\documentclass[12pt,a4paper,twoside]{book}
\usepackage[top=3.5cm, bottom=3.5cm,left=3.2cm,right=2.5cm]{geometry}

\usepackage[english]{babel}
\usepackage[a-1b]{pdfx}
\usepackage{hyperref}
\usepackage{amsmath,amssymb} 
\usepackage{amsthm}
\usepackage{mathabx}
\usepackage{bm} 
\usepackage{color}
\usepackage{tikz-cd}
\usetikzlibrary{matrix,arrows,decorations.pathmorphing}
\usepackage{mathtools}
\usepackage{graphicx}
\usepackage{subcaption}
\usepackage{dsfont}
\usepackage{cite}
\usepackage[bulletsep]{collref}
\usepackage{tensor}
\usepackage{fancyhdr}
\usepackage{enumerate}
\usepackage{multirow}
\usepackage{rotating}
\usepackage{titlesec}
 
\hypersetup{
	bookmarksnumbered=true, 
	hidelinks,
	}

\usepackage{fancyhdr}
\allowdisplaybreaks[3]

\numberwithin{equation}{chapter}

\usepackage[font=small,labelfont=bf,width=0.85\textwidth]{caption}

\let\oldbfseries=\bfseries
\let\oldmdseries=\mdseries
\let\oldnormalfont=\normalfont
\renewcommand{\bfseries}{\oldbfseries\boldmath}
\renewcommand{\mdseries}{\oldmdseries\unboldmath}
\renewcommand{\normalfont}{\oldnormalfont\unboldmath}

\makeatletter
\newlength{\apb@width}
\newcommand{\autoparbox}[2][c]{\settowidth{\apb@width}{#2}\parbox[#1]{\apb@width}{#2}}

\makeatother

\makeatletter
\def\mr@ignsp#1 {\ifx\:#1\@empty\else #1\expandafter\mr@ignsp\fi}%
\newcommand{\multiref}[1]{\begingroup
\xdef\mr@no@sparg{\expandafter\mr@ignsp#1 \: }%
\def\mr@comma{}%
\@for\mr@refs:=\mr@no@sparg\do{\mr@comma\def\mr@comma{,}\ref{\mr@refs}}%
\endgroup}
\makeatother

\newcommand{\hypref}[2]{\ifx\href\asklfhas #2\else\href{#1}{#2}\fi}

\renewcommand{\eqref}[1]{(\multiref{#1})}

\newcommand{\nn}{\nonumber}

\DeclareMathOperator{\tr}{tr}                               
\DeclareMathOperator{\str}{str}                               
\newcommand{\unit}{\mathbf{1}}                              
\newcommand{\comm}[2]{[#1,#2]}                              
\newcommand{\grSO}{\mathrm{SO}}                             
\renewcommand{\O}{\mathcal{O}}                              
\newcommand{\AdS}{\mathrm{AdS}}
\newcommand{\EAdS}{\mathrm{EAdS}}
\newcommand{\MM}{\mathrm{M}}

\newcommand{\spl}{\sigma ^+}
\newcommand{\smi}{\sigma ^-}
\newcommand{\zb}{{\bar z}}   
\newcommand{\id}{{\mathrm{id}}}  
\newcommand{\J}{\mathrm{J}}
\newcommand{\jay}{\mathrm{j}}
\newcommand{\f}{\mathbf{f}}

\newcommand{\Dbizz}{\mathcal{D}}
\newcommand{\plexp}{\overleftarrow{\mathrm{P}} \hspace*{-1.4mm} \exp}
\newcommand{\prexp}{\overrightarrow{\mathrm{P}} \hspace*{-1.4mm} \exp}

\newcommand{\delC}{\widehat\delta}
\newcommand{\master}{\widehat \delta}
\newcommand{\masterj}{\widehat j}
\newcommand{\masterQ}{\widehat Q}
\newcommand{\chargeC}{\widehat Q}

\newcommand{\chargeY}{Q}

\renewcommand{\a}{\alpha}
\renewcommand{\b}{\beta}
\newcommand{\g}{\gamma}

\newcommand{\da}{{\dot{\alpha}}}
\newcommand{\db}{{\dot{\beta}}}
\newcommand{\dg}{{\dot{\gamma}}}
\newcommand{\la}{\lambda}

\newcommand{\eps}{\epsilon}
\newcommand{\s}{\sigma}
\newcommand{\bs}{\widebar{\sigma}}

\newcommand{\BSi}{\widebar \Sigma}
\newcommand{\btheta}{\widebar\theta}
\newcommand{\bTheta}{\widebar\Theta}
\newcommand{\bt}{\widebar \theta}
\newcommand{\dt}{\dot \theta}
\newcommand{\dbt}{\dot{\widebar{\theta}}}
\newcommand{\vart}{\vartheta}
\newcommand{\bvart}{\widebar{\vartheta}}
\newcommand{\weta}{\widebar \eta}

\newcommand{\A}[1]{A^{(#1)}}
\newcommand{\AAp}{A^{(1)+(3)}}
\newcommand{\AAm}{A^{(1)-(3)}}
\newcommand{\commwedge}[2]{
	\big [ #1 \, 
	\underset{\smash{\raisebox{0.6em}{,}}}{\smash{\raisebox{.2em}{$\wedge$}}} \, 
	#2 \big ] }

\newcommand{\Qb}{{\widebar{Q}}}
\newcommand{\Sb}{{\widebar{S}}}

\ifx\genfrac\sdflkaj\else\fi
\newcommand{\sfrac}[2]{{\textstyle\frac{#1}{#2}}}
\newcommand{\half}{\sfrac{1}{2}}
\newcommand{\ihalf}{\sfrac{i}{2}}
\newcommand{\thirrd}{\sfrac{1}{3}}
\newcommand{\quarter}{\sfrac{1}{4}}

\newcommand{\alg}[1]{\mathfrak{#1}}
\newcommand{\grp}[1]{\mathrm{#1}}

\newcommand{\ft}[2]{{\textstyle\frac{#1}{#2}}}

\def\l<{\langle}\def\r>{\rangle}

\newcommand{\lrbrk}[1]{\left(#1\right)}
\newcommand{\bigbrk}[1]{\bigl(#1\bigr)}

\newcommand{\lreval}[1]{\left.#1\right|}

\newcommand{\Nfour}{\mathcal{N}\!=4}
\newcommand{\Tr}{\mathop{\mathrm{Tr}}}
\newcommand{\sign}{\mathop{\mathrm{sign}}}

\newcommand{\eqn}[1]{(\ref{#1})}

\newcommand{\dx}{{\dot x}}
\newcommand{\ddx}{{\ddot x}}

\newcommand*{\diff}{{\mathrm d}} 

\newcommand\doilink[2]{\href{http://dx.doi.org/#1}{#2}}
\newcommand\arxivlink[2]{\href{http://arxiv.org/abs/#1}{#2}}

\theoremstyle{definition} 
\newtheorem{definition}{Definition}[chapter]
\theoremstyle{plain}

\titleformat{\chapter}[display]
  {\normalfont\huge\bfseries}{\chaptertitlename\ \thechapter}{20pt}{\huge}
\titlespacing*{\chapter}
  {0pt}{20pt}{30pt}

\title{Symmetries of Maldacena--Wilson Loops from Integrable String Theory}
\author{Hagen M\"unkler}

\begin{document}  
\pagenumbering{roman}
\raggedbottom


\begin{center}
\thispagestyle{empty}
\LARGE{\textsc{Symmetries of Maldacena--Wilson Loops from Integrable String Theory}} \\     
\vspace{.5cm}
\LARGE{\textbf{Dissertation}} \\  
\large{zur Erlangung des akademischen Grades} \\
\large{doctor rerum naturalium} \\
\large{im Fach Physik} \\[.1cm]
\small{Spezialisierung: Theoretische Physik}

\vspace{0.5cm}

\includegraphics[width=35mm]{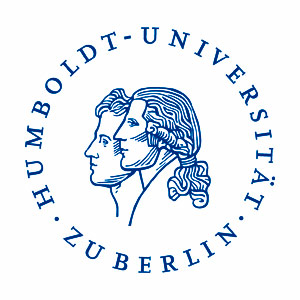}\\

\vspace{0.5cm}

\large{eingereicht an der \\ Mathematisch-Naturwissenschaftlichen Fakult\"at \\ der Humboldt-Universit\"at zu Berlin}

\vspace{0.5 cm}

\large{ von \\ \textbf{Herrn M. Sc. Hagen M{\"u}nkler}} \\[0.1cm] 

\vspace{0.5cm}
\end{center}
\begin{flushleft}

\large{Pr{\"a}sidentin der Humboldt-Universit\"at zu Berlin: \\
	Prof. Dr. Sabine Kunst}\\
\vspace{0.5cm}
\large{Dekan der Mathematisch-Naturwissenschaftliche Fakult\"at: \\
	Prof. Dr. Elmar Kulke}

\begin{tabular}{@{}l l}\\
Gutachter: & 1. Prof. Dr. Jan Plefka \\[.2cm]
 		   & 2. Dr. Valentina Forini  \\[.2cm]
 		   & 3. Prof. Dr. Gleb Arutyunov \\
 & \\
Disputation: & \hspace{4.5mm} 11. September 2017 \\
\end{tabular}
\end{flushleft}  

\newpage
\thispagestyle{empty} 
$ \mbox{ }$

\newpage
\pagestyle{plain}
\begin{center}
\Large \textbf{Abstract}
\end{center}
\vspace*{5mm}

This thesis discusses hidden symmetries within $\Nfour$ supersymmetric Yang--Mills theory or its AdS/CFT dual, string theory in $\AdS_5 \times \mathrm{S}^5$. Here, we focus on the Maldacena--Wilson loop, which is a suitable object for this study since its vacuum expectation value is finite for smooth contours and the conjectured duality to scattering amplitudes provides a conceptual path to transfer its symmetries to other observables. Its strong-coupling description via minimal surfaces in $\AdS_5$ allows to construct the symmetries from the integrability of the underlying classical string theory. This approach has been utilized before to derive a strong-coupling Yangian symmetry of the Maldacena--Wilson loop and describe equiareal deformations of minimal surfaces in $\AdS_3$. These two findings are connected and extended in the present thesis.   

In order to discuss the symmetries systematically, we first discuss the symmetry structure of the underlying string model. The discussion can be generalized to the discussion of generic symmetric space models. For these, we find that the symmetry which generates the equiareal deformations of minimal surfaces in $\AdS_3$ has a central role in the symmetry structure of the model: It acts as a raising operator on the infinite tower of conserved charges, thus generating the spectral parameter, and can be employed to construct all symmetry variations from the global symmetry of the model. It is thus referred to as the master symmetry of symmetric space models. Additionally, the algebra of the symmetry variations and the conserved charges is worked out. 

For the concrete case of minimal surfaces in $\AdS_5$, we discuss the deformation of the four-cusp solution, which provides the dual description of the four-gluon scattering amplitude. This marks the first step toward transferring the master symmetry to scattering amplitudes. Moreover, we compute the master and Yangian symmetry variations of generic, smooth boundary curves. The results leads to a coupling-dependent generalization of the master symmetry, which constitutes a symmetry of the Maldacena--Wilson loop at any value of the coupling constant. Our discussion clarifies why previous attempts to transfer the deformations of minimal surfaces in $\AdS_3$ to weak coupling were unsuccessful. We discuss several attempts to transfer the Yangian symmetry to weak or arbitrary coupling, but ultimately conclude that a Yangian symmetry of the Maldacena--Wilson loop seems not to be present.  

The situation changes when we consider Wilson loops in superspace, which are the natural supersymmetric generalizations of the Maldacena--Wilson loop. Substantial evidence for the Yangian invariance of their vacuum expectation value has been provided at weak coupling and the description of the operator as well as its weak-coupling Yangian invariance were subsequently established in parallel to the work on this thesis. We discuss the strong-coupling counterpart of this finding, where the Wilson loop in superspace is described by minimal surfaces in the superspace of type IIB superstring theory in $\AdS_5 \times \mathrm{S}^5$. The comparison of the strong-coupling invariance derived here with the respective generators at weak coupling shows that the generators contain a local term, which depends on the coupling in a non-trivial way.

Additionally, we find so-called bonus symmetry generators. These are the higher-level recurrences of the superconformal hypercharge generator, which does not provide a symmetry itself. We show that these symmetries are present in all higher levels of the Yangian. 
\newpage

\begin{center}
\LARGE \textbf{Zusammenfassung}
\end{center}
\vspace*{5mm}

In der vorliegenden Arbeit werden versteckte Symmetrien innnerhalb der $\Nfour$ supersymmetrischen Yang--Mills Theorie oder der nach der AdS/CFT Korrespondenz dualen Beschreibung durch eine String-Theorie in $\AdS_5 \times \mathrm{S}^5$ besprochen. 
Dabei be\-trach\-ten wir die Maldacena--Wilson Schleife, die sich f{\"u}r diese Untersuchungen besonders eignet, da ihr Vakuum-Erwartungswert f{\"u}r glatte Kurven nicht divergiert und die vermutete Dualit{\"a}t zu Streuamplituden wenigstens konzeptionell eine M{\"o}glichkeit bietet, etwaige Symmetrien zu anderen Observablen zu {\"u}bertragen. 
Ihre Beschreibung durch Minimalfl{\"a}chen in $\AdS_5$ erlaubt es, Symmetrien mithilfe der Integrabilit{\"a}t der zugrunde liegenden klassischen String-Theorie zu konstruieren.
Dieser Zugang wurde bereits in der Herleitung der Yang'schen Symmetrie der Maldacena--Wilson Schleife bei starker Kopplung sowie in der Beschreibung von Deformationen gleiches Fl{\"a}cheninhalts von Minimalfl{\"a}chen in $\AdS_3$ verwendet. 
Diese beiden Ergebnisse werden in der vorliegenden Arbeit miteinander verbunden und erweitert. 

Im Sinne einer systematischen Herangehensweise besprechen wir zun{\"a}chst die Symmetriestruktur der zugrunde liegenden String-Theorie. Diese Diskussion l{\"a}sst sich auf die Diskussion von String-Theorien in symmetrischen R{\"a}umen verallgemeinern. 
Dabei zeigt sich, dass die Symmetrie, welche  die Deformationen gleiches Fl{\"a}cheninhalts in 
$\AdS_3$ erzeugt, in der Symmetriestruktur dieser Modelle eine zentrale Rolle einnimmt: 
Sie wirkt als Aufsteige-Operator auf den unendlich vielen erhalten Ladungen und generiert somit den Spektralparameter. 
Weiterhin l{\"a}sst sie sich anwenden, um ausgehend von der globalen Symmetrie s{\"a}mtliche Symmetrien des zugrunde liegenden Modells zu konstruieren. 
Sie wird daher als die Master-Symmetrie dieser Modelle bezeichnet. Zus{\"a}tzlich wird die Algebra der Symmetrie-Variationen sowie der erhaltenen Ladungen ausgearbeitet. 

F{\"u}r den konkreten Fall von Minimalfl{\"a}chen in $\AdS_5$ diskutieren wir die Deformation der Minimalfl{\"a}chenl{\"o}sung f{\"u}r den Fall eines lichtartigen Vierecks. 
Diese liefert die duale Beschreibung der Streuamplitude f{\"u}r vier Gluonen. Damit unternehmen wir einen ersten Schritt zur {\"U}bertragung der Master-Symmetrie auf Streuamplituden. 
Weiterhin berechnen wir die Variation der Randkurven der Minimalfl{\"a}chen unter der Master- und Yang'schen Symmetrie für allgemeine, glatte Randkurven. 
Das Ergebnis dieser Rechnung f{\"u}hrt auf eine Verallgemeinerung der Master-Symmetrie zu einer Variation, die von der Kopplungskonstanten abh{\"a}ngt und f{\"u}r beliebige Werte der Kopplungskonstanten eine Symmetrie der Maldacena--Wilson Schleife darstellt. 
Unsere Diskussion erkl{\"a}rt das Scheitern vorheriger Versuche, die entsprechende Symmetrie im Spezialfall von Minimalfl{\"a}chen in $\AdS_3$ zu schwacher Kopplung zu {\"u}bertragen. 
Wir besprechen verschiedene Ans{\"a}tze, die Yang'sche Symmetrie zu schwacher oder beliebiger Kopplung zu {\"u}bertragen, schlussfolgern aber letztendlich, dass eine Yang'sche Symmetrie der Maldacena--Wilson Schleife nicht vorzuliegen scheint. 

Die Situation {\"a}ndert sich, wenn wir Wilson Schleifen in Superr{\"a}umen betrachten. 
Diese sind die nat{\"u}rlichen supersymmetrischen Erweiterungen der Maldacena--Wilson Schleife. 
F{\"u}r die Yang'sche Invarianz ihres Vakuum-Erwartungswerts wurden wichtige Anhaltspunkte gefunden und sowohl die Beschreibung dieser Operatoren als auch der Beweis der Yang'schen Invarianz bei schwacher Kopplung wurden parallel zur Arbeit an der vorliegenden Dissertation vervollst{\"a}ndigt.
Wir diskutieren das Gegenst{\"u}ck zu diesem Ergebnis bei starker Kopplung. Dort wird die Wilson Schleife durch eine Mini\-mal\-fl{\"a}che beschrieben, welche im Superraum der Superstring-Theorie vom Typ IIB in $\AdS_5 \times \mathrm{S}^5$ liegt. 
Der Vergleich der bei starken Kopplung etablierten Invarianz mit den entsprechenden Generatoren bei schwacher Kopplung zeigt, dass die Symmetrie-Generatoren einen lokalen Anteil enthalten, der auf nicht-triviale Weise vom Wert der Kopplungskonstanten abh{\"a}ngt. 

Zus{\"a}tzlich finden wir sogenannte Bonus-Symmetrien. Diese sind die analogen Ge\-neratoren in den h{\"o}heren Ordnungen zum Hyperladungs-Generator, der selbst keine Symmetrie darstellt. Wir zeigen, dass diese Symmetrien in allen h{\"o}heren Ordnungen der Yang'schen Algebra vorliegen.

\newpage
\begin{center}
\LARGE \textbf{Publications}
\end{center}
\vspace*{10mm}
This thesis describes the continuation of the research begun in the publication \cite{Muller:2013rta}, 
\begin{itemize}
	\item[{[1]}] D.~M{\"u}ller, H.~M{\"u}nkler, J.~Plef\-ka, J.~Pollok and K.~Za\-rembo,
	\textit{``{Yangian Symmetry of smooth Wilson Loops in $\mathcal{N} = $ 4 super
  	Yang-Mills Theory}''}, \\ 
  	\textsf{\doilink{10.1007/JHEP11(2013)081}{JHEP~1311,~081~(2013)}},
    \texttt{\arxivlink{1309.1676}{arxiv:1309.1676}}, 
\end{itemize}
which was included in the author's master's thesis. It is based on the peer-reviewed publications \cite{Munkler:2015gja,Munkler:2015xqa,Klose:2016uur,Klose:2016qfv} of the author in different collaborations, 
\begin{itemize}
	\item[{[2]}] H.~M{\"u}nkler and J.~Pollok, 
	\textit{``{Minimal surfaces of the ${{AdS}}_{5}\times {S}^{5}$ superstring and
  		the symmetries of super Wilson loops at strong coupling}''}, \\
	\textsf{\doilink{10.1088/1751-8113/48/36/365402}{J.~Phys.~A48,~365402~(2015)}},
	\texttt{\arxivlink{1503.07553}{arxiv:1503.07553}}, 
    \item[{[3]}] H.~M{\"u}nkler,
	\textit{``{Bonus Symmetry for Super Wilson Loops}''}, \\
	\textsf{\doilink{10.1088/1751-8113/49/18/185401}{J.~Phys.~A49,~185401~(2016)}},
	\texttt{\arxivlink{1507.02474}{arxiv:1507.02474}},
    \item[{[4]}] T.~Klose, F.~Loebbert and H.~M{\"u}nkler,
	\textit{``{Master Symmetry for Holographic Wilson Loops}''}, 
	\textsf{\doilink{10.1103/PhysRevD.94.066006}{Phys.~Rev.~D94,~066006~(2016)}},
	\texttt{\arxivlink{1606.04104}{arxiv:1606.04104}}.
    \item[{[5]}] T.~Klose, F.~Loebbert and H.~M{\"u}nkler,
	\textit{``{Nonlocal Symmetries, Spectral Parameter and Minimal Surfaces in
  	AdS/CFT}''},
	\textsf{\doilink{10.1016/j.nuclphysb.2017.01.008}{Nucl.~Phys.~B916,~320~(2017)}},
	\texttt{\arxivlink{1610.01161}{arxiv:1610.01161}}.
\end{itemize}
The results of the publication \cite{Dorn:2012cn}, 
\begin{itemize}
	\item[{[6]}] H.~Dorn, H.~M{\"u}nkler and C.~Spielvogel,
	\textit{``{Conformal geometry of null hexagons for Wilson loops and scattering
  	amplitudes}''},
	\textsf{\doilink{10.1134/S1063779614040066}{Phys.~Part.~Nucl.~45,~692~(2014)}},
	\texttt{\arxivlink{1211.5537}{arxiv:1211.5537}},
\end{itemize}
are not included in this thesis.



\newpage
\setcounter{tocdepth}{1}
\tableofcontents


\chapter{Introduction} 
\pagestyle{fancy}
\newcounter{oldpage}
\setcounter{oldpage}{\value{page}}
\pagenumbering{arabic}
\setcounter{page}{\theoldpage}

Our current understanding of the microscopic world is based on Yang--Mills gauge theories \cite{Yang:1954ek}, which describe the fundamental interactions between elementary particles. The gauge theory descriptions of the electromagnetic, weak nuclear and strong nuclear force are combined in the Standard Model of particle physics, which has the gauge group $\grp{SU}(3) \times \grp{SU}(2) \times \grp{U}(1)$. The Standard Model describes all known elementary particles and since the discovery of the Higgs boson \cite{Aad:2012tfa,Chatrchyan:2012xdj,Aad:2015zhl} all particles which it predicts have been observed.  

Despite its great success in the description of scattering experiments, it does not explain all of the observed phenomena that a theory of all fundamental interactions should explain. Apart from the lack of a description of gravity, it does for example neither predict the observed neutrino oscillations or the neutrino masses inferred from them, nor does it provide appropriate candidate particles for the dark matter needed to explain astrophysical observations.  

The incompleteness of the Standard Model is however not the only challenge theoretical physicist face in the description of gauge theories. A different challenge lies in the mathematical problem of obtaining predictions from these models and concerns gauge theories in general. Indeed, our understanding of gauge theories in the full range of energy scales is disappointingly limited, since we have to refer to perturbation theory in order to obtain results. In the case of quantum chromodynamics (QCD), which describes the strong nuclear force, this method only provides reliable results in the case of high-energy collisions where the running coupling constant is small. For other phenomena --- or even the time spans shortly after the scattering events, when the scattered constituents again combine into hadrons --- we must rely on numerical results in combination with experimental data. 

The most promising approach to reach an analytic understanding beyond perturbation theory is the study of a particular class of gauge theories, which allow for exact results. The prime example for such a theory is $\Nfour$ supersymmetric Yang--Mills (SYM) theory, an $\grp{SU}(N)$ gauge theory  which one may view as a theoretical laboratory for QCD, with which it shares a similar field content. The similarity to QCD can be exemplified by the possibility to compute certain tree-level QCD scattering amplitudes within $\Nfour$ SYM theory or by the fact that parts of the anomalous dimensions appearing in the study of infrared singularities can be transferred also at higher loop orders. 
While $\Nfour$ SYM theory was already discussed in the late 70's 
\cite{Brink:1976bc,Gliozzi:1976qd}, the interest in it grew considerably after the advent of the AdS/CFT correspondence \cite{Maldacena:1997re}. The conjectured correspondence relates the four-dimensional (conformal) gauge theory to a superstring theory with target space 
$\AdS_5 \times \mathrm{S}^5$. It connects the strong-coupling regime of the gauge theory to the weak-coupling regime of the string theory and vice versa and can thus be employed to gain insights in either theory at values of the coupling constant that are otherwise inaccessible. 

The appeal of $\Nfour$ SYM theory is further raised by the availability of two additional methods which allow to look beyond the perturbative curtain: localization and integrability. Either of these methods can be applied to derive exact results. The technique of localization is based on the supersymmetry of the theory and can be applied to reduce the path integrals appearing for certain observables to ordinary integrals, see reference \cite{Pestun:2016zxk} for a recent review. In this way, one can for example derive an exact result for the circular Wilson loop \cite{Pestun:2007rz,Zarembo:2016bbk}, which we encounter later on in this thesis.     

In the planar limit, where one sends the Yang--Mills coupling constant $g$ to zero and the parameter $N$ of the gauge group to infinity in such a way that the 't Hooft coupling constant 
$\lambda = g^2 N$ is kept fixed, the theory appears to be integrable. Integrability is not conceptually tied to the presence of supersymmetry, although it does appear within a supersymmetric theory in the present case. Integrable structures were first observed within $\Nfour$ SYM theory in the study of spectral problem, which could be reformulated as an integrable spin chain \cite{Minahan:2002ve}. This reformulation allowed for spectacular progress \cite{Beisert:2003tq,Beisert:2003yb} in the study of two-point functions within $\Nfour$ SYM theory, see also \cite{Beisert:2010jr} for an overview. 

It is a common belief that the availability of exact results is linked to the presence of an underlying hidden symmetry. Such symmetries have indeed been observed for different objects in $\mathcal{N}=4$ SYM theory, e.g.\ the symmetries of the dilatation operator \cite{Dolan:2003uh}, which are relevant for the solution of the spectral problem, as well as the dual superconformal or Yangian symmetry of scattering amplitudes \cite{Drummond:2008vq,Drummond:2009fd}.

Much like the study of the integrable structures itself, the investigation of the associated symmetry structures is typically a case-by-case study, although an interesting attempt has been made recently \cite{Beisert:2017pnr} to study the symmetry of the action or equations of motion of $\Nfour$ SYM theory directly. Here, we aim at finding hidden symmetries within $\Nfour$ SYM theory in a systematic way. The object of study for this investigation is the Maldacena--Wilson loop, which is a specific generalization of the Wilson loop considered in generic Yang--Mills theories and naturally appears in $\Nfour$ SYM theory. The Wilson loop is a central object in any gauge theory, but possibly even more so in $\Nfour$ SYM theory and we will see shortly, why it is a particularly suitable observable for the investigation of hidden symmetries.  

Concretely, the Maldacena--Wilson loop over a contour $\gamma$ is given by 
\cite{Maldacena:1998im,Rey:1998ik} 
\begin{align}
W(\gamma) = \frac{1}{N} \, \tr \plexp 
	\left( i \int _\gamma \left( A_\mu \, \diff x^\mu + i \, \Phi_I \lvert\dx \rvert n^I 
	\right) \right) .
\end{align}
Here, the gauge fields $A_\mu$ couple to the curve $\gamma$, which is described by the parametrization $x^\mu (\s)$, whereas the scalar fields $\Phi_I$ of $\Nfour$ SYM theory couple to a six-vector $n^I(\s)$, which has unit length and thus describes a point on $\mathrm{S}^5$. We note that for space-like contours, the Maldacena--Wilson loop is no longer a phase 
\cite{Drukker:1999zq} due to the appearance of the factor $i$ in front of the scalars. 
A remarkable aspect of the Maldacena--Wilson loop is that, due to cancellations between the gauge and scalar fields, its expectation value is finite for smooth contours. This is a welcome property for the study of symmetries, since it implies that potential symmetries are not overshadowed by renormalization effects. 

Another intriguing feature of the (Maldacena--)Wilson loop%
\footnote{Since the contours under consideration are light-like in this case, the Maldacena--Wilson and the ordinary Wilson loop are the same object.}
in $\Nfour$ SYM theory is the conjectured duality to scattering amplitudes 
\cite{Alday:2007hr,Brandhuber:2007yx,Drummond:2008aq}, which relates certain scattering amplitudes to Wilson loops over specific, light-like polygons. Due to the cusps of these contours, the Wilson loops are divergent, corresponding to the infrared divergences of the scattering amplitudes. This implies that any symmetry found for smooth Maldacena--Wilson loops could become anomalous for these contours. Nonetheless, the correspondence provides a conceptual path to transfer any symmetry found for the Maldacena--Wilson loop to other observables within 
$\Nfour$ SYM theory. 

In the strong-coupling limit, the Maldacena--Wilson loop is described by the area of a minimal surface ending on the conformal boundary of $\AdS_5$, 
\begin{align}
\left \langle W(\gamma) \right \rangle \overset{\lambda \gg 1}{=}
	\exp \left( - \ft{\sqrt{\la}}{2 \pi} A_\mathrm{ren}(\gamma) \right) \, .
\end{align}
Here, we have restricted to the case where the vector $n^I$ is constant, the general case will be discussed later on. In order to describe the boundary value problem, we use so-called Poincar{\'e} coordinates $(X^\mu , y)$ for $\AdS_5$. In these coordinates, the metric is given by 
\begin{align*}
\diff s^2 = \frac{\diff X^\mu \, \diff X_\mu + \diff y \, \diff y}{y^2} \, ,
\end{align*}
and the conformal boundary, which corresponds to infinity in $\AdS_5$, is given by the Minkowski space located at $y=0$. The minimal surface is then specified by the boundary conditions
\begin{align*}
X^\mu ( \tau = 0 , \sigma ) &= x^\mu (\sigma) \, , &
y ( \tau = 0 , \sigma ) &= 0 \, .
\end{align*}
The strong-coupling description of the Maldacena--Wilson loop is particularly suitable for the study of hidden symmetries, since the minimal surface is described by classical string theory in $\AdS_5$, which is known to be integrable. The symmetries of the string theory induce symmetry transformations of the boundary curve, which can then be studied also at weak coupling. 

This approach has been discussed in reference \cite{Muller:2013rta}, where it was shown that the  Maldacena--Wilson loop is Yangian invariant in the strong-coupling limit. The weak-coupling side of this finding was reported on also in reference \cite{Muller:2013rta} as well as the author's master's thesis \cite{Master}. A Yangian invariance of the one-loop expectation value of the Maldacena--Wilson loop could not be established, but it was observed that an extension of the Maldacena--Wilson loop into a non-chiral superspace is Yangian invariant. Interestingly, a similar situation was observed for the lambda-deformations of minimal surfaces in Euclidean $\AdS_3$, which were found by Kruczenski and collaborators in references \cite{Ishizeki:2011bf,Kruczenski:2013bsa}. These deformations were studied at weak coupling by Dekel in reference \cite{Dekel:2015bla}, where it was also found that they are not symmetries of the one-loop expectation value of the Maldacena--Wilson loop. 

The aim for this thesis is to complete the picture sketched above. In particular, we establish the connection between the Yangian symmetry and the lambda-deformations and lift the discussion of the strong-coupling Yangian symmetry to the superspace Wilson loop. The field theory description of this operator as well as its Yangian symmetry at the one-loop level were completed in a parallel line of research in references \cite{Beisert:2015jxa,Beisert:2015uda}. 

The thesis is structured as follows. We first discuss the foundations for the research presented in this thesis in chapter \ref{chap:Basics}. Here, we mainly focus on the underlying symmetry structures and the Maldacena--Wilson loop. A short introduction to $\Nfour$ SYM theory and the AdS/CFT correspondence is also provided.

We then turn to the study of the Maldacena--Wilson loop at strong coupling. The discussion of the symmetries of minimal surfaces in $\AdS_5$ naturally generalizes to a class of spaces known as symmetric spaces and so we discuss the symmetry structures for minimal surfaces or strings in these spaces in general in chapter \ref{chap:SSM}. In addition to a review of the integrability of symmetric space models, this chapter presents the research published in references \cite{Klose:2016qfv,Klose:2016uur}, which was carried out in collaboration with Thomas Klose and Florian Loebbert. In particular, the relation between the Yangian symmetry and Kruczenski's lambda deformations is established. In fact, we observe that the symmetry behind the lambda deformation is fundamental for the integrability structure of symmetric space models and can be employed to construct all other symmetry variations and hence we refer to it as the \emph{master symmetry}. The connections to the literature on symmetric space models, where parts of the results had already been discussed, are discussed as well. 

The application of the symmetries derived for generic symmetric space models to minimal surfaces in $\AdS_5$ is discussed in chapter \ref{chap:MinSurf}. Also this chapter is based on references \cite{Klose:2016qfv,Klose:2016uur}, although it includes a discussion of the large master symmetry transformations of some analytically-known minimal surfaces, which was not published before. Moreover, we discuss the variations of the boundary curves that follow from the symmetries discussed before. 

With the symmetry variations of the contours established, we address the question, if and how these symmetries can be extended to weak or even arbitrary values of the 't~Hooft coupling constant 
$\lambda$ in chapter \ref{chap:CarryOver}. The variation of the boundary curve obtained in the previous chapter explains why the same transformation was not observed to be a symmetry also at weak coupling in reference \cite{Dekel:2015bla}. It turns out, however, that the master symmetry variation can be generalized to a coupling-dependent variation which does provide a symmetry of the Maldacena--Wilson loop at any value of the coupling constant $\lambda$. Furthermore, we discuss the continuation of the Yangian symmetry generators from strong to weak coupling. In their original form as employed in reference \cite{Muller:2013rta}, the Yangian generators are not cyclic and this finding alone predicts that they are not symmetries of the one-loop expectation value of the Maldacena--Wilson loop. While there exists a possibility to adapt the generators in such a way that they become cyclic \cite{Chicherin:2017cns}, also the adapted generators do not form symmetries of the one-loop expectation value of the Maldacena--Wilson loop, such that the result of reference \cite{Muller:2013rta} still holds.   

This concludes the discussion of the Maldacena--Wilson loop and we turn to the discussion of Wilson loops in superspace. In this case, the underlying symmetry algebra is the superconformal algebra 
$\alg{psu}(2,2 \vert 4)$, such that potential Yangian symmetry generators are automatically cyclic due to the vanishing of the dual Coxeter number of $\alg{psu}(2,2 \vert 4)$. As in the field-theory construction of reference \cite{Muller:2013rta}, the strong-coupling description of the Wilson loop in superspace arises from the supersymmetric extension of the respective description of the Maldacena--Wilson loop. Instead of minimal surfaces in $\AdS_5 \times \mathrm{S}^5$, we thus consider minimal surfaces of the superstring in $\AdS_5 \times \mathrm{S}^5$, i.e.\ in the superspace appearing in the description of the Green-Schwarz superstring in this space \cite{Metsaev:1998it}. Appropriate boundary conditions for these minimal surfaces which generalize the conformal boundary of $\AdS_5$ have been discussed in reference \cite{Ooguri:2000ps}. 

The symmetry variations of the boundary curves again follow from the integrability of the string theory that describes the minimal surface in the bulk space \cite{Bena:2003wd}, which we discuss in chapter \ref{chap:Semi-SSM}. Here, we generalize the discussion to a class of models known as semisymmetric space models. In addition to reviewing the construction of the conserved charges, we also introduce the master symmetry for these models, which was formulated for the pure spinor superstring in reference \cite{Chandia:2016ueo}. Moreover, we construct an infinite tower of so-called bonus symmetry charges. Inspired by the master symmetry, we find a more elegant approach than the one described in reference \cite{Munkler:2015xqa} by the author. The results, however, remain unaltered.

The application to minimal surfaces in the $\AdS_5 \times \mathrm{S}^5$-superspace is considered in chapter \ref{chap:SurfaceSuperspace}, where we construct the expansion of the minimal surface around the boundary curve in order to derive the Yangian symmetry generators for the superspace Wilson loop at strong coupling. This chapter describes the results published in reference \cite{Munkler:2015gja}, which were obtained in collaboration with Jonas Pollok.  

The thesis is concluded by a summary of the results and an outlook on their implications for possible future works in chapter \ref{chap:conclusion}. Technical aspects of some of the topics which are important for our discussion are collected in appendices 
\ref{app:Spinor}-\ref{app:densities}. 

\chapter{Symmetries, Fields and Loops}
\label{chap:Basics}

This chapter provides a more detailed introduction to the aforementioned concepts and objects which form the foundation of the research on which this thesis reports. The focus lies on the discussion of the Maldacena--Wilson loop as well as the various symmetry structures that will be interesting with respect to it. As a prerequisite for the discussion of the Maldacena--Wilson loop, brief introductions to $\Nfour$ supersymmetric Yang--Mills theory as well as the AdS/CFT correspondence are also given. 

\section{Symmetries}
\label{sec:Int_Symm}

Symmetries are one of the most fundamental concepts of theoretical physics and have served as an important guiding principle in the construction of new theories. Since the research described in this thesis focuses on symmetry structures, it seems fitting to begin with the basic symmetry structures that underlie our discussion. 

\subsection{Conformal Symmetry}
We first discuss conformal transformations of $d$-dimensional Minkowski space $\mathbb{R}^{(1,d-1)}$ both for infinitesimal and large transformations. The discussion of the large conformal transformations and their singularities leads us to discuss the concept of the conformal compactification, which will also be important in the discussion of the AdS/CFT correspondence. 

Conformal transformations are generalized isometries, which leave the angle between two vectors invariant while generically changing their length. Stated more formally, a conformal transformation is a map $f: U \subset \mathbb{R}^{(1,d-1)} \to \mathbb{R}^{(1,d-1)}$, which satisfies $f^\ast \eta = e^{2 \sigma} \eta$, or in coordinates
\begin{align}
\frac{\partial f(x) ^\rho}{\partial x^\mu} \, 
	\frac{\partial f(x) ^\sigma}{\partial x^\nu} \, \eta _{\rho \sigma} =
	e^{2 \sigma(x)} \, \eta_{\mu \nu} \, , 
\label{Def:Conf_Trans}	
\end{align} 
where $\sigma(x)$ denotes an arbitrary, smooth function and $\eta$ denotes the mostly-plus metric of $d$-dimensional Minkowski space. We will see below that generic conformal transformations have singularities in $\mathbb{R}^{(1,d-1)}$, and we have thus restricted their definition to an open subset $U$ of $\mathbb{R}^{(1,d-1)}$. In order to find the conformal transformations of Minkowski space we first consider infinitesimal transformations. These are described by vector fields $\xi$ which satisfy the conformal Killing equation
\begin{align}
2 \left( \partial_\mu \, \xi_\nu + \partial_\nu \, \xi_\mu \right) 
	= \left( \partial_\rho \, \xi ^\rho \right) \eta_{\mu \nu} \, .
\label{eqn:Conf_Killing}	
\end{align}
Here, we have already taken a trace to express the arbitrary function appearing in the infinitesimal version of equation \eqref{Def:Conf_Trans} in terms of the vector field $\xi$. The large transformations associated to these vector fields are given by the flows associated to them:
To each vector field $\xi$ we can associate a set of integral curves $\gamma_{x_0} ^\xi$, which are defined by
\begin{align}
\partial_ \tau \, \gamma_{x_0} ^\xi (\tau) &= \xi (  \gamma_{x_0} ^\xi (\tau) ) \, , & 
\gamma_{x_0} ^\xi (0) &= x_0 \, .   
\end{align}    
The flow of the vector field $\xi$ is then given by the map
\begin{align}
\sigma ^\xi ( \tau , x_0 ) =  \gamma_{x_0} ^\xi (\tau) \, , 
\end{align} 
and for fixed values of $\tau$ it gives a diffeomorphism between open subsets of Minkowski space. The conformal Killing equation ensures that these diffeomorphisms are conformal transformations. We note that it may not be possible to extend the interval on which the integral curves are defined to all of $\mathbb{R}$ and this would in turn restrict the domain of the flows. This behavior is related to the singularities of the conformal transformations we encounter below. 

Let us however stay with the infinitesimal transformations for a moment. From the conformal Killing equation one may show that for $d>2$ the vector fields $\xi(x)$ are polynomials with maximal degree two and based on this finding it is easy to see that any conformal Killing field can be written as
\begin{align}
\xi = a^\mu \, p_\mu 
	+ \omega ^{\mu \nu} \, m_{\mu \nu} 
	+ s \, d 
	+ c^\mu \, k_\mu \, .
\end{align}
Here, we have introduced the following basis of vector fields for the conformal algebra
\begin{align}
\begin{aligned}
	p_\mu (x) &= \partial_\mu \, , \\
	d(x) &= x^\mu \partial_\mu \, ,
\end{aligned}
&&
\begin{aligned}
	 m_{\mu \nu} (x) &= x_\mu \partial_\nu - x_\nu \partial_\mu \, , \\
	k_\mu (x)  &= \left( x^2 \delta_\mu ^\nu - 2 x_\mu x^\nu \right) \partial_\nu  \, . 
\end{aligned}
\label{conf_basis}
\end{align}
The vector fields $p_\mu$ and $m_{\mu \nu}$ generate translations and Lorentz-transformations and span the Lie algebra of the Poincar{\'e} group. The non-vanishing commutators between these generators are given by
\begin{align}
\begin{aligned}
\left[ m_{\mu \nu}, m_{\rho \sigma} \right] &=  
	\eta_{\mu \sigma} \, m_{\nu \rho} + \eta_{\nu \rho} \, m_{\mu \sigma} 
	- \eta_{\mu \rho} \, m_{\nu \sigma} - \eta_{\nu \sigma} \, m_{\mu \rho}\, , \\
	\left[ m_{\mu \nu}, p_{\lambda} \right] &= 
	\eta_{\nu \lambda} \, p_{\mu}  -  \eta_{\mu \lambda} \, p_{\nu} \, .
\end{aligned}	
\end{align}
In addition to the Poincar{\'e} generators, we have the dilatation generator $d$ as well as the generators $k_\mu$ of special conformal transformations. With these generators included, we have the additional commutation relations
\begin{align}
\begin{aligned}
	\left[d , p_\mu \right] &= - p_\mu \, , \\
	\left[d , k_\mu \right] &=  k_\mu \, ,
\end{aligned}
&&
\begin{aligned}
	\left[ m_{\mu \nu}, k_{\lambda} \right] &= 
	\eta_{\nu \lambda} \, k_{\mu}  -  \eta_{\mu \lambda} \, k_{\nu} \, , \\
	\left[p_\mu , k_\nu \right] &= 2 m_{\mu \nu} - 2 \eta_{\mu \nu} \,  d \, . 
\end{aligned}
\end{align}
Here, we have again only written out the non-vanishing commutators. A noteworthy aspect of the above commutation relations is that the commutator with $d$ is diagonal in the given basis, i.e.\ we have
\begin{align}
\left[ d , t_a \right] = \mathrm{dim}(t_a) \, t_a \, ,  
\label{d_comm}
\end{align}
where $t_a$ is one of the basis elements \eqref{conf_basis} and the dimensions of the generators are given by
\begin{align}
\dim ( p_\mu ) &= -1 \, , &
\dim ( k_\mu ) &= 1 \, , &
\dim ( m_{\mu \nu} ) &= 0 \, .
\end{align}
This aspect will prove to be crucial in the explicit construction of coset representatives in chapter \ref{chap:MinSurf}. One may show that the conformal algebra specified by the above commutation relations is isomorphic to the Lie algebra $\alg{so}(2,4)$.

We employ this isomorphism in order to construct a matrix representation of the conformal algebra, which will be used in chapter \ref{chap:MinSurf}. We note that we choose the basis 
$\lbrace T_a \rbrace$ for this representation in such a way that we have the commutation relations
\begin{align*}
[T_a , T_b ] = \mathbf{f} \indices{_{ba} ^c} \, T_c
	= f\indices{_{ab} ^c} \, T_c \, , 
\end{align*}
where $\mathbf{f} \indices{_{ab} ^c}$ denote the structure constants of the conformal Killing fields introduced above, $[t_a , t_b ] = \mathbf{f} \indices{_{ab} ^c} t_c$, and the above relation defines the structure constants $f\indices{_{ab} ^c}$ of the matrix generators. The reason for choosing this difference in the commutation relation will become clear in chapter \ref{chap:MinSurf}, where we obtain the above conformal Killing fields from a coset construction involving the basis $\lbrace T_a \rbrace$. 
We begin by defining the generators in the fundamental representation of $\grp{SO}(2,d)$, for which we note 
\begin{align}
\left( M_{I J} \right) ^\alpha {} _\beta 
	= \eta_{I \beta} \, \delta _J ^\alpha - \eta_{J \beta} \, \delta _I ^\alpha \, .
\end{align}
Here, the metric is given by $\eta_{IJ} = \mathrm{diag}(-1,1, \ldots , 1 , -1 )$ and the indices $I,J,\alpha$ and $\beta$ take values in $\lbrace 0, 1 , \ldots , d+1 \rbrace$. The above matrices satisfy the commutation relations
\begin{align}
\left[M_{IJ} , M_{KL} \right] = \eta_{I K} M_{J L} - \eta_{I L} M_{J K} + \eta_{J L} M_{I K} - \eta_{J K} M_{I L}  \, . 
\label{comm:MIJ}
\end{align}
We employ the trace to introduce a metric on the Lie algebra, which is given by
\begin{align}
\left \langle M_{IJ} , M_{KL} \right \rangle
	= \tr \left( M_{I J} M_{K L} \right) 
	= 2\, \eta_{I L} \, \eta_{J K} - 2\, \eta_{I K} \, \eta_{J L} \, .
\label{metric:MIJ}
\end{align}
The basis elements introduced above may be related to the usual basis elements of the conformal algebra by 
\begin{align}
P_\mu &= M_{\mu, N+1} - M_{\mu, N} \, , &
K_\mu &= M_{\mu, N+1} + M_{\mu, N} \, , &
D &= - M_{N, N+1} \, .
\end{align}
Here, the index $\mu$ extends from 0 to $d-1$. Apart from the commutation relations \eqref{comm:MIJ}, the non-vanishing commutators are given by
\begin{alignat}{3}
\left[ D, P_{\mu} \right] &=  P_{\mu} \, ,& \qquad  
\left[ M_{\mu \nu}, P_{\lambda} \right] &= \eta_{\mu \lambda} P_{\nu}  
	- \eta_{\nu \lambda} P_{\mu}   \, ,& \qquad 
\left[ P_{\mu}, K_{\nu} \right] &=  2 \eta_{\mu \nu} \, D  - 2 M_{\mu \nu} \, , \nn \\
\left[ D, K_{\mu} \right] &= - K_{\mu}\, ,& \qquad  
\left[ M_{\mu \nu}, K_{\lambda} \right] &= \eta_{\mu \lambda} K_{\nu} 
	-  \eta_{\nu \lambda} K_{\mu}   \, .   \label{conf_algebra_0}
\end{alignat}
In addition to \eqref{metric:MIJ}, we note the remaining non-vanishing elements of the metric 
\begin{align}
\tr \left(P_\mu \, K_\nu \right) &= 4 \, \eta_{\mu \nu} \, , & 
\tr \left( D \, D \right) &= 2 \, .
\label{eqn:Metric_0}
\end{align}

We can introduce a $\mathbb{Z}_2$-grading on the conformal algebra by considering the automorphism 
\begin{align}
\Omega(X) = K X K^{-1} \, , \quad \text{where} \quad 
K = \mathrm{diag}(1, \ldots ,1,-1) \, .
\end{align}
The $\mathbb{Z}_2$-grading will be important for the coset construction discussed in the following chapters. Here, we note that $\Omega$ is an involution, $\Omega ^2 = \id$, and hence it has eigenvalues $\pm 1$. We can then decompose the algebra into the eigenspaces, 
\begin{align}
\alg{h} &= \left \lbrace X \in \alg{so}(2,d) : \Omega(X) = X \right \rbrace \, , &
\alg{m} &= \left \lbrace X \in \alg{so}(2,d) : \Omega(X) = - X \right \rbrace \, .
\end{align}
Since $\Omega$ is an automorphism, the above decomposition gives a $\mathbb{Z}_2$-grading, i.e.\ we have
\begin{align}
\left[ \alg{h} , \alg{h} \right] &\subset \alg{h} \, , & 
\left[ \alg{h} , \alg{m} \right] &\subset \alg{m} \, , &
\left[ \alg{m} , \alg{m} \right] &\subset \alg{h} \, .
\end{align}
In terms of the generators introduced above, the subspaces $\alg{h}$ and $\alg{m}$ are given by
\begin{align}
\alg{h} &= \mathrm{span} \left \lbrace M_{\mu \nu} , P_\mu - K_\mu \right \rbrace  \, , &
\alg{m} &= \mathrm{span} \left \lbrace P_\mu + K_\mu , D \right \rbrace \, .
\end{align}

The fundamental representation of the Euclidean conformal group 
$\mathrm{SO}(1,d+1)$ can be constructed in the same way, and the corresponding relations follow by replacing $\eta_{\mu \nu} \to \delta_{\mu \nu}$.

Let us then turn to the discussion of the large conformal transformations. It is easy to see that the dilatation $d$ generates the scaling transformation $x^\mu \mapsto e^{2s} x^\mu$. The transformations associated to the generators $k_\mu$ can be obtained by combining the inversion map
\begin{align}
I(x)^\mu = \frac{x^\mu}{x^2}  
\end{align}  
with translations. One may show by explicit calculation that $I(x)$ is a conformal transformation and hence conclude that so is its concatenation with translations. This leads us to consider the special conformal transformations
\begin{align}
K_c &= I \circ T_c \circ I \, , &
K_c (x) ^\mu &= \frac{x^\mu + x^2 c^\mu}{1 + 2 c \cdot x + c^2 x^2} \, , 
\end{align}
where $T_c(x)^\mu = x^\mu + c^\mu$ denotes a translation by $c^\mu$. Expanding the above transformation for small $c$ shows that it is indeed the large transformation associated to the generators $k_\mu$. We note that the special conformal transformations become singular when $x$ approaches the light-cone described by
\begin{align}
0 = 1 + 2 c \cdot x + c^2 x^2 = c^2 \left( x + \frac{c}{c^2} \right) ^2  , 
\end{align}
i.e.\ the light-cone centered at $-c/c^2$. The special conformal transformations are hence not well-defined and thus do not strictly form a subgroup of the diffeomorphism group of 
$\mathbb{R}^{(1,3)}$. In order to address this problem, one considers a conformal compactification of Minkowski space, which we discuss below following references
\cite{Hawking:1973uf,book:Scho,Dorn:Lecture_Notes}. The basic idea is to include infinity in the space in a way that is compatible with the conformal structure. This construction allows to lift the relation between the conformal algebra and $\alg{so}(2,4)$ to the respective symmetry groups and it will also be important for our discussion of the AdS/CFT correspondence.   

\begin{figure}[t]
\centering
\includegraphics[width=120mm]{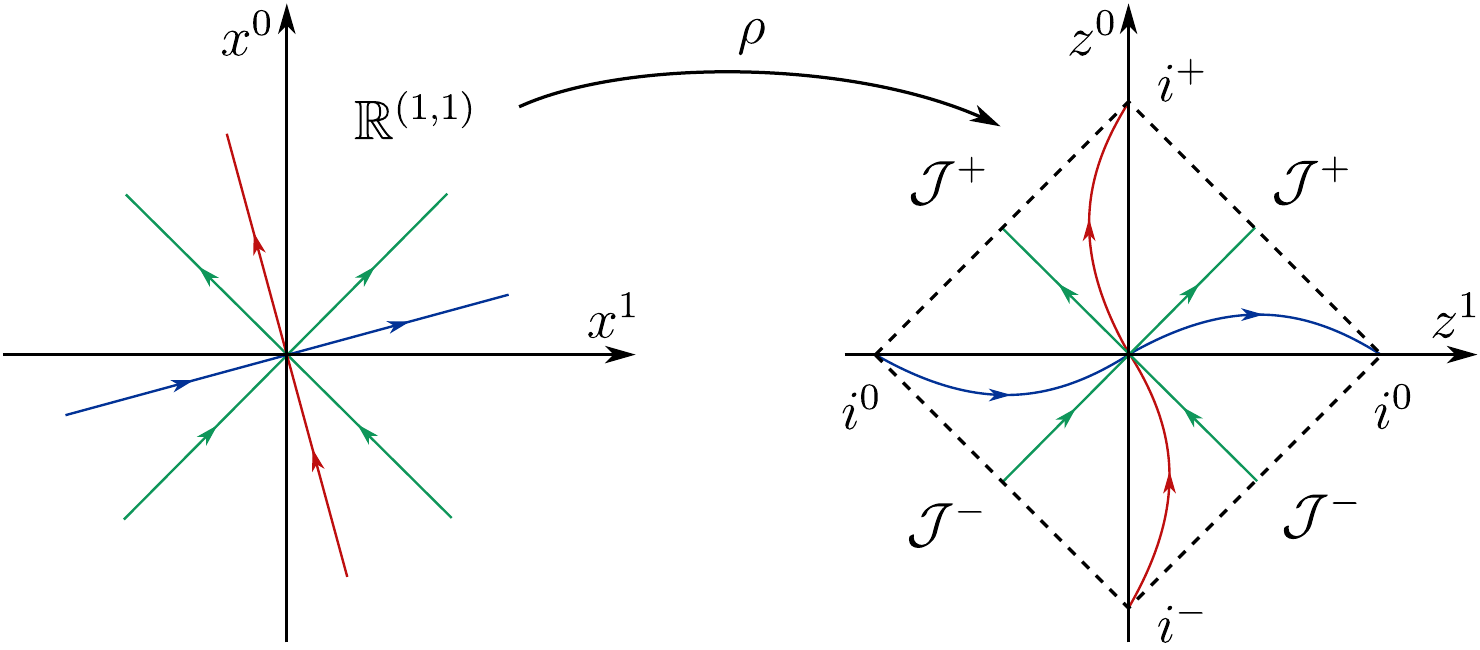}
\caption{Penrose diagram of $\mathbb{R}^{(1,1)}$. The red, green and blue lines correspond to time-, light- and space-like geodesics, respectively. Their images in the Penrose diagram approach 
the time-like past and future infinities $i^-$ and $i^+$, the light-like past and future infinities 
$\mathcal{J}^-$ and $\mathcal{J}^+$ and the space-like infinity $i^0$. }
\label{fig:Penrose}
\end{figure}

In order to gain some intuition for the idea of the conformal compactification, we discuss the two-dimensional Minkowski space $\mathbb{R}^{(1,1)}$ first. We treat this space as an example for the higher-dimensional spaces we are interested in and hence we leave out the conformal transformations that additionally appear in two-dimensional spacetime and only consider the conformal transformations discussed above, which appear for any dimension of the spacetime. In order to discuss infinity in spacetime one typically employs the Penrose diagrams introduced in reference \cite{Penrose:1962ij} and we begin by constructing the diagram for two-dimensional Minkowski spacetime. We map the coordinates $(x^0 , x^1)$ of Minkowski space $\mathbb{R}^{(1,1)}$ to the region $ \lbrace (z^0 , z^1) \in \mathbb{R}^{(1,1)} : - \pi < z^0 \pm z^1 < \pi \rbrace$ by setting 
\begin{align}
\begin{aligned}
z^1 &= \arctan \left( x^1 + x^0 \right) + \arctan \left( x^1 - x^0 \right) \, , \\
z^0 &= \arctan \left( x^1 + x^0 \right) - \arctan \left( x^1 - x^0 \right) \, , 
\end{aligned}
\end{align}
The metrics on the two spaces are then related by 
\begin{align}
(\diff x^1)	^2 - ( \diff x^0 )^2 
	&= \frac{ (\diff z^1) ^2 - (\diff z^0)^2 }{
	4 \cos ^2  \ft{z^1 + z^0}{2} \cos ^2  \ft{z^1 - z^0}{2} } \, , 
\label{map:conf}	
\end{align}
such that the above mapping is conformal. Note that the metric induced by the mapping diverges upon approaching the boundary of the Penrose diagram, which is given by the edges 
$z^1 \pm z^0 = \pi$ and  $z^1 \pm z^0 = - \pi$. The divergence has to appear since these edges correspond to infinity in Minkowski space. With infinity captured in a finite domain, we can identify the different parts of the causal structure. For this purpose, we consider the images of time-, light- and space-like geodesics of Minkowski space in the Penrose diagram, cf.\ figure \ref{fig:Penrose}. The images of these geodesics approach different parts of the boundary of the Penrose diagram and we discriminate between the time-like past and future infinities $i^-$ and $i^+$, the light-like past and future infinities $\mathcal{J}^-$ and $\mathcal{J}^+$ and the space-like infinity $i^0$. The light-like past and future infinities correspond to the edges of the boundary, the time- and space-like infinities are given by the cusp points. 

To construct the conformal compactification, we drop the conformal factor in equation \eqref{map:conf} and consider the closure of the image of Minkowski space, which is the entire region $ \lbrace (z^0 , z^1) \in \mathbb{R}^{(1,1)} : - \pi \leq z^0 \pm z^1 \leq \pi \rbrace$. However, in order to have a conformal compactification, we have to be able to analytically continue the conformal transformations of the original spacetime to the boundary. In order to see which conditions this requirement entails, let us consider the special conformal transformations, which are the only conformal transformations that map points inside Minkowski space to infinity and vice versa. 

We consider the special conformal transformation $K_c$ and take $y$ to lie on the critical light-cone, i.e.\ $\left(y + c /c^2 \right)^2 = 0$. For $x_\epsilon = y + \epsilon v$, we find  
\begin{align*}
K_c (x_\epsilon) ^\mu &= \frac{1}{2 \epsilon} \, 
	\frac{y^\mu + c^\mu y^2}{c^2 (v y) + v c} 
	+ \frac{v^\mu + 2 c^\mu (y v) }{2 \left( c^2 (vy) + vc \right) } 
	+ \O (\epsilon) \, ,  
\end{align*} 
where we have assumed that $v$ is not light-like (so that $x_\epsilon$ is not on the critical light-cone) and $y$ is not the center $-c/c^2$ of the critical light-cone. We observe that for $\epsilon \to 0$, $K_c (x_\epsilon)$ is asymptotically a straight line and a short calculation reveals that it is light-like. We thus conclude that for 
$\epsilon \to 0^+$ and $\epsilon \to 0^-$, $K_c (x_\epsilon)$ approaches opposite points in the light-like past and future infinity. For the case $y= - c/c^2$, we get
\begin{align*}
K_c (x_\epsilon) ^\mu &= \frac{1}{\epsilon} \, 
	\frac{c^2 v^\mu - 2 (vc) c^\mu}{c^4 v^2} + \frac{1}{c^2} \, , 
\end{align*}
which is again a straight line and we find that the direction is either time-like or space-like, depending on the sign of $v^2$. Thus we find that $K_c (x_\epsilon)$ approaches the time- and space-like infinities $i^0$, $i^-$ and $i^+$. 

In order for analytic continuations of the special conformal transformations to exist, the different points approached by $K_c(x_\varepsilon)$ on the conformal boundary upon approaching the same point $y$ on the critical light-cone have to be identified with each other. Schematically, we thus have the identification
\begin{align*}
\includegraphics[width=20mm]{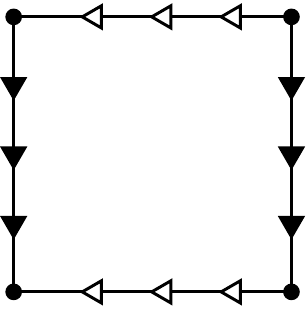} 
\end{align*} 
and turn the Penrose diagram into a torus $\mathrm{S}^1 \times \mathrm{S}^1$. 

In order to generalize the above construction to $d$-dimensional Minkowski space 
$\mathbb{R}^{(1,d-1)}$, we map it into the (periodically identified) Einstein static universe
$\mathrm{ESU}_d$, which is the product manifold $\mathrm{S}^{1} \times \mathrm{S}^{d-1}$. We use an angular coordinate $z^0 \in [-\pi,\pi)$ for the one-dimensional sphere and embedding coordinates $(z^1, \ldots , z^d)$ for $\mathrm{S}^{d-1}$, such that the metric is given by 
\begin{align}
\diff s^2 _{\mathrm{ESU}} = - (\diff z^0)^2 +  \diff \Omega_{d-1}  ^2 \, .
\end{align}
If we iteratively introduce spherical coordinates in $\mathrm{S}^{d-1}$ by
\begin{align}
z^d &= \cos \vartheta \, , & 
z^j &= \sin \vartheta \, w^j \, , &
\sum \limits_{j=1} ^{d-1} (w^j)^2 = 1 \, ,
\end{align}
we obtain the metric
\begin{align}
\diff s^2 _{\mathrm{ESU}} = - (\diff z^0)^2 + \diff \vartheta ^2 
	+ \sin ^2 \vartheta \, \diff \Omega_{d-2}  ^2 \, .
\end{align}
We introduce spherical coordinates for the space-like part of $\mathbb{R}^{(1,d-1)}$ as well, i.e.\ we introduce the radius $r = \sqrt{(x^1)^2 + \ldots (x^{d-1})^2}$ and obtain the metric 
\begin{align}
\diff s^2 _{\mathrm{MS}} =  - (\diff x^0)^2 + \diff r ^2 
	+ r^2 \, \diff \Omega_{d-2}  ^2 \, .
\end{align}
A conformal mapping of Minkowski space into $\mathrm{ESU}_d$ is then given by
\begin{align}
\begin{aligned}
z^0 &= \arctan \left( r + x^0 \right) - \arctan \left( r - x^0 \right) \, , \\
\vartheta &= \arctan \left( r + x^0 \right) + \arctan \left( r - x^0 \right) \, , 
\end{aligned}
\end{align}
and identifying the respective coordinates for the $(d-2)$-dimensional spheres 
$\mathrm{S}^{d-2}$. Since we have $r>0$, this mapping does not cover $\mathrm{ESU}_d$ entirely but only the region described by
\begin{align}
- \pi &< z^0 < \pi \, , &
z^d + \cos z^0 > 0 \, .
\end{align}
Inverting the above mapping gives the relations
\begin{align}
x^0 &= \frac{\sin z^0}{\cos z^0 + z^d} \, , &
x^j &= \frac{z^j}{\cos z^0 + z^d} \, , 
\end{align}
and we see that we can extend this map to the region $z^d + \cos z^0 \neq 0$ to obtain a double cover of Minkowski space. Alternatively, we can identify antipodal points on the two spheres by setting
\begin{align*}
z^0 &\sim z^0 \pm \pi \, , & z^j &\sim - z^j \, , & z^d &\sim - z^d \, ,
\end{align*}
and obtain a one-to-one mapping to Minkowski space with the conformal boundary identified as the set of points described by $z^d + \cos z^0 = 0$. Below, we describe an equivalent construction of the conformal compactification. The one described above is of particular interest for the discussion of the conformal boundary of Anti-de Sitter space $\AdS_d$. 

We consider the light-cone in projective space also known as the Dirac cone, cf.\ reference \cite{Dirac:1936fq}, where it was first considered. We identify points $w \in \mathbb{R}^{(2,d)}$ by the equivalence relation
\begin{align}
w \sim w^\prime \quad \Leftrightarrow \quad \exists \ell \neq 0: \; \;
	w = \ell w^\prime \, ,
\label{identify}	
\end{align}
and consider the image of the null cone,
\begin{align}
\mathrm{N}^{(1,d-1)} = \left. \left \lbrace w \neq 0 \, : \, 
	w_{-1} ^2 - w_0 ^2 + w_1 ^2 + \ldots + w_d ^2 = 0 \right \rbrace \middle / \! \sim 
	\right.  \, ,
\end{align}
under the projection onto the equivalence classes. The conformal map to Minkowski space can be written as
\begin{align}
x^\mu = \frac{w^\mu}{w^{-1} + w^d} \, .
\end{align}
Note that the right-hand side is scale-invariant, such that it is indeed well-defined on the equivalence classes $[w] \in \mathrm{N}^{(1,d-1)}$. The region corresponding to $w^{-1} + w^d \neq 0$ is mapped to Minkowski space in one-to-one fashion and $\mathrm{N}^{(1,d-1)}$ is indeed identified as the conformal compactification in Minkowski space, such that the region $w^{-1} + w^d = 0$ corresponds to the conformal boundary. The conformal transformations of Minkowski space can be analytically continued there, cf.\ reference \cite{book:Scho} for more details.   

Using the above construction, we can map isometries $\Lambda \in \grp{SO}(2,d)$ of $\mathbb{R}^{(2,d)}$ to maps $\Phi _\Lambda : \mathrm{N}^{(1,d-1)} \to \mathrm{N}^{(1,d-1)}$, which correspond to conformal transformations in Minkowski space. In fact, these are the sought-after analytic continuations and they clarify the relation between the conformal group of  Minkowski space and 
$\grp{SO}(2,d)$. Due to the equivalence relation \eqref{identify}, we note that $\Phi_{\Lambda} = \Phi_{-\Lambda}$ and hence the conformal group is identified as%
\footnote{The subscript $+$ denotes the connected component of the unit element.} 
$\grp{SO}_+(2,d)$ or $\grp{SO}_+(2,d)/\lbrace \pm E_{d+2} \rbrace$, respectively, depending on whether or not $d$ is even. 

\subsection{\texorpdfstring{$\mathcal{N}\!=4$}{N=4} Superconformal Symmetry}

We now turn to the discussion of the $\mathcal{N}\!=4$ supersymmetric extension of the conformal algebra $\alg{so}(2,4)$. We follow the approach of reference \cite{Weinberg}, where the superconformal algebra is obtained by including Poincar{\'e} supersymmetry generators in the conformal algebra and repeatedly applying the Jacobi identity in order to deduce the other generators and commutators. It is also instructive to discuss the fundamental representation of the superconformal algebra in terms of $\left(4 \vert 4 \right) \times \left(4 \vert 4 \right)$ supermatrices as in reference \cite{Arutyunov:2009ga,Beisert:2010kp}. As a prerequisite for our discussion of strings or minimal surfaces in semisymmetric spaces, we also discuss the fundamental representation of the superconformal algebra below. More details on our conventions and the choice of generators are collected in appendix \ref{app:u224}. 

The supersymmetry generators carry spinor indices and in order to be able to write the commutation relations in a compact fashion it is convenient to use spinor indices for the generators of the conformal algebra as well. Using the conventions introduced in appendix \ref{app:Spinor}, we note the following generators:
\begin{align}
p^{\da \a} &= \s ^{\mu \, \da \a} \, p_\mu \, , &
k_{\a \da} &= \bs ^\mu _{\a \da} \, k_\mu \, , &
m \indices{_\a ^\b} &= (\s ^{\mu \nu} ) \indices{_\a ^\b} \, m_{\mu \nu} \, , &
\widebar{m} \indices{^\da _\db} &= ( \bs ^{\mu \nu} ) \indices{^\da _\db} \, m_{\mu \nu} \, .
\end{align}
The commutation relations of these generators can be carried over straight-forwardly from the commutation relations given above by applying the spinor identities provided in appendix \ref{app:Spinor}. The commutation relations of the generators $m$ and $\widebar{m}$ can be written out conveniently by noticing that they only depend on the spinor indices and their position, 
\begin{align}
\begin{aligned}
\big[ m \indices{_\a ^\b}  \, , \,  J_\g \big] &= 
	- 2i \, \delta ^\b _\g J_\a + i \delta ^\b _\a J_\g \, , \\
\big[ \widebar{m} \indices{^\da _\db}  \, , \, J^\dg \big] &= 
	- 2i \, \delta ^\dg _\db J^\da + i \delta ^\da _\db J^\dg \,
\end{aligned}
&&
\begin{aligned}
\big[ m \indices{_\a ^\b} \, , \, J^\g \big] &= 
	2i \, \delta ^\g _\a J^\b - i \delta ^\b _\a J^\g \, , \\
\big[ \widebar{m} \, \indices{^\da _\db} \, , \, J_\dg \big] &= 
	2i \, \delta ^\da _\dg J_\db - i \delta ^\da _\db J_\dg	\, .
\end{aligned}
\label{spinor_transf}
\end{align}
Additionally, we note the commutator
\begin{align}
\big[ p^{\da \a} \, , \, k_{\b \db} \big] = 
	2i \, \delta ^\da _\db \, m \indices{_\b ^\a}
	-2i \, \delta ^\a _\b \, \widebar{m} \indices{^\da _\db} 
	+ 4 \, \delta ^\da _\db \, \delta ^\a _\b \, d \, .
\end{align}
We then include the supersymmetry generators $ q \indices{_A ^\a}$ and $\widebar{q}^{\, \da A}$, whose anti-commutator gives the translation generator, 
\begin{align}
\big \lbrace q \indices{_A ^\a} \, , \, \widebar{q}^{\, \da B} \big \rbrace 
 	&= -2i \, \delta ^B _A \, p^{\da \a} \, .
\label{susy_comm} 	
\end{align}  
Here, we consider an extension that does not contain central charges $Z^{AB}$, i.e.\ we have
\begin{align}
\big \lbrace q \indices{_A ^\a} \, , \, q \indices{_B ^\b} \big \rbrace &= 0 \, , &
\big \lbrace \widebar{q}^{\da A} \, , \, \widebar{q}^{\db B} \big \rbrace &= 0 \, .
\end{align}
The transformation of the spinor charges with respect to the $\alg{sl}(2,\mathbb{C})$ generators 
$m \indices{_\a ^\b}$ and $\widebar{m} \, \indices{^\da _\db}$ can be read off from equation \eqref{spinor_transf} and we note that the supersymmetry generators have half of the dimension of $p$, i.e.\ we have the commutation relations \eqref{d_comm} with
\begin{align}
\mathrm{dim} (q) = \mathrm{dim} ( \widebar{q} ) = -\frac{1}{2} \, .
\end{align}
An additional set of superconformal generators arises from calculating the commutator with 
$k_{\a \da}$, which gives 
\begin{align}
\big[ k_{\a \da} \, , \, q \indices{_A ^\b} \big] &= 
	+ 2i \, \delta^\b _\a \, \widebar{s}_{A \da} \, , & 
\big[ k_{\a \da} \, , \, \widebar{q} ^{\, \db A}  \big] &= 
	- 2i \, \delta^\db _\da \, s \indices{_\a ^A} \, .
\label{comm_kq}	
\end{align}
These generators are the counterparts of the supersymmetry generators associated to $k$, 
\begin{align}
\big \lbrace s^A _\a \, , \, \widebar{s}_{B \da} \big \rbrace &= 
	-2i \, \delta ^A _B \, k_{\a \da} \, , 
\end{align}
and hence we note that in addition to the commutation relations \eqref{spinor_transf} they have half of the dimension of $k$, 
\begin{align}
\mathrm{dim} (s) = \mathrm{dim} ( \widebar{s} ) = \frac{1}{2} \, .
\end{align}
In analogy to the commutation relations \eqref{comm_kq}, we note the commutators
\begin{align}
\big[ p^{\da \a} \, , \, s \indices{_\b ^A} \big] &= 
	- 2i \, \delta^\a _\b \, \widebar{q}^{\, \da A} \, , & 
\big[ p^{\da \a} \, , \, \widebar{s}_{A \db} \big] &= 
	+ 2i \, \delta^\da _\db \, q \indices{_A ^\a} \, .
\end{align}
The non-vanishing commutators between the supersymmetry and special superconformal generators 
are given by
\begin{align}
\begin{aligned}
\big \lbrace q \indices{_A ^\a} \, , \, s \indices{_\b ^B} \big \rbrace &= 
	-2i \, \delta ^B _A \, m \indices{_\b ^\a} 
	- \delta ^\a _\b \, r \indices{^B _A} 
	- 2 \, \delta ^B _A \, \delta ^\a _\b \, ( d + c ) \, , \\
\big \lbrace \widebar{q}^{\, \da A} \, , \, \widebar{s}_{B \db} \big \rbrace &= 
	-2i \, \delta ^A _B \, \widebar{m} \, \indices{^\da _\db} 
	- \delta ^\da _\db \, r \indices{^A _B} 
	+ 2 \, \delta ^A _B \, \delta ^\da _\db \, (d - c) \, .
\end{aligned}	
\end{align}
Here, $c$ is the central charge of the superconformal algebra $\alg{su}(2,2 \vert 4 )$, which commutes with all other generators. Representations of the above algebra, in which the central charge is absent, $c=0$, are denoted by $\alg{psu}(2,2\vert4)$. The above anti-commutator moreover produces the R-symmetry generators $r \indices{^A _B}$, which correspond to the R-symmetry group $\grp{SU}(4)$. As for the Lorentz generators, their commutation relations with the other generators of the algebra only depend on the set of indices and their positions, 
\begin{align}
\big[ r \indices{^A _B} \, , \, J^C \big] &= 
	- 4 \, \delta ^C _B J^A + \delta ^A _B J^C \, , &  
\big[ r \indices{^A _B} \, , \, J_C \big] &= 
	4 \, \delta ^A _C J_B - \delta ^A _B J_C \, .
\end{align}
The commutation relations also show that $\sum _A r \indices{^A _A} = 0$, such that we indeed have 15 linearly independent R-symmetry generators for the R-symmetry group $\grp{SU}(4)$. A crucial aspect of the superconformal algebra is noted in reference \cite{Weinberg}: The R-symmetry generators form a part of the superconformal algebra and are not merely outer automorphisms of it. They are thus necessarily symmetries of a superconformally invariant action, which is not generically the case for super Poincar{\'e} algebras.   

It is a difficult task in general to extend the representation \eqref{conf_basis} of the conformal algebra in terms of vector fields on Minkowski space to a representation of the superconformal algebra on a superspace containing this space as the bosonic base. In fact, reference \cite{Muller:2013rta} failed to do this correctly. We will construct such a representation explicitly in chapter \ref{chap:SurfaceSuperspace}, where it follows automatically from the superstring coset model. 

We now turn to the discussion of the fundamental representation of $\alg{u}(2,2 \vert 4)$, for which we follow reference \cite{Arutyunov:2009ga}. The representation is based on $( 4 \vert 4 )$ supermatrices, i.e.\ matrices
\begin{align}
N = \begin{pmatrix}
	m & \theta \\
	\eta & n
\end{pmatrix} \, , 
\end{align}
for which the entries of the off-diagonal blocks $\theta$ and $\eta$ are Gra{\ss}mann odd%
\footnote{Here, we follow the conventions of reference \cite{Arutyunov:2009ga}. One may also introduce supermatrices without referring to Gra{\ss}mann odd numbers, cf.\ references 
\cite{Cornwell,Beisert:2010kp}.}
numbers. The set of all such supermatrices satisfying the reality condition
\begin{align}
N = \begin{pmatrix}
	m & \theta \\ \eta & n
\end{pmatrix} = \begin{pmatrix}
	-H\, m^\dag H^{-1} & -H\, \eta^\dag   \\ -\theta^\dag H^{-1} & - n ^\dag
\end{pmatrix} = - \begin{pmatrix}
	H & 0 \\ 0 & \unit_4 
\end{pmatrix} N^\dag \begin{pmatrix}
H^{-1} & 0 \\ 0 & \unit_4 
\end{pmatrix}
\label{reality_constraint}
\end{align}
is denoted by $\alg{u}(2,2 \vert 4)$ and we note that the lower left block gives a representation of $\alg{u}(4)$. For the complex conjugation of the Gra{\ss}mann odd numbers, we note the conventions
\begin{align}
\left( c \, \theta \right)^\ast &= c ^\ast \, \theta ^\ast \, , & 
\theta ^{\ast \, \ast} &= \theta \, , &
\left( \theta_1 \, \theta_2 \right) ^\ast &= \theta_2 ^\ast \, \theta_1 ^\ast \, , 
\end{align}
where $c$ denotes an ordinary complex number and $\theta$ a Gra{\ss}mann odd number. These conventions ensure that commutators of supermatrices satisfying the reality constraint 
\eqref{reality_constraint} again satisfy \eqref{reality_constraint}. The matrix $H$ is given by
\begin{align}
H = \begin{pmatrix}
0 & \unit_2  \\ \unit_2 & 0  
\end{pmatrix} ,
\end{align}
and since it has split signature, we note that the upper right blocks of the matrices $N$ form a representation of $\alg{u}(2,2)$. The choice of the matrix $H$ differs from the one made in reference \cite{Arutyunov:2009ga} and is better adapted to the choice of generators that are typically used on the field theory side, cf.\ also reference \cite{Beisert:2010kp}. The different choices for the matrix $H$ are related by a unitary transformation.

We can introduce a $\mathbb{Z}_4$-grading on $\mathfrak{u}(2,2 \vert 4)$ by using the map
\begin{align}
N \mapsto \Omega(N) &= - \mathcal{K} \, N^{\mathrm{st}} \, \mathcal{K}^{-1} \, , &
\mathcal{K} &= \begin{pmatrix}
	K & 0 \\ 0 & K
\end{pmatrix} , &
K = \begin{pmatrix} 
	-i \sigma ^2 & 0 \\ 0 & -i \sigma^2 
	\end{pmatrix} ,
\end{align}
Here, $N^{\mathrm{st}}$ denotes the super-transpose
\begin{align}
N^{\mathrm{st}} = \begin{pmatrix}
	m^\mathrm{t} & - \eta^\mathrm{t} \\ \theta^\mathrm{t} & n^\mathrm{t} 
\end{pmatrix} ,  
\end{align}
and we note that $\Omega ^4 = \id$. The map $\Omega$ gives an automorphism of 
$\alg{gl}(4 \vert 4 )$ and introduces a $\mathbb{Z}_4$-grading by decomposing 
$\alg{gl}(4 \vert 4 )$ into its eigenspaces, 
\begin{align}
\begin{aligned}
\mathfrak{gl}(4 \vert 4) &=
	\mathfrak{gl}(4 \vert 4)^{(0)} \oplus \mathfrak{gl}(4 \vert 4)^{(2)} 
	\oplus \mathfrak{gl}(4 \vert 4)^{(1)} \oplus \mathfrak{gl}(4 \vert 4)^{(3)} \, , \\
\mathfrak{gl}(4 \vert 4)^{(k)} &= \left \lbrace N \in \mathfrak{gl}(4 \vert 4) \, : \, 
	\Omega(N) = i^k N \right \rbrace .
\end{aligned}
\end{align}
We can project any element of $\mathfrak{gl}(4\vert4)$ onto the eigenspace for the eigenvalue $i^k$ with the projectors
\begin{align}
P^{(k)}(N) =  N^{(k)} = \quarter \left( N + i^{3k} \Omega(N) +  i^{2k} \Omega^2(N) +  i^{k} \Omega^3(N) \right)  . 
\end{align}
While the automorphism $\Omega$ clearly does not map $\alg{u}(2,2 \vert 4)$ to itself, the projectors $P^{(k)}$ do, and we can hence carry over the above $\mathbb{Z}_4$-grading to 
$\alg{u}(2,2 \vert 4)$, i.e.\ we have
\begin{align}
\begin{aligned}
\mathfrak{u}(2,2\vert4) &= \mathfrak{g}^{(0)} \oplus \mathfrak{g}^{(2)} 
	\oplus \mathfrak{g}^{(1)} \oplus \mathfrak{g}^{(3)} \, , & \quad
\mathfrak{g}^{(k)} &= \left\lbrace  P^{(k)}(N) \, \vert \, N \in \mathfrak{u}(2,2\vert4) \right \rbrace  , \\
\left[ \mathfrak{g}^{(k)} \, , \, \mathfrak{g}^{(l)} \right] &\subset \mathfrak{g}^{(k+l) \, \mathrm{mod} 4 }  \,.
\end{aligned}
\end{align}
In appendix \ref{app:u224}, we introduce a basis of generators for $\alg{u}(2,2 \vert 4)$. As already for the conformal algebra discussed above, the generators are chosen in such a way that they satisfy the commutation relations
\begin{align*}
\left[ T_a , T_b \right] = \mathbf{f} \indices{_{ba} ^c} \, T_c 
	= f \indices{_{ab} ^c} \, T_c \, , 
\end{align*}
where $\mathbf{f} \indices{_{ab} ^c}$ denote the structure constants of the generators introduced in the beginning of this subsection, 
$\left[ t_a , t_b \right] = \mathbf{f} \indices{_{ab} ^c} \, t_c \, $. Here, we only note two of the generators, the central charge $C$ and the hypercharge generator $B$, which are given by
\begin{align}
C &= \frac{1}{2} \left( \begin{array}{c c} 
	\unit_4  & 0 \\ 
	0 & \unit_4  \\
	\end{array} \right) \, , &
B &=  - \frac{1}{2} \left( \begin{array}{c c} 
	0  & 0 \\ 
	0 & \unit_4  \\
	\end{array} \right)  \, .
\end{align}
The algebra $\alg{su}(2,2 \vert 4)$ is reached by restricting to elements with vanishing supertrace,
\begin{align*}
\str (N) = \tr (m) - \tr (n) = 0 \, ,
\end{align*}
which corresponds to leaving out the hypercharge generator $B$. A representation in which additionally the central charge $C$ vanishes is denoted by $\alg{psu}(2,2 \vert 4 )$. In these two cases, the bosonic subalgebras are given by 
$\alg{su}(2,2) \oplus \alg{su}(4) \oplus \alg{u}(1)$ or  
$\alg{su}(2,2) \oplus \alg{su}(4)$, respectively. 

We can employ the supertrace to introduce a metric $G_{a b} = \str (T_a T_b)$ on the algebra. The metric satisfies the symmetry property 
$G_{a b} = \left(-1 \right) ^{\lvert a \rvert} G_{b a}$, 
where $\lvert a \rvert = \mathrm{deg}(T_a)$ denotes the Gra{\ss}mann degree of a homogeneous basis element, i.e.\ $\lvert a \rvert = 0$ or 1 for an even or odd generator, respectively. The components of the metric for a set of generators of $\alg{u}(2,2 \vert 4)$ are collected in appendix \ref{app:u224}. Here, we note that 
\begin{align}
\str ( B \, C ) = 1 
\end{align} 
and all other components of the metric that involve the generators $B$ or $C$ vanish. The metric is hence degenerate for the superalgebra $\alg{su}(2,2 \vert 4)$, but not for 
$\alg{u}(2,2 \vert 4)$ or $\alg{psu}(2,2 \vert 4)$.

\subsection{Yangian Symmetry}

A typical feature of many integrable models is the appearance of Yangian symmetry, which can be viewed as a generalization of Lie algebra symmetries. The Yangian $\mathrm{Y}[\alg{g}]$ over a simple Lie algebra $\alg{g}$ was introduced by Drinfeld in references \cite{Drinfeld:1985rx,Drinfeld:1986in} and has played an important role in the study of integrable systems since. While it was first mainly studied in the context of integrable two-dimensional field theories, cf.\ e.g.\ references \cite{Bernard:1990jw,MacKay:1992he,Bernard:1992ya}, it has also been encountered within 
$\mathcal{N}\!=4$ supersymmetric Yang--Mills theory both in the context of anomalous dimensions or the associated integrable spin chains \cite{Minahan:2002ve,Beisert:2003tq,Beisert:2003yb,Dolan:2003uh,Dolan:2004ps} as well as for scattering amplitudes \cite{Drummond:2009fd,Bargheer:2009qu,Beisert:2010gn,CaronHuot:2011kk}. 

Below, we give a short and introductory review to the so-called first realization of the Yangian, which is based on references \cite{Ferro:2011ph,Loebbert:2016cdm} as well as the author's master's thesis \cite{Master}, where the reader may find an accessible account of some of the algebraic prerequisites which are not elaborated on below. Other interesting accounts of Yangian symmetry can be found in references \cite{MacKay:2004tc,Beisert:2010jq,Torrielli:2010kq,Torrielli:2011gg} as well as \cite{Rocen}, which contains many technical details.

The Yangian $\mathrm{Y}[\alg{g}]$ is an infinite-dimensional extension of the underlying Lie algebra $\alg{g}$ and we can organize the generators in levels, beginning with the level zero, which is spanned by the generators $\J_a ^{(0)}$ of $\alg{g}$. The full algebra can be obtained by additionally specifying the generators $\J_a ^{(1)}$, which span the level one. The higher levels can then be obtained from repeated commutators of the level-1 generators. 

In order to discuss the algebraic structure of the Yangian, it is perhaps simplest not to consider the completely abstract setting right away. Rather, we consider an $N$-site space, which could arise from e.g.\ a spin chain or a color-ordered amplitude describing the interaction of $N$ particles. This has the advantage that the meaning of the product of two generators is clear and we do not have to introduce an enveloping algebra to define it. At each site $i$, we have a representation of the underlying Lie algebra $\alg{g}$ in terms of generators $\J_{a, i}$, for which we note the commutation relations
\begin{align}
\big[ \J_{a, i} \, , \,  \J_{b, k} \big] = \delta_{ik} \, 
	\mathbf{f} \indices{_{ab} ^c} \, \J_{c, i} \, .
\end{align} 
In this situation, the level-0 and level-1 Yangian generators typically have the form
\begin{align}
\J_a ^{(0)} &= \sum \limits _{i=1} ^N \J_{a, i} \, , &
\J_a ^{(1)} &= \mathbf{f} \indices{_a ^{cb}} \sum \limits _{i<k}  
	\J_{b, i} \, \J_{c, k} \, ,
\label{Yangian_Gen}		
\end{align}
and a simple calculation shows that they obey the commutation relations
\begin{align}
\big[ \J_a ^{(0)} \, , \,  \J_b ^{(0)} \big] &= 
	\mathbf{f} \indices{_{ab} ^c} \, \J_c ^{(0)} \, , &
\big[ \J_a ^{(0)} \, , \,  \J_b ^{(1)} \big] &= 
	\mathbf{f} \indices{_{ab} ^c} \, \J_c ^{(1)} \, . 
\label{Yangian_Comm}	
\end{align}
The commutators of two level-1 generators are more involved and contain the level-2 generators. We see directly that it contains terms which act on three sites. A crucial aspect of the above generators is that they obey the Serre relation%
\footnote{The brackets denote the symmetrization or anti-symmetrization of the enclosed indices, specifically we define 
$X_{(i_1 \ldots i_n)} = \ft{1}{n!} \sum _{\sigma \in S_n} 
	X_{i_{\sigma(1)} \ldots i_{\sigma(n)}}$
as well as 
$X_{[i_1 \ldots i_n]} = \ft{1}{n!} \sum _{\sigma \in S_n} \mathrm{sign}(\sigma)
	X_{i_{\sigma(1)} \ldots i_{\sigma(n)}}$.
We note moreover that in the case of $\alg{g}=\alg{su}(2)$, the Serre relation is replaced by another relation, which is otherwise implied, cf.\ reference \cite{MacKay:2004tc}.}
\begin{align}
\f \indices{^d _{[ab}} \big[ \J^{(1)} _{c]} , \J^{(1)} _d \big] 
	&= \frac{1}{12} \,  
	\f \indices{_{ag} ^d} \,\f \indices{_{bh} ^e} \,
	\f \indices{_{ck} ^f} \, \f ^{ghk} \,  
	\J ^{(0)}_{(d} \, \J ^{(0)}_e \, \J ^{(0)}_{f)} .  
\label{Serre1}	
\end{align}
This was shown in reference \cite{Dolan:2004ps} in the case of the underlying Lie algebra being given by $\alg{su}(N)$. Let us elaborate on this relation for a moment. It is instructive to compare the Yangian algebra to the polynomial algebra $\alg{g}[u]$ over the Lie algebra $\alg{g}$. An appropriate basis for the polynomial algebra is given by the generators
$J_a ^{(n)} = u^n J_a$, where $J_a$ are the generators of $\alg{g}$. For these generators, we note the commutation relations
\begin{align}
\big[ J_a ^{(n)} ,  J_b ^{(m)} \big] = \mathbf{f} \indices{_{ab} ^c} \, J_c ^{(n+m)} \, .
\end{align}
The polynomial algebra is sometimes referred to as half of a loop algebra, since above we are assuming $m$ and $n$ non-negative. For the above algebra, we note that the left-hand side of the Serre relation \eqref{Serre1} vanishes due to the Jacobi identity. In this sense, the Serre relation is sometimes referred to as a generalized Jacobi identity and the Yangian algebra may be viewed as a deformation of the universal enveloping algebra of the polynomial algebra. The Serre relation constrains the construction of higher-grade elements of the Yangian algebra. If we define an element of grade two by%
\footnote{Here, $\alg{c}$ denotes the dual Coxeter number, which arises in the contraction
	$\mathbf{f} \indices{_a ^{bc}} \mathbf{f} \indices{_{cb} ^{d}} = 2 \alg{c} \delta ^d _a$.} 
\begin{align*}
\J_a^{(2)} = \frac{1}{2 \alg{c}} \, \mathbf{f} \indices{_a ^{b c}} \, 
	\big[ \J_c^{(1)} ,  \J_b^{(1)} \big] \, , 
\end{align*}
the commutator of the level-1 generators can be expressed as
\begin{align*}
\big[ \J_b^{(1)} , \J_c^{(1)} \big] = 
	\mathbf{f} \indices{_{ b c} ^d} \, \J_d^{(2)} + X_{b c} \, ,
\end{align*}
where $\mathbf{f} \indices{_a ^{c b}} \, X_{b c} = 0$. The imposition of the Serre relation then uniquely determines $X_{b c}$, cf.\ reference \cite{MacKay:2004tc} for details.

Abstractly, one may define the Yangian algebra as the algebra generated by $\J_a ^{(0)}$ and 
$\J_a ^{(1)}$, such that the commutation relations \eqref{Yangian_Comm} and the Serre relation \eqref{Serre1} hold true. The explicit construction is similar to the construction of the universal enveloping algebra of a Lie algebra, which we review now. 

The idea behind this construction is to embed a Lie algebra into an algebra in such a way that the Lie bracket and the algebra product are compatible, i.e.\ we have
\begin{align}
\iota \left( [A,B] \right) = \iota (A) \otimes \iota(B) - \iota(B) \otimes \iota(A) \, ,
\label{UEA_Id}
\end{align}
where $A,B$ are elements of the Lie algebra and $\iota$ denotes the inclusion map from the Lie algebra into the universal enveloping algebra and $\otimes$ denotes the algebra product. The universal enveloping algebra can be constructed by appropriately identifying elements in the tensor algebra
\begin{align}
T( \alg{g} ) = \bigoplus \limits _{n \geq 0} \alg{g} ^{\otimes n} \, .
\label{TensorAlg}
\end{align} 
Here, $\alg{g} ^{\otimes n}$ denotes the $n$-fold tensor product of $\alg{g}$ and we have set 
$\alg{g} ^{\otimes 0} = \mathbb{R}$, assuming we are considering a real Lie algebra. The algebra product in the tensor algebra is simply given by the tensor product. The Lie algebra is naturally embedded in the tensor algebra by simply mapping it to the copy of $\alg{g}$ in the direct sum in equation \eqref{TensorAlg}. The appropriate way to achieve the identification \eqref{UEA_Id} is to factor out a two-sided ideal $\mathrm{I}$, which contains all elements of the form
\begin{align*}
[x,y] - x \otimes y + y \otimes x \, ,
\end{align*}
with $x,y \in \alg{g}$. A two-sided ideal $\mathrm{I}$ is a subspace of the algebra for which the multiplication with algebra elements from either side lies again in $\mathrm{I}$. This structure is needed to ensure that when we carry over the algebra structure of $T( \alg{g} )$ to the factor algebra $T( \alg{g} ) / \mathrm{I}$ the definitions do not depend on the representatives of the equivalence classes. Here, we could write it explicitly as the set of all linear combinations of elements of the form 
\begin{align*}
r \otimes \left( [x,y] - x \otimes y + y \otimes x \right) \otimes s
\end{align*}
with $r,s \in T( \alg{g} )$ and with $x,y \in \alg{g}$ as before. We have thus simply enforced the desired relation \eqref{UEA_Id} by factoring out $\mathrm{I}$ and note that the universal enveloping algebra is precisely the factor algebra $T( \alg{g} ) / \mathrm{I}$. 

The Yangian $\mathrm{Y}[\alg{g}]$ can be constructed similarly by considering an enveloping algebra of $\alg{g} \oplus \mathrm{span} \lbrace \J_a ^{(1)}\rbrace$ and enforcing the commutation relations \eqref{Yangian_Comm} as well as the Serre relation \eqref{Serre1} e.g.\ in the way we have discussed above. In this construction, the product of level-0 generators on the right-hand side of the Serre relation is just given by the product in the enveloping algebra. 

An important aspect of the Yangian algebra is that it can be equipped with a Hopf algebra structure with the coproduct 
$\Delta:\mathrm{Y}[\alg{g}] \to \mathrm{Y}[\alg{g}] \otimes \mathrm{Y}[\alg{g}]$ satisfying the relations
\begin{align}
\begin{aligned}
\Delta \big( \J_a ^{(0)} \big) &= 
	\J_a ^{(0)} \otimes \unit + \unit \otimes \J_a ^{(0)} \, , \\ 
\Delta \big( \J_a ^{(1)} \big) &= 
	\J_a ^{(1)} \otimes \unit + \unit \otimes \J_a ^{(1)} 
	+ \frac{1}{2} \, \mathbf{f} \indices{_a ^{cb}} \, \J_b ^{(0)} \otimes \J_c ^{(0)} \, .
\end{aligned}	
\end{align}
A Hopf algebra contains more structure than the coproduct specified above, the reader is invited to consult reference \cite{Loebbert:2016cdm,Master} for a short introduction to Hopf algebras and more details on the specific Hopf algebra structure of the Yangian. Here, we will be satisfied with just pointing out the implications of the above coproduct structure. Along with the trivial coproduct 
$\Delta (\unit) = \unit \otimes \unit$, the above definitions completely specify the coproduct on 
$\mathrm{Y}[\alg{g}]$. This is due to the facts that the coproduct is required to be an algebra morphism in a Hopf algebra and that we can obtain any element of the Yangian as linear combinations of products of the level-0 and level-1 generators. The algebra constraints \eqref{Yangian_Comm} and \eqref{Serre1} are compatible with this requirement. It is shown explicitly in reference \cite{Rocen} that when we require the relations \eqref{Yangian_Comm}, we must also include the Serre relation \eqref{Serre1} in order for $\Delta$ to become an algebra morphism.

The coproduct also motivates the level-1 generators given in equation \eqref{Yangian_Gen}. If we begin with some representation of the Yangian on a single-site space, we may employ the coproduct to obtain a representation on a two-site space. The two-site representation of the level-1 generators would then contain the contribution 
\begin{align*}
\mathbf{f} \indices{_a ^{cb}} \, \J_b ^{(0)} \otimes \J_c ^{(0)} 
\end{align*}
and in fact the level-1 generator in equation \eqref{Yangian_Gen} is obtained by summing these contributions over all pairs of two sites. 

We note that the Yangian algebra is invariant under the map $\J_a^{(1)} \mapsto \J_a^{(1)} + \alpha \J_a^{(0)}$. This is easy to see for the commutation relation \eqref{Yangian_Comm} and follows from the Jacobi identity for the Serre relation \eqref{Serre1}. For the $N$-site space discussed above, we can more generally add a local contribution of the form
\begin{align}
\J_{a , \mathrm{lo}}^{(1)} = \sum \limits _{i=1} ^N v_i \, \J_{a,i}      
\end{align}    
to the level-1 generators $\J_a^{(1)}$ without altering the Yangian algebra relations. We will see in chapter \ref{chap:CarryOver} that this local contribution can be employed to control the starting-point dependence of the level-1 generators, cf.\ reference \cite{Chicherin:2017cns}.

\section{\texorpdfstring{$\mathcal{N}\!=4$}{N=4} Supersymmetric Yang--Mills Theory}

We introduce $\mathcal{N}\!=4$ supersymmetric Yang--Mils (SYM) theory, which is the unique gauge theory in four dimensions with this amount of supersymmetry. This is the maximal amount of supersymmetry in four dimensions, since any theory with $\mathcal{N} \, > 4$ would have to contain particles with spin $s > 1$ and would hence not be renormalizable. 

A convenient way to derive the action of $\mathcal{N}\!=4$ SYM theory is given by considering the dimensional reduction of ten-dimensional $\mathcal{N}\!=1$ SYM theory to four dimensions. This approach was first described in reference \cite{Brink:1976bc}. The ten-dimensional gauge theory contains the gauge field $A_m$ and a ten-dimensional Majorana--Weyl spinor $\Psi$. We take the gauge group to be $\grp{SU}(N)$, such that all fields take values in the Lie algebra $\alg{su}(N)$, 
\begin{align}
A_m &= A_m ^a \, t^a \, , &
\Psi &= \Psi ^a \, t^a \, . 
\end{align} 
Here, the generators $t^a$ denote a basis of $\alg{su}(N)$, for which we choose the convention
$2 \tr ( t^a \, t^b ) = \delta ^{ab}$. We note the expressions for the covariant derivative and the field strength,
\begin{align}
D_m \Psi &= \partial_m \Psi - i \left[ A_m , \Psi \right] \, , &
F_{mn} &= \partial_m A_n - \partial_n A_m - i \left[A_m , A_n \right] .
\end{align}
The conventions are chosen in such a way that the fields have classical mass dimensions
$\left[ A \right] = 1$ and $\left[ \Psi \right] = 3/2$, whereas the Yang--Mills coupling constant  $g_{10}$ in ten dimensions has dimension $[ g_{10}] = -3$. The action then takes the form
\begin{align}
S = \frac{1}{g_{10} ^2} \int \diff ^{10} x \, 
	\tr \left( - \half \, F_{mn} \, F^{mn} + i \bar{\Psi} \, \Gamma^m D_m \, \Psi \right) . 
\label{10daction}	
\end{align}
Here, the matrices $\Gamma_m$ are ten-dimensional Dirac matrices, which satisfy the Clifford algebra for $\mathbb{R}^{(1,9)}$, 
\begin{align}
\left \lbrace \Gamma ^m , \Gamma ^n \right \rbrace 
	= - 2 \eta ^{mn} \, \unit \, .
\end{align}
Note that we are working with the mostly-plus convention
$\eta = \mathrm{diag}(-, + , \ldots , +)$. Apart from the gauge invariance under the transformations
\begin{align}
A_m \; \; & \mapsto \; \; U(x) \left( A_m + i \, \partial_m \right) U(x) ^\dag \, , &
\Psi \; \; & \mapsto \; \; U(x) \Psi U(x) ^\dag \, ,
\label{gaugetransf}
\end{align}
the action is invariant under the supersymmetry transformations
\begin{align}
\delta _\xi A_m &= i \, \bar{\xi} \, \Gamma_m \, \Psi \, , & 
\delta _\xi \Psi &= \ihalf F_{mn} \, \Gamma ^{mn} \, \xi \, .
\label{10dSUSY}
\end{align}
The supersymmetry parameter $\xi$ is a constant, ten-dimensional Majorana--Weyl spinor and the matrices $\Gamma _{mn}$ are defined by 
$\Gamma _{mn} = \ihalf \left[ \Gamma_m , \Gamma _n \right]$. 

The dimensional reduction to four dimensions is obtained by demanding that the fields only depend on the coordinates $x^\mu$ of $\mathbb{R}^{(1,3)} \subset \mathbb{R}^{(1,9)}$. The remaining volume integrals in the action, $V = \int \diff x^4 \ldots \diff x^9$, are absorbed by a redefinition of the coupling constant, $g = V^{-1/2} g_{10}$, which is then dimensionless. The independence of the fields on the coordinates $x^4$ to $x^9$ implies in particular that 
\begin{align}
\partial_m A_n(x) &= 0 \, , &
\partial_m \Psi(x) &= 0 \, , &
\partial_m U(x) &= 0 \, , 
\end{align} 
for $m$ taking values in $\lbrace 4 , 5 , \ldots , 9 \rbrace$. Since also the gauge transformations only depend on the first four coordinates, we note that the fields
\begin{align}
\Phi _I &= A_{I+3} \, , \qquad I \in \lbrace 1 , 2 , \ldots , 6 \rbrace ,
\end{align}
no longer transform as gauge fields, but transform simply in the adjoint representation,
\begin{align}
\Phi_I \; \; & \mapsto \; \; U(x) \Phi_I U(x) ^\dag \, .
\end{align}
These are scalar fields from the four-dimensional viewpoint, i.e.\ with respect to the Lorentz group in $\mathbb{R}^{(1,3)}$. The discussion of the spinor fields is facilitated by choosing an appropriate representation of the ten-dimensional Clifford algebra, which discriminates naturally between the four- and six-dimensional spinor indices. For the choice introduced in appendix \ref{app:Spinor}, the left-handed Weyl spinor $\Psi$ takes the form
\begin{align}
\Psi = \big( \Psi \indices{_\a ^A} , 0 , 0 , \tilde{\Psi} \indices{_A ^\da} \big)^T  .
\end{align}
The Majorana condition then implies that 
$\Psi \indices{_\a ^A} = ( \tilde{\Psi} _{A \da} ) ^\ast$. The original spinor indices in ten dimensions are split up into the four-dimensional spinor indices $\a, \da$ and the R-symmetry indices labeled by $A \in \lbrace 1 , \ldots , 4 \rbrace $ above. They correspond to the subgroup
$\grp{SO}(6)\! \sim \! \grp{SU} (4)$ of the ten-dimensional Lorentz group. 
From the four-dimensional viewpoint we are thus considering a set of four four-dimensional Majorana spinors. 

In order to write out the action \eqref{10daction} in terms of the fields 
$\big( A_\mu , \Phi_I ,  \Psi \indices{_\a ^A} , \tilde{\Psi} \indices{_A ^\da} \big)$, it is convenient to employ the matrices 
$\Sigma ^{I}$ and $\widebar{\Sigma}{}^I$, which are introduced in appendix \ref{app:Spinor} as the blocks of the six-dimensional Dirac matrices, to associate R-symmetry indices to the scalar fields,
\begin{align}
\Phi ^{AB} &= \frac{1}{\sqrt{2}} \, \Sigma ^{I \, AB} \, \Phi_I \, , &
\widebar{\Phi}_{AB} &=  \frac{1}{\sqrt{2}} \, \BSi^I _{AB} \, \Phi_I
	= \frac{1}{2} \, \epsilon_{ABCD} \, \Phi ^{CD} \, . 
\end{align}
The contractions of the different indices are related by the identity
\begin{align}
\widebar{\Phi}_{AB} \, \Phi ^{AB} = 2 \, \Phi^I \, \Phi^I \, .
\end{align}
Writing out the action \eqref{10daction} in terms of the fields discussed above, we find the action of $\mathcal{N}\!=4$ supersymmetric Yang--Mils theory in four dimensions to be given by
\begin{align}
S &= \frac{1}{g^2} \int \diff ^4 x \tr \Big(  
	- \half \, F_{\mu \nu} F^{\mu \nu} + \half \left( D_\mu \widebar{\Phi}_{A B} \right) 
	\left( D^\mu \Phi^{A B} \right) 
	+ \ft{1}{8} \, \left[ \, \widebar{\Phi}_{A B} , \widebar{\Phi}_{C D} \right]  
	\left[ \Phi^{A B}  , \Phi^{C D} \right] \nn \\
	& \quad + 2 i \, \tilde{\Psi}_{A \, \da} \,  
	D^{\da \b}	
	\, \Psi \indices{_\b ^A} 
	+ \sqrt{2}   \, \tilde{\Psi}_{A \, \da} \,  
	\big[  \Phi^{A B} , \tilde{\Psi}  \indices{_B ^\da }  \big] 
	- \sqrt{2} \, \Psi^{\a A} \,  
	\left[ \widebar{\Phi}_{A B} ,  \Psi \indices{ _\a ^B} \right] \Big) . 
\label{4daction}
\end{align}
The above action inherits the invariance under the supersymmetry transformations \eqref{10dSUSY} from the ten-dimensional theory, which appears as $\mathcal{N}\!=4$ supersymmetry after decomposing the ten-dimensional spinor into four-dimensional spinors as for the gluino fields above. In addition to the Poincar{\'e} and supersymmetry invariance, the action is classically invariant under the scale transformations
\begin{align*}
x \; \; & \mapsto \; \; e^{-s} \, x \, , & 
A \; \; & \mapsto \; \; e^{s} \, A \, , &  
\Phi \; \; & \mapsto \; \; e^{s} \, \Phi \, , &  
\Psi \; \; & \mapsto \; \; e^{3s/2} \, \Psi \, , &    
\end{align*}   
and indeed also invariant under conformal transformations, such that the $\mathcal{N}\!=4$ Poincar{\'e} supersymmetry is extended to the superconformal symmetry 
$\alg{psu}(2,2 \vert 4)$. However, the conformal symmetry of a theory is often broken by the introduction of a renormalization scale $\mu$ in the quantum theory. The observables of the theory then become scale-dependent due to the scale dependence of the parameters of the theory. The scale dependence of the coupling constant $g$, which is the only parameter of the theory for fixed $N$, is described by the beta function 
\begin{align}
\beta = \mu \, \frac{\diff g}{\diff \mu} \, .
\end{align}
The beta function of $\mathcal{N}\!= 4$ SYM theory is believed to vanish to all orders in perturbation theory and also non-perturbatively, cf.\ references \cite{Sohnius:1981sn,Mandelstam:1982cb,Howe:1983sr,Brink:1982pd,Brink:1982wv,Sohnius:1985qm}. This implies that the conformal symmetry of the classical theory also holds true for the quantum theory, which thus still has the full $\alg{psu}(2,2 \vert 4 )$-invariance. 

Since the coupling constant is not running, the theory is described by two freely tunable parameters: the coupling constant $g$ and the parameter $N$ describing the dimension of the gauge group $\grp{SU}(N)$. The two parameters are often combined to the 't Hooft coupling constant 
\begin{align}
\lambda = g^2 N \, , 
\end{align}
such that observables can be described in the double expansion in $\lambda$ and $N^{-1}$. In this thesis, we will mainly consider the limit of large $N$, in which we send 
$N \to \infty$ and $g \to 0$ in such a way that the 't Hooft coupling constant $\lambda$ is kept fixed \cite{tHooft:1973alw}. This limit is known as the planar limit, since the dominant diagrams in this limit are planar in the double-line notation. It is in this limit that the aforementioned integrable structures in $\Nfour$ supersymmetric Yang--Mills theory appear. 

\section{The AdS/CFT Correspondence}
\label{sec:AdS/CFT}

Another important aspect of $\mathcal{N}\!=4$ supersymmetric Yang--Mills theory is the conjectured duality to type IIB superstring theory on $\AdS_5 \times \mathrm{S}^5$, which was proposed in references \cite{Maldacena:1997re} and further elaborated in references 
\cite{Gubser:1998bc,Witten:1998qj}.
This duality provided the first concrete realization of 't Hooft's idea of a string/gauge duality
\cite{tHooft:1973alw}, which was based on the finding that the $1/N$-expansion in a large-$N$ field theory can be viewed as a genus expansion for the discrete surfaces arising from the theory's Feynman diagrams. 
Below, we give a brief overview over the conjectured duality. For a detailed introduction, the reader is referred to the reviews \cite{Aharony:1999ti,DHoker:2002nbb,Plefka:2005bk} or the textbook \cite{Nastase:2015wjb}.

\subsection{Anti-de Sitter Space}

The duality between $\mathcal{N}\!=4$ supersymmetric Yang--Mills theory and superstring theory on 
$\AdS_5 \times S^5$ is called holographic, since the field theory is considered to live in the (conformal) boundary of Anti-de Sitter space. We discuss this relation explicitly below.   

Anti-de Sitter space $\AdS_{d+1}$ can be introduced as the hyperquadric in $\mathbb{R}^{(2,d)}$  defined by
\begin{align}
- Z_{-1} ^2 -  Z_0 ^2 
	+  Z_1  ^2 + \ldots +  Z_d  ^2 = - R^2 \, , 
\label{AdS:def}	
\end{align}
where $R$ is the radius of $\AdS_{d+1}$, which we will set to $R=1$ in our below discussion of different coordinates systems of $\AdS_{d+1}$ and the conformal compactification. A frequently-used coordinate system is formed by the global coordinates, which are introduced by
\begin{align}
Z^{-1} &= \cosh \rho \, \cos \tau \, , &
Z^0 &= \cosh \rho \, \sin \tau \, , &
Z^j &= \sinh \rho \, v^j \, , 
\end{align}
where $v^j$ are embedding coordinates for a $(d-1)$-dimensional sphere $\mathrm{S}^{d-1}$, i.e.\ they satisfy the constraint $\sum _{j=1} ^d v_j ^2 = 1$. In these coordinates, the AdS-metric is given by 
\begin{align}
\diff s^2 _{\AdS} &= \diff Z^\mu \diff Z_\mu - \diff Z^{-1} \diff Z^{-1} + 
	\diff Z^d \diff Z^d \nn \\
	&= - \cosh^2 \rho \, \diff \tau ^2 
	+ \diff \rho ^2 + \sinh ^2 \rho \, \diff \Omega_{d-1} ^2 \, .
\end{align}
Alternatively, one often uses Poincar{\'e} coordinates $(X^\mu ,y)$, which are given by
\begin{align}
Z^\mu &= \frac{X^\mu}{y} \, , &
Z^{-1} + Z^d  &= \frac{1}{y} \, , &
Z^{-1} - Z^d  &= \frac{X^\mu X_\mu + y^2}{y} \, ,
\end{align}
where $\mu$ takes values in $\lbrace 0 , 1 , \ldots , d-1 \rbrace$. It is easily checked that this parametrization satisfies the embedding relation \eqref{AdS:def} and for the induced metric, we have
\begin{align}
\diff s _\AdS ^2 = 	\frac{\diff X^\mu \diff X_\mu + \diff y^2 }{y^2} \, . 
\end{align}
Note that the Poincar{\'e} patch given by $y>0$ covers only one half of $\AdS_N$, which is described by $Z^{-1} + Z^d > 0$. 

For the construction of the conformal compactification, we initially work with global coordinates and map $\AdS_{d+1}$ into the Einstein static universe $\mathrm{ESU}_{d+1}$, where we again introduce spherical coordinates for the space-like part by setting
\begin{align}
z^{d+1} &= \cos \vartheta \, , &
z^j &= \sin \vartheta \, u^j \, .
\end{align}
The map from $\AdS_{d+1}$ into the Einstein static universe $\mathrm{ESU}_{d+1}$ is then described by 
\begin{align}
\tau &= z^0 \, , &
\sinh \rho &= \tan \vartheta \, ,  & 
v^j &= u^j \, ,
\end{align}
and one may show that it is conformal by direct calculation. We see that the conformal boundary is assumed at the equator $\vartheta = \pi/2$ of the sphere 
$\mathrm{S}^d$, which corresponds to the Einstein static universe 
$\mathrm{ESU}_d$ of one dimension less. In our above discussion of the conformal compactification, we have seen that $\mathrm{ESU}_d$ gives a double cover of $d$-dimensional Minkowski space. More precisely, we have mapped Minkowski space to the region $\cos z^0 + z^d > 0$, which covers one half of this space. 

The relation between $\AdS_{d+1}$ and its conformal boundary is clearest when we use Poincar{\'e} coordinates. These cover the region $Z^{-1} + Z^d > 0$, which corresponds to the region
\begin{align}
\cos z^0 + \frac{\tan \vartheta}
	{\sin \vartheta \, \sqrt{1 + \tan ^2 \vartheta }} \, z^d > 0    
\end{align}  
in the Einstein static universe $\mathrm{ESU}_{d+1}$. In the boundary limit 
$2 \vartheta \to \pi ^{-}$, we thus approach the region $\cos z^0 + z^d > 0$, in which we have mapped Minkowski space. The boundary limit $y \to 0$ in Poincar{\'e} coordinates thus approaches the Minkowski space located at $y=0$ as the form of the metric suggests.  

The Wick rotation to Euclidean $\AdS_{d+1}$ is subtle but again one reaches the conclusion that in Poincar{\'e} coordinates one approaches a flat Euclidean space $\mathbb{R}^d$ in the boundary limit. A discussion of the Euclidean case can be found e.g.\ in reference \cite{Witten:1998qj}.

\subsection{The Correspondence}

The correspondence between $\Nfour$ SYM theory and type IIB superstring theory on $\AdS_5 \times 
\mathrm{S}^5$ was proposed based on the consideration of $N$ parallel D3 branes, which extend along a $(3+1)$-dimensional plane in $(9+1)$-dimensional Minkowski space, and the study of the low-energy limit of this system. The branes are separated by a distance $r$ and the low-energy limit can be considered as the limit of taking $(r , \alpha ^\prime) \to 0$ in such a way that the ratio between them is kept fixed \cite{Maldacena:1997re}; this procedure is known as the Maldacena limit.

The system can be viewed in two different ways. In one approach, the low-energy limit leads to two decoupled systems, free type IIB supergravity in the ten-dimensional bulk space and $\Nfour$ supersymmetric Yang--Mills theory on the (3+1)-dimensional brane. 
The second approach is based on an insight of Polchinski \cite{Polchinski:1995mt} and again leads to two decoupled systems. The first is again given by free type IIB supergravity in the the ten-dimensional bulk space, the second is the geometry near the horizon of the D3 branes, which turns out to be given by $\AdS_5 \times \mathrm{S}^5$.

The conjecture then arises from identifying the second systems appearing in the two descriptions, concretely the conjecture states the duality:  

\begin{center}
$\Nfour$ supersymmetric $\grp{SU}(N)$ Yang--Mills theory \\ 
in four-dimensional Minkowski space \\[.2cm]
$\longleftrightarrow$ \\[.2cm]
type IIB superstring theory on $\AdS_5 \times \mathrm{S}^5$ with equal\\
radii $R$ and $N$ corresponding to the flux of the \\ 
five-form Ramond--Ramond field strength on 
$\mathrm{S}^5$.
\end{center}

An immediate consistency check of this identification is to note that both systems have an underlying $\grp{PSU}(2,2 \vert 4)$-symmetry. We have seen this above for $\Nfour$ supersymmetric Yang--Mills theory; for the string theory on $\AdS_5 \times \mathrm{S}^5$ it arises from the inclusion of supersymmetry in the isometry algebra 
$\alg{so}(2,4) \oplus \alg{so}(6) \simeq \alg{su}(2,2) \oplus \alg{su}(4)$ of $\AdS_5 \times \mathrm{S}^5$.  

The parameters of the two theories are related by
\begin{align}
g^2 &= 4 \pi \, g_s ^2 \, , &
2 \pi T = \sqrt{\lambda} = \frac{R^2}{\alpha ^\prime} \, , 
\end{align}
where $g_s$ denotes the string coupling constant and $T$ the string tension. We note that only the combination $R^2 / \alpha ^\prime$ appears as a parameter of the string theory. In supergravity calculations it can be convenient to set the radii of $\AdS_5$ and $\mathrm{S}^5$ to one and we will also do this below. 

In its strongest form, the conjecture is valid for the full parameter region of both theories, but also the restriction to limiting cases is interesting to study. Considering the planar limit implies $g_s \to 0$, such that the splitting and joining of strings is suppressed and we are hence considering free string theory. In this case, a small gauge theory coupling constant $\lambda$ corresponds to a strongly curved background in string units and a weakly curved background corresponds to a strongly coupled gauge theory. 
This makes the conjecture both hard to test and powerful, since it allows insights into strongly coupled gauge theory. 

The precise relations between the different objects of the string and gauge theory form the so-called AdS/CFT dictionary; for these relations the reader is referred to the above-mentioned reviews. Below, we only discuss the dual description of the Maldacena--Wilson loop, which we introduce in the next section.  

\section{The Maldacena--Wilson loop}
\label{sec:MWL}

We are now in a position to introduce the Maldacena--Wilson loop operator in $\Nfour$ supersymmetric Yang--Mills theory. We begin by introducing Wilson loops in Yang-Mills theories and obtain the Maldacena--Wilson loop from the dimensional reduction of light-like Wilson loops in the ten-dimensional $\mathcal{N}\! = 1 $ supersymmetric Yang--Mills theory. As an example for a test of the AdS/CFT correspondence, we discuss the Maldacena--Wilson loop over the circle, which has been exactly calculated on the gauge theory side. 

\subsection{Wilson loops in gauge theories}

The Wilson loop was first introduced by Kenneth Wilson in reference \cite{Wilson:1974sk} in the study of quark confinement using gauge theory on a lattice. Here, we introduce the Wilson loop from general considerations of gauge invariance following reference \cite{Dorn:1986dt}. Consider for example a quark field $\psi$ in the fundamental representation evaluated on two points 
$x,y \in \mathbb{R}^{(1,3)}$. We cannot compare the fields directly, since the difference does not transform appropriately under gauge transformations
\begin{align}
\psi(x) \; \; &\mapsto \; \; U(x) \psi(x) \, , &
\psi(y) \; \; &\mapsto \; \; U(y) \psi(y) \, .
\end{align}
This problem is similar to the comparison of tangent vectors at different points on a manifold and requires to introduce a parallel transport or Wilson line in the context of Yang--Mills theories. We consider a curve $\gamma$, denoted also by $x(\sigma)$, from $y$ to $x$ and introduce the Wilson line $V_\gamma$ from the requirement that it be covariantly constant along $\gamma$, i.e.\ it satisfies 
$\dx ^\mu D_\mu V_\gamma = 0$ or more explicitly
\begin{align}
\partial_\sigma \, V_\gamma(x(\sigma), y) 
	= i \dx^\mu (\sigma) A_\mu (x(\sigma)) \, V_\gamma(x(\sigma), y) \, .
\label{def:WilsonLine}
\end{align}
Along with the initial condition $V_\gamma(y, y)= \unit$, the above differential equation completely determines $V$ due to the uniqueness theorem for ordinary differential equations. Under a gauge transformation 
$A_\mu \mapsto A_\mu ^\prime = U \left( A_\mu + i \partial_\mu \right) U^{-1}$, 
the Wilson line transforms as
\begin{align}
V_\gamma (x,y) \; \; &\mapsto \; \; V_\gamma ^\prime (x,y) = U(x) V_\gamma(x,y) U^{-1}(y) \, .
\end{align}
In order to prove this behavior, we only need to show that $V_\gamma ^\prime$ satisfies the gauge transformed version of the definition \eqref{def:WilsonLine}, which follows from a short calculation:
\begin{align*}
\partial_\sigma V_\gamma ^\prime (x(\sigma),y) &= 
	U(x(\sigma)) \left[ i \dx^\mu(\sigma) A_\mu (x(\sigma)) 
	- \dx^\mu(\sigma) \partial_\mu \right] U^{-1} (x(\sigma)) \, 
	V_\gamma ^\prime(x(\sigma),y) \\
	&= i \dx^\mu(\sigma) A^\prime _\mu (x(\sigma)) \, 
	V_\gamma ^\prime(x(\sigma),y) \, .
\end{align*}
Noticing that $V_\gamma ^\prime(y,y) = \unit$ concludes the proof. With the gauge transformation of the Wilson line established, we note that we can now compare the fields 
$\psi(x)$ and $V_\gamma (x,y) \psi(y)$, since they transform in the same way under gauge transformations. Moreover, if we have e.g.\ scalar fields $\Phi$ in the adjoint representation as in $\mathcal{N}\!=4$ supersymmetric Yang--Mills theory, we can construct non-local gauge invariant operators such as
\begin{align*}
\tr \left( \Phi (x) V_\gamma (x,y) \Phi(y) V_{\gamma} ^{-1} (y,x) \right) . 
\end{align*}
We can also construct a non-local gauge invariant operator from the Wilson line itself. For a closed curve $\gamma$, we note that the Wilson line transforms as
\begin{align}
V_\gamma (x,x) \; \; \mapsto \; \; U(x) V_\gamma (x,x) U(x)^{-1} \, ,  
\label{W_gauge}
\end{align}
and we define the gauge-invariant Wilson loop as
\begin{align}
\mathrm{W}(\gamma) = \frac{1}{N} \, \tr \left(  V_\gamma (x,x) \right) \, .
\end{align}
The normalization factor $N^{-1}$ ensures that in a $\grp{U}(N)$ or $\grp{SU}(N)$ gauge theory, the trivial loop over a constant curve gives $\mathrm{W}(\gamma)=1$. We note that the gauge transformation behavior \eqref{W_gauge} allows to construct other gauge invariants than the trace, since all eigenvalues of $V_\gamma (x,x)$ are invariant. Considering other combinations of eigenvalues amounts to studying the Wilson loop in different representations of the gauge group. Here, we focus on the Wilson loop in the fundamental representation, where the gauge field, while it transforms in the adjoint representation, takes values in the fundamental representation. 

A better-known expression for the Wilson loop is obtained by rewriting the defining equation \eqref{def:WilsonLine} as an integral equation, 
\begin{align}
V_\gamma (x(\sigma),y) = \unit + i \int \limits _0 ^\sigma \diff \sigma_1 \, 
	\dx_1 ^\mu A_\mu (x_1) V_\gamma (x_1,y) \, ,
\end{align}   
where we have abbreviated $x(\sigma_1) = x_1$. Plugging this recursion into itself, we obtain the formal solution
\begin{align}
V_\gamma (x(\sigma),y) = \plexp \left( i \int _0 ^\sigma \diff \sigma_1 \, 
	\dx_1 ^\mu A_\mu (x_1) \right) , 
\end{align}
where the arrow indicates that in the expansion of the path-ordered exponential, greater values of $\sigma$ are ordered to the left. For the Wilson loop we thus have the expression
\begin{align}
\mathrm{W}(\gamma) = \frac{1}{N} \, \tr 
	\plexp \left( i \int _\gamma \diff \sigma \, \dx^\mu A_\mu (x) \right) ,
\end{align}
which also allows us to carry out perturbative calculations. 

Physically, we may interpret the Wilson loop to describe the insertion of heavy external quarks into the theory. In order to motivate this interpretation, we consider the path integral in a  
$\grp{U}(1)$ gauge theory with a source term specified by
\begin{align}
J^\mu(y) = \int \diff \sigma \, \dx^\mu (\sigma) \, \delta ^{(4)} (y - x(\sigma)) \, . 
\end{align}
The partition function including the source term is then given by
\begin{align}
Z[J] = \int \left[ \diff A \right] \, \exp \left( i S 
	+ i \int \diff ^4 x \, J^\mu A_\mu \right) ,
\end{align}
and we find that
\begin{align}
\left \langle \mathrm{W}(\gamma) \right \rangle = \frac{Z[J]}{Z[0]} \, .
\end{align}
Here, $\gamma$ denotes the contour appearing in the prescription for the current $J^\mu$. A particularly interesting result is obtained for a rectangular contour $\gamma_{R,T}$ with side length $T$ in the time direction and $R \ll T$ in some spatial direction, for which one finds
\begin{align}
\left \langle \mathrm{W}(\gamma_{R,T} ) \right \rangle \simeq e^{- T \, V(R)} \, , 
\end{align}
where $T$ is considered to be asymptotically large. The above conclusion is based on the fact that the path integral for large Euclidean times is dominated by the ground state energy and that for $R \ll T$ we may neglect the spatial parts of the loop, such that we are considering a quark-antiquark pair separated by a distance $R$. We note that, while we have only motivated the above result for a $\grp{U}(1)$ gauge theory, it also holds in non-Abelian Yang--Mills theory, cf.\ e.g.\ reference \cite{Smit:2002ug} for more details. The calculation of the expectation value of the Wilson loop is thus crucial in the study of confinement, which is the problem Wilson originally addressed in reference \cite{Wilson:1974sk}. We note that in a conformal field theory scale invariance requires that the expectation value is of the form $e^{-T/R}$, such that we obtain the Coulomb potential. 

The perturbative calculation of the expectation value of the Wilson loop leads to divergences which require renormalization. These divergences were noted first in reference \cite{Polyakov:1979ca}, the renormalization properties were established in references \cite{Dotsenko:1980wb,Brandt:1981kf}. We discuss the divergences for the one-loop approximation of the expectation value, which is given by
\begin{align}
\left \langle \mathrm{W}(\gamma) \right \rangle = 
	1 - \frac{g^2 (N^2-1)}{16 \pi^2 N} \int \limits _0 ^L \diff \sigma_1 \, \diff \sigma_2 \, 
	\frac{\dx_1 \dx_2 }{(x_1-x_2)^2} + \O ( g^4) \, ,
\end{align}
Here we have taken the gauge group to be given by $\grp{SU}(N)$ and have plugged in the gauge field propagator in Feynman gauge, 
\begin{align}
\left \langle A_\mu ^a (x_1) A^\nu _b (x_2) \right \rangle 
	= \frac{g^2}{4 \pi^2} \, \frac{\eta_{\mu \nu} \, \delta ^{ab}}{(x_1-x_2)^2} 
\end{align}
In the following, we will demand that the parametrization satisfy $\dx^2 = 1$ and we have indicated the use of such a parametrization above by using the integration boundaries 0 and $L$. We employ a cut-off regularization to discuss the divergence of the one-loop contribution and find 
\begin{align}
\int \limits _0 ^L \diff s \int \limits _{-s} ^{L-s} \diff t \, 
	\frac{\dx(s) \dx(s+t)}{[x(s+t)-x(s)]^2 + a^2 } 
	& = \int \limits _0 ^L \diff s \int \limits _{-s} ^{L-s} \diff t \, 
	\frac{1}{t^2 + a^2} \, + ( \text{finite}) \nn \\
	&= \frac{\pi L}{a} + ( \text{finite}) \, .
\label{Linear_Div}	
\end{align}
This linear divergence appears in all orders of perturbation theory and 
$G(\gamma) = \left \langle \mathrm{W}(\gamma) \right \rangle$ 
can be renormalized as
\begin{align}
G (\gamma)
	= e^{-cL/a} \, G_\mathrm{ren}(\gamma)  \, ,
\end{align}
which we may interpret as a mass renormalization of the test particle prescribed by the Wilson loop. It was further noted in reference \cite{Polyakov:1979ca} that the Wilson loop has additional divergences for cusped contours. In order to discuss these divergences, we switch to dimensional regularization where the above linear divergence is absent. The cusp divergence depends only on the intersection angle at the cusp and we can hence compute the divergence for the intersection of two straight lines intersecting at a hyperbolic angle $\rho$. The relevant integral for the one-loop contribution is given by%
\footnote{In dimensional regularization, the momentum space propagators are unaltered, but the Fourier transformation is carried out in $D=4 - 2 \epsilon$ dimensions. This leads to the alteration of the two-point function noted here, cf.\ e.g.\ reference \cite{Erickson:2000af}.}
\begin{align*}
\int \limits _0 ^L 
	\frac{\diff \sigma_1 \, \diff \sigma_2 \, \cosh(\rho)}{\left[\sigma_1 ^2 + \sigma_2 ^2 
	+ 2 \sigma_1 \sigma_2 \cosh(\rho) \right] ^{1 - \epsilon} }
	&= \int \limits _0 ^L \frac{\diff \ell}{\ell ^{1 - 2 \epsilon}}
	\int \limits _0 ^1  \frac{\diff z \, \cosh(\rho)}{z^2 + \bar{z}^2 
	+ 2 z \bar{z} \cosh(\rho)} + \O (\epsilon^0) \\
	&= \frac{\rho \coth(\rho)}{2 \epsilon} + \O (\epsilon^0) \, .
\end{align*}
Here, we have used the substitution $\sigma_1 = \ell z$, 
$\sigma_2 = \ell \bar{z} = \ell (1-z)$ in order to capture the divergence in the scale integral over $\ell$. The cusp divergence is renormalized multiplicatively through a $\rho$-dependent $Z$-factor
\begin{align}
G_R (\gamma) = Z(\rho) \, G(\gamma) \, ,
\end{align}
where we have omitted the dependence on the regulator which the quantities appearing on the right-hand side have. The renormalization prescription for the Wilson loop was completed in reference 
\cite{Brandt:1981kf}, where they additionally discussed the case of self-intersecting Wilson loops. These are still renormalized multiplicatively, but now the $Z$-factor mixes between correlators of Wilson loops taken over the same contour with different orderings around the intersection point.

The anomalous dimensions associated to the cusp and cross divergences are known as the cusp and cross or soft anomalous dimension and are of phenomenological relevance in the description of infrared divergences of scattering amplitudes. Intuitively, we can understand the connection as follows: If an outgoing quark emits a soft gluon of zero momentum, it will not recoil and follow the straight-line trajectory that also describes the associated Wilson line which accounts for the acquired phase factor. The connection between soft singularities of scattering amplitudes and Wilson loops was established soon after the study of Wilson loops was initiated \cite{Frenkel:1984pz,Korchemsky:1993uz} and is still widely applied today, see e.g.\ reference \cite{Magnea:2014vha} for a recent review or \cite{White:2015wha} for a pedagogical introduction. We note that the cusp anomalous dimension has been calculated to two loops 
\cite{Korchemsky:1987wg} and more recently to three loops \cite{Grozin:2015kna} in QCD. In $\mathcal{N}\!=4$ supersymmetric Yang--Mils theory, it is known up to four loops \cite{Makeenko:2006ds,Correa:2012nk,Henn:2013wfa}. 

\subsection{The Maldacena--Wilson loop}
Let us now turn to $\mathcal{N}\!=4$ supersymmetric Yang--Mills theory, where one considers a generalization of the Wilson loop known as the Maldacena--Wilson loop. Maldacena's original motivation was based on considering $(N+1)$ five-branes and separating one of them from the others and thus studying the Higgs mechanism for the symmetry breaking 
$\grp{U}(N+1) \to \grp{U}(N) \times \grp{U}(1)$. For this reasoning, the reader is referred to the original papers \cite{Maldacena:1998im,Rey:1998ik} or the textbook \cite{Nastase:2015wjb}. 
Here, we motivate the Maldacena--Wilson loop by referring to the dimensional reduction and considerations of supersymmetry. 

We begin by considering the Wilson loop in ten-dimensional $\mathcal{N}\!=1$ supersymmetric Yang--Mills theory which is given by
\begin{align}
\mathrm{W}(\gamma) = \frac{1}{N} \, \tr \plexp 
	\left( i \int _\gamma A_m \, \diff x^m \right) .
\end{align}
We note now that if the ten-dimensional curve $\gamma$ is light-like, the linear divergence of the Wilson loop discussed in equation \eqref{Linear_Div} is absent since the length of the curve vanishes. Choosing a light-like contour also has implications for supersymmetry. Consider the supersymmetry variation of the Wilson loop, which we find using the field variation \eqref{10dSUSY} to be given by
\begin{align}
\delta_\xi \mathrm{W}(\gamma) =
	- \frac{1}{N} \, \tr \overleftarrow{\mathrm{P}}  \left[
	\int \diff \sigma  \left( \widebar{\Psi} \, \dx^m \Gamma_m \, \xi \right)
	\exp \left( i \int A_m \, \diff x^m \right) \right] \, . 
\end{align} 
The light-likeness of $\dx^m$ implies that the matrix coupling the supersymmetry parameter $\xi$ to the fermionic fields squares to zero,
\begin{align}
\left( \dx^m \Gamma_m \right) ^2 = \half \dx^m \dx^n \, 
	\left \lbrace \Gamma_m , \Gamma_n \right \rbrace = 0 \, ,
\end{align}
such that its rank is at most half of its dimension. Hence, if we allow the supersymmetry variation to be local for a moment, we find at least eight linearly independent supersymmetry parameters $\xi (\sigma)$ for which the supersymmetry variation vanishes. The property of being locally supersymmetric carries over to the counterpart of the light-like Wilson loop in the dimensionally reduced theory, which is the Maldacena--Wilson loop 
\begin{align}
W(\gamma) = \frac{1}{N} \, \tr \plexp 
	\left( i \int _\gamma \left( A_\mu \, \diff x^\mu + i \, \Phi_I \lvert\dx \rvert n^I 
	\right) \right) .
\end{align}
Here, $n^I$ is a six-dimensional unit vector, which can in general depend on the curve parameter 
$\sigma$. This ensures that the constraint of light-like tangent vectors in ten dimensions is satisfied, 
\begin{align}
\dx ^m (\sigma) = \left( \dx^\mu(\sigma) , i \, n^I (\sigma) \lvert \dx (\sigma) \rvert \right) 
\quad \Rightarrow \quad
\dx ^m  \dx _m 
	= \dx^2 - \lvert \dx \rvert ^2 = 0 \, . 
\end{align}
Note that here we have defined
\begin{align}
\lvert \dx \rvert = \begin{cases}
		\sqrt{\dx^2} \quad &\text{if} \; \; \dx^2 \geq 0 \, , \\
		i \, \sqrt{\lvert \dx^2 \rvert} \quad &\text{if} \; \; \dx^2 < 0 \, , 
	\end{cases}
\end{align}
such that the Maldacena--Wilson loop is only a phase if $\dx^\mu$ is time-like. For a space-like tangent vector, we cannot obtain a light-like vector in ten dimensions by adding space directions and we are thus forced to consider the additional components imaginary.    

As pointed out above, the Maldacena--Wilson loop inherits the local supersymmetry property%
\footnote{We note that if $\dx^m$ has imaginary components, 
it is not possible to find solutions to $\dx^m \Gamma_m \, \xi = 0$, which satisfy the Majorana condition for spinors in ten dimensions. For more details, the reader is referred to reference \cite{Master}. }
of the ten-dimensional Wilson loop. Of course, the action is not invariant under local supersymmetry variations and supersymmetry only has consequences for the expectation value of the Maldacena--Wilson loop, if we are able to find constant supersymmetry parameters for which the supersymmetry variation vanishes. The simplest case in which this is possible, is the straight line for which the Maldacena--Wilson loop is a 1/2 BPS object. This implies that its expectation value is finite and does not receive quantum corrections, 
\begin{align}
\left \langle W ( 
	\raisebox{1mm}{\includegraphics[width=2.2ex]{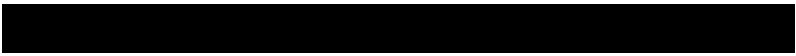}} 
	)  \right \rangle = 1 \, . 
\end{align}
We can apply this result to convince ourselves that the linear divergences of Wilson loops are indeed absent for smooth Maldacena--Wilson loops. These divergences arise from the limit where all integration points along the Wilson line are close to each other. In this limit, the curvature of the considered curve is not relevant and the finiteness hence carries over to generic smooth curves. 

It was noted in reference \cite{Zarembo:2002an} that also for other curves than the straight line, some of the supersymmetry can be preserved globally, leading to 1/4, 1/8 or 1/16 BPS operators, depending on the amount of supersymmetry which can be preserved. The construction involves a coupling between the $\mathrm{S}^5$-vectors $n^I(\sigma)$ and the contour $x^\mu (\sigma)$. Further classes of contours were found in reference \cite{Drukker:2007dw} by considering also superconformal variations of the Maldacena--Wilson loop. A classification of loops, for which at least one supersymmetry can be preserved, was obtained in \cite{Dymarsky:2009si,Cardinali:2012sy}. 

For the expectation value of the Maldacena--Wilson loop, we find 
\begin{align}
\left \langle W(\gamma \right \rangle =
	1 - \frac{\lambda \left( 1 - N^{-2} \right)}{16 \pi ^2} \, 
	\int \diff \sigma_1 \, \diff \sigma_2 \, 
	\frac{\dx_1 \dx_2 - n_1 n_2 \, \lvert \dx_1 \rvert \lvert \dx_2 \rvert }
	{(x_1 - x_2)^2} + \O (\lambda ^2) \, , 
\label{W1loop}	
\end{align}
Here, we have used the result 
\begin{align}
\left \langle \Phi _I ^a (x_1) \, \Phi _J ^b (x_2) \right \rangle 
	= \frac{g^2}{4 \pi^2} \,  
	\frac{\delta_{IJ} \, \delta ^{ab}}{(x_1-x_2)^2} \, ,
\end{align}
for the scalar propagator in order to find the one-loop contribution to the above expectation value. We can observe the finiteness of the expectation value for the one-loop contribution by considering the integrand in the limit $\sigma_2 \to \sigma_1$. Since $n^2 = 1 $ implies that 
$n \cdot \dot{n} = 0$, we find that the denominator is of order $(\sigma_2 - \sigma_1 )^2$, such that the integrand is indeed finite in this limit.  

\subsection{The Strong-Coupling Description}
The strong-coupling description of the Maldacena--Wilson loop was obtained in reference \cite{Maldacena:1998im}. On the string theory side of the correspondence, the Maldacena--Wilson loop is described by the string partition function, with the string configuration bounded by the Wilson loop contour on the conformal boundary of $\AdS_5$. In the limit of large $\lambda$, the partition function is dominated by the classical action and the AdS/CFT prescription for the Maldacena--Wilson loop at strong coupling is given by
\begin{align}
\left \langle W(\gamma) \right \rangle \overset{\lambda \gg 1}{=}
	\exp \left( - \ft{\sqrt{\la}}{2 \pi} A_\mathrm{ren}(\gamma) \right) \, .
\label{W:Strong}	
\end{align}
Here, $A_\mathrm{ren}(\gamma)$ denotes the area of the minimal surface ending on the contour 
$\gamma$, which is situated at the conformal boundary. The boundary value problem is simplest to describe in Poincar{\'e} coordinates $(X^\mu , y)$, for which we recall the metric
\begin{align}
\diff s^2 = \frac{\diff X^\mu \, \diff X_\mu + \diff y \, \diff y}{y^2} \, ,
\end{align}
and note that the conformal boundary corresponds to $y=0$. For suitably chosen coordinates $\tau$ and $\sigma$, we thus impose the boundary conditions
\begin{align}
X^\mu ( \tau = 0 , \sigma ) &= x^\mu (\sigma) \, , &
y ( \tau = 0 , \sigma ) &= 0 \, .
\end{align}
The area of the minimal surface is obtained from the area functional or Nambu--Goto action
\begin{align}
A = \int \diff \tau \, \diff \sigma \, \sqrt{\det \left( \Gamma_{ij} \right) } \, ,
\end{align}
where $\Gamma_{ij} = y^{-2} \left( \partial_i X^\mu \, \partial_j X_\mu 
	+ \partial_i y \, \partial_j y \right)$ is the induced metric on the surface. The minimal surface described by the above boundary conditions is divergent due to the divergence of the AdS-metric. We regulate the area of the minimal surface by introducing a cut-off $\varepsilon$ for the $y$-direction, i.e.\ we integrate only over the region $y \geq \varepsilon$. Since the minimal surface leaves the conformal boundary perpendicularly due to the divergence of the AdS-metric, we can identify the divergence to be given by $L(\gamma)/\varepsilon$ and we note that the AdS/CFT prescription \eqref{W:Strong} contains the renormalized minimal area
\begin{align}
A_\mathrm{ren}(\gamma) = \lim \limits _{\varepsilon \to 0} \left \lbrace A(\gamma) \big \vert_{y \geq \varepsilon} - \frac{L(\gamma)}{\varepsilon} \right \rbrace \, .  
\label{def:Aren_0}
\end{align}
Let us point out again that the Maldacena--Wilson loop over a smooth contour is finite and does not require renormalization. The above renormalization of the area of the minimal surface stems from considering the Legendre transformation with respect to the loop variables coupling to the scalar fields, see reference \cite{Drukker:1999zq} for more details. 

The above description is restricted to the case of constant $n^I$. For the general case of $n^I (\sigma)$ describing a closed curve on $\mathrm{S}^5$, the strong-coupling description contains a minimal surface in $\AdS_5 \times \mathrm{S}^5$, which is bounded by $x^\mu(\s)$ in the conformal boundary of $\AdS_5$ and $n^I(\sigma)$ in $\mathrm{S}^5$. 

\subsection{The Circular Maldacena--Wilson loop}
As an example, we now discuss the Maldacena--Wilson loop over the circle. This contour is of particular interest, since the expectation value of the Maldacena--Wilson loop can be calculated exactly on the gauge theory side, which allows to compare with the AdS/CFT prediction at strong coupling. The circular Maldacena--Wilson loop is a 1/2 BPS operator. The supersymmetry variations discussed above are not sufficient to establish this fact, rather one also needs to include special superconformal variations to find linear combinations which leave the operator invariant \cite{Drukker:2007dw}.     

The minimal surface for the circular Wilson loop in Euclidean space was obtained soon after the AdS/CFT proposal in reference \cite{Berenstein:1998ij}. Assuming that the sections of the minimal surface at constant $y$ are still circular, one may use the following parametrization of the minimal surface:
\begin{align}
X^\mu (r , \sigma ) &= \left( r \cos \sigma , r \sin \sigma \right) \, , &
y &= y(r) \, . 
\label{SurfCircleInitial}
\end{align} 
The area of the surface is then given by
\begin{align}
A = \int \diff r \, \diff \sigma \, \frac{r \sqrt{1+ y^\prime (r) ^2}}{y(r)^2} \, , 
\end{align}
such that we have the equations of motion
\begin{align}
\partial_r \left( \frac{r \, y^\prime (r)}{y(r) ^2 \sqrt{1+ y^\prime (r) ^2} } \right) 
	+ \frac{2r \, \sqrt{1+ y^\prime (r) ^2}}{y(r)^3} = 0 \, .
\label{Circle:EOM}	
\end{align}
Due to the large amount of symmetry and the particular parametrization used above, the equations of motion have reduced to an ordinary differential equation for the function $y(r)$. For the circle of radius 1, we note the boundary condition $y(1)=0$. There is, however, a simpler way to find the minimal surface for the circle than to solve the above equation, which was used in reference \cite{Berenstein:1998ij}. Note that the inversion map on $\mathbb{R}^2$, 
\begin{align*}
I(x)^\mu = \frac{x^\mu}{x^2} \, ,
\end{align*}
maps the circle to a straight line and vice versa. Specifically, we consider the boundary curves
\begin{align*}
x(\sigma) &= \left( \cos \sigma , \sin \sigma + 1 \right) \, , &
I(x(\sigma)) &= \left( \frac{\cos \sigma}{2 ( 1 + \sin \sigma)}  , \frac{1}{2} \right) \, .
\end{align*} 
The inversion map can be extended to the AdS-isometry 
\begin{align}
I_\AdS (X,y) &= \left( \frac{X^\mu}{X^2 + y^2} \, , \, \frac{y}{X^2 + y^2} \right) ,
\end{align}
which can be used to map the (formal) minimal surface attached to the straight line to the one attached to the circle. Writing the minimal surface for the straight-line%
\footnote{It is straight-forward to check that this intuitive solution indeed solves the equations of motion.}
as
\begin{align}
X^\mu (\tau , \sigma) &= \left( \sigma , \ft{1}{2} \right) \, , &
y(\tau , \sigma ) &= \tau \, ,  
\end{align}
we obtain the surface
\begin{align}
X^\mu (\tau , \sigma ) &= \left( \frac{4 \sigma}{1 + 4 ( \sigma ^2 + \tau ^2 )} \, , 
	\frac{1 - 4 (\sigma ^2 + \tau ^2) }{1 + 4 ( \sigma ^2 + \tau ^2 ) } \right)  \, , &
y (\tau , \sigma) = \frac{4 \tau}{1 + 4 ( \sigma ^2 + \tau ^2 )} \, .	
\label{Surf:Circle1}
\end{align}
Here, we have employed a translation by $(0,-1)$ in addition to the inversion such that the circle is centered around the origin. The parametrization obtained above is not well suited for the further study of this minimal surface. We note that the surface described by equation 
\eqref{Surf:Circle1} satisfies the equation
\begin{align}
X^2 + y^2 = 1 \, . 
\end{align}
For the parametrization \eqref{SurfCircleInitial} employed at the beginning of our discussion, we thus find
\begin{align}
y(r) = \sqrt{1 - r^2} \, ,
\end{align}
which indeed solves the equations of motion \eqref{Circle:EOM}. An often-used parametrization is given by
\begin{align}
X_1 (\tau , \sigma) &= \frac{\cos \sigma}{\cosh \tau} \, , &
X_2 (\tau , \sigma) &= \frac{\sin \sigma}{\cosh \tau} \, , & 
y (\tau , \sigma) &= \tanh \tau \, . &  
\label{sol:circle:0} 
\end{align} 
For this parametrization, the induced metric is conformal to the flat metric such that it solves the equations of motion in conformal gauge, if one chooses to work with a Polyakov rather than Nambu--Goto action. 

In order to calculate the area of the minimal surface, we introduce a cut-off at 
$y= \varepsilon$, which corresponds to $r= \sqrt{1 - \varepsilon^2}$. Then, we get
\begin{align}
A_{\mathrm{ren}} \left( 
	\raisebox{-.9mm}{\includegraphics[height=2.2ex]{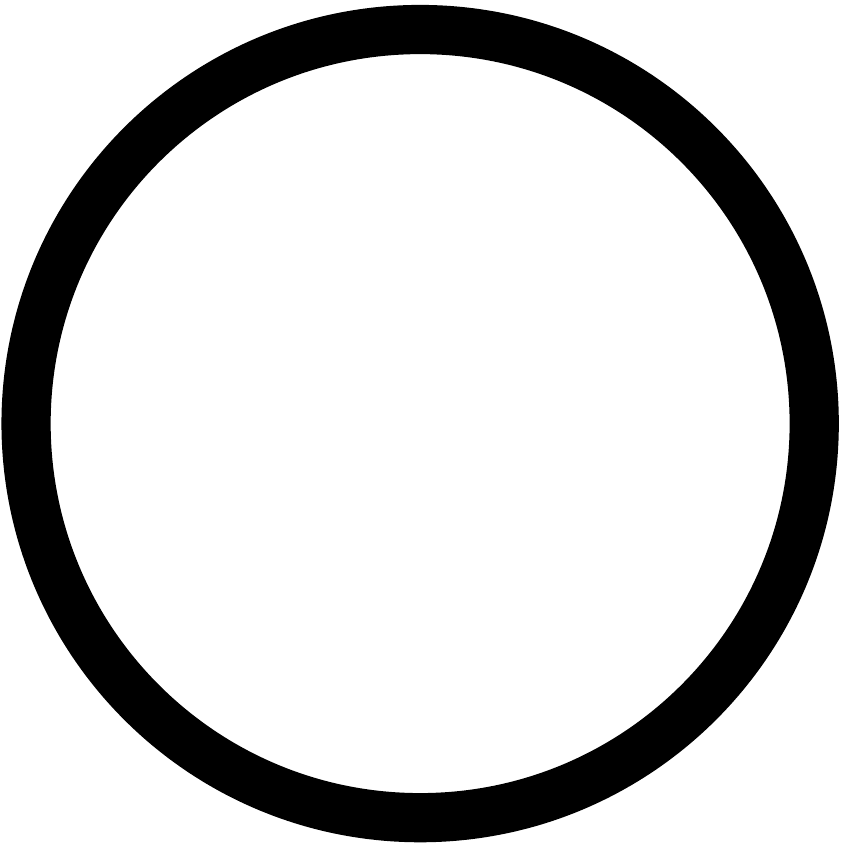}}   
	\right)
	= \lim \limits _{\varepsilon \to 0} 
	\bigg \lbrace \int \limits _0 ^{2 \pi} \diff \sigma 
	\int \limits _0 ^{\sqrt{1-\varepsilon^2}} 
	\frac{r \, \diff r}{\left( 1 - r^2 \right)^{3/2} } \, 
	- \frac{2 \pi}{\varepsilon}  \bigg \rbrace
	= - 2 \pi \, .
\end{align} 
We have thus found that the circular Maldacena--Wilson loop has the following asymptotic behavior at strong coupling:
\begin{align}
\left \langle W  \left( 
	\raisebox{-.9mm}{\includegraphics[height=2.2ex]{circle.pdf}}   
	\right) 
	\right \rangle 
	\overset{\lambda \gg 1}{=}  
	e^{\sqrt{\lambda}} .
\label{eqn:Pred}	
\end{align}
Remarkably, the expectation value of the circular Maldacena--Wilson loop has been calculated exactly on the gauge theory side, beginning with the calculation of reference \cite{Erickson:2000af}, which we sketch below. 

\begin{figure}[t]
\centering
\subcaptionbox{Double-Propagator}
[.3\linewidth]{\includegraphics[width=.25\linewidth]{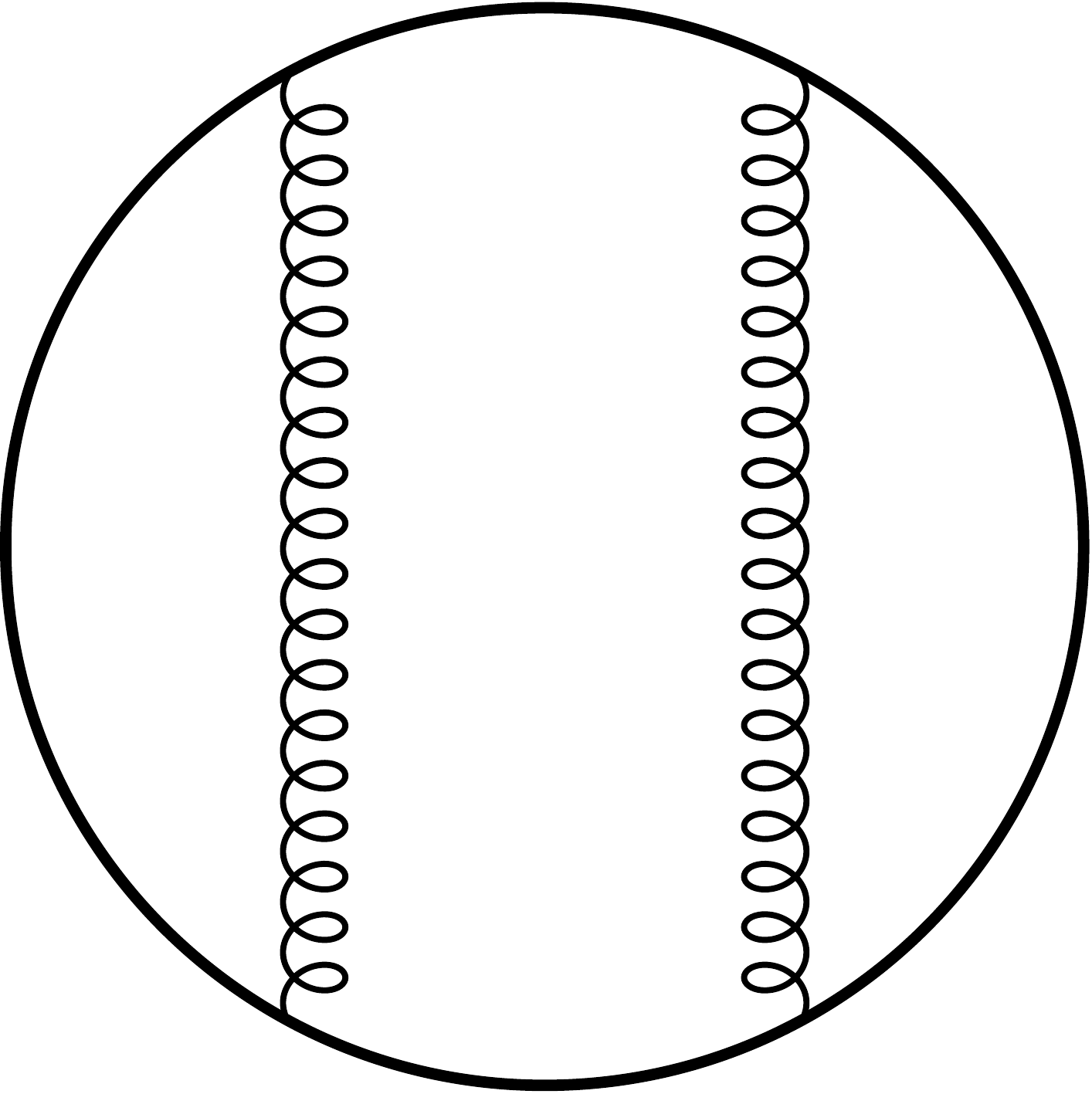}}
\subcaptionbox{Self-Energy} 
[.3\linewidth]{\includegraphics[width=.25\linewidth]{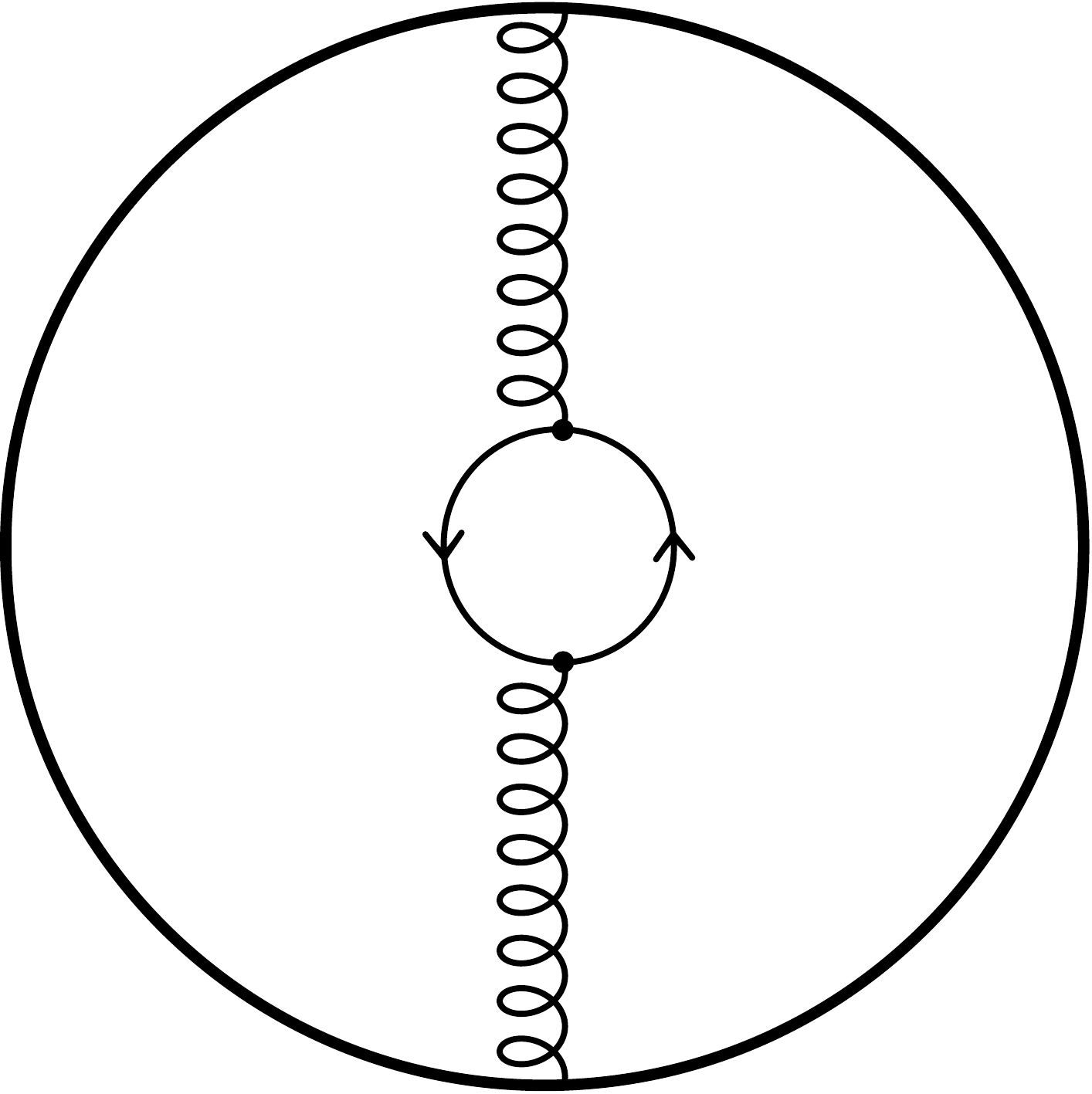}}
\subcaptionbox{Three-Vertex} 
[.3\linewidth]{\includegraphics[width=.25\linewidth]{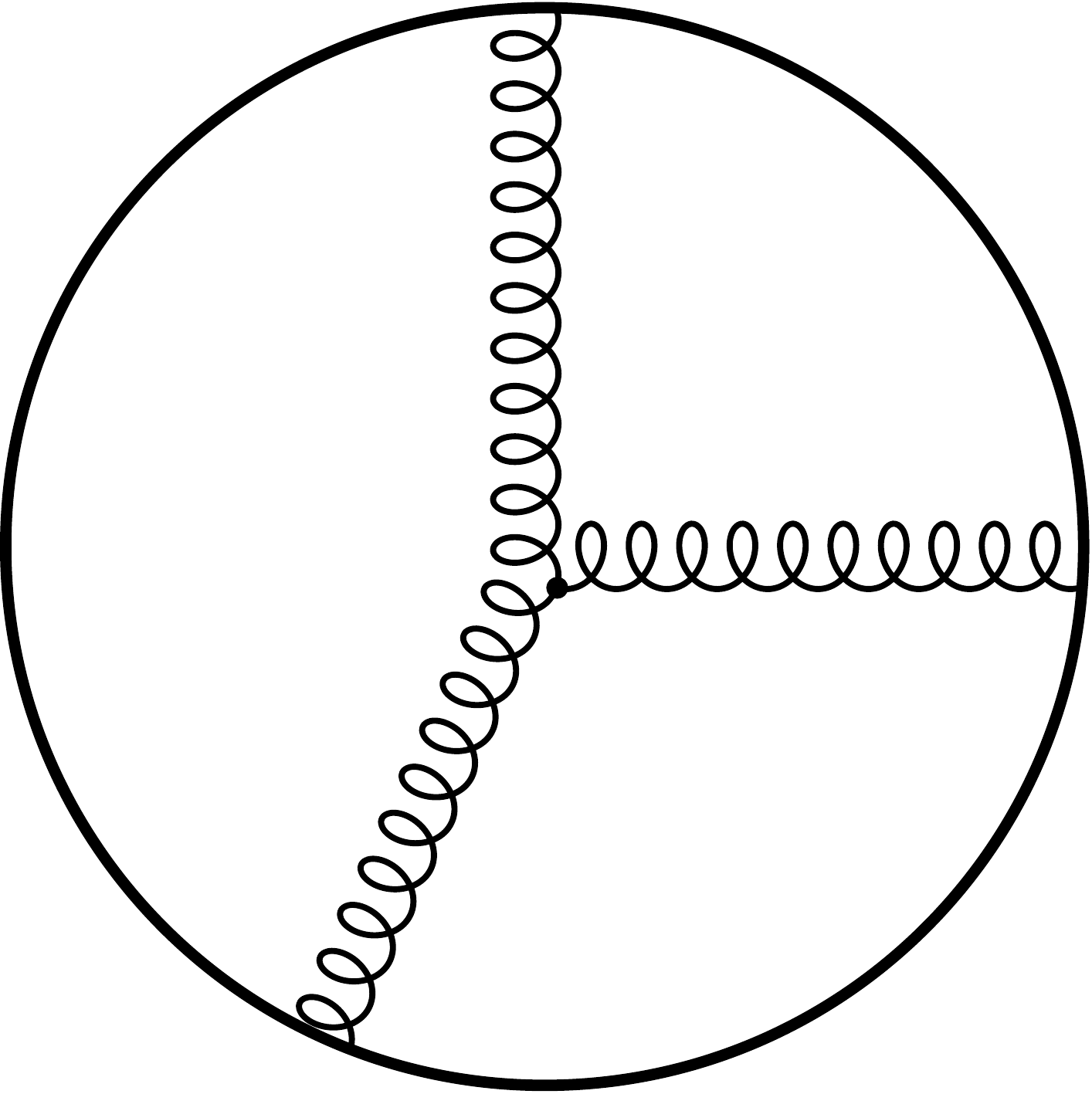}}
\caption{Examples of the double-propagator, self-energy and three-vertex diagrams appearing in the two-loop calculation of the expectation value of the Maldacena--Wilson loop. }
\label{fig:2-loop-diagrams}
\end{figure}

Let us first consider the one-loop order of the expectation value \eqref{W1loop}. For the circle parametrized by 
$x(\sigma) = ( \cos \sigma , \sin \sigma)$ and constant $n^I$, we find
\begin{align*}
\frac{\dx_1 \dx _2 - \vert \dx_1 \vert \vert \dx_2 \vert}{(x_1 - x_2 ) ^2}
	= \frac{\cos \sigma_1 \cos \sigma _2  + \sin \sigma_1 \sin \sigma_2 - 1 }
	{2 - 2 ( \cos \sigma_1 \cos \sigma _2  + \sin \sigma_1 \sin \sigma_2) }
	= - \frac{1}{2} \, .
\end{align*}
This finding renders the integral appearing in the calculation of the one-loop coefficient trivial and we obtain (in the planar limit)
\begin{align}
\left \langle W  \left( 
	\raisebox{-.9mm}{\includegraphics[height=2.2ex]{circle.pdf}}   
	\right) \right \rangle
	= 1 + \frac{\lambda}{8} + \O ( \lambda ^2 ) \, .	
\end{align}
At the two-loop order, the calculation becomes more intricate, since in addition to the diagrams consisting of two gluon or scalar propagators, diagrams with three-vertices and the self-energy correction to the propagators appear, see figure \ref{fig:2-loop-diagrams}. These diagrams are divergent and require regularization. The regularization scheme used in reference \cite{Erickson:2000af} is known as the dimensional reduction scheme. It is a version of dimensional regularization which is particularly well-suited for supersymmetric theories. Here, $\mathcal{N} \! =  4$ supersymmetric Yang--Mills theory is viewed as the theory obtained from dimensionally reducing ten-dimensional $\mathcal{N}=1$ supersymmetric Yang--Mills theory to D dimensions. The regularized theory hence has a D-component vector field $A_\mu ^a$ as well as $10 - \mathrm{D}$ scalar fields $\Phi _I ^a$, see reference \cite{Siegel:1979wq} for more details. The expectation value of the Maldacena--Wilson loop is finite for smooth contours and indeed one observes that the divergences of the self-energy and three-vertex diagrams cancel each other for generic (smooth) contours. The calculation in reference \cite{Erickson:2000af} showed that these diagrams cancel each other exactly for the circular Wilson loop in $\mathrm{D}=4$ dimensions.

The authors of reference \cite{Erickson:2000af} then conjectured that these cancellations occur at all loop orders. This conjecture can be interpreted by again referring to the conformal mapping of the straight line to the circle. For the straight line the gluon and scalar propagators ending on the line cancel each other individually since $\dx_1 \parallel \dx_2$. Moreover, since the expectation value is identically 1, all contributions containing interaction vertices cancel each other at any given loop order. The transformation mapping the straight line to the circle is anomalous, since it maps one point on the circle to infinity. Indeed, the gluon and scalar propagators do no longer cancel each other for the circle as we have seen above. The conjecture then states that the cancellations between the diagrams containing interaction vertices do carry over to the circular Maldacena--Wilson loop. 

Given this conjecture, the expectation value of the circular Maldacena--Wilson loop can be calculated from diagrams containing propagators ending on the circle, such as diagram (a) in figure \ref{fig:2-loop-diagrams}. Moreover, if we consider the planar theory, the propagators do not cross each other. In order to calculate these diagrams, it is sensible to combine the gluon and scalar contributions as in the one-loop contribution, such that each propagator contributes a factor of 
\begin{align*}
- \frac{\dx_1 \dx _2 - \vert \dx_1 \vert \vert \dx_2 \vert}{(x_1 - x_2 ) ^2} = \frac{1}{2} \, .
\end{align*}
The color factors are obtained by applying the identity $2\, T^a T^a = N \, \unit$ and we note that the ordered 2n-fold integral over the interval $[ 0 , 2 \pi]$ contributes a factor of
\begin{align*}
\frac{(2 \pi)^{2n}}{(2n)!} \, .
\end{align*}
Each individual diagram at the $n$-th loop order thus contributes the factor
\begin{align}
\frac{1}{2^n} \, \frac{(2 \pi)^{2n}}{(2n)!} \, \frac{g^2}{4 \pi ^2} 
	\left( \frac{N}{2} \right) ^n = \frac{\lambda^n}{4 ^n \, (2n)!} \, .
\label{eqn:Diagramfactor}	
\end{align}
We need hence only count all possible rainbow-like diagrams consisting of $n$ propagators, which are not crossing each other. Any diagram of this type can be drawn in the following form:
\begin{align*}
\includegraphics[width=82mm]{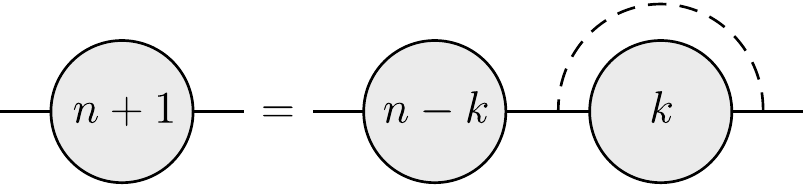}
\end{align*}
Here, the grey blob denotes a generic rainbow-like diagram containing $n$ propagators. It is then easy to see that the number $A_n$ of the rainbow-like diagrams satisfies the recursion relation
\begin{align}
A_{n+1} &= \sum \limits _{k=0} ^n A_{n-k} \, A_k \, , &
A_0 &= 1 \, .
\label{eqn:recursion}
\end{align}
This recursion relation may be solved by finding a generating function
\begin{align}
f(z) = \sum \limits _{n=0} ^\infty A_n \, z^n \, .
\end{align}
In terms of this generating function, the recursion relation \eqref{eqn:recursion} translates to the functional equation
\begin{align}
f(z) ^2 &= \frac{f(z) - 1 }{z} \, , &
f(0) &= 1 \, ,
\end{align}
which is solved by
\begin{align}
f(z) = \frac{1 - \sqrt{1-4z}}{2z} = \sum \limits _{n=0} ^\infty
	\frac{(2n)!}{(n+1)! \, n!} \, z^n \, .
\end{align}
We have hence found the number of rainbow-like diagrams containing $n$ propagators to be given by
\begin{align}
A_n = \frac{(2n)!}{(n+1)! \, n!}  \, .
\end{align}
It was noted in reference \cite{Erickson:2000af} that $A_n$ can also be calculated from a matrix model introduced in reference \cite{Brezin:1977sv}. Combining the above finding with the factor \eqref{eqn:Diagramfactor} contributed by each individual diagram, we find (assuming the conjecture of reference \cite{Erickson:2000af} holds true)
\begin{align}
\left \langle W  \left( 
	\raisebox{-.9mm}{\includegraphics[height=2.2ex]{circle.pdf}}   
	\right) \right \rangle
	= \sum \limits _{n=0} ^\infty 
	\frac{\lambda ^n}{4^n \, (n+1)! \, n!} 
	= \frac{2}{\sqrt{\lambda}} \, \mathrm{I}_1 \big( \sqrt{\lambda} \, \big) \, .
\end{align}
Here, $\mathrm{I}_1$ is the modified or hyperbolic Bessel function $\mathrm{I}_\nu$ of the first kind for the value $\nu = 1$, cf.\ e.g.\ reference \cite{FunctionAtlas} for more details on this function. The asymptotic expansion for large $\lambda$ is given by
\begin{align}
\left \langle W  \left( 
	\raisebox{-.9mm}{\includegraphics[height=2.2ex]{circle.pdf}}   
	\right) \right \rangle
	\overset{\la \gg 1}{=} \sqrt{\frac{2}{\pi}} \, 
	\frac{e^{\sqrt{\lambda}}}{\lambda^{3/4}} \, , 
\end{align}
which agrees with the AdS/CFT prediction \eqref{eqn:Pred} within the limits of its accuracy. The calculation of the circular Maldacena--Wilson loop was extended by Drukker and Gross \cite{Drukker:2000rr} to include all non-planar corrections. They studied the anomaly arising from the singular mapping of the straight line to the circle also relying on the conjecture that all diagrams containing interaction vertices cancel against each other. The conjecture was later proven by Pestun \cite{Pestun:2007rz}, who used localization techniques to reduce the calculation of the circular Maldacena--Wilson loop to a matrix model calculation. 

At the strong coupling side, the one-loop string correction to the circular Wilson loop was also considered \cite{Drukker:2000ep,Kruczenski:2008zk}, but a mismatch to the localization result was observed, cf.\ e.g.\ reference \cite{Vescovi:2016zzu} for more details. In this light, it is interesting to consider the ratio between the circular 1/2 BPS Maldacena--Wilson loop and a 1/4 BPS Maldacena--Wilson loop known as the latitude Wilson loop, for which some of the potential ambiguities of the string one-loop calculation drop out. The mismatch between the localization result and the string correction observed there \cite{Forini:2015bgo,Faraggi:2016ekd} was recently resolved in reference \cite{Forini:2017whz}. 

\subsection{The Duality to Scattering Amplitudes}

In the description of Wilson loops in generic Yang--Mills theories, we have touched upon the relation between the ultraviolet (cusp) divergences of Wilson loops and the infrared divergences of scattering amplitudes. In $\mathcal{N}\!=4$ supersymmetric Yang--Mills theory, the connection between scattering amplitudes and Wilson loops goes even further, cf.\ the reviews 
\cite{Alday:2008yw,Henn:2009bd} on which the below discussion is mainly based. The first signs for the conjectured duality between these observables were found by Alday and Maldacena in reference \cite{Alday:2007hr}, who found that gluon scattering amplitudes at strong coupling are described by the area of certain minimal surfaces. Concretely, the boundary curves for these minimal surfaces are given by polygons with light-like edges, where the cusp points are related to the momenta of the gluons by
\begin{align}
x_{i+1} - x_i = p_i \, .  
\end{align} 
The strong-coupling description of the scattering amplitudes is thus the same as for the Wilson loop over the polygon with cusp points $x_i$. The calculation of the minimal surface in the four-cusp case lead to a result in agreement with the BDS-ansatz for the four-gluon amplitude, which is a conjecture for color-ordered, maximally helicity violating n-gluon amplitudes that was put forward in reference \cite{Bern:2005iz} based on a 3-loop calculation. 

Let us shortly explain the nature of the duality in more detail. We are considering maximally helicity violating (MHV) gluon amplitudes in the planar limit of $\mathcal{N}\!=4$ supersymmetric Yang--Mills theory, where one considers color-ordered amplitudes, see references \cite{Dixon:1996wi,Henn:2014yza} for an introduction. For MHV amplitudes, two of the gluons have one helicity while all other gluons have the opposite helicity. In this case, the same function of the helicity variables appears at all loop orders, and the amplitude can be written as
\begin{align}
A_n = A_n ^{\mathrm{tree}} \, M_n \, ,
\end{align}  
where $M_n$ is a function only of the momentum invariants $(p_i+p_j)^2$ and the helicity information is contained in the tree-level amplitude $A_n ^{\mathrm{tree}}$. The conjectured duality states \cite{Brandhuber:2007yx} that the function $M_n$ is related%
\footnote{Since the duality relates ultraviolet and infrared divergent quantities, both the regularization parameters $\epsilon_{\mathrm{UV}}$ and $\epsilon_{\mathrm{IR}}$ and the renormalization constants $\mu_{\mathrm{UV}}$ and $\mu_{\mathrm{IR}}$ have to be related to each other. This can in general be done in such a way that the divergent pieces of the amplitude and the Wilson loop match, cf.\ e.g.\ reference \cite{Henn:2009bd} for a more detailed explanation.}
to the expectation value $\left \langle W_n \right \rangle$ of the Wilson loop over the associated polygon by
\begin{align}
\left \langle W_n \right \rangle = M_n + d \, .
\end{align}
Here, $d$ is a constant which does not depend on the kinematical information. This conjecture was tested further in references \cite{Drummond:2007cf,Drummond:2008aq}, where it was found to hold true, while both the Wilson loop and the scattering amplitude begin to deviate from the BDS-ansatz starting from six gluons and two loops.
\begin{figure}[th]
\centering
\includegraphics{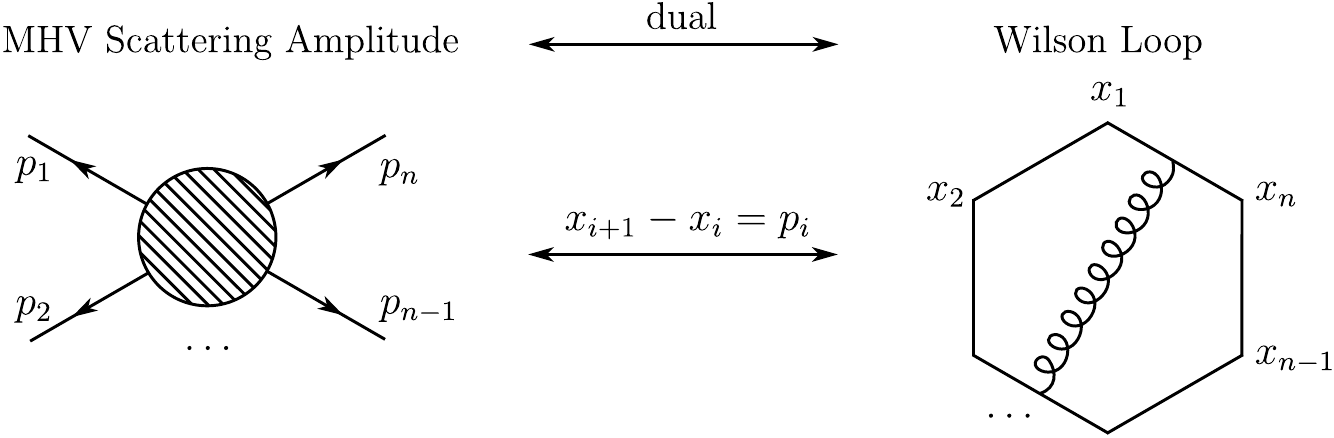}
\caption{Graphical representation of the duality between Wilson loops and scattering amplitudes.}
\label{fig:Duality}
\end{figure}

The appearance of these deviations can be understood from symmetry considerations, which one may motivate from the above duality. From the viewpoint of the scattering amplitudes, the conformal symmetry of the Wilson loop appears as a dual conformal symmetry in the dual variables $x_i$; this symmetry had indeed been obtained in reference \cite{Drummond:2006rz,Drummond:2007aua}. The dual conformal symmetry is very restrictive in the case of four- and five-point scattering amplitudes, for which one cannot construct conformally invariant combinations of the dual variables $x_i$. Starting from six points however, one may consider the conformally invariant cross-ratios
\begin{align*}
u_{ijkl} = \frac{(x_i - x_j)^2 (x_k - x_l) ^2 }{(x_i - x_k)^2 (x_j - x_l) ^2} \, ,
\end{align*}
and both Wilson loop and scattering amplitude contain a functional dependence on the respective cross-ratios, which the BDS ansatz does not account for. 

The amplitudes of $\mathcal{N}\!=4$ supersymmetric Yang--Mills theory can be organized conveniently by introducing an auxiliary superspace containing the Gra{\ss}mann odd variables $\eta_i ^A$ for each external leg of the amplitude, see reference \cite{Henn:2014yza} for more details. The original amplitudes can then be obtained as the coefficients in the $\eta$-expansion of the superamplitudes. It was noted in reference \cite{Drummond:2008vq} that the dual conformal symmetry extends to a dual superconformal symmetry of the color-ordered superamplitudes, see also the review \cite{Drummond:2010km}. The combination of the dual with the ordinary superconformal symmetry yields a Yangian symmetry of tree-level superamplitudes, which was observed in reference \cite{Drummond:2009fd}. 

Attempts have been made to extend the duality between Wilson loops and MHV amplitudes to a duality between superamplitudes and corresponding supersymmetric extensions of light-like Wilson loops \cite{CaronHuot:2010ek,Mason:2010yk}. However, at the one-loop order the proposed correspondence seems to fail due to singularities appearing for the supersymmetric extensions of the light-like Wilson loop, cf.\ references \cite{Belitsky:2011zm,Belitsky:2012nu,Beisert:2012xx}. 

Still, the combination of the Yangian symmetry found for the scattering amplitudes with the conjectured dualities hints that a Yangian symmetry could be present for (Maldacena--)Wilson loops or possibly supersymmetric extensions of these. Indeed it was found that an extension of smooth Maldacena--Wilson loops into smooth Wilson loops in superspace possesses a Yangian symmetry \cite{Muller:2013rta,Beisert:2015jxa,Beisert:2015uda}. We will discuss the strong-coupling counterpart of this object as well as its symmetries in chapter \ref{chap:SurfaceSuperspace}. 

\chapter{Symmetric Space Models}
\label{chap:SSM}

We have seen above that the Maldacena--Wilson loop is described by the string action in 
$\AdS_5 \times \mathrm{S}^5$ at strong coupling and it thus inherits the symmetries of the underlying string theory, which is known to be classically integrable. One can employ the classical integrability of the theory to derive symmetries of the minimal surface problem; this approach was applied in reference \cite{Muller:2013rta} to derive Yangian Ward identities for the Maldacena--Wilson loop at strong coupling. 

Apart from finding strong-coupling symmetries, we are interested in transferring these symmetries from strong coupling to weak or even arbitrary coupling. Since it is a priori unclear for which symmetries this is possible and what formulation is suitable for it, we discuss different approaches to construct the symmetries of string theory in $\AdS_5$. In fact, our discussion is not limited to Anti-de Sitter spaces but extends to a class of spaces known as symmetric spaces, which can be described by a specific class of coset spaces. The group-theoretic approach makes the integrability of the theory and the associated symmetries particularly transparent without needing to refer to the specific underlying space.   

The explicit application of the symmetries found for these models to minimal surfaces in 
$\AdS_5$ is discussed in chapter \ref{chap:MinSurf}. In the present chapter, the fact that the focus lies on minimal surfaces rather than closed strings appears in the assumption that the surface is simply connected, which is crucial for the discussion of conserved charges. We discuss the differences to the theory of closed strings or two-dimensional field theory at various points. 

The present chapter consists partly of an extensive review of the integrability of symmetric space models and partly of the original results obtained in references 
\cite{Klose:2016uur,Klose:2016qfv}, which highlight the role of the \textit{master symmetry} in the construction of the symmetries. We begin by discussing the geometric aspects of symmetric spaces in section \ref{sec:SymmSpaces} and go on to discuss the string action in a symmetric space as well as its integrability in sections \ref{sec:Action} and \ref{sec:Integrability}. 
We then turn to the discussion of the master symmetry in section \ref{sec:Master}. This symmetry was discussed in reference \cite{Ishizeki:2011bf,Kruczenski:2013bsa} as the spectral parameter deformation of minimal surfaces in Euclidean $\AdS_3$ and much earlier in reference \cite{Eichenherr:1979ci} in the context of two-dimensional field theories. In the latter reference it was applied to derive an infinite set of conserved charges, thus establishing the model's integrability. We initially focus on the deformations of minimal surfaces and establish the properties of these transformations. We then go on to show in section \ref{sec:IntCompl} that in addition to the construction of conserved charges, one can also apply the master symmetry to construct one-parameter families of symmetry variations, justifying its name. The algebra relations of the symmetry variations so obtained as well as the Poisson algebra of the conserved charges are established in section \ref{sec:Algebra}.

\section{Symmetric Spaces}
\label{sec:SymmSpaces}
A general symmetric space $\MM$ can be described as a homogeneous or coset space 
$\grp{G} / \grp{H}$ with appropriate Lie groups $\grp{G}$ and $\grp{H}$. The condition of being a symmetric space is then entailed in an algebraic restriction which the Lie algebras
$\alg{g}$ and $\alg{h}$ obey. This algebraic approach to the study of symmetric spaces goes back to the work of {\'E}lie Cartan \cite{Cartan:1926,Cartan:1927}, who applied it to give a classification of Riemannian symmetric spaces. It provides an extremely efficient description of symmetric spaces which is widely applied in the literature. However, in order to explain what is symmetric about symmetric spaces, we will initially take a geometric viewpoint. Here, we are not aiming for a complete and rigorous mathematical introduction to the subject of symmetric spaces such that most proofs are omitted below. For a complete mathematical introduction, the reader is referred to reference \cite{Helgason:2001} and in particular reference \cite[chapters 8-11]{ONeill:1983}, on which the account given below is based. 

In a rough sense, a symmetric space is a space for which the set of isometries is large enough to make the space regular. To reach a more precise description, let us first consider the example of $n$-dimensional Euclidean space $\mathbb{R}^n$. Apart from rotations and translations, the point reflections $R_a$ with respect to a certain fixed point $a \in \mathbb{R}^n$ are isometries. The set of point reflections $R_a: x \mapsto 2a - x$ generates all translations, since the concatenation of two point reflections is given by $R_a \circ R_b : x \mapsto x + 2(a-b)$. The generalization of the point reflections to manifolds leads us to the concept of a symmetric space.

\begin{definition}
We call a smooth, connected semi-Riemannian manifold $\MM$ a \textit{symmetric space} if for each $p \in \MM$ there is an involutive isometry $\xi_p$ (i.e.\ an isometry satisfying $\xi_p ^2 = \id _\MM$) with isolated fixed point $p$. The isometry $\xi_p$ is called the \textit{global symmetry} of $\MM$ at $p$.
\end{definition}

Since $\xi_p$ is an involution, it follows that 
$(\diff \xi_p)_p : T_p \MM \to T_p \MM$ is also an involution and hence it must have eigenvalues 
$\pm 1$. Remembering that a geodesic is entirely specified by its tangent vector in a given point, we realize that if $(\diff \xi_p)_p$ has an eigenvalue $+1$, then the geodesic through $p$ which is tangent to the respective eigenvector is mapped to itself by $\xi_p$, in contradiction to $p$ being an isolated fixed point. Hence we have 
$(\diff \xi_p)_p = - \id _ {T_p \MM}$, such that $\xi_p$ reverses all geodesics passing through $p$. In this way, the global symmetries provide a natural generalization of the point reflections of Euclidean space. 

As an example, we consider the three-dimensional sphere $\mathrm{S}^3$, which we describe in embedding coordinates as
\begin{align*}
\mathrm{S}^3 = \lbrace (X_1 , X_2 , X_3 , X_4 ) \in \mathbb{R}^4 :
	X_1^2 + X_2^2 + X_3^2 + X_4^2 = 1 \rbrace .
\end{align*}
The global symmetry at the pole 
$ N = (0,0,0,1)$
is given by the map
\begin{align}
\xi_N (X_1 , X_2 , X_3 , X_4 ) =  (- X_1 , -X_2 , -X_3 , X_4 ) \, ,
\label{globalsymmetryS3}
\end{align}
which is clearly an involution on $\mathrm{S}^3 $ and has isolated fixed points $N$ and $-N$.

Spaces of constant curvature such as the example above are symmetric spaces. While the converse is not true in general, the condition of being a symmetric space implies that the covariant derivative of the curvature tensor vanishes, 
$\nabla \mathcal{R} \equiv 0 $, where $\nabla$ denotes the Levi--Civita connection on $\MM$. The vanishing of the covariant derivative of the curvature tensor in turn only implies that $\MM$ is locally symmetric, which means that for each $p \in \MM$ there is a neighbourhood $U_p$ of $p$ on which a map with the properties of $\xi_p$ can be defined.  

With the geometric definition of a symmetric space discussed, we turn to the representation of a symmetric space as a coset space $\grp{G}/\grp{H}$. In order for M to be a homogeneous space, the action of the set of isometries of $\MM$ must be transitive. We can argue that this is the case similarly to the example of Euclidean space, where we found that the set of point reflections generates translations. The appropriate generalization of translations to manifolds is given in the following definition. 

\begin{definition}
An isometry $\xi: \MM \to \MM$ is called a \textit{transvection} along a geodesic $\gamma$, provided that
\begin{enumerate}[(i)]
\item $\xi$ is a translation along $\gamma$, that is 
	$\xi(\gamma(s)) = \gamma(s+c)$ for all $s \in \mathbb{R}$ and some $c \in \mathbb{R}$, 
\item $\diff \xi$ gives parallel translation along $\gamma$, i.e.\ if $v \in T_{\gamma(s)} \MM$, then 
	$\diff \xi _{\gamma(s)} (v) \in T_{\gamma(s+c)} \MM$ is the parallel translate of $v$ along $\gamma$. 
\end{enumerate}
\end{definition}

The global symmetries of M at $p$ generate transvections in the same way the point reflections of Euclidean space generate translations. For a given geodesic $\gamma$, consider the isometry 
$\xi _{\gamma(c)} \circ \xi _{\gamma(0)}$ constructed from the global symmetries of M at $\gamma(c)$ and 
$\gamma(0)$. We have already seen that $\xi_p$ reverses the geodesics through $p$, i.e.\ 
$\xi _{\gamma(0)} (\gamma(s)) = \gamma(-s)$, and since $c$ is the midpoint of $[-s, s + 2c]$, we have
$\xi _{\gamma(c)} \circ \xi _{\gamma(0)} (\gamma(s)) = \gamma(s + 2c)$. The second property above can be proven similarly. 

Since M is connected, two points $p$ and $q$ can be connected by a (broken) geodesic. The composition of the transvections along the smooth pieces of the geodesic then provides an isometry which maps $p$ to $q$. We have thus seen that the isometry group $I(\MM)$ acts transitively on M. One can then show that also the identity component $I_0(\MM)$ of the isometry group $I(\MM)$ acts transitively on M. We then define 
$\grp{G}= I_0(\MM)$ and take H to be the isotropy or stabilizer group of some point $p \in \MM$,   
\begin{align}
\grp{H} = \lbrace g \in \grp{G} : g \cdot p = p \rbrace .
\end{align} 
Concerning the manifold structures of M and $\grp{G}/\grp{H}$, we note that since H is a closed subgroup of G, there is a unique way to equip $\grp{G}/\grp{H}$ with a manifold structure, such that the canonical projection becomes smooth. With this structure, the map $j: \grp{G}/\grp{H} \to \MM$, $g\grp{H} \mapsto g \cdot p$ becomes a diffeomorphism. 

For the example of the three-dimensional sphere $\mathrm{S}^3$, the identity component of the isometry group is $\grp{SO}(4)$ and the isotropy group of the pole $N=(0,0,0,1)$ is given by
\begin{align}
H = \left \lbrace \begin{pmatrix}
	A & 0 \\ 0 & 1 
	\end{pmatrix} : A \in \grp{SO}(3) \right \rbrace
	\simeq \grp{SO}(3) \, .
\end{align}
We can hence identify the sphere $\mathrm{S}^3$ with the coset space $\grp{SO}(4)/\grp{SO}(3)$. 

We have seen above that we can represent any symmetric space as a coset space $\grp{G}/\grp{H}$. The additional structure provided by the global symmetry has profound consequences for the Lie algebras $\alg{h}$ and $\alg{g}$ of H and G. Consider the map  $\sigma(g) = \xi_p \cdot g \cdot \xi_p$, where $\xi_p$ is the global symmetry of $\MM = \grp{G}/\grp{H}$ at the fixed point of the action of H. Since $\xi_p = \xi_p ^{-1}$, we find that $\sigma: G \to G$ is an involutive automorphism of G. Moreover, one can show that the set 
\begin{align*}
G_\sigma = \lbrace g \in G : \sigma (g) = g \rbrace
\end{align*}
of fixed points of $\sigma$ is a closed subgroup of G in such a way that 
$( G_\sigma ) _0 \subset H \subset G_\sigma$. The derivative of $\sigma$ at the unit element $e$ gives rise to a Lie algebra automorphism $\Omega: \alg{g} \to \alg{g}$, 
\begin{align}
\Omega (X) = \diff \sigma_e (X) = \frac{\diff}{\diff t} \, 
	\sigma \left( \exp \left( t X \right) \right) \vert _{t=0} \, .
\end{align}
It follows that $\Omega ^2 = \mathrm{id}_{\alg{g}}$ and thus $\Omega$ has eigenvalues $\pm 1$. We can identify the Lie algebra $\alg{h}$ with the eigenspace of $\Omega$ for the eigenvalue $+1$, 
\begin{align}
\alg{h} = \left \lbrace X \in \alg{g} \, : \, \Omega (X) = X \right \rbrace .
\end{align}
In order to prove the above relation, suppose first that $X \in \alg{h}$. Since $\sigma$ acts as the identity on H, we have $\diff \sigma _e (X) = X$. Conversely, suppose that $\Omega (X) = X$. We consider the one-parameter subgroup $\alpha_t (X) = \exp ( t X)$. Then the set 
$\sigma ( \alpha_t (X) )$ is also a one-parameter subgroup with the same initial velocity, 
\begin{align*}
\frac{\diff}{\diff t} \sigma ( \alpha_t (X) ) \vert _{t=0}
	= \Omega(X) = X \, ,   
\end{align*}  
such that $\sigma ( \alpha_t (X) ) = \alpha_t (X)$. Hence we have 
$\alpha_t (X) \in ( G_\sigma ) _0 \subset H$, such that $X \in \alg{h}$. The Lie algebra $\alg{g}$ can then be written as the direct sum of $\alg{h}$ and the subspace
\begin{align}
\alg{m} = \left \lbrace X \in \alg{g} \, : \, \Omega (X) = - X \right \rbrace  .
\end{align}
The projections onto the components $\alg{h}$ and $\alg{m}$ are given by
\begin{align}
P_{\alg{h}} (X) = \half \left(X + \Omega(X) \right) \, , \qquad P_{\alg{m}} (X) = \half \left(X - \Omega(X) \right) \, .
\label{def:projectors}
\end{align}    
Since $\Omega$ is a Lie algebra automorphism, it follows that the decomposition 
$\alg{g} = \alg{h} \oplus \alg{m}$ gives a $\mathbb{Z}_2$-grading on $\mathfrak{g}$,
\begin{align}
\left[ \alg{h} , \alg{h} \right] &\subset \alg{h} \, , & 
\left[ \alg{h} , \alg{m} \right] &\subset \alg{m} \, , &
\left[ \alg{m} , \alg{m} \right] &\subset \alg{h} \, .
\label{eqn:algebra-relations}
\end{align}
These are the sought-after algebraic relations for the Lie algebra 
$\alg{g}= \alg{h} \oplus \alg{m}$ that guarantee that the coset $\grp{G}/\grp{H}$ becomes a symmetric space. 

Let us once more return to our example of the three-dimensional sphere $\mathrm{S}^3$.
The global symmetry given in equation \eqref{globalsymmetryS3} induces the automorphism 
$\sigma (g) = K g K$ with 
$K = \mathrm{diag}\left( -1 , -1, -1, 1 \right)$. Then we have
\begin{align}
\grSO(4)_\sigma = \left \lbrace \begin{pmatrix}
g & 0 \\ 0 & 1
\end{pmatrix} : g \in \grSO(3) \right \rbrace = \grp{H} \simeq \grSO(3) \, ,
\end{align}
as before. The Lie algebra automorphism $\Omega = \diff \sigma_e $ is then given by 
$\Omega(X)= K X K $ and we find
\begin{align}
\alg{h} = \left \lbrace \begin{pmatrix}
X & 0 \\ 0 & 0
\end{pmatrix} : X \in \alg{so}(3) \right \rbrace \simeq \alg{so}(3) \, , \qquad \alg{m} = \left \lbrace \begin{pmatrix}
0 & v \\ -v^T & 0
\end{pmatrix} : v \in \mathbb{R}^3 \right \rbrace .
\end{align}
By direct matrix multiplication one may then show that the relations \eqref{eqn:algebra-relations} are indeed satisfied.

So far, we have seen that a symmetric space can be identified with a coset space G/H and studied the implications which the global symmetries of M have on G. Rather than having a metric only on M, it is more convenient to work with a metric on G such that we can express all quantities through coset representatives. The subspace $\alg{m}$ of the Lie algebra $\alg{g}$ corresponds to the tangent space $T_p \MM$. If we identify $\alg{m}$ as usual with the set of subspaces of tangent spaces obtained by left-multiplication in G, the correspondence between $\alg{m}$ and the tangent spaces of M extends to arbitrary points. 

If we now fix an Ad(H)-invariant\footnote{For some $h \in \grp{H}$, the map $\mathrm{Ad}_h: \alg{g} \to \alg{g}$ is defined as the differential of the conjugation map $C_h(g) = h g h^{-1}$.} scalar product on $\alg{m}$, it corresponds to a metric on M for which the G-action gives isometries. A natural choice for a scalar product like this is given by the Killing metric B on G, which is given by the trace in the adjoint representation: With 
$\mathrm{ad}_X(Y) = [X,Y]$, we have
\begin{align}
\mathrm{B}(X,Y) = \tr \left( \mathrm{ad}_X \, \cdot \mathrm{ad}_Y \right) \, . 
\end{align} 
The Killing metric is in fact invariant under all automorphisms of G, such that we have in particular 
\begin{align}
\mathrm{B}(X,Y) = \mathrm{B}(\Omega(X), \Omega (Y) ) \, \quad \forall \, X,Y \in \alg{g} \, ,   
\end{align} 
which implies that $\alg{h}$ and $\alg{m}$ are perpendicular, $\alg{h} \perp \alg{m}$. If the Lie algebra $\alg{g}$ is simple, the Killing metric is proportional to the trace in the fundamental representation,
\begin{align}
\mathrm{B}(X,Y) = c \, \tr \left( X \, Y \right) .
\end{align}
In the case of $\AdS_N$, we are dealing with a simple Lie algebra $\alg{g}$ and in the general discussion of symmetric space models below, we will employ the fundamental representation with the scalar product on $\alg{g}$ given by the trace. This will simplify the notation considerably as we can e.g.\ use direct matrix multiplication instead of the differential of some group action. We need not, however, assume that we are working with a simple Lie-algebra 
$\alg{g}$. It is sufficient for us that the trace gives a non-degenerate scalar product, which is invariant under the automorphism $\Omega$; this can also be seen directly from the form of 
$\Omega$ discussed above.  

\begin{figure}[t]
\centering
\includegraphics[width=100mm]{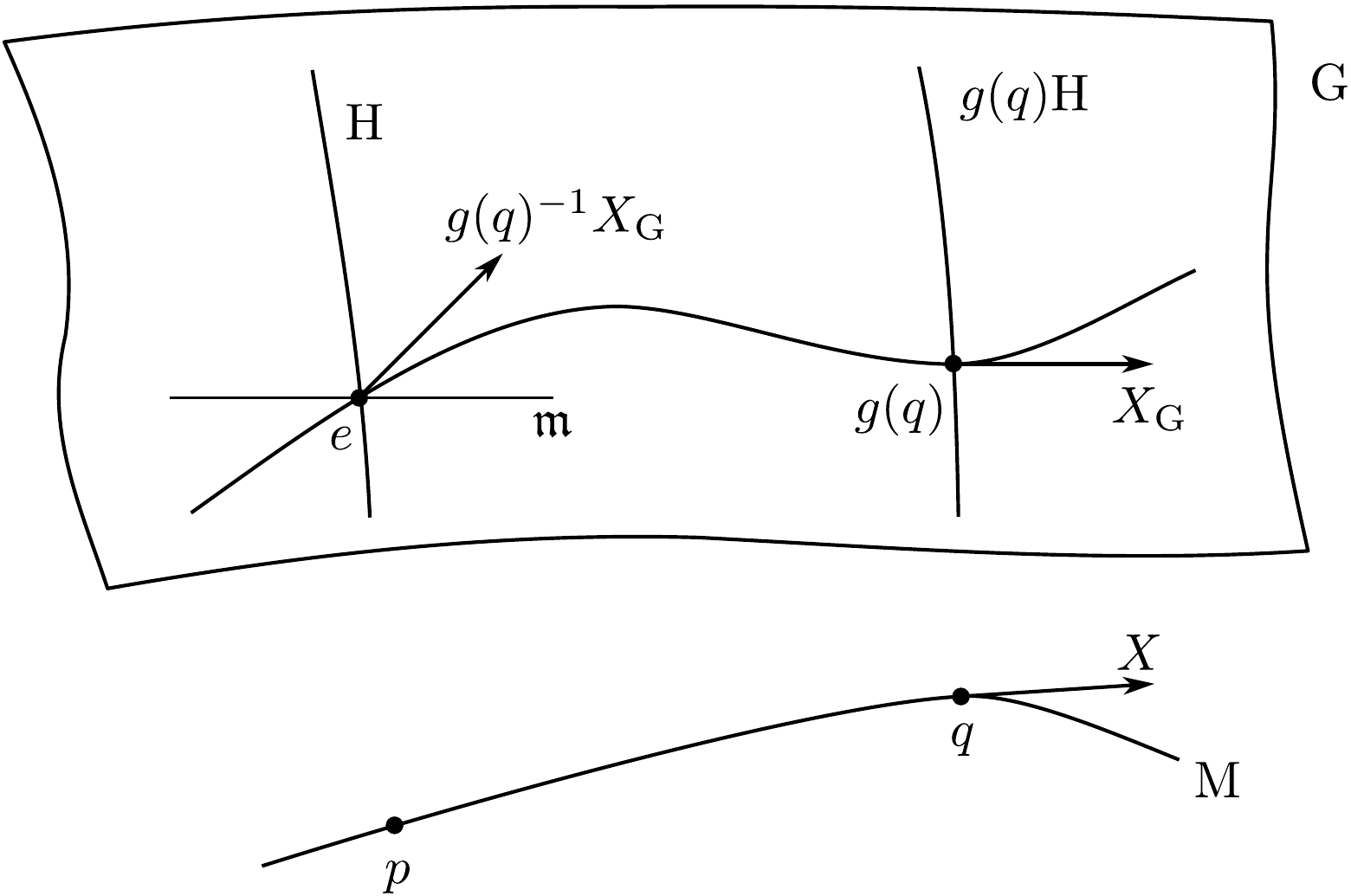}
\caption{Depiction of the coset space G/H.}
\label{fig:Coset}
\end{figure}

With a metric on G established, we are now in a position to work directly with group elements without referring to the original space M. Concretely, for each $q \in \MM$ we choose a coset representative $g(q)$ in a smooth way. A natural choice for the set of coset representatives is given by exponentiating the subspace $\alg{m} \subset \alg{g}$, but we are free to 
right-multiply with elements of H at each point. Indeed, a different choice of coset representatives will prove to be convenient in the coset description of $\AdS_5$, which we discuss in chapter \ref{chap:MinSurf}. If we now consider two tangent vectors $X,Y \in T_q \MM$, then by virtue of our choice of coset representatives we have corresponding tangent vectors $X_{\grp{G}} , Y_{\grp{G}}  \in T_{g(q)} \grp{G}$, see also figure \ref{fig:Coset}. Since the metric on G is left-invariant by construction, we can express it in terms of the scalar product on $T_e \grp{G}$. The scalar product of $X$ and $Y$ can then be expressed in terms of the corresponding quantities on G as
\begin{align}
\left \langle X , Y \right \rangle_q = \tr \left( 
	P_{\alg{m}} \left( g(q)^{-1} X_{\grp{G}} \right) \,
	P_{\alg{m}} \left(  g(q)^{-1} Y_{\grp{G}} \right)  \right) . 
\end{align}
In particular, for a parametrization $q(\tau,\sigma)$ of a surface on M, we have a corresponding parametrization $g(\tau,\sigma)$ on G and the scalar product of the tangent vectors $\partial_\tau q$ and $\partial_\sigma q$ can be expressed in terms of $g(\tau,\sigma)$ as
\begin{align}
\left \langle \partial_i q , \partial_j q \right \rangle_q = \tr \left( 
	P_{\alg{m}} \left( g^{-1} \partial_i g \right) \,
	P_{\alg{m}} \left( g^{-1} \partial_j g \right)  \right) . 
\end{align}
This allows us to describe the string action for the surface described by $q(\tau,\sigma)$ entirely in terms of the corresponding parametrization $g(\tau,\sigma)$, which will be the starting point of our discussion in the next section. 

\section{The String Action}
\label{sec:Action}

We now turn to the discussion of the string action or area functional in a symmetric space.
In the last section, we found that we can describe a surface in $\MM = \grp{G}/\grp{H}$ by a map $g(\tau,\sigma)$ from the worldsheet $\Sigma$ to G. 
The relevant quantities for the action are the Maurer--Cartan current 
$U_i = g^{-1} \partial_i g$ as well as its projections
\begin{align}
a_i &= P_{\alg{m}} \left( g^{-1} \partial_i g \right)  , &
A_i &= P_{\alg{h}} \left( g^{-1} \partial_i g \right)  .
\end{align}
In terms of these currents, the Polyakov action is given by
\begin{align}
S[g,h] = - \frac{T}{2} \, \int _\Sigma \diff \tau \wedge \diff \sigma \, 
	\sqrt{h} \, h^{ij} \, \tr \big( a_i \, a_j \big) .
\label{actionindex}	
\end{align}
Here, we introduced the Euclidean worldsheet metric $h_{ij}$ and denoted $h= \det(h_{ij})$ and the inverse of the metric $h^{i j}$. Moreover, we have set 
$(\sigma^1 , \sigma ^2 )$ = $(\tau , \sigma)$. The boundary curve of the minimal surface is situated at $\tau=0$ and we have periodic boundary conditions in the 
$\sigma$-variable. The worldsheet metric is subject to the Virasoro constraints
\begin{align}
\tr \big( a_i \, a_j \big) - \half h_{i j} \, h^{k l} \tr \big( a_k \, a_l \big) = 0 \, .
\end{align}
If we fix conformal gauge and introduce a complex worldsheet coordinate $z = \sigma + i \tau$, the Virasoro constraints take the form
\begin{align}
\tr \big( a_z\, a_z \big) = \tr \big( a_\zb \, a_\zb \big) &= 0 \, . 	
\end{align} 
It is customary in string theory to fix conformal gauge, but in the study of minimal surfaces one may work with the Nambu-Goto action and not all minimal surface solutions are necessarily given in a conformal parametrization. We will hence avoid fixing a gauge condition for the parametrization of the surface and work with a general parametrization. This can be achieved without creating additional complications by employing the language of differential forms on the worldsheet. An introduction to the use of differential forms in the case at hand as well as a collection of helpful identities is provided in appendix \ref{app:Hodge}. In calculating with these differential forms it is important to remember that their components take values in a Lie algebra $\alg{g}$ and hence do not commute. For two general one-forms $\omega$ and $\rho$ we have e.g.\
\begin{align}
\omega \wedge \rho + \rho \wedge \omega = \left[ \omega_i , \rho _j \right] \, 
	\diff \sigma^i \wedge \diff \sigma^j \, . 
\end{align}
The worldsheet metric is incorporated in the Hodge star operator, which maps a $k$-form to a $(2-k)$-form, since the worldsheet is two-dimensional. On one-forms, the Hodge star operator acts as
\begin{align}
\ast \, \diff \sigma^i = \sqrt{ h } \, 
	h^{i j} \, \epsilon_{j k} \, \diff \sigma ^k .
\end{align}
Here, we fix the convention $\eps_{\tau \sigma} = \eps_{1 2} = 1$ for the Levi-Civita symbol and note the identities 
\begin{align}
\ast \ast w &= - \omega \, , &
\omega \wedge \ast \rho &= - \ast \omega \wedge \rho \, , 
\label{HodgeStarRel}
\end{align} 
for two general one-forms $\omega$ and $\rho$. The notation using differential forms and the Hodge star operator is related to index notation by
\begin{align*}
\omega \wedge \ast \rho & =  \sqrt{h} \, h^{i j} \omega_i \, \rho_j 
	\left( \diff \tau \wedge \diff \sigma \right) \, , &  
\diff \ast \omega &= \partial_i \big( \sqrt{h} h^{i j} \omega_j \big) 
	\left( \diff \tau \wedge \diff \sigma \right) \, . 	
\end{align*}
We can hence write the Polyakov action \eqref{actionindex} in a compact way as
\begin{align}
S[g,h] = - \frac{T}{2} \, \int _\Sigma \tr \left( a \wedge \ast a \right) .
\label{CosetAction}
\end{align}
The variation of the action with respect to $g$ gives the equations of motion. Varying the embedding $q(\tau , \sigma)$ with a fixed set of coset representatives would not give a completely general $\delta g$, but since the choice of coset representatives does not carry any physical information allowing for a general $\delta g$ does not imply additional restrictions. Since the subspaces $\alg{h}$ and $\alg{m}$ are perpendicular to each other, we have 
$\delta \tr \left( a \wedge \ast a \right) = 2 \tr \left( \delta U \wedge \ast a \right)$
and using that 
$\delta U = \diff ( g^{-1} \delta g ) + [ U, g^{-1} \delta g ]$, 
we obtain the variation of the action as
\begin{align}
\delta S & = - T \int_\Sigma \tr \left[ \lrbrk{ \diff ( g^{-1} \delta g ) 
	+ [ U, g^{-1} \delta g ] } \wedge \ast a  \right] \nn \\
         & = T  \int_\Sigma \tr \left[ \lrbrk{ \diff \ast a 
         + \ast a \wedge U + U \wedge \ast a } g^{-1} \delta g \right] 
         - T \int _{\partial \Sigma} \tr \left( g^{-1} \delta g \ast a \right)  \, ,  
\end{align}
where we have used equation \eqref{TraceTrick} and integrated by parts. For a minimal surface, we keep the boundary fixed and thus we demand that the variation $g^{-1} \delta g$ takes values in $\alg{h}$ at the boundary, such that the boundary term above vanishes. Inserting $U = A+a$, we see that the $a$-part drops out of the first term, which leaves us with the equations of motion
\begin{align}
\diff \ast a + \ast a \wedge A + A \wedge \ast a &= 0 \, . 
\label{EOM}
\end{align}
We have constructed the coset model to have a gauge redundancy associated to the isotropy group H, which we will call the gauge group from now on. In terms of the functions $g(\tau , \sigma)$, the gauge transformations are given by right multiplication of $g$ with local elements $R(\tau, \sigma) \in \grp{H}$. Then, the Maurer--Cartan form transforms as 
\begin{align*}
U \;\;\mapsto\;\; R^{-1} U R + R^{-1} \diff R
\end{align*}
and since $[\alg{h} , \alg{h} ] \subset \alg{h}$ and  
$[\alg{h} , \alg{m} ] \subset \alg{m}$, we can conclude that its components transform as
\begin{align*}
& A \;\;\mapsto\;\; R^{-1} A R + R^{-1} \diff R \, , &
& a \;\;\mapsto\;\; R^{-1} a R \, ,
\end{align*} 
such that the action is invariant, as it should be. The global G-Symmetry acts on $g$ by left-multiplication with a constant $L \in \grp{G}$. The Maurer--Cartan form then transforms as 
\begin{align*}
g^{-1}\diff g \mapsto g^{-1}L^{-1}\diff(Lg) = g^{-1}\diff g \, , 
\end{align*}
such that the action is invariant. Of course, this symmetry was built in by choosing G to be the isometry group of M. In writing the action in the form \eqref{CosetAction}, we have already applied the G-invariance of the metric on G, which then appears as the invariance of the Maurer--Cartan form under left-multiplication. 

In order to derive the Noether current associated to this symmetry, we consider the infinitesimal transformation 
\begin{align}
\delta g = \epsilon g 
\label{eq:Liesymmetry}
\end{align}
with $\epsilon \in \alg{g}$. We can derive the Noether current by introducing a local parameter in the variation, which can be achieved here by allowing $\epsilon$ to depend on the worldsheet coordinates. Then, since for constant $\epsilon$ we are considering a symmetry variation, the variation of the action can be written in the form 
$\delta S = \int \ast j \wedge \diff \epsilon$.  
If we impose the equations of motion, any variation of the action will vanish such that the current $j$ has to be conserved, $\diff \ast j = 0$. In the present case, the variation of $g$ implies $\delta U = g^{-1} \diff \epsilon g$ and thus the variation of the action is given by  
\begin{align}
\delta S = T \int \tr ( \ast a \wedge g^{-1} \diff\epsilon g ) 
  	= T \int \tr (   g\ast a g^{-1} \wedge  \diff\epsilon).
\end{align}
We have thus found the Noether current and the associated charge to be given by
\begin{align} 
j &= -2 g a g^{-1} \, , &
Q &= \int \ast j \, .
\label{eqn:Noethercurrent}
\end{align}
We can convince ourselves that the Noether current is indeed conserved by employing equation \eqref{dconjg} to find 
\begin{align}
\diff \ast j = -2 g \left( \diff \ast a 
	+ A \wedge \ast a + \ast a \wedge A \right) g^{-1} = 0 \, , 
\label{jcons}
\end{align}
where we have inserted the equations of motion \eqref{EOM}. A crucial feature of symmetric space models is that the Noether current is also flat, i.e.\ 
$\diff j + j \wedge j = 0$. This follows from the flatness of the Maurer--Cartan form, which is flat by construction,
\begin{align*}
\diff U = \diff \left( g^{-1} \diff g \right) 
	= \diff g^{-1} \wedge \diff g 
	= - g^{-1} \diff g \wedge g^{-1} \diff g
	= - U \wedge U \, .
\end{align*}
For the projections $A$ and $a$, this implies the relations 
\begin{align}
\diff A + A \wedge A +  a \wedge a &= 0 \, , &
\diff a + a \wedge A +  A \wedge a &= 0 \, . 
\label{Uflatcomp}
\end{align}
Here, we made use of the fact that the combination $(a \wedge A +  A \wedge a)$ is given by a commutator of the components of $A$ and $a$ and employed the grading 
\eqref{eqn:algebra-relations} of $\alg{g}$ to obtain the projections. 
The flatness of the Noether current $j$ then follows from a short calculation,   
\begin{align}
\diff j + j \wedge j = -2 g \left( \diff  a 
	+ A \wedge  a +  a \wedge A + 2 a \wedge a 
	- 2 a \wedge a  \right) g^{-1} = 0 \, .  
\end{align}
It allows to construct an infinite hierarchy of conserved charges and hence indicates the integrability of the model. We review the construction of the infinite tower of conserved currents and charges in the next section. 

\section{Integrability and Conserved Charges}
\label{sec:Integrability}
Given a flat and conserved current, an infinite tower of conserved charges may be obtained from an iterative procedure introduced by Br{\'e}zin, Itzykson, Zinn--Justin and Zuber (BIZZ) in reference
\cite{Brezin:1979am}. In reviewing their construction we begin by defining the covariant derivative $\Dbizz$ which contains the Noether current $j$ as a connection and acts on $\alg{g}$-valued functions $f$ and one-forms $\omega$ as
\begin{align} \label{eqn:Dbizz}
\Dbizz f &= \diff f - f j \, , &
\Dbizz \, \omega &= \diff \omega + \omega \wedge j \, .
\end{align} 
With these definitions, we find the identities
\begin{align}
\Dbizz \, \Dbizz f &= \diff ^2 f - \diff f \wedge j - f \diff j + (\diff f - f j) \wedge j  
	= -f \left( \diff j + j \wedge j \right) = 0 \, ,
\end{align}
as well as
\begin{align}
\diff \ast \Dbizz f &= \diff \ast \diff f - \diff f \wedge \ast j - f \diff \ast j 
	= \diff \ast \diff f + \ast \diff f \wedge j
	= \Dbizz \ast \diff f \, .
\end{align}
Here, we have applied that the Noether current is conserved, $\diff \ast j = 0$, such that the above relation only holds on-shell. Let us now suppose that we have a conserved current $\tilde{j}^{(n)}$ which can be written as $\tilde{j}^{(n)} = \Dbizz \chi^{(n-1)}$ for some function $\chi^{(n-1)}$. Since $\tilde{j}^{(n)}$ is conserved, we can calculate a potential defined by $\tilde{j}^{(n)} = \ast \diff \chi^{(n)}$. Given this potential, we define a higher current by 
\begin{align}
 \tilde{j}^{(n+1)} = \Dbizz \chi^{(n)} \, .
\end{align}
The crux of the construction is that the new current is conserved,
\begin{align}
  \diff \ast \tilde{j}^{(n+1)} = \diff \ast \Dbizz \chi^{(n)} = \Dbizz \ast \diff \chi^{(n)} = \Dbizz \tilde{j}^{(n)} = \Dbizz \, \Dbizz \chi^{(n-1)} = 0 \, . 
\end{align}
We can hence construct an infinite set of conserved currents beginning with $\tilde{j}^{(0)} = -j$ which can indeed be expressed as $\tilde{j}^{(0)} = \Dbizz \chi^{(-1)}$ with $\chi^{(-1)} = \unit$. The first two currents obtained from this recursion take the form
\begin{align}
 \tilde{j}^{(0)} &= -j \, , &
 \tilde{j}^{(1)} &= \ast j - \chi^{(0)} j \, , &
 \tilde{j}^{(2)} &= j + \chi^{(0)} \ast j - \chi^{(1)} j \, .
\label{eq:bizzcurrents}
\end{align}

A different method to construct an infinite tower of conserved charges is to introduce a Lax connection, which here is a family of flat connections parametrized by a spectral parameter $u \in \mathbb{C}$. We can obtain such a connection by considering the ansatz
\begin{align}
L_u = A + \alpha \, a + \beta \, \ast a \, , 
\end{align}
and fixing the coefficients from the requirement that $L_u$ is flat when the equations of motion are enforced. Using the flatness condition \eqref{Uflatcomp} for $a$ as well as the equations of motion \eqref{EOM}, we find
\begin{align}
\diff L_u + L_u \wedge L_u &= \diff A + A \wedge A 
	+ \left( \alpha ^2 + \beta ^2 \right) a \wedge a \, . 
\end{align}
The comparison with the flatness condition for $A$ then shows that we need to require
$ \alpha ^2 + \beta ^2 = 1 $ in order for $L_u$ to be flat. We then have the Lax connection
\begin{align} 
L_u = A + \frac{1-u^2}{1+u^2} \, a - \frac{2u}{1+u^2} \ast a 
	= A + \cos (\theta) \, a + \sin (\theta) \, \ast a \, . 
\label{eqn:definition-Lax}
\end{align} 
Here, we have introduced two different parametrizations by $u$ and $\theta$, which are related to each other by 
\begin{align}
e^{i\theta} = \frac{1-iu}{1+iu}.
\label{eq:theta}
\end{align}
Depending on the context, either the $u$- or the $\theta$-parametrization will be more convenient. The undeformed case is reached by setting $u$ or $\theta$ to zero, respectively. We note that the $\mathfrak{h}$-part of the Maurer--Cartan current remains unaltered, while the transformation of the $\mathfrak{m}$-part is reminiscent of a worldsheet rotation, which is particularly apparent using the $\theta$-parametrization and in conformal gauge, where we can write the deformation as 
\begin{align} 
a \;\; \mapsto \;\; a_\theta  = e^{-i\theta} a_z \, \diff z 
	+ e^{i\theta} a_{\bar{z}} \, \diff \bar{z} \, .
\label{eqn:deform-a}
\end{align}
This transformation is not an honest rotation, however, since the $\alg{h}$-valued part is not transformed and the argument of $a_z$ is kept fixed. 

In order to derive conserved charges in this approach, we consider the monodromy over the Lax connection $L$, which is given by
\begin{align}
T_u = \prexp \left( \int _\gamma L_u \right) \, . 
\label{monodromy:L}
\end{align} 
Here, $\gamma$ is a closed curve wrapping the worldsheet, which we typically assume to lie at constant $\tau$ and cover the whole period of the $\sigma$-coordinate.  

The conserved charges obtained from the BIZZ recursion can be extracted from $T_u$ by expanding around the undeformed case $u=0$. However, in the form given above, $T_u$ is completely non-local at $u=0$, which complicates the expansion. We can circumvent this issue by considering the transformed Lax connection 
\begin{align}
\ell_u = g L_u g^{-1} - \diff g \, g^{-1} = g (L_u - U) g^{-1} \, .
\end{align}
The flatness of $\ell_u$ follows directly from the flatness of $U$ and $L_u$. Making use of equation \eqref{dconjg}, we have
\begin{align*}
\diff \ell_u &= 
	g \left(U \wedge U - L_u \wedge L_u + U \wedge (L_u-U)  
	+ (L_u-U) \wedge U \right) g^{-1} \\
	&= - g \left( (L_u -U) \wedge (L_u -U) \right) g^{-1} 
	= - \ell_u \wedge \ell_u \, .
\end{align*}
In terms of the Lax connection $\ell_u$, we find that $T_u$ is given by
\begin{align}
T_u &= g_0 \,  t_u \,  g_0^{-1} \, , & 
t_u &= \prexp \left( \int _\gamma \ell_u \right)  ,
\end{align}
where $g_0$ is the value of $g$ at the base point $z_0$ of the curve over which we consider the monodromy. For a proof of the above transformation behaviour, the reader is referred to appendix \ref{app:Hodge}, where we discuss flat connections and path-ordered exponentials in more detail.
The transformation is similar to the gauge transformation of a Wilson loop and it is sometimes referred to as a gauge transformation, although strictly it is not a gauge transformation in this context since $g$ does not take values in the gauge group H. Inserting the explicit form of $L_u$ from \eqref{eqn:definition-Lax}, we see that the Lax connection $\ell_u$ represents a flat deformation of the Noether current $j$:
\begin{align} 
\ell_u = \frac{u}{1+u^2} \big (u\, j +  \ast j \big)\, .
\label{eqn:deformed-Noether}
\end{align}
In particular, the transformed connection $\ell_u$ vanishes at $u=0$ such that we may expand around $u=0$ to find
\begin{align}
t_u = \unit + u  \int_\gamma \ast j  + \mathcal{O}(u^2)  \, . 
\end{align} 
In order to compare the charges obtained from the BIZZ recursion to the ones obtained from the expansion of $t_u$, it is convenient to consider the quantity $\chi_u$ which is defined by
\begin{align}
\diff \chi_u &= \chi_u \ell _u \, , &
\chi_u (z_0) &= \unit \, .
\label{eqn:dchi}  
\end{align}
The above relation is known as the auxiliary linear problem for the connection $\ell_u$. 
The existence of a solution follows from the flatness of $\ell_u$, which implies that
\begin{align*}
\diff ^2 \chi _u = \diff \left( \chi_u \ell_u \right) 
	= \chi_u \left( \ell_u \wedge \ell_u + \diff \ell _u \right) = 0 \, .
\end{align*}
We can then apply the Poincar{\'e} lemma to infer that a unique solution exists. To be more precise, applying the lemma requires that the underlying space is contractible, which implies in particular that it is simply connected. In the case of closed strings, where the string embedding is not simply connected, we may work on the worldsheet $\Sigma$ extending over a single period. This prohibits considering multiply wound contours and leads to a unique solution. If the connection $\ell_u$ were not flat, a solution would not exist. We could then restrict the defining relation \eqref{eqn:dchi} to a curve and solve the resulting ordinary differential equation along the curve. The solution can formally be written as the path-ordered exponential of the contour-integral over the connection as for the monodromy above. In the case of a flat connection, the existence of a solution of equation \eqref{eqn:dchi} shows that the monodromy becomes path independent. This holds for any two contours which can be deformed continuously into each other. Since the minimal surfaces are simply connected, we can in particular shrink any closed contour into a point and hence the monodromy is trivialized, 
$t_u = \unit$, in this case. This implies in particular that all conserved charges vanish. 

The quantity $\chi_u$, however, also allows to extract the conserved currents. We can then convince ourselves that the conserved currents derived from the auxiliary linear problem for $\ell_u$ are the same as the ones obtained from the BIZZ recursion. This was established in reference \cite{Wu:1982jt}. Concretely, the quantity $\chi_u$ is the generating function for the BIZZ potentials $\chi^{(n)}$ introduced above, 
\begin{align} 
\chi_u = \sum_{n=0}^\infty \chi^{(n-1)} u^n \, .
\label{eqn:Taylor-chi}
\end{align}
In order to prove this relation, we express the BIZZ recursion as a recursion relation for the potentials, 
\begin{align}
\ast \diff \chi^{(n+1)} = \Dbizz  \chi^{(n)} \, . 
\label{eqn:BIZZ-recursion}
\end{align}
It is then easy to see that the defining equation \eqref{eqn:dchi} implies the above recursion relation. We invert \eqref{eqn:deformed-Noether} to obtain
\begin{align} \label{eqn:jfroml}
j = \ell_u - u^{-1} \ast \ell_u  \quad \Rightarrow \quad \ast \ell_u = u ( \ell_u - j ) \, , 
\end{align}
which shows that $\chi_u$ satisfies
\begin{align}
\ast \diff \chi_u = \chi_u \, \ast \ell_u = u \, \chi_u \left( \ell_u - j \right) = u \, \left( \diff\chi_u - \chi_u j \right) = u \, \Dbizz \chi_u \, .
\end{align}
Expanding this equation in powers of $u$ results in the BIZZ recursion \eqref{eqn:BIZZ-recursion}. Noting that $\ell_{u=0} = 0$ implies $\chi_{u=0} = \unit$ concludes the proof. However, there is a subtle difference between the charges obtained from the BIZZ recursion and the ones obtained from the expansion of the monodromy $t_u$, which will become clear in the following discussion.  

We have seen above that the monodromy is trivial in the case of minimal surfaces, since they are simply connected, such that all conserved charges obtained from the expansion of the monodromy vanish. In the case of closed strings, the situation is more complicated. The difference becomes particularly apparent when considering the charge
\begin{align}
\tilde{Q}^{(1)} = \int \ast j^{(1)} 
= - \int \left( \chi^{(0)} \ast j + j \right) .
\label{BIZZQ1}
\end{align}
For simplicity, we restrict to conformal gauge and consider the contour to be at constant $\tau$. Then we have
\begin{align}
\partial_\tau \tilde{Q}^{(1)} 
	= j^{(1)} _\sigma (\tau , 2 \pi ) - j^{(1)} _\sigma (\tau , 0 ) 
	= \left( \chi ^{(0)} (\tau , 2 \pi ) - \chi ^{(0)} (\tau , 0 ) \right) 
	j_\sigma (\tau , 0 ) \, , 
\label{Conservation:BIZZ}
\end{align}
where we have used the periodicity of $j_\sigma$ in the last step. We observe that in order for 
$\tilde{Q}^{(1)}$ to be a conserved charge, the function $\chi^{(0)}(\tau , \sigma)$ has to be periodic as well. This corresponds to the vanishing of the Noether charge 
\begin{align*}
Q^{(0)} = \int \ast j \, , 
\end{align*}
which happens for the minimal surfaces but not in general in the case of closed strings as we discussed above. 

However, one can still extract an infinite set of conserved charges for the closed strings using the expansion of the monodromy. Here, the difference to the charges obtained from the BIZZ recursion becomes important. We have seen above that the expressions for the charges derived from both formalisms in terms of the Noether current $j$ and the potentials $\chi^{(n)}$ are the same. The difference between the charges is due to the initial conditions we impose on the functions $\chi^{(n)}$. For example in the case of the charge $\tilde{Q}^{(1)}$, 
\begin{align*}
\tilde{Q}^{(1)} = - \int \left( \chi^{(0)} \ast j - j \right) \, , 
\end{align*}
we impose the initial conditions 
\begin{align*}
\chi_{\mathrm{BIZZ}} ^{(0)} (\tau_0 , 0) &= 0 \, , &
\chi_{\mathrm{mon}} ^{(0)} (\tau, 0) &= 0 \, ,
\end{align*}
for the BIZZ recursion and the expansion of the monodromy, respectively. Here, $\tau_0$ correspond to some fixed point, at which we impose the initial condition in determining the potential 
$\chi_{\mathrm{BIZZ}} ^{(0)}$. In contrast, the integration in calculating $\tilde{Q}^{(1)}$ is taken at constant $\tau$ and the initial condition we impose for $\chi_{\mathrm{mon}} ^{(0)}$ changes accordingly. The two versions of $\chi^{(0)}$ are then related by
\begin{align*}
\chi_{\mathrm{mon}} ^{(0)} (\tau, \sigma ) = \chi_{\mathrm{BIZZ}} ^{(0)} (\tau , \sigma) 
	- \int _{(\tau_0,0)} ^{(\tau,0)} \ast j \, , 
\end{align*}
such that the level-1 charges are related by
\begin{align*}
\tilde{Q}_{\mathrm{mon}} ^{(1)} (\tau)  = \tilde{Q}_{\mathrm{BIZZ}} ^{(1)} (\tau) 
	- \Big( \int  _{(\tau_0,0)} ^{(\tau,0)} \ast j \Big) Q^{(0)} \, ,
\end{align*}
which implies that 
\begin{align}
\partial _\tau \, \tilde{Q}_{\mathrm{mon}} ^{(1)} (\tau) = \left[Q^{(0)} \, , \, j_\sigma (\tau , 0 ) \right] .
\end{align}
The above relation differs from the respective relation \eqref{Conservation:BIZZ} for the BIZZ construction by the appearance of the commutator, which is crucial for the construction of the conserved charges in the case of the closed string. In order to discuss this construction, we consider the $\tau$-derivative of the monodromy $t_u$, which can be obtained from noting that $t_u$ is the monodromy for the auxiliary linear problem, i.e.\
\begin{align}
\chi_u (\tau , 2 \pi ) = \chi_u (\tau, 0) \, t_u \, .
\end{align}
Here, we have used that $t_u$ is constructed from the solution of the auxiliary linear problem along 
$\gamma$ and that multiplying with $\chi_u (\tau, 0)$ ensures the appropriate initial condition. We can now take the $\tau$-derivative to find
\begin{align}
\chi_u (\tau, 0) \, t_u \, \ell_{u , \tau} (\tau , 0) 
 	= \chi_u (\tau, 0) \, \ell_{u , \tau} (\tau , 0) \, t_u
 	+ \chi_u (\tau, 0) \, \partial_\tau t_u \, , 
\end{align}
which implies the \textit{evolution equation} for the monodromy $t_u$, 
\begin{align}
 \partial_\tau t_u (\tau)  = \left[t_u (\tau) \, , \, \ell_{u , \tau} (\tau , 0) \right] \, .
\end{align}
Equations of this type are well-known in the study of integrable systems, where they appear for the Lax pair. We can formally solve the above relation by noting that the monodromies for different values of $\tau$ are related by a similarity transformation, 
\begin{align}
t_u (\tau) = S_u(\tau , \tau_0) ^{-1}  \, t_u (\tau_0) \, S_u(\tau , \tau_0) \, , 
\end{align}  
where $ S_u(\tau , \tau_0)$ is defined by the relations
\begin{align}
\partial_\tau S_u(\tau, \tau_0 ) &= S_u(\tau, \tau_0 ) \, \ell_{u , \tau} (\tau , 0) \, , & 
S_u(\tau_0 , \tau_0 ) &= \unit \, .
\end{align}
This shows that the spectrum of $t_u (\tau)$ is $\tau$-independent, and we can hence employ the $u$-dependent eigenvalues of $t_u$ to construct the conserved charges. The expansion of the monodromy $t_u$ or its eigenvalues is the typical approach to construct an infinite tower of conserved charges in an integrable string theory and is important in the construction of the spectral curves employed in integrability calculations on either side of the AdS/CFT correspondence, cf.\ e.g.\ references \cite{Kazakov:2004qf,Kazakov:2004nh,Beisert:2004ag,SchaferNameki:2004ik,Beisert:2005bm}.

Let us conclude by noting that in a two-dimensional field theory the vanishing of the currents at spatial infinity implies that $t_u$ (taken from $-\infty$ to $+\infty$) is conserved. The study of minimal surfaces is thus in some sense more similar to the two-dimensional field theory than to string theory.  

\section{Master Symmetry}
\label{sec:Master}
In this section, we introduce the \emph{master symmetry}, which maps a solution $g$ of the equations of motion to another solution $g_u$ of the equations of motion. Such transformations are often called B{\"a}cklund or dressing transformations. However, the precise relation between the master symmetry and the B{\"a}cklund transformations considered in the literature remains to be made explicit. The starting point for this analysis was the observation made in references\cite{Ishizeki:2011bf,Kruczenski:2013bsa}, which constructed spectral parameter deformations for minimal surfaces in Euclidean $\AdS_3$ based on a Pohlmeyer reduction of the corresponding string theory. The symmetry had however been discussed earlier in reference \cite{Eichenherr:1979ci} for general symmetric space models, where it was used to construct an infinite set of conserved charges. In fact, the master symmetry can be applied to derive an infinite set of symmetry variations as well; this is discussed in section \ref{sec:IntCompl}. Here, we study the map $g \mapsto g_u$ and discuss its properties. 

The symmetry is based on the observation that the action, the equations of motion and the Virasoro constraints remain unaltered upon replacing $U \mapsto L_u$. For the action, this follows immediately by replacing
\begin{align}
a \; \; \mapsto \; \; a_u = \frac{1-u^2}{1+u^2} \, a 
	- \frac{2u}{1+u^2} \, \ast a \, ,
\end{align}
which implies that
\begin{align}
S_u = - \frac{T}{2} \int \tr ( a_u \wedge \ast a_u ) 
	= - \frac{T}{2}\, \frac{(1-u^2)^2 + 4 u^2}{(1+u^2)^2} 
	\int  \tr \left( a \wedge \ast a \right) = S \, .
\end{align}
In the case of the Virasoro constraints, we restrict to conformal gauge and employ relation 
\eqref{eqn:deform-a} to find
\begin{align}
\tr \left( a_{\theta,z} \, a_{\theta,z} \right) &= 
	e^{-2i \theta} \tr \left( a_{z} \, a_{z} \right) = 0 \, , & 
\tr \left( a_{\theta,\zb} \, a_{\theta,\zb} \right) &= 
	e^{2i \theta} \tr \left( a_{\zb} \, a_{\zb} \right) = 0 \, .	
\end{align}
The invariance of the equations of motion follows by making use of the flatness condition \eqref{Uflatcomp}: 
\begin{align}
\diff \ast a_u + \ast a_u \wedge A_u + A_u \wedge \ast a_u =& \frac{1-u^2}{1+u^2} \left( \diff \ast a + \ast a \wedge A + A \wedge \ast a \right) = 0 \, .
\end{align}
We can carry over the transformation $U \mapsto L_u$ to the fundamental fields $g$ by imposing the differential equation
\begin{align} 
g_u^{-1} \diff g_u &= L_u \, , &
g_u (z_0) &= g (z_0) \, ,
\label{eqn:def-deformed-sol}
\end{align}
where we have have demanded that the transformation has a fixed point $z_0$ on the worldsheet.   We note that in order to think of the deformation \eqref{eqn:deform-a} as a symmetry transformation of physical solutions, we need to impose a reality condition on $L_u$ which leads to the restriction that $u \in \mathbb{R}$.
In the following, we will refer to this transformation as the \emph{master symmetry} due to its property to map conserved charges to conserved charges and to generate infinite towers of nonlocal symmetries of the model.  

A convenient description of the transformed solution is obtained by writing the deformed solution as the left-multiplication of $g$ by some $\mathrm{G}$-valued function $\chi_u$ that mediates between the original and the transformed solution. A quick analysis shows that this function is indeed given by $\chi_u$. The transformation described by equation 
\eqref{eqn:def-deformed-sol} can thus be rewritten as 
\begin{align}
g_u &= \chi_u \, g \, , 
&
\diff \chi_u &= \chi_u \, \ell _u \, ,  
&
\chi_u(z_0) &= \unit \, .
\label{eqn:deformed-sol-chi}
\end{align}
In this form, the transformation was described for general symmetric space models in reference \cite{Eichenherr:1979ci}.  

The infinitesimal version of the master symmetry transformation is thus given by
\begin{align} 
\delC g = \dot{g}_u \vert _{u=0} = \dot{\chi}_u \vert_{u=0} \, g 
	= \chi^{(0)} g \, ,
\label{eqn:C-variation}
\end{align}
where we introduced the dot to denote $u$-derivatives. The symbol $\chi^{(0)}$ represents the coefficient of the linear term in the Taylor expansion \eqref{eqn:Taylor-chi}, which satisfies
\begin{align}
\diff \chi ^{(0)} =  \ast j \, .
\label{eqn:dchi0}
\end{align}
We have thus identified the function $\chi^{(0)}$, which generates the master symmetry as the potential of the G-symmetry Noether current. We note that the master symmetry acts on the Maurer--Cartan form $U$ as
\begin{align}
\master \, U = - \ast a = - \left( P_\alg{m} \circ \ast \right) U \, .
\end{align}
It hence reproduces the map $\Sigma$, which was employed in reference \cite{Beisert:2012ue} to construct the Lax connection of symmetric space models by exponentiation. There, it was observed that the exponentiation construction can be extended to a wider class of models. 

Before turning to the discussion of infinitesimal symmetries, we consider the properties of large master symmetry transformations. Changing the base point where the initial condition is imposed from $z_0$ to $z_1$ corresponds to a global $\mathrm{G}$-transformation of the solution from the left by $\chi_u(z_1)^{-1}$. This can be shown by using the uniqueness of the solution to equation \eqref{eqn:deformed-sol-chi}. If we denote by $\chi_u ^{z_1}$ the function defined by the differential equation
\begin{align}
\diff \chi_u ^{z_1} &= \chi_u ^{z_1} \, \ell_u \, , &
\chi_u ^{z_1} (z_1 ) &= \unit \, , 
\label{eqn:dchiz1}
\end{align}
and likewise for $\chi_u ^{z_0}$, 
we see that both $\chi_u ^{z_1}$ and 
$\chi_u^ {z_0} (z_1)^{-1} \chi_u ^{z_0}$ satisfy relation \eqref{eqn:dchiz1}. The uniqueness of the solution to this equation then leads us to conclude that the two solutions are equal.  

We can study the relation of the master symmetry to $\grp{G}$-symmetries as well as the concatenation of two master symmetry transformations in a similar fashion. For this purpose, it is convenient to denote the map $g \mapsto g_u$ by $g_u = M_u (g)$. 
Let us first note that the master symmetry commutes with the $\grp{G}$-symmetry transformations of the model, i.e.\ that we have
\begin{align}
M_u(L g) = L M_u(g) \, .
\end{align}
This may be concluded from the fact that the Maurer--Cartan current $U = g^{-1} \diff  g$ is invariant under $g \mapsto L  g$ using again the uniqueness of the solution to \eqref{eqn:def-deformed-sol}, once a boundary condition is specified.

The structure appearing for the concatenation of two master symmetry transformations is particularly clear in terms of the angular spectral parameter $\theta$ introduced in equation \eqref{eq:theta}. If we take $L_{\theta_1}$ to be the Maurer--Cartan current of a deformed solution $g_{\theta_1}$ and calculate the Lax connection for a different angle $\theta_2$, we obtain the Lax connection 
\begin{align*}
\left( L_{\theta_1} \right) _{\theta_2} = 
	A + \cos \theta_2 \left( \cos \theta_1 \, a + \sin \theta_1 \ast a \right)
	+ \sin \theta_2 \ast \left( \cos \theta_1 \, a + \sin \theta_1 \ast a \right)
	= L_{\theta_1 + \theta_2} \, .
\end{align*}
This structure can again be carried over to the solutions $g_\theta$ by using the uniqueness of the solution of the defining relation \eqref{eqn:def-deformed-sol}, for which we correspondingly obtain the relation
\begin{align}
\left( M_{\theta_1} \circ M_{\theta_2} \right) (g) = M_{\theta_1 + \theta_2} (g) \, .
\end{align}
In order to express this relation in terms of the parameter $u$, $ M_{u_1} \circ M_{u_2} = M_{u_3}$, we need to solve
\begin{align}
e^{i \theta_1 + i \theta_2} = \frac{1-iu_1}{1+iu_1}\frac{1-iu_2}{1+iu_2} = \frac{1-iu_3}{1+iu_3} = e^{i\theta_3}
\end{align}
to find the following composition rule for the spectral parameter $u$:
\begin{align} \label{eqn:addition-theorem}
u_3 = u_1 \oplus u_2 = \frac{u_1+u_2}{1-u_1 u_2} \; .
\end{align}
From this rule, we obtain a formula for the variation of the deformed solution $g_u$,
\begin{align} \label{eqn:addition-theorem-infinitesimal}
  \master g_u = \lreval{ \frac{\diff}{\diff u'} \, g_{u\oplus u'} }_{u'=0} 
  = \frac{\diff g_u}{\diff u} \lreval{ \frac{\diff}{\diff u'} \frac{u+u'}{1-u u'} }_{u'=0} 
  = (1+u^2) \frac{\diff}{\diff u} \, g_u \; .
\end{align}
We can translate this relation into an expression for the variation of $\chi_u$ under $\delC$, 
\begin{align}
 \master \left( \chi_u \, g \right) = (1+u^2) \frac{\diff}{\diff u} \, \left( \chi_u \, g \right)
 \quad \Rightarrow \quad 
 \delC \chi_u = \left(1 + u^2 \right) \frac{\diff}{\diff u} \chi_u - \chi_u \cdot \chi^{(0)} \, . \label{eqn:delchi}
\end{align}

Let us conclude again by commenting on the master symmetry in the case of closed strings. As we have noted before, the key difference there is that the solution $\chi_u$ of the auxiliary linear problem is no longer periodic. The deformed solutions $g_u$ would thus typically violate the periodic boundary conditions. In other words, applying a master symmetry transformation breaks open the closed strings. However, this finding does not exclude the possibility that the master symmetry generates a recursion on the conserved charges obtained from the eigenvalues of the monodromy also for the case of closed strings. 

\section{Master Symmetry of Principal Chiral Models}
\label{sec:PCM}

We take a short excursion and discuss the master symmetry for principal chiral models, which will be employed in the discussion of certain minimal surfaces in $\AdS_3$. Principal chiral models are a well-known class of integrable models. In the case of a string theory, the target space is given by a group manifold $\grp{G}$. The basic variables of a principal chiral model are matrices $g_p(\tau , \sigma )$ taking values in some representation of a Lie Group $\grp{G}$. We define the left and right Maurer--Cartan forms to be given by
\begin{align}
U_p ^r &= g_p ^{-1} \diff g_p \, , & U_p ^l &= - \diff g_p g_p ^{-1} \, .    
\end{align}  
Both currents are flat by construction, $\diff U_p ^{r/l} + U_p ^{r/l} \wedge U_p ^{r/l} = 0$. The classical principal chiral model is defined by the action
\begin{align}
S = - \frac{T}{2} \int \tr \big( U_p ^r \wedge \ast U_p ^r \big) 
	= - \frac{T}{2} \int \tr \big( U_p ^l \wedge \ast U_p ^l \big) \, .   \label{eqn:PCM_action}
\end{align}
The action is invariant under left- and right-multiplication of $g$ by elements of $\grp{G}$ and the currents $U_p ^r$ and $U_p ^l$ can be identified as the respective Noether currents. The equations of motion for the action \eqref{eqn:PCM_action} are given by
\begin{align}
\diff \ast U_p  ^r = 0 \quad \Leftrightarrow \quad \diff \ast U_p  ^l = 0 \, .
\end{align}
Both currents can be deformed to obtain a Lax connection, which is flat if the equations of motion are satisfied,
\begin{align}
L_{p\, u} ^{r/l} = \frac{u^2}{1+u^2} \,  U_p ^{r/l} + \frac{u}{1+u^2} \, \ast U_p ^{r/l} \, . \label{eqn:Lax_PCM}
\end{align}
For the symmetric space model, the master symmetry transformation is obtained from deforming $g$ to $g_u$ in such a way that the Maurer--Cartan current associated to $g_u$ is the Lax connection. If one defines the master transformation for the principal chiral model in the same way, one does not obtain a symmetry of the action since
\begin{align*}
\tr \big( L_{p\, u} ^r \wedge \ast L_{p\, u} ^r \big) \neq \tr \big( U_p ^r \wedge \ast U_p ^r \big) \, .
\end{align*}
In order to obtain the appropriate definition for the master symmetry transformation, we consider a specific symmetric space model, which reproduces the principal chiral model on G. This connection was explained in reference \cite{Eichenherr:1979ci} and is based on considering a symmetric space model with Lie group $\grp{G}_s = \grp{G} \times \grp{G}$. The basic variables are matrices $g_s$, which we represent as
\begin{align*}
g_s = \begin{pmatrix}
g_1 & 0 \\ 0 & g_2
\end{pmatrix}  .
\end{align*} 
Here, $g_1$ and $g_2$ take values in G. We introduce an automorphism $\sigma: \grp{G}_s \to \grp{G}_s$ given by
\begin{align}
\sigma \left( g_s \right) &= M g_s M^{-1} = \begin{pmatrix}
g_2 & 0 \\ 0 & g_1 
\end{pmatrix} , & 
M &= \begin{pmatrix}
0 & \unit \\ \unit & 0 
\end{pmatrix} .
\end{align}
The set of fixed points of $\sigma$ is the diagonal subgroup 
\begin{align*}
\grp{H} = \left \lbrace \begin{pmatrix}
g & 0 \\ 0 & g 
\end{pmatrix} \, : g \in \grp{G} \right \rbrace \, ,
\end{align*}
which appears as the gauge group in this context. We have
\begin{align*}
a_s = \frac{1}{2} \begin{pmatrix}
U_1 - U_2 & 0 \\ 0 & U_2 - U_1 
\end{pmatrix} \, , 
\end{align*}
and correspondingly the Lagrangian of the symmetric space model is given by
\begin{align*}
\mathcal{L}_s = \tr \big( a_s \wedge \ast a_s \big) = \frac{1}{2} \tr \big( \left(U_1 - U_2 \right) \wedge \ast \left(U_1 - U_2 \right) \big) \, .
\end{align*}
We can hence identify the symmetric space model with a principal chiral model by setting
\begin{align}
g_p = g_2 g_1 ^{-1} \, .
\end{align}
This leads to $U_p ^r = g_1 \left( U_2 - U_1 \right) g_1 ^{-1}$ and thus we have
\begin{align}
\mathcal{L}_p = \tr \big( U_p^r \wedge \ast U_p ^r \big) = 2 \, \mathcal{L}_s \, .
\end{align}
For symmetric space models, large master symmetry transformations can be formulated as
\begin{align*}
g_{s , u} &= \chi_u \cdot g_s \, , &  \diff \chi_u  &= \chi_u  \ell_u \, .
\end{align*}
Here, $\ell_u$ is the Lax connection
\begin{align}
\ell_u &= \left( \frac{u^2}{1+u^2} + \frac{u}{1+u^2} \,  \ast \right) 
	\left( - 2 g_s \, a_s \, g_s ^{-1} \right)  \\
	 &= \left( \frac{u^2}{1+u^2} + \frac{u}{1+u^2} \,  \ast \right) \!
	 \begin{pmatrix}
		g_1 \left(U_2 - U_1 \right) g_1 ^{-1} & 0 \\
		0 & g_2 \left(U_1 - U_2 \right) g_2 ^{-1} 
	\end{pmatrix} \!
	= \! \begin{pmatrix}
		L^r _p & 0 \\
		0 & L^l _p
	\end{pmatrix} . \nn
\end{align}
Correspondingly we have
\begin{align*}
g_{1 \, u} &= \chi ^r _u \cdot g_1 \, , & g_{2 \, u} &= \chi ^l _u \cdot g_2 \, , 
\end{align*}
where $\chi ^{r/l}_u $ are defined by the auxiliary linear problems
\begin{align*}
\diff \chi ^r_u  &= \chi _u ^r \cdot L^r _{p \, u} \, , & \diff \chi _u ^l &= \chi_u ^l \cdot L^l _{p \, u} \, .
\end{align*}
The master symmetry transformation for the principal chiral model is thus given by
\begin{align}
g_{p \, u} = \chi _u ^l \cdot g_p \cdot \chi ^r_u  {} ^{-1} \, ,
\end{align}
and for the associated variation we have
\begin{align}
\delC  g_p = \chi ^{l , (0)} \cdot g_p - g_p \cdot \chi ^{r , (0)} \, ,
\end{align}
where $\chi ^{r/l , (0)}$ are the potentials for the left and right Noether currents, 
\begin{align}
\chi ^{r/l , (0)} = \int \ast U_p ^{r/l} \, .  
\end{align}

\section{Integrable Completion}
\label{sec:IntCompl}
In section \ref{sec:Integrability}, we have outlined different methods to obtain an infinite tower of conserved charges for symmetric space models. In this section, we discuss another option, namely to employ the master symmetry discussed above to construct towers of conserved charges, which is the approach taken in reference  \cite{Eichenherr:1979ci}. We derive the relation between these charges and the ones obtained from the BIZZ recursion below. Moreover, we demonstrate explicitly that the master symmetry can be employed to deform \emph{any symmetry variation} $\delta_0$ into a one-parameter family of symmetries $\delta_{0,u}$. Given some symmetry variation $\delta_0$ with associated conserved current $j_0$, we can then apply the master symmetry to deform both the symmetry variation $\delta_0$ and the associated conserved current $j_0$ into one-parameter families of variations $\delta_{0,u}$ and conserved currents $j_{0,u}$, respectively. It is then natural to ask whether the deformed variations and currents are related by the Noether procedure. However, since the deformed variations are typically on shell and an off-shell continuation is not always known, we can establish this relation only formally. In a sense which we will make more explicit later on, we have the following picture:  
\begin{displaymath}
\begin{tikzcd}[column sep=7em,row sep=7em]
    \delta_0 \arrow{r}{\displaystyle{\text{Noether}}} 
    \arrow{d}[sloped,below]{\displaystyle{\text{Completion}}} 
    & Q_0 \arrow{d}[sloped,above]{\displaystyle{\text{Master}}} \\ 
    \delta_{0,u}  \arrow{r}[swap]{\displaystyle{\text{Noether}}} & Q_{0,u}       
\end{tikzcd}
\end{displaymath}
Since the parameter $u$ generated by the master symmetry is the spectral parameter underlying the integrability of the model, we refer to this procedure as the \emph{integrable completion} of the symmetry $\delta_0$ and its associated current $j_0$, respectively.
In particular, this procedure applies to the master symmetry itself, which results in a one-parameter family of master symmetries $\master_u$ with associated conserved charges.

\subsection{Generic Symmetries}
\label{sec:GenSymm}
We begin by discussing the integrable completion of a generic symmetry. In particular, we show how to obtain a one-parameter family of symmetry variations from a given symmetry variation $\delta_0$.  

\paragraph{Completion of conserved currents and charges.}
The symmetry variation of a Noether current must also be a conserved current. This is due to the fact that the equations of motion are invariant under the variation and that the conservation of the Noether current is equivalent to the equations of motion. In particular this holds for the master variation $\master$ applied to any conserved current $j_0$, i.e.\ we have the conservation equation
\begin{align}
\diff \ast \master j_0 = 0 \, ,
\end{align} 
which implies that the charge 
\begin{align}
\master Q_0=\int  \ast \master j_0
\end{align}
associated with this current is time-independent, at least if the current $\master j_0$ is again periodic, which we have seen to be the case for minimal surfaces. Instead of acting multiple times with the master symmetry variation, we can apply a large master symmetry transformation to the respective currents and charges by making the replacement $g \mapsto g_u=\chi_u g$ in the definition of any current $j_0$ or charge $Q_0$, respectively. This gives a one-parameter family of conserved currents and charges,
\begin{align}
j_0 \mapsto j_{0,u} &= j_0 \vert_{g \mapsto g_u} \, , &
	Q_0 \mapsto Q_{0,u} &= Q_0 \vert_{g \mapsto g_u} \, .
\label{eq:onepjJ}
\end{align}
We will refer to the above currents and charges as the integrable completions of the current $j_0$ or the charge $Q_0$, respectively.

\paragraph{Completion of Symmetry Variations.}
For any given variation $\delta_0 g$ of the field $g$, we introduce a one-parameter family of variations $\delta_{0,u} g$, which we define by 
\begin{align}
\delta_{0,u} g = \chi_u^{-1} \delta_0 ( \chi_u g) .
\label{eq:defvariation}
\end{align}
Below, we show that if $\delta_0 g$ is a symmetry of the equations of motion, then 
$\delta_{0,u} g$ will be a symmetry of the equations of motion as well, i.e.\ we show that
\begin{align}
\diff \ast \delta_0 j = 0 \quad \Rightarrow \quad 
\diff \ast \delta_{0,u} j = 0 \, .
\end{align} 
We note that the quantity $\chi_u$ is inherently on shell since its definition requires the Lax connection $\ell_u$ to be flat, which is equivalent to the equations of motion. Consequently, we are not in a position to study the invariance of the action and rather study the equations of motion. 

We begin by deriving a necessary and sufficient criterion for a given variation $\delta g$ to be a symmetry. We define $\eta = \delta g\, g^{-1}$ so that we can write the variation as 
$\delta g = \eta g$ and the induced change of the Maurer--Cartan form as 
$\delta U = g^{-1} \diff \eta \, g$. In order to calculate the variation of the Noether current, it is convenient to write it in the form
\begin{align}
j = - 2 g P_{\alg{m}}(U) g^{-1} = g \bigbrk{\Omega(U) - U } g^{-1} \, .
\end{align}
Since $\Omega$ is a linear map on $\alg{g}$, its action and the variation commute and we find
\begin{align}
\delta j =  -\diff \eta - [j, \eta ] + g \, \Omega \bigbrk{ g^{-1} \diff \eta \, g } g^{-1} \, .
\label{eqn:delta-j}
\end{align} 
Hence, we have a symmetry of the equations of motion, $\diff \ast \delta j = 0$, if and only if
\begin{align}
\diff \ast \bigbrk{ \diff \eta + [j, \eta ] } = \diff \bigbrk{ g \, \Omega \bigbrk{ g^{-1} \ast \diff \eta \, g } g^{-1} } \, .
\label{eqn:step}
\end{align}
In order to reach a more convenient form, we need to rewrite the right hand side of the above relation. Abbreviating $\omega = g^{-1} \ast \diff \eta g$, we have
\begin{align*}
\diff \omega = g^{-1} \diff \ast \diff \eta g - U \wedge \omega - \omega \wedge U
\end{align*}
and we find
\begin{align*}
\diff \left( g \, \Omega(\omega) g^{-1} \right) 
	= g \left( U \wedge \Omega(\omega) + \Omega(\omega) \wedge U 
	+ \Omega \left( g^{-1} \diff \ast \diff \eta g - U \wedge \omega - \omega \wedge U \right)
	\right) g^{-1} \, .
\end{align*}
We can rewrite this further by noting that since 
$ \omega \wedge \rho + \rho \wedge \omega$ leads to a commutator and $\Omega$ is an involutive automorphism on $\alg{g}$, we have
\begin{align*}
U \wedge \Omega (\omega) + \Omega (\omega)  \wedge U
	= \Omega \left( \Omega(U) \wedge \omega
	+ \omega  \wedge \Omega(U) \right) , 
\end{align*}
such that 
\begin{align*}
\diff \left( g \, \Omega(\omega) g^{-1} \right) 
	&= g\, \Omega \left( g^{-1} \left( \diff \ast \diff \eta 
	+ j \wedge \ast \diff \eta + \ast \diff \eta \wedge j \right) g \right) g^{-1} \\
	& =  g\, \Omega \left( g^{-1} \diff \ast \left( 
	\diff \eta + \left[ j , \eta \right] \right) g \right) g^{-1} \, ,
\end{align*}
where we have used the conservation of $j$ in the last step to find
\begin{align*}
\diff \ast \left[ j , \eta \right] = - \ast j \wedge \diff \eta - \diff \eta \wedge \ast j 
	= j \wedge \ast \diff \eta + \ast \diff \eta \wedge j \, .
\end{align*}
We have thus rewritten the condition \eqref{eqn:step} as
\begin{align}
 g^{-1} \diff \ast \lrbrk{ \diff \eta + \left[ j , \eta \right]  } g
 =  \Omega \bigbrk{ g^{-1} \diff \ast \lrbrk{ \diff \eta + \left[ j , \eta \right]  } g} \, ,
\label{CriterionStep}
\end{align}
which states that
\begin{align}
  g^{-1} \diff \ast \lrbrk{ \diff \eta + \left[ j , \eta \right]  } g \in \alg{h} \, .
 \label{eqn:criterion}
\end{align}
This is the sought-after necessary and sufficient condition for $\delta g = \eta g$ to be a symmetry of the model. In most of the cases we consider, this condition is actually satisfied in the stronger form
\begin{align}
  \diff \ast \lrbrk{ \diff \eta + \left[ j , \eta \right] } = 0 \, .
 \label{eqn:criterion-strong}
\end{align}

With this criterion at hand, we return to the variation \eqref{eq:defvariation} which is given by $\delta_{0,u} g = \chi_u^{-1} \delta_0 ( \chi_u g)$. By Leibniz' rule we obtain
\begin{align}
\delta_{0,u} g = \delta_0 g + (\chi_u^{-1} \delta_0 \chi_u ) g \, .
\end{align}
Hence, the total variation splits into a part $\delta_0 g$, which is a symmetry  by assumption, and the part $\delta'_{0,u} g = \eta g$ with $\eta = \chi_u^{-1} \delta_0 \chi_u$. We now demonstrate that also the second part is a symmetry by showing that \eqref{eqn:criterion-strong} is satisfied. We begin by noting that
\begin{align} 
\diff \eta = \diff (\chi_u^{-1} \delta_0 \chi_u) 
	= -\chi_u^{-1} \diff \chi_u \, \chi_u^{-1} \delta_0 \chi_u 
	+ \chi_u^{-1} \delta_0 \diff \chi_u 
	= \delta_0 \ell_u + [\eta,\ell_u] \, ,
\label{eqn:conjug-deta}
\end{align}
where we used $\diff \chi_u = \chi_u \ell_u$ as given in \eqref{eqn:deformed-sol-chi}. Then, it follows that
\begin{align}
\diff \eta + [ j , \eta ] 
	= \delta_0 \ell_u + [\eta,\ell_u-j] 
	= \delta_0 \ell_u + \frac{1}{u} [\eta,\ast\ell_u]  \, ,
\label{eq:zwischenschritt}
\end{align}
using equation \eqref{eqn:jfroml}, 
$\ast \ell_u = u ( \ell_u - j )$, in the last step. Taking the divergence of this equation yields the terms
\begin{align} 
\diff \ast \left( \diff \eta + [ j , \eta ] \right)
   &= \diff \ast \delta_0 \ell_u - \frac{1}{u} \diff [\eta,\ell_u] \nn \\
   &= \diff \ast \delta_0 \ell_u 
   - \frac{1}{u} \left( \diff\eta\wedge\ell_u + \ell_u\wedge\diff\eta \right) 
   - \frac{1}{u} [\eta,\diff\ell_u] \, . 
\label{eqn:conjug-test1}
\end{align}
Within the middle terms on the right hand side of the above relation, we replace $\diff\eta$ again with the help of \eqref{eqn:conjug-deta}, which gives
\begin{align}\label{eq:expthree}
  \diff\eta\wedge\ell_u + \ell_u \wedge \diff\eta & = \delta_0 \ell_u \wedge \ell_u + \ell_u \wedge \delta_0 \ell_u + [\eta,\ell_u] \wedge \ell_u + \ell_u \wedge [\eta,\ell_u] \nn \\
  &= \delta_0 \lrbrk{ \ell_u \wedge \ell_u } + [\eta,\ell_u \wedge \ell_u ] 
   = - \delta_0 \, \diff \ell_u - [\eta,\diff\ell_u ]\,.
\end{align}
Using once more that $\ast \ell_u = u ( \ell_u - j )$ or equivalently
$\ell_u + u \ast \ell_u = \ast j$, we then find
\begin{align}
\diff \ast \lrbrk{ \diff \eta + [ j , \eta ] } 
	= \diff \, \delta_0 \Big( \ast \ell_u + \frac{1}{u} \, \ell_u \Big) 
	= \frac{1}{u} \, \diff \ast \delta_0 j = 0 \, .
\end{align}
Thus, the condition \eqref{eqn:criterion-strong} is satisfied and the integrable completion \eqref{eq:defvariation} of a symmetry variation $\delta_0$ indeed furnishes a one-parameter family of symmetry variations.

Let us note here that the only symmetry variation which only satisfies the criterion 
\eqref{eqn:criterion} rather than the stronger form \eqref{eqn:criterion-strong} is the master symmetry variation itself, for which we have
\begin{align}
g^{-1} \diff \ast \lrbrk{ \diff \chi^{(0)} + \left[ j , \chi^{(0)}  \right]  } g 
	= - 4 \, a \wedge a \in \alg{h} \, .
\end{align}

\paragraph{Noether procedure and on-shell symmetries.}
We would now like to derive conserved charges, which are associated to the nonlocal symmetry transformations discussed above.  Let us point out that due to the definition of $\chi_u$ all of the higher symmetry transformations are inherently on shell. Carrying out Noether's procedure strictly would require to continue the symmetry variations to off-shell symmetries of the action. This was done in references \cite{Dolan:1980kz,Dolan:1981fq,Hou:1981hn} for the Yangian-like symmetries of principal chiral models and it seems plausible that it could also be done for symmetric space models. Here, we will be satisfied with deriving on-shell expressions for conserved currents. 
Let us clarify, how these currents are related to the currents one would derive from a (hypothetical) off-shell continuation of the underlying symmetry via Noether's procedure.   

Suppose we had found a way to extend the symmetry transformations discussed above off shell. This would involve finding off-shell expressions for the quantity $\chi_u$, which can e.g.\ be done as in references \cite{Dolan:1980kz,Hou:1981hn} by fixing specific paths from any point on the worldsheet 
(or spacetime in their case)
to a common starting point and by defining $\chi_u$ to be the solution of $\diff \chi_u = \chi_u \ell_u$ along this path. This implies that the continued $\chi_u$ satisfies
\begin{align*}
\diff \chi_u = \chi_u \ell _u + f_u \, , 
\end{align*}
where $f_u$ is some one-form which vanishes on-shell. By assumption, the variation of the Lagrangian can be written as a total derivative, $\delta \mathcal{L} = \diff \ast k $. Hence, $k$ represents the contribution to the Noether current which would follow from the off-shell symmetry.

Since we do not have an off-shell continuation of the above symmetries at hand, we simply perform a formal calculation where we use $\diff \chi_u = \chi_u \ell_u$, but we will not use the equations of motion otherwise. For the symmetries we consider, one can show that in this way we obtain $\delta \mathcal{L} = \diff \ast k^\prime$. That is, $k^\prime$ represents the on-shell contribution to the current following from our formal procedure.
It is then clear that $\diff \ast \left( k - k ^\prime \right)$ will be proportional to $f_u$, which vanishes on-shell. Hence, $\left( k - k ^\prime \right)$ vanishes on shell up to the usual freedom in reading off $\ast k$ from $\diff \ast k$. We thus see that the conserved current we derive would agree on shell with the Noether current one would find if one continued the symmetry off shell and carried out Noether's procedure. 

\subsection{Yangian Symmetry}

In this subsection we discuss the first nontrivial example of the above integrable completion via the master symmetry. In particular, we demonstrate that the completion of the Lie algebra symmetry \eqref{eq:Liesymmetry} of symmetric space models yields a tower of nonlocal Yangian symmetries. 


\paragraph{Completion of conserved currents and charges.}
The application of the master symmetry variation to the conserved current $j$ introduced in \eqref{eqn:Noethercurrent} yields
\begin{align}
\master j = -2 \ast j + [\chi^{(0)}, j ],
\label{MasterJ}
\end{align}
which is indeed a nonlocal conserved current. The corresponding charge 
\begin{align}
Q^{(1)} := \int \ast \master \,j =2 \int j + \int \limits_{\sigma_1 < \sigma_2} [\ast j_1, \ast j_2 ] 
\label{MasterQ}
\end{align}
takes the standard form of a level-1 Yangian charge. We will demonstrate in section \ref{sec:Algebra} that these charges indeed obey the commutation relations of the Yangian algebra. Higher conserved currents and charges can be constructed by repeated application of $\master$. However, since we know the large transformation generated by $\master$, we need not carry out this cumbersome procedure. Exponentiating the variation $\master$ essentially transforms $g$ into $g_u$ and all derived quantities like the Noether current $j$ transform accordingly. The higher charges to be constructed from $Q$ by repeated application of $\master$ should thus be contained in the one-parameter family of conserved charges obtained from the transformed solutions $g_u$, 
\begin{align}
\label{def:chargeY}
Q _u &= \int \ast j_u \, ,  
&
j_u &= - 2 g_u \, a_u \, g_u ^{-1} \, = \frac{1-u^2}{1+u^2} \, \chi_u \, j \, \chi_u ^{-1} - \frac{2u}{1+u^2} \, \chi_u \, \ast j \, \chi_u ^{-1} \, . 
\end{align}
A more precise relation can be established by applying equation \eqref{eqn:addition-theorem-infinitesimal} to obtain a recurrence relation for the coefficients of the Taylor expansion of $Q_u$. Since $\Omega$ is a linear map on $\alg{g}$ the variation $\delC$ acts on $a_u = P_\alg{m} ( g_u ^{-1} \diff g_u) $ in the same way as on $g_u$, such that 
\begin{align}
\master \, Q _u= (1+u^2) \frac{\diff}{\diff u} Q_u \, .
\end{align}
The relation takes a simpler form for the angular spectral parameter $\theta$ introduced in \eqref{eq:theta} via the relation $e^{i\theta} = \frac{1-iu}{1+iu}$. In terms of this parameter, the relation reads
\begin{align}
\master \, Q_\theta= \frac{\diff }{\diff \theta} Q_\theta  \, ,
\end{align}
which makes manifest that the master symmetry generates the spectral parameter. Defining the charges $Q^{(n)}$ to be the coefficients in the Taylor expansion 
\begin{align*}
Q_\theta = \sum \limits _{n=0} ^{\infty}  \frac{\theta^n}{n!} \, Q ^{(n)} \, ,
\end{align*}
we find the recurrence relation
\begin{align}
\master\, Q^{(n)}= Q^{(n+1)}.
\end{align}

\paragraph{Relation to BIZZ charges.} 
It is evident from comparing the expressions \eqref{MasterQ}  
and \eqref{BIZZQ1} that the charges obtained from the application of the master symmetry and the BIZZ recursion differ from each other. However, since both charges are based on the quantity $\chi_u$, one should expect that they are related to each other. In fact, such a relation can be established using the variation $\delC$. 
Since the charges themselves vanish, we compare the expressions for the conserved currents integrated over open contours. The generating functional for the BIZZ currents is given by 
$\tilde{j}_u = \ast \diff \chi_u $ and hence the function 
$\chi_u$ describes the integrals over open contours. 
The relation to the respective integral over $j_u$ is given by
\begin{align}
\left(1 + u^2 \right) \dot{\chi}_u (z, \bar{z}) \chi_u ^{-1} (z, \bar{z}) = \int \limits _{z_0} ^z \ast j_u \, . \label{eqn:chiandjt}
\end{align}
Note first that the two sides of this equation are the same for $u=0$ since $\diff \chi^{(0)}= \ast j$. In order to prove equality for any value of $u$, we can employ the variation $\delC$ to construct a recurrence relation for the Taylor coefficients on either side of equation \eqref{eqn:chiandjt}. We have already seen that 
\begin{align}
\delC \, j_u = \left( 1 + u^2 \right) \frac{\diff}{\diff u} \, j_u 
\end{align}
and a simple application of equation \eqref{eqn:delchi} shows that 
\begin{align*}
\delC \left[ \left(1 + u^2 \right) \dot{\chi}_u  \chi_u ^{-1} \right] = \left(1 + u^2 \right) \frac{\diff}{\diff u} \left[ \left(1 + u^2 \right) \dot{\chi}_u  \chi_u ^{-1} \right] \, , 
\end{align*}
which proves the relation \eqref{eqn:chiandjt} and hence shows that the charges $Q^{(n)}$ carry the same information as those obtained from the BIZZ procedure.

\paragraph{Completion of symmetry variations.}

It is in general not straightforward to obtain nonlocal symmetries associated to nonlocal charges. However, the integrable completion of the symmetry variation 
$\delta_{\epsilon}$ provides a natural candidate for the symmetry variations associated to the charges discussed above. In order to find the variation 
$\delta_{\epsilon , u}$, we need to calculate the variation $\delta_\epsilon \chi_u$. We can find this variation e.g.\ from noting that since the left-multiplication $g \mapsto L g$ induces $g_u \mapsto L g_u$, we have $\chi_u \mapsto L \chi_u L^{-1}$ and hence
$\delta_\epsilon \chi_u = \left[ \epsilon , \chi_u \right]$. The integrable completion of the symmetry variation $\delta_\epsilon$ is thus given by
\begin{align}
\delta_{\epsilon,u}  \, g = \chi_u ^{-1} \delta_\epsilon \left( \chi_u \, g \right) 
= \chi_u ^{-1} \epsilon \chi_u \, g 
= \eta _{\epsilon,u} \, g .
\label{eq:Yangvars}
\end{align}
These variations were considered in reference \cite{Schwarz:1995td}, 
where they were uplifted from known symmetry variations of principal chiral models
\cite{Dolan:1980kz,Dolan:1981fq,Devchand:1981wy,Wu:1982jt}. 

\paragraph{Noether procedure for Yangian symmetry.}
We now turn to the derivation of the conserved current for the nonlocal symmetries of Yangian type, which are given by \eqref{eq:Yangvars}.
As for the derivation of the $\grp{G}$-symmetry Noether current, we allow $\epsilon$ to vary over the worldsheet. We thus find the variation of the Maurer--Cartan current to be
\begin{align}
\delta_{\epsilon,u} U 
= g^{-1} \left( \left[ \eta_{\epsilon,u} , \ell_u \right] + \chi_u ^{-1} \diff \epsilon \, \chi_u \right) g\,.
\end{align}
For the variation of the action%
\footnote{For brevity, we drop the factor containing the string tension $T$ in the calculations of the Noether currents.}
we obtain
\begin{align}
\delta_{\epsilon,u} S 
	&= \int \tr \left( \ast j \wedge \left( \left[\eta_{\epsilon,u} , \ell_u \right] 
	+ \chi_u ^{-1} \diff \epsilon \, \chi_u \right) \right) \nn \\
	&= - \int \tr \left( \left( \ast j \wedge \ell_u + \ell_u \wedge \ast j \right) \eta_{\epsilon,u} 
	- \chi_u  \ast j \,  \chi_u ^{-1} \wedge \diff \epsilon \right) \nn \\
	& =  - \int \tr \left( \frac{2 u}{1+u^2} ( j \wedge j ) \, \eta_{\epsilon,u} 
	- \chi_u  \ast j \,  \chi_u ^{-1} \wedge \diff \epsilon \right) \nn \\
	&= \int \tr \left( \frac{2 u}{1+u^2} \left(\chi_u \,  \diff j \,  \chi_u ^{-1}  \right) \epsilon 
	+ \chi_u  \ast j \,  \chi_u ^{-1} \wedge \diff \epsilon \right) , 
\end{align}
where we have inserted the explicit expression \eqref{eqn:deformed-Noether} for $\ell_u$. Note now that 
\begin{align}
\diff \left( \chi_u \,  j \,  \chi_u ^{-1} \right) 
= \chi_u \left( \ell_u \wedge j + j \wedge \ell_u + \diff j \right) \chi_u ^{-1} 
= \frac{1-u^2}{1+u^2} \, \chi_u \,  \diff j \,  \chi_u ^{-1} \, .   
\end{align}
Consequently we have
\begin{align}
\delta_{\epsilon,u}S = \int \tr \left( \frac{2u}{1-u^2} \, \diff \left( \chi_u \,  j \,  \chi_u ^{-1} \right) \, \epsilon + \chi_u  \ast j \,  \chi_u ^{-1} \wedge \diff \epsilon \right) \, ,
\end{align}
and dropping boundary terms, we find
\begin{align}
\delta_{\epsilon,u}  S = \int \tr \left( \chi_u \left(\ast j + \frac{2u}{1-u^2} \, j \right) \chi_u ^{-1} \wedge \diff \epsilon \right).
\end{align}
The Noether current associated to the nonlocal symmetry $\delta_{\epsilon,u} $ is thus given by
\begin{align}
\mathtt{j}_u = \chi_u \left( j - \frac{2u}{1-u^2} \, \ast j \right) \chi_u ^{-1} \, .
\label{eqn:jY}
\end{align}
Comparing this expression with the $\grp{G}$-symmetry Noether current \eqref{def:chargeY} of the transformed solution, we find the relation
\begin{align}
\mathtt{j}_u = \frac{1+u^2}{1-u^2} \, j_u \, .
\end{align}

\subsection{Master Symmetry}
We consider the second nontrivial example of an integrable completion via the master symmetry
and apply the completion to the master variation $\master$ itself, yielding a one-parameter family of master transformations $\master_u$ and associated charges $\chargeC_u$.

\paragraph{Completion of conserved currents and charges.}

We begin by deriving the conserved current for the master symmetry, for which we utilize the same method as before and introduce a coordinate-dependent transformation parameter $\rho=\rho(z,\bar{z})$ into the variation, $\delC g = \rho \chi^{(0)} g$, such that we can read off a conserved current 
$\masterj$ from $\delC S = \int \ast \masterj \wedge \diff \rho$. As a first step we note the variation
\begin{align}
\delC \,U = g^{-1} \diff(\rho\chi^{(0)}) g = -2 \rho \ast a + \diff \rho \, g^{-1} \chi^{(0)} g \, ,
\end{align}
where we used equations \eqref{eqn:dchi0} and \eqref{eqn:Noethercurrent}. The first term leaves the action invariant while the second term produces
\begin{align*}
  \delC S =  2 \int \tr ( \diff \rho \, g^{-1} \chi^{(0)} g \wedge \ast a )
          = -2 \int \tr ( g \ast a g^{-1} \chi^{(0)} ) \wedge \diff \rho
          = \int \tr ( \ast j \chi^{(0)} ) \wedge \diff \rho \, ,
\end{align*}
such that we find the conserved current 
\begin{align}
\masterj = \tr \big( j \chi^{(0)} \big) . 
\label{eq:bbcurrent}
\end{align}
The associated conserved charge takes the form
\begin{align}
\masterQ = \int \ast \, \masterj  = \frac{1}{2} \tr(Q Q)\, ,
\label{eq:chargeC}
\end{align}
and is recognized as the quadratic Casimir of $\mathrm{G}$.
The interesting aspect of this result is not that $\tr(Q Q)$ is conserved, which is obvious since $Q$ is conserved, but the fact that there is a symmetry transformation that has the Casimir as a conserved charge. 

We now consider the integrable completion of the conserved charge associated to the master symmetry. Acting with $\master$ on the current $\masterj$ of \eqref{eq:bbcurrent} gives the conserved master current of level one:
\begin{align}
\master \; \masterj = 
	\tr \big( j \big( 2 \chi^{(1)} - \chi^{(0)\, 2} \big) - 2 \ast j \chi^{(0)}  \big) \, .
\end{align}
The structure of the conserved quantities, however, turns out to be more transparent if one considers the charges directly. Applying the master symmetry variation $\master$ the conserved charge $\chargeC$ defined in \eqref{eq:chargeC} gives the conserved master charge of level one:
\begin{align}
\label{eq:lev1master}
\chargeC ^{(1)} = \tr \big( Q \, Q ^{(1)} \big) \, . 
\end{align}
Employing a large master transformation provides the generating function 
\begin{align}\label{eq:masterchargeu}
\masterQ_u = \sfrac{1}{2} \tr \big( Q_u Q_u \big) \, ,
\end{align}
and switching to the angular spectral parameter $\theta$, we again find the relation
\begin{align*}
\delC \chargeC_\theta = \frac{\diff}{\diff \theta} \chargeC_\theta \, 
\end{align*}
such that the Taylor coefficients $\chargeC ^{(n)}$ of $\chargeC_\theta$ satisfy the recurrence relation
\begin{align}
\delC \, \chargeC ^{(n)} = \chargeC ^{(n+1)} \, .
\end{align}

\paragraph{Completion of symmetry variations.}
Employing relation \eqref{eqn:delchi} for the action of the master symmetry $\master$ on 
$\chi_u$, we find the integrable completion of the master symmetry to be given by 
\begin{align}
\master_u g = \frac{1}{1+u^2} \, \chi_u ^{-1} \master \left( \chi_u \, g \right) =  \chi_u ^{-1} \dot{\chi}_u \, g \, . \label{eq:mastervars}
\end{align}
Again, similar variations were considered by Schwarz in reference \cite{Schwarz:1995td}, who did not, however, consider the master symmetry $\master$ itself. Curiously, some Yangian-like variations $\delta_{\epsilon , u}$ can be obtained from the above variations. Shifting the base point of $\chi_u$ amounts to left-multiplying $\chi_u$ by $\chi_u (z_1)^{-1}$, 
$\chi_u ^\prime = \chi_u (z_1)^{-1} \cdot \chi_u$, 
such that we have
\begin{align}
\master _u  ^{\, \prime} = \master _u + \delta_{\epsilon , u} \, , 
\end{align}
with 
$\epsilon = - \dot{\chi}_u (z_1) \, \chi_u (z_1)^{-1}$. 

\paragraph{Noether procedure.}
We now derive the Noether current associated to the one-parameter family of master symmetries:
\begin{align}
\delC_{u} g = \rho \, \eta _u \, g = \rho\,\chi_u^{-1} \dot{\chi}_u \, g \, .
\end{align}
Here, we have again introduced a coordinate-dependent transformation parameter $\rho$ in order to derive the Noether current. Making use of $\delC U = g^{-1} (  \diff \eta_u  \rho + \eta_u \diff \rho ) g$ we find the variation of the action to be
\begin{align}
\delC S = \int \left \lbrace \tr \left( \ast j \wedge \diff \eta _u \right) \rho + \tr \left( \ast j \eta _u \right) \wedge \diff \rho\right \rbrace \, .
\end{align}
By using that $\diff \eta _u= [ \eta _u , \ell ] + \dot{\ell}$, we can recast
\begin{align*}
\tr \left( \ast j \wedge \diff \eta_u\right) 
& = - \tr \left( \left( \ast j \wedge \ell + \ell \wedge \ast j \right) \eta_u \right) + \tr \left( \ast j \wedge \dot{\ell} \right) \\
&= \frac{2u}{1+u^2} \tr \left( \diff j \, \eta_u + \frac{1}{1+u^2} \ast j \wedge j \right) \, ,  
\end{align*}
and comparing with
\begin{align*}
\diff \tr \left( j \eta_u \right) = \frac{1-u^2}{1+u^2} \, \tr \left( \diff j \, \eta_u+ \frac{1}{1+u^2} \, \ast j \wedge j \right) 
\end{align*}
yields
\begin{align*}
\delC S = \int \left \lbrace \frac{2u}{1-u^2} \, \diff \tr \left( j \, \eta_u \right) \rho+ \tr \left( \ast j \, \eta_u \right) \wedge \diff\rho \right \rbrace 
= \int \tr \left[ \left( \ast j + \frac{2u}{1-u^2} \, j \right) \eta_u \right] \wedge \diff \rho \, .
\end{align*}
From this we read off the Noether current 
\begin{align}
\widehat{\mathtt{j}} = \tr \left( \mathtt{j}_u \, \dot{\chi}_u \chi_u^{-1} \right)
= \frac{1+u^2}{1-u^2} \tr \left( j_u \, \dot{\chi}_u \chi_u^{-1} \right) \, .
\end{align}
By virtue of equation \eqref{eqn:chiandjt}, we conclude that
\begin{align}
\widehat{\mathtt{j}}_u \left( z , \bar{z} \right) = \frac{1}{1-u^2} \tr \left( j_u \left( z , \bar{z} \right) \int _{z_0} ^z \ast j_u \right) \, , 
\end{align}
such that the Noether charge is identified as
\begin{align}
\int \ast \widehat{\mathtt{j}}_u = \frac{1}{1-u^2} \, \chargeC_u \, .
\end{align}
Note that $\master_u$ yields the charge $\chargeC_u$ up to a $u$-dependent factor.

\subsection{Spacetime or Worldsheet Symmetries}
The principle to conjugate a known symmetry $\delta_0$ of the model with $\chi_u$ can also be applied to spacetime or worldsheet symmetries. A general worldsheet symmetry 
$\delta_0=\delta_\text{WS}$ is described by
\begin{align}
\delta_\text{WS} \,g = b^i (\tau , \sigma) \partial_i g \, ,
\end{align}
where the specific form of $b^i$ depends on the specific symmetry.
The conjugation with $\chi_u$ then leads to the variations
\begin{align}
\label{eqn:conjvar}
\delta_{\text{WS},u} \,g = b^i \, l_{u , i} g 
	= b^i  \left( \frac{u^2}{1+u^2} \, j_i - \frac{u}{1+u^2} \, 
	\sqrt{h} \, \eps_{i j} \, h^{j k} j_k \right) g \, ,
\end{align}
where for convenience we write the currents $j$ and $\ast j$ in terms of their components.
Note now that
\begin{align}
j_i \cdot g = - 2 g P_\alg{m} \left( g^{-1} \partial_i g \right) g^{-1} g 
	= - 2 \partial_i g + 2 g P_\alg{h} \left( g^{-1} \partial_i g \right) \, .
\end{align}
We thus observe that the variations \eqref{eqn:conjvar} are merely $u$-dependent linear combinations of the original worldsheet symmetries and gauge transformations.

\begin{table}
\begin{center}
\renewcommand{\arraystretch}{1.6}
\setlength{\tabcolsep}{1.5mm}
\begin{tabular}{| c | l || c | c | c |}
	\cline{3-5}
	\multicolumn{2}{c||}{} & Level-0 & Level-1 & Completion \\
	\hline \hline
	\multirow{2}{*}{{
	\begin{turn}{270} \hspace*{-4mm} Yangian  \end{turn}
	}} 
	& Variation & $\delta_\epsilon g =\epsilon g$
	& $\delta_\epsilon^{(1)}g=\comm{\epsilon}{\chi^{(0)}}g$ 
	& $\delta_{\epsilon,u} g =  \chi_u ^{-1} \epsilon \chi_u \, g  $ \\ \cline{2-5}
    & Charge & $Q^{(0)}\equiv Q = \int \hspace*{-1mm} \ast j$ 
    & $Q^{(1)} =   \int _{<}
    [\ast j_1, \ast j_2 ] + 2 \int \hspace*{-1mm} j $ 
    & $Q_u= \int \hspace*{-1mm} \ast j_u$   \\ \hline \hline 
    \multirow{2}{*}{{
	\begin{turn}{270} \hspace*{-4mm} Master  \end{turn}
	}} 
	& Variation & $\master g=\chi^{(0)} g$ 
	& $\master^{(1)} g=[2\chi^{(1)}-(\chi^{(0)})^2]g$ 
	& $\master_u g=\chi_u ^{-1} \dot{\chi}_u \, g$ \\ \cline{2-5}
    & Charge 
    & $\chargeC ^{(0)} =\half \tr \big( \chargeY \, \chargeY  \big)$ 
    & $\chargeC ^{(1)} = \tr \big( \chargeY \, \chargeY ^{(1)} \big)$ 
    & $\chargeC_u = \sfrac{1}{2} \tr \big( \chargeY_u \chargeY_u\big)$ \\ \hline 
\end{tabular}
\caption{Summary of the symmetries and charges obtained from the integrable completion. The symbol $\int_{<}$ denotes the ordered double-integral.} 
\label{tab:overview}
\end{center}
\end{table}

We conclude this section with a summary of the symmetry variations and charges found from the integrable completion of the Lie algebra and master symmetry, see also table \ref{tab:overview}. The integrable completion of the G-symmetry yields the nonlocal Yangian-like symmetries 
\begin{align}
\delta_{\epsilon,u} g =  \chi_u ^{-1} \epsilon \chi_u \, g \, .
\end{align}
The associated conserved charges are (up to a $u$-dependent factor) given by
\begin{align}
Q _u &= \int \ast j_u \, ,  
&
j_u &= - 2 g_u \, a_u \, g_u ^{-1} \, = \frac{1-u^2}{1+u^2} \, \chi_u \, j \, \chi_u ^{-1} - \frac{2u}{1+u^2} \, \chi_u \, \ast j \, \chi_u ^{-1} \, , 
\end{align}
which are the charges obtained from the integrable completion of the G-symmetry Noether charge $Q$. 
From the integrable completion of the master symmetry we obtain the higher master symmetry variations 
\begin{align}
\master_u g=\chi_u ^{-1} \dot{\chi}_u \, g \, , 
\end{align}
for which we find (up to a $u$-dependent factor) the charges obtained from the integrable completion of the Casimir charge of the G-symmetry,  
\begin{align}
\chargeC_u = \sfrac{1}{2} \tr \big( \chargeY_u \chargeY_u\big) \, .
\end{align}

\section{The Symmetry Algebra}
\label{sec:Algebra}
We now turn to the discussion of the algebra of the symmetries obtained in the last section. We discuss the algebra of the symmetry variations in section \ref{sec:Variations} and the Poisson algebra of the Noether charges in section \ref{sec:Poissbrack}. In the latter discussion, we restrict to the case of a 2-dimensional field theory in order to make contact to the results of reference \cite{MacKay:1992he}, upon which the discussion builds. 

\subsection{The Algebra of the Variations}
\label{sec:Variations}
We provide the commutation relations for the variations $\delta_{\epsilon, u}$ and $\master_u$ given in \eqref{eq:Yangvars} and \eqref{eq:mastervars}, respectively. The commutation relations of similar nonlocal symmetries were derived by Schwarz \cite{Schwarz:1995td} and we follow the methods explained there. Note however, that Schwarz' discussion does not include the symmetry $\master\equiv\master^{(0)}$.

Let us begin by considering two generic variations $\delta_1$, $\delta_2$, which are the infinitesimal variations associated to some transformations $g \mapsto F_i ^{t} (g) = f_i ^{t} (g) \cdot g $ ($i=1,2$). Taking $t$ to be infinitesimal, we have the variations
\begin{align*}
\delta_i  g = \dfrac{\diff}{\diff t} \, F_i ^t (g) \big \vert _{t=0} =  \dfrac{\diff}{\diff t} \, f_i ^t (g) \big \vert _{t=0} \cdot g = \phi_i(g) \cdot g \, .
\end{align*}
The concatenation of two such variations is given by
\begin{align*}
\delta_1 \left( \delta_2 g \right) 
= \dfrac{\diff}{\diff t_2} \left( \phi_1 \left( f_2 ^{t_2}(g) \cdot g \right) \cdot f_2 ^{t_2}(g) \cdot g \right)_{t_2=0} 
= \phi_1(g) \cdot \phi_2(g) \cdot g + \left(\delta_2 \, \phi_1 \right) \cdot g \, , 
\end{align*}
and hence we have the commutator
\begin{align}
\left[ \delta_1 \, , \, \delta_2 \right] g = \left( \left[ \phi_1  ,  \phi_2 \right] + \delta_2 \phi_1 - \delta_1 \phi_2 \right) g \, .
\label{eqn:comm_generic}
\end{align}
\paragraph{The Variations of $\chi_u$.}
All arising terms of the form $\delta_i \phi_j$ can be inferred from the variations $\delta_{\epsilon_1, u_1} \chi_{u_2}$ and $\master _{u_1} \chi_{u_2}$. To reduce the amount of writing, we introduce the abbreviations
\begin{align*}
 \delta_{\epsilon_i, u_i} g &= \delta _i \, g = \eta _i \, g =  \chi _i  ^{-1} \epsilon_i \, \chi _i \, g \, , &  \master_{u_i} g &= \psi_i g = \chi_i^{-1} \dot{\chi}_i g \, .
\end{align*}
The variations are constructed from the solution $\chi_u$ to the auxiliary linear problem
\begin{align}
\diff \chi_u &= \chi_u \, \ell_u \, , &
\chi_u ( z_0 , \bar{z}_0 ) &= \unit \, ,  
\label{eqn:auxiliary-problem}
\end{align}
and their variations can be derived from the varied auxiliary linear problem 
\begin{align}
\chi_u ^{-1} \left( \diff \left( \delta  \chi_u \right) - \left( \delta  \chi_u \right) \ell_u \right) &= \delta   \ell_u \, , &
\delta  \chi_u (z_0, \bar{z}_0) &= 0 \, . 
\label{eqn:varied-auxiliary-problem}
\end{align}
We first consider the variation $\delta_1 \chi_2$ and begin by calculating $\delta_1 \ell_2 $. We recall that $\ell_i = \left(1 + u_i^2 \right) ^{-1} \left( u_i^2 \, j +  u_i \ast j \right)$ and use equation \eqref{eqn:delta-j} to find
\begin{align*}
\delta_1 j & = - \diff \eta_1 - \left[ j , \eta_1 \right] + g \, \Omega \left( g^{-1} \diff \eta_1 g \right) g^{-1} \, .
\end{align*}
We can further simplify this expression by noting that 
$\diff \eta_1 = [\eta_1 , \ell_1 ]$ and rewriting
\begin{align*}
g\, \Omega( g^{-1} [j , \eta_1] g ) g^{-1} 
	= -2 g \, \Omega(  [a , g^{-1} \eta_1 g ] ) g^{-1} 
	= 2 g [a ,\Omega( g^{-1} \eta_1 g) ] g^{-1} 
	= - [j , \tilde{\eta}_1] .
\end{align*}
Here, we introduced the abbreviation $\tilde{\eta} = g \, \Omega \left( g^{-1} \eta g \right) g^{-1} $. Note that the above relation extends to $\ell_1$, since it is given by a linear combination of $j$ and $\ast j$. Thus we find that
\begin{align}
\delta_1 j & = \left[ \ell_1 - j , \eta _1 \right] + \left[ \ell_1 , \tilde{\eta} _1 \right] 
=  \frac{1}{1+ u_1^2} \left(  \left[- j +  u_1 \ast j , \eta_1 \right] + \left[u_1^2 \, j + u_1 \ast j , \tilde{\eta}_1 \right] \right) \, ,
\end{align}
which implies that the variation $\delta_1 \ell_2 $ may be written as
\begin{align}
\delta_1 \ell_2 &= \frac{1}{1+u_2^2} \left( u_2 ^2 \, \delta_1 j + u_2 \ast \delta_1 j \right) \nn \\
&= \frac{u_2}{u_1-u_2} \left[ \ell_2 - \ell_1 , \eta_1 \right] + \frac{u_1 u_2}{1 + u_1 u_2} \left[ \ell_2 + \ell_1 - j , \tilde{\eta}_1 \right] \, .
\end{align}
We now construct the solution to equation \eqref{eqn:varied-auxiliary-problem} from the ansatz
\begin{align*}
\delta_1 \chi_2 &= \gamma _1 \left(\delta_1 \chi_2 \right)_1 + \gamma _2 \left(\delta_1 \chi_2 \right)_2 \, , & 
\left(\delta_1 \chi_2 \right)_1 &= \chi_2 \eta_1 - \epsilon_1 \chi_2  \, , & \left(\delta_1 \chi_2 \right)_2 &= \chi_2 \tilde{\eta}_1 - \epsilon_1 ^\prime \chi_2 \, .
\end{align*}
Here, $\epsilon^\prime$ is defined by
\begin{align*}
\epsilon ^\prime &= \tilde{\eta}( z_0 , \bar{z}_0 )  = g_0 \, \Omega \left( g_0 ^{-1} \epsilon g_0 \right) g_0 ^{-1} \, , &
g_0 &= g ( z_0 , \bar{z}_0 ) \, ,
\end{align*}
such that both $\left(\delta_1 \chi_2 \right)_1$ and $\left(\delta_1 \chi_2 \right)_2$ satisfy the boundary condition $\delta_1 \chi_2 (z_0, \bar{z}_0) = 0$. By using similar arguments as in the discussion preceding equation \eqref{CriterionStep}, we find that
\begin{align}
\diff \tilde{\eta}_1 = g \, \Omega \left(g^{-1}\left( \diff \eta_1 + \left[j, \eta_1 \right] \right) g \right) g^{-1} = \left[\ell_1 - j , \tilde{\eta}_1 \right] \, , 
\end{align}
and thus we obtain 
\begin{align*}
\chi_2 ^{-1} \left( \diff \left( \delta_1  \chi_2 \right) - \left( \delta_1  \chi_2 \right) \ell_2 \right) = \gamma _1 \left[ \ell_2 - \ell_1 , \eta_1 \right] + \gamma_2 \left[ \ell_2 + \ell_1 - j , \tilde{\eta}_1 \right] \, , 
\end{align*}
from which we can read off the solution to equation \eqref{eqn:varied-auxiliary-problem} as 
\begin{align}
\delta_1 \chi_2 = \frac{u_2}{u_1-u_2} \left( \chi_2 \eta_1 - \epsilon_1 \chi_2 \right) + \frac{u_1 u_2}{1+ u_1 u_2} \left( \chi_2 \tilde{\eta}_1 - \epsilon_1 ^\prime \chi_2 \right) \, .
\label{eqn:delta1chi2}
\end{align}
We proceed similarly for the calculation of $\master_{u_1} \chi_2$ and begin by calculating $\master_{u_1} \ell_2$. Using equation \eqref{eqn:delta-j} as well as the relation 
$\diff \psi = [ \psi , \ell ] + \dot{\ell} $, we have
\begin{align}
\master_{u_1} \,  j = \big[\ell_1 -j , \psi_1 \big] +  \big[\ell_1 , \tilde{\psi}_1 \big] - 2 \, \dot{\ell}_1 \, , 
\end{align} 
abbreviating  $\tilde{\psi} = g \, \Omega \left( g^{-1} \psi g \right) g^{-1}$ as before. The above relation implies that
\begin{align*}
\master_{u_1} \, \ell_2 = \frac{u_2}{u_1 - u_2} \, \big[\ell_2 - \ell_1 , \psi_1 \big] + \frac{u_1 u_2}{1 + u_1 u_2} \, \big[\ell_2 + \ell_1 - j , \tilde{\psi}_1 \big] - \frac{2}{1+ u_2^2} \left( u_2 ^2 \, \dot{\ell}_1 + u_2 \ast \dot{\ell}_1 \right) .
\end{align*}
Making use of the relation 
\begin{align}
\diff \tilde{\psi} = g \Omega \left( g^{-1} \left( \diff \psi + \left[ j , \psi \right] \right) g \right) g^{-1} = \big[ \ell - j , \tilde{\psi} \big] - \dot{\ell} \, ,
\end{align}
one may then show that the defining relation \eqref{eqn:varied-auxiliary-problem} for $ \master_{u_1} \chi_2$ is solved by
\begin{align}
\master_{u_1} \, \chi_2 = \frac{u_2}{u_1 - u_2} \, \chi_2 \psi_1 + \frac{u_1 u_2}{1 + u_1 u_2} \, \chi_2 \tilde{\psi}_1 - \frac{u_2(1+u_2^2)}{(u_1 - u_2)(1+ u_1 u_2) } \, \chi_2 \psi_2 \, . 
\label{eqn:deltaVchi} 
\end{align}
Note that the boundary condition $ \master_{u_1}  \chi_2 ( z_0 ) = 0$ is automatically satisfied since we have $\psi_i  ( z_0 ) = 0$. 

\paragraph{The Commutators.}
The above results allow us to compute the commutators of the variations $\delta_\epsilon ^{(u)}$ and $\master_u$. In order to compute the commutator $\left[\delta_1 , \delta_2 \right]$ we note that 
\begin{align*}
\delta_1 \eta_2 = \left[ \eta_2 , \chi_2 ^{-1} \delta_1 \chi_2 \right] &= \frac{u_2}{u_1-u_2}  \left( \left[\eta_2 , \eta_1 \right] - \chi_2 ^{-1} \left[\epsilon_2 , \epsilon_1 \right] \chi_2 \right) \\
& + \frac{u_1 u_2}{1 + u_1 u_2} \left( \left[\eta_2 , \tilde{\eta}_1 \right] - \chi_2 ^{-1} \left[\epsilon_2 , \epsilon_1 ^\prime \right] \chi_2 \right) \, ,
\end{align*}
such that 
\begin{align}
\left[\delta_1 , \delta_2 \right] g = \left( \left[\eta_1 , \eta_2 \right] + \delta_2 \eta_1 - \delta_1 \eta_2 \right) g 
=  \frac{u_1 \, \delta_{\left[\epsilon_1 , \epsilon_2 \right], u_1}  - u_2 \, \delta_{\left[\epsilon_1 , \epsilon_2 \right], u_2} }{u_1-u_2} \, g& \nn \\
\qquad + \frac{u_1 u_2 \left( \left[\eta_1 , \tilde{\eta}_2 \right] - \left[\eta_2 , \tilde{\eta}_1 \right] - \delta_{[\epsilon_1 , \epsilon_2 ^\prime ], u_1}  + \delta_{[\epsilon_2 , \epsilon_1 ^\prime ], u_2}\right)}{1+u_1 u_2} \, g& \, .
\end{align}
Noting that
\begin{align}
\left( \left[\eta_1 , \tilde{\eta}_2 \right] - \left[\eta_2 , \tilde{\eta}_1 \right] \right) g = g P_{\alg{h}} \left[ g^{-1} \eta_1 g , \Omega \left(g^{-1} \eta_2 g \right) \right]  ,
\label{eqn:gaugetrans}
\end{align}
we thus have
\begin{align}
\Big[ \delta_{\epsilon_1, u_1}  , \delta_{\epsilon_2, u_2}  \Big] 
&= \frac{ 
	u_1 \, \delta_{\left[\epsilon_1 , \epsilon_2 \right],u_1} 
	- u_2 \, \delta_{\left[\epsilon_1 , \epsilon_2 \right], u_2}
	}{u_1-u_2}  
	+ \frac{u_1 u_2}{1 + u_1 u_2}  
	\left( \delta_{[\epsilon_2 , \epsilon_1 ^\prime ], u_2} 
	- \delta_{[\epsilon_1 , \epsilon_2 ^\prime ], u_1} \right), 
\label{eqn:comm1}
\end{align}
up to gauge transformations. In order to compute the commutator $\left[ \master_{u_1},  \master_{u_2} \right]$, we employ equation \eqref{eqn:deltaVchi} to find
\begin{align*}
\master_{u_1} \psi_2 &= - \chi_2 ^{-1} \big( \master_{u_1} \chi_2 \big) \psi_2 + \chi_2 ^{-1} \frac{\partial}{\partial u_2} \big( \master_{u_1} \chi_2 \big)   \nn \\
&= \frac{u_2}{u_2 - u_1} \big[ \psi_1 , \psi_2 \big] 
- \frac{u_1 u_2}{1 + u_1 u_2} \big[ \tilde{\psi}_1 , \psi_2 \big] 
- \frac{u_2 (1 + u_2 ^2 ) }{(u_1 - u_2 ) (1+ u_1 u_2)} \, \dot{\psi}_2  \\
&  - \left( \frac{\partial}{\partial u_2} \, \frac{u_2}{u_2 - u_1} \right) \psi_1 
+ \left( \frac{\partial}{\partial u_2} \, \frac{u_1 u_2}{1 + u_1 u_2} \right) \tilde{\psi}_1 
- \left( \frac{\partial}{\partial u_2} \, \frac{u_2 (1 + u_2 ^2 ) }{(u_1 - u_2 ) (1+ u_1 u_2)} \right) \psi_2 \, .
\end{align*}
Using equations similar to \eqref{eqn:gaugetrans} and noting that
\begin{align}
\tilde{\psi}_1 \, g = g \, \Omega \left( g^{-1} \psi_1 g \right) 
= - \psi_1 \, g  + g \, 2 P_{\alg{h}} \left( g^{-1} \psi_1 g \right) 
= - \master_{u_1} g + \delta_h g \, , 
\label{eqn:gaugetrans_comm}
\end{align}
we find the commutator
\begin{align}
\Big[ \master _{u_1} , \master _{u_2}  \Big] 
	&= \sum \limits _{i=1} ^2 \frac{(1 + u_i^2) 
	\left( u_i  \partial_{u_i} + 1 \right) \master _{u_i}}
	{(u_1 - u_2) (1 + u_1 u_2)} +
	\frac{2 (1+u_1^2)(1+u_2^2)
	\big(u_2  \master _{u_2} - u_1 \master _{u_1} \big)}
	{(u_1-u_2)^2 (1+ u_1 u_2)^2} ,
\label{eqn:comm3}
\end{align}
where we have again left out the gauge transformations.
In order to compute the commutator $\left[ \master_{u_1}, \delta_{\epsilon_2, u_2} \right]$, we use equations \eqref{eqn:delta1chi2} and \eqref{eqn:deltaVchi}, to find the variations
\begin{align}
\master_{u_1} \eta_2 &= 
\frac{u_2}{u_1 - u_2} \, \big[ \eta_2, \psi_1 \big] 
+ \frac{u_1 u_2}{1 + u_1 u_2} \, \big[ \eta_2 , \tilde{\psi}_1 \big] 
- \frac{u_2 (1 + u_2^2)}{(u_1 - u_2) (1 + u_1 u_2)} \, \partial_{u_2} \eta_2 \, , \\
\delta_{\epsilon_2, u_2} \psi_1 &=
\frac{u_1}{u_2 - u_1} \, \big[ \psi_1 , \eta_{\epsilon_2, u_2} \big]
+ \frac{u_1 u_2}{1 + u_1 u_2} \big[ \psi_1 , \tilde{\eta}_{\epsilon_2, u_2}, \big] \\
& \; + \left( \frac{\partial}{\partial u_1}  \, \frac{u_1}{u_2 - u_1} \right) \left( \eta_{\epsilon_2, u_2} - \eta_{\epsilon_2, u_1} \right) 
+ \left( \frac{\partial}{\partial u_1}  \, \frac{u_1 u_2}{1 + u_1 u_2} \right) \left( \tilde{\eta}_{\epsilon_2, u_2} - \eta_{\epsilon ^\prime _2, u_1} \right) \, . \nn
\end{align} 
By making use of identities similar to \eqref{eqn:gaugetrans} and \eqref{eqn:gaugetrans_comm} we then obtain the commutator
\begin{align}
\Big[ \master _{u_1} , \delta_{\epsilon, u_2}  \Big]  
&=  \frac{u_2 \big( \delta_{\epsilon, u_2}  - \delta_{\epsilon, u_1}  \big)}{(u_1 - u_2 )^2} 
- \frac{u_2 \big( \delta_{\epsilon, u_2}  + \delta_{\epsilon^\prime, u_1} \big) }{(1+ u_1 u_2 )^2} 
+ \frac{u_2 (1+ u_2 ^2 ) \, \partial_{u_2} \delta_{\epsilon, u_2} }{( u_1 - u_2 ) ( 1 + u_1 u_2)  } .
\label{eqn:comm2}
\end{align}

\paragraph{Expanded Algebra.}
The underlying algebra takes a more intuitive form after performing an expansion around $u=0$. We define the coefficients $\delta_{\epsilon}^{(n)}$ and $\master ^{(n)}$ by
\begin{align}
\delta_{\epsilon, u} &= \sum \limits _{n=0} ^\infty u^n \, \delta_{\epsilon}^{(n)} \, , 
& 
\master_{u} &= \sum \limits _{n=0} ^\infty u^n \, \master ^{(n)} \,,
\end{align}
and we set $\delta_{\epsilon}^{(n)} = 0 = \master ^{(n)}$ for  $n < 0 \, $.
The expansion of all commutators can be performed by making repeated use of the identity 
\begin{align*}
u_1 ^{n+1} - u_2 ^{n+1} = \left( u_1 - u_2 \right) \sum \limits _{k=0} ^n u_1 ^{n-k} \, u_2 ^k \, . \qquad 
\end{align*}
Expanding the commutation relations of the Yangian-type symmetries leads to the relations
\begin{align}
\Big[ \delta_{\epsilon_1}^{(n)} , \delta_{\epsilon_2}^{(m)} \Big] &= 
\begin{cases}
\delta_{[\epsilon_1,\epsilon_2]}^{(m+n)} & \text{if} \quad n=0 \vee m=0, \\[3mm]
\delta_{[\epsilon_1,\epsilon_2]}^{(m+n)} + \left(-1 \right) ^{n}  \, \delta_{[\epsilon_2 , \epsilon_1 ^\prime ]} ^{(m-n)} -  \left(-1 \right) ^{m}  \, \delta_{[\epsilon_1 , \epsilon_2 ^\prime ]} ^{(n-m)} 
& \text{if} \quad n,m \neq 0 \, .
\end{cases} 
\label{eqn:comm1_expanded} 
\end{align}
The first term represents the commutation relations of a loop algebra, which is the symmetry algebra of principal chiral models. The additional terms can be simplified if we fix the condition $g_0 = \unit$, such that $\epsilon^\prime = P_\alg{h} \epsilon - P_\alg{m} \epsilon$. 
Discriminating the cases $\epsilon_i \in \alg{h}$ and $\epsilon_i \in \alg{m}$ one then reaches the following commutation relations for $n,m \neq 0$:
\begin{align}
\Big[ \delta_{\epsilon_1}^{(n)} , \delta_{\epsilon_2}^{(m)} \Big] &= 
\begin{cases} 
\delta_{[\epsilon_1,\epsilon_2]}^{(m+n)} - \left(-1 \right) ^{n}  \, \delta_{[\epsilon_1 , \epsilon_2 ]} ^{(m-n)} -  \left(-1 \right) ^{m}  \, \delta_{[\epsilon_1 , \epsilon_2]} ^{(n-m)} 
& \quad \epsilon_1 \in \alg{h} , \, 
\epsilon_2 \in \alg{h}   \, , \\[3mm]
\delta_{[\epsilon_1,\epsilon_2]}^{(m+n)} - \left(-1 \right) ^{n}  \, \delta_{[\epsilon_1 , \epsilon_2 ]} ^{(m-n)} +  \left(-1 \right) ^{m}  \, \delta_{[\epsilon_1 , \epsilon_2]} ^{(n-m)} 
& \quad \epsilon_1 \in \alg{h} , \, 
\epsilon_2 \in \alg{m}   \, , \\[3mm]
\delta_{[\epsilon_1,\epsilon_2]}^{(m+n)} + \left(-1 \right) ^{n}  \, \delta_{[\epsilon_1 , \epsilon_2 ]} ^{(m-n)} -  \left(-1 \right) ^{m}  \, \delta_{[\epsilon_1 , \epsilon_2]} ^{(n-m)} 
& \quad \epsilon_1 \in \alg{m} , \, 
\epsilon_2 \in \alg{h}   \, , \\[3mm]
\delta_{[\epsilon_1,\epsilon_2]}^{(m+n)} + \left(-1 \right) ^{n}  \, \delta_{[\epsilon_1 , \epsilon_2 ]} ^{(m-n)} +  \left(-1 \right) ^{m}  \, \delta_{[\epsilon_1 , \epsilon_2]} ^{(n-m)} 
& \quad \epsilon_1 \in \alg{m} , \, 
\epsilon_2 \in \alg{m}   \, .
\end{cases} 
\label{eqn:comm1_double_expanded} 
\end{align}
If the additional terms with varying signs were absent, the symmetry algebra would be half of a loop algebra $\widehat{\grp{G}}$. Due to the dependence of the signs on the discrimination between $\alg{h}$ and $\alg{m}$, Schwarz denotes the symmetry of symmetric space models by 
$\widehat{\grp{G}}_\grp{H}$ in reference \cite{Schwarz:1995td}.

For the commutator of the higher master and Yangian-like variations, we find
\begin{align}
\left[ \master ^{(n)} , \delta_{\epsilon}^{(m)} \right] &= -m \, \delta_{\epsilon} ^{(m+n+1)} + (-1)^m m \, \delta_{\epsilon^\prime} ^{(n+1-m)} - (-1)^n m \, \delta_{\epsilon} ^{(m-n-1)} 
\, . 
\label{eqn:comm2_expanded} 
\end{align}
We observe that all of the higher master variations commute with the generators $\delta_{\epsilon}$ of the G-symmetry. Moreover, the raising operator structure observed for the action of the master variation $\master$ on the conserved charges is also present. Given the variations 
$\delta_\epsilon ^{(1)}$ and $\master$ all of the higher Yangian-like variations follow from the above commutation relations.   

The commutator of two higher master variations can be shown to take the form
\begin{align}
\left[ \master ^{(n)} , \master ^{(m)} \right] &= 
	(n-m) \, \master ^{(n+m+1)} 
	+ (-1)^m (n-3m-2) \, \master ^{(n-m-1)} \nn \\
	& \quad - (-1)^n (m-3n-2) \,  \master ^{(m-n-1)} \, .
\label{eqn:comm3_expanded} 
\end{align}
Note that the knowledge of the first two generators $\master$ and $\master ^{(1)}$ is sufficient to construct all of the higher master symmetry generators as it is the case for the Yangian-like variations $\delta_\epsilon$ and $\delta_\epsilon^{(1)}$ as well.   

The algebra of the nonlocal symmetries of symmetric space models has been discussed by Schwarz in reference \cite{Schwarz:1995td}, who obtained the symmetry transformations by generalizing the known nonlocal symmetries of principal chiral models, which had been obtained in references 
\cite{Dolan:1980kz,Dolan:1981fq,Devchand:1981wy,Wu:1982jt}. 
He described two types of nonlocal symmetries. The first one corresponds to the Yangian-like symmetry 
$\delta_{\epsilon, u}$. The second is named ``Virasoro-like'' by Schwarz, since the algebraic properties of its generators resemble those of linear combinations of Virasoro generators. It is similar to the higher master variations $\master_u$.  

In order to put Schwarz' Virasoro symmetry into the context of our discussion, we note that he uses the following variation for the Virasoro-like symmetries (translated to the conventions used within this thesis)
\begin{align}\label{eq:Schwcompare}
\delta_{\mathrm{V}, u} \, g &= \left( (1+u^2) \chi_u ^{-1} \dot{\chi}_u - \chi^{(0)} \right) g  
= \chi_u ^{-1} \big(\, \master \chi_u \big) g 
=: \eta  _\mathrm{V, u} \, g \, .
\end{align}
The above symmetry action is the natural symmetric space generalization of the nonlocal symmetries of the principal chiral model discussed in the same paper. Let us compare this to the variation \eqref{eq:mastervars}, which is given by
\begin{align}\label{eq:wecompare}
\master_u g =  \chi_u ^{-1} \dot{\chi}_u \, g =: \eta _u \, g\,.
\end{align}
We see that the two variations are related by
\begin{align}
\delta_{\mathrm{V}, u} \, g = \big( ( 1 + u^2 ) \master_u - \master \, \big) g \, .
\end{align}
The master symmetry $\delC$ is thus absent in the discussion of Schwarz since the variation $\delta_{\mathrm{V}, u} $ becomes trivial in the limit $u \to 0$. Note also that the variation $\master$ cannot be extracted for $u \in \mathbb{C}$ at $u^2 = -1$ since $\chi_u$ has poles at these points. In fact, there is a simple argument which shows that $\master$ is not contained in the family of variations given by \eqref{eq:Schwcompare}. Since the variation $\delta_{\mathrm{V}, u} $  is of the form $\chi_u ^{-1} \delta_0 \chi_u$, the proof given in section \ref{sec:GenSymm} shows that
\begin{align}
\diff \ast \left( \diff \eta  _\mathrm{V,u} + \left[ j , \eta  _\mathrm{V,u} \right] \right) = 0 \, ,
\end{align}
which can also be seen from the proof given in reference \cite{Schwarz:1995td}. In contrast, the variation $\master$ only satisfies the necessary condition \eqref{eqn:criterion}, since
\begin{align}
g^{-1} \diff \ast \lrbrk{ \diff \chi^{(0)} + \left[ j , \chi^{(0)}  \right]  } g = - 4 \, a \wedge a \in \alg{h} \, .
\end{align}
In the case of the principal chiral model, Schwarz gives an interpretation of the analogue of the algebra given by the commutation relations \eqref{eqn:comm3_expanded}. He shows that the respective generators can be related to half of a Virasoro algebra by considering suitable linear combinations and thus refers to this symmetry as ``Virasoro-like''. In the case of symmetric space models and for the variations $\master^{(n)}$, a similar relation to the Virasoro algebra seems not to be present.  


\subsection{The Algebra of the Charges}
\label{sec:Poissbrack}

In this section, we discuss the Poisson algebra of the conserved charges associated to the nonlocal symmetries of symmetric space models. We find that up to ambiguous boundary terms, which commonly appear in the study of such Poisson algebras \cite{Luscher:1977rq}, the Poisson algebra of the charges $\chargeY^{(n)}$ is given by the classical analogue of a Yangian algebra. Similar results hold for principal chiral and Gross--Neveu models \cite{MacKay:1992he}. Since the form of the current algebra for symmetric space models is close to the one of principal chiral models, the analysis is simplified and many of the results of reference \cite{MacKay:1992he} can be transferred. As the charges associated to the master symmetry turn out to be compositions of the Yangian charges, their Poisson algebra is inherited from the latter.
For convenience, the calculations within this subsection are performed in components instead of differential forms. Also, in order to be compatible with reference \cite{MacKay:1992he}, we switch to a field theory point of view in this subsection. That is, we view the symmetric space model to describe the internal degrees of freedom of a two-dimensional field theory rather than the target space of a string theory. In particular, the conserved charges are integrated over an infinite line rather than a closed cycle. In order to emphasize this point, we also switch the notation from $(\tau,\sigma)$ to $(t,x)$ within this subsection.

The Poisson algebra of the conserved charges of symmetric space models can be obtained from the respective algebra for principal chiral models, which was derived in reference \cite{MacKay:1992he}. We have introduced principal chiral models in section \ref{sec:PCM}. The action can be written as 
\begin{align}
S = -\frac{T}{2} \int \tr \left( j \wedge \ast j \right) ,
\label{PCMAction}
\end{align}
where $j = g^{-1} \diff g$ is the Noether current corresponding to the G-symmetry of left-multiplication. The components of the the Noether current, 
\begin{align*}
j (t,x) = j_0 ^{a} (t , x ) \, T_a \, \diff t  +  j_1 ^{a} (t , x ) \, T_a \, \diff x ,
\end{align*}
are constrained by the fact that the current is both flat and conserved, 
\begin{align}
\partial_t \, j_0 ^{a} + \partial_x \, j_1 ^{a} &= 0 \, , &
\partial_t \, j_1 ^{a} - \partial_x \, j_0 ^{a} 
+ f \indices{^a _{bc}} \, j_0 ^{b} \, j_1 ^{c} &= 0 \, .
\end{align}
Here, $f \indices{^a _{bc}}$ denote the structure constants of the generators $T_a$, 
\begin{align*}
\left[ T_a , T_b \right] = f \indices{_{ab} ^c} \, T_c \, .
\end{align*}
The Poisson brackets for the components of the Noether current were derived in reference \cite{Faddeev:1987ph} by regarding $q^a = j_1 ^a$ as generalized coordinates. Due to the flatness of the Noether current, the time derivatives of the coordinates can be expressed as
\begin{align}
\partial _t q^a = \partial_x j_0 ^a + f \indices{^a _{bc}} q^b j_0 ^c 
	= \left( \mathbf{D} j_0 \right) ^a \, , 
\end{align}
and the conjugate momenta can be identified as
\begin{align}
\pi ^a  = - \left( \mathbf{D}^{-1} j_0 \right) ^a \, .
\end{align}
The Poisson algebra for the Noether current can then be derived from the canonical Poisson brackets for the variables $\pi^a$ and $q^a$ and is found as
\begin{align}
\left \lbrace j_0 ^{a} (t,x) \, , \,  j_0 ^{b} (t,y) \right \rbrace_{\mathrm{PCM}} 
	&= \mathbf{f} ^{ab} {} _c \, j_0 ^c (t,x) \, \delta (x-y) \, , \nn \\
\left \lbrace j_0 ^{a} (t,x) \, , \,  j_1 ^{b} (t,y) \right \rbrace_{\mathrm{PCM}} 
	&= \mathbf{f} ^{ab} {} _c \, j_1 ^c (t,x) \, \delta (x-y) 
	+ G^{a b} \, \partial_x \delta(x-y) \, , \\
\left \lbrace j_1 ^{a} (t,x) \, , \,  j_1 ^{b} (t,y) \right \rbrace_{\mathrm{PCM}} &= 0 \, . \nn
\end{align}
Here, $G^{ab} = \tr \left( T^a T^b \right)$ denotes the metric on the group G. 
Moreover, following the conventions introduced in section \ref{sec:Int_Symm}, we are using the structure constants of the symmetry variations here, 
\begin{align}
\left[ t_a , t_b \right] &= \mathbf{f} _{ab} {} ^c \, t_c \, , & 
\mathbf{f} _{ab} {} ^c &= f _{ba} {} ^c \, .
\end{align}
The metric $G_{ab}$, which we use lower the group indices, is the same for the generators $T_a$ and $t_a$. 

It is tempting to transfer the above algebra for the Noether currents to symmetric space models,  since the action of these models can also be written in the form \eqref{PCMAction} employing the Noether current of symmetric space models, which is flat and conserved as well. Note however, that in the case of symmetric space models the Noether current $j$ has $\mathrm{dim}(\alg{g})$ components corresponding to only $\mathrm{dim}(\alg{m})$ degrees of freedom, such that the components are not independent and should hence not be considered as a set of generalized coordinates. The above Poisson algebra for the components of the Noether current can hence not be transferred directly to symmetric space models. 

The Poisson algebra of the Noether currents for symmetric space models was derived in a different approach in reference \cite{Forger:1991cm}. They found the Poisson-brackets
\begin{align}
\left \lbrace j_0 ^{a} (t , x)\, , \, 
	j_0 ^{b} (t , y) \right \rbrace_{\mathrm{SSM}} &= 
	\mathbf{f} ^{ab} {} _c \, j_0 ^c (t , x) \, \delta (x - y) \, , \\
\left \lbrace j_0 ^{a} (t , x) \, , \,  
	j_1 ^{b} (t , y) \right \rbrace_{\mathrm{SSM}} &= 
	\mathbf{f}  ^{ab} {} _c \, j_1 ^c (t , x)  \, \delta (x - y)
	+ k^{a b} (t , y) \, \partial_x \delta (x - y) \, , \\
\left \lbrace j_1 ^{a} (t , x)\, , \,  
	j_1 ^{b} (t , y) \right \rbrace_{\mathrm{SSM}} &= 0 \, .
\end{align}
Here, the quantity $k^{ab}$ is given by
\begin{align}
k ^{ab} = \tr \big( T^a k \big( T^b \big) \big) = \tr \big( P_\alg{m} \big( g^{-1} T^a g \big) 
P_\alg{m} \big( g^{-1} T^b g \big) \big) \, ,
\end{align}
with $k: \alg{g} \to \alg{g}$ being a map from $\alg{g}$ to itself defined by
\begin{align}
k &= \mathrm{Ad}(g) \circ P_\alg{m} \circ \mathrm{Ad}(g)^{-1} \, , &
k (X) &= g P_\alg{m} \left( g^{-1} X g \right) g^{-1} \, . 
\end{align}
The map $k$ is related to the Noether current $j$ by the relations
\begin{align}
\mathrm{ad}(j_\mu) &= k \circ \mathrm{ad}(j_\mu) + \mathrm{ad}(j_\mu) \circ k \, , &
j_{\mu} ^c \, f_c {} ^{a b} &=  j_{\mu} ^c \left(  f_c {} ^{ad} k_d {} ^b -  f_c {} ^{bd} k_d {} ^a \right)
\, , 
\label{kidentity}
\end{align}
which follow from a short calculation:
\begin{align*}
& \left( k \circ \mathrm{ad}(j_\mu) + \mathrm{ad}(j_\mu) \circ k \right) (X) 
= g P_\alg{m} \left( g^{-1} \left[ j_\mu , X \right] g \right) g^{-1} + \left[ j_\mu , g P_\alg{m} \left( g^{-1} X g \right) g^{-1} \right] \\
& \hspace*{20mm} = g P_\alg{m} \left( \left[ -2 a_\mu , g^{-1} X g \right]  \right) g^{-1} + g \left[ -2 a_\mu , P_\alg{m} \left( g^{-1} X g \right) \right] g^{-1} \\
& \hspace*{20mm} = g \left( \left[ -2 a_\mu , P_\alg{h} \left( g^{-1} X g \right) + P_\alg{m} \left( g^{-1} X g \right) \right] \right) g^{-1}
= \left[ j_\mu , X \right] \, .
\end{align*}
The Poisson algebra closes if one includes $k^{ab}(t,x)$. The additional Poisson brackets take the form
\begin{align}
\left \lbrace j_0 ^{a} (t,x) \, , \, k^{b c} (t,y) \right \rbrace_{\mathrm{SSM}} &= 
	\left( \mathbf{f} ^{ab} {} _d \, k^{d c} (t,x) 
	+ \mathbf{f} ^{ac} {} _d \, k^{d b} (t,x) \right) \delta ( x - y ) \, , \nn \\
\left \lbrace j_1 ^{a} (t,x) \, , \, k^{b c} (t,y) \right \rbrace_{\mathrm{SSM}} &= 0 \, , \\
\left \lbrace k^{ab} (t,x) \, , \, k^{c d} (t,y) \right \rbrace_{\mathrm{SSM}} &= 0 \, . \nn
\end{align}

\paragraph{Yangian charges.}
Given the above Poisson brackets for the components of the Noether current, we turn to the calculation of the Poisson brackets of the conserved charges $\chargeY^{(n)}$, for which we note the explicit expressions
\begin{align}\label{eq:Yangcharg}
\chargeY^{(0)\, a} &= \int \limits _{-\infty} ^\infty \diff x \, j_0 ^a (t, x) \, , \nn \\ 
\chargeY^{(1)\, a} &= \mathbf{f} ^a {}_{cb} \int \limits _{-\infty} ^\infty \diff x_1 \diff x_2 \,  \theta(x_2 - x_1) \, j_0 ^b (x_1) j_0 ^c (x_2)
+ 2 \int \limits _{-\infty} ^\infty  \diff x \, j_1 ^a (x) \, .
\end{align}
We now demonstrate that the Poisson algebra of these charges represents the classical counterpart of a Yangian algebra, i.e.\ that the charges satisfy the relations
\begin{align}\label{eq:Yangcomms}
\left \lbrace \chargeY^{(0)\, a} \, , \, \chargeY^{(0)\, b} \right \rbrace 
	&= \, \mathbf{f} ^{ab} {} _c \, \chargeY^{(0)\, c} \, , & 
\left \lbrace \chargeY^{(0)\, a} \, , \, \chargeY^{(1)\, b} \right \rbrace 
	&= \, \mathbf{f} ^{ab} {} _c \, \chargeY^{(1)\, c} \, ,
\end{align}
as well as the classical counterpart of the Serre relations:
\begin{align}
f_d {} ^{[ab} \left \lbrace  \chargeY ^{(1) c]} \, , \, \chargeY ^{(1) d}\right \rbrace 
	= \half \, \mathbf{f} ^a {} _{ip} \, \mathbf{f}^b {} _{jq} \,
	\mathbf{f}^c {}_{kr} \, \mathbf{f}^{ijk} \,
	 \chargeY ^{(0) p} \, \chargeY ^{(0) q} \, \chargeY ^{(0) r}  .
\label{Serre}
\end{align}
For the Gross--Neveu and the principal chiral model, the Serre relations for the conserved charges were shown by MacKay \cite{MacKay:1992he}. Due to the similarity of the Poisson brackets for the principal chiral and symmetric space model, the most part of the calculation for the symmetric space model has already been performed there. In order to show that the result can be transferred, we only need to evaluate the difference in the Poisson brackets, which can be done by employing the following notation: 
\begin{align}
\left \lbrace j_0 ^{a} (t,x) \, , \,  j_0 ^{b} (t,y) \right \rbrace_{\mathrm{SSM-PCM}} &= 0 \, , \nn \\
\label{brack:ssm-pcm}
\left \lbrace j_0 ^{a} (t,x) \, , \,  j_1 ^{b} (t,y) \right \rbrace_{\mathrm{SSM-PCM}} &=  \left( k^{ab} (t,y) - G^{a b} \right) \, \partial_x \delta(x-y) \, , \\
\left \lbrace j_1 ^{a} (t,x) \, , \,  j_1 ^{b} (t,y) \right \rbrace_{\mathrm{SSM-PCM}} &= 0 \, . \nn
\end{align}
We begin by studying the Poisson bracket between a level-0 and a level-1 charge. The calculation of this bracket gives rise to boundary terms, which depend on the precise way in which the upper and lower integration boundaries are taken to infinity in \eqref{eq:Yangcharg}, cf.\ reference \cite{Luscher:1977rq}. They arise for both the principal chiral and symmetric space model from integrating out the $\partial_x \delta(x-y)$ contributions. The problem stems from the fact that both $G^{ab}$ and $k^{ab}$ are not suitable test functions as they do not vanish when $x$ approaches infinity. 

In order to keep the discussion general for the moment, we consider the charges $\chargeY^{(0)}$ and $\chargeY^{(1)}$ with the following boundaries:
\begin{align}
\chargeY^{(0)\, a} &= 
\int  \limits _{-L_1} ^{L_2}  \diff x j_0 ^a (x) 
\\
\chargeY^{(1)\, a} &= \mathbf{f} ^a {}_{cb} 
\int \limits _{- L_3} ^{L_4} \diff x_1 \diff x_2 \,  \theta(x_2 - x_1) j_0 ^b (x_1) \,  j_0 ^c (x_2)
+ 2 
\int  \limits _{-L_5} ^{L_6}  \diff x\,  j_1 ^a (x) \, .
\end{align}
Since the Poisson bracket of two 0-components of the Noether current $j$ is the same as in the case of the principal chiral model (see \eqref{brack:ssm-pcm}), the Lie algebra commutator for the level-0 charges follows trivially. For the Poisson bracket of a level-0 with a level-1 charge we find 
\begin{align}
& \left \lbrace \chargeY^{(0)\, a} \, , \, \chargeY^{(1)\, b} \right \rbrace _{\mathrm{SSM-PCM}} = 
\int \limits _{-L_1} ^{L_2}  \diff x 
\int  \limits _{-L_5} ^{L_6}  \diff y 
\left( k^{ab} (y) - G^{a b} \right) \, \partial_x \delta(x-y) \nn \\
& \qquad = \int  \limits _{-L_5} ^{L_6}  \diff y 
\left( k^{ab} (y) - G^{a b} \right) \left( \delta(L_2-y) - \delta(-L_1 -y) \right) \nn \\
& \qquad = \left( k^{ab} (L_2) - G^{a b} \right) \theta (L_6 - L_2 ) - 
\left( k^{ab} (-L_1) - G^{a b} \right) \theta (L_5 - L_1 ) .
\end{align}
The result shows the difference between the boundary terms for the principal chiral and symmetric space model. In the case of the principal chiral model, it has been noted in reference \cite{MacKay:1992he} that the ambiguous boundary terms disappear if one sets the upper and lower boundaries equal, $L_1 = L_2$ and $L_5 = L_6$. This prescription is not sufficient in the case of the symmetric space model, since $k^{ab} (L_1)$ generically differs from $k^{ab} (-L_1)$. For the boundary terms to disappear --- and the Yangian algebra to be satisfied --- we have to require $L_1 > L_5$ and $L_2 > L_6$. 

In order to study the Serre relations \eqref{Serre} next, let us now turn to the Poisson bracket of two level-1 charges $\chargeY^{(1)\, a}$. Making use of equation \eqref{brack:ssm-pcm},
 we find
\begin{align}
& \left \lbrace \chargeY^{(1)\, a} , \chargeY^{(1)\, b} \right \rbrace _{\mathrm{SSM-PCM}} 
= \int \limits _{-L_3} ^{L_4} \diff x_1 \diff x_2 \, \theta(x_2-x_1) 
\int \limits _{-L_5} ^{L_6} \diff x_3 \times \nn \\
& \times \Big[
\mathbf{f} ^{a} {}_{dc} \left \lbrace j_0 ^c (x_1) j_0 ^d (x_2) 
, j_1 ^b (x_3) \right \rbrace   _{\mathrm{SSM-PCM}}
+ \mathbf{f} ^{b} {}_{dc} \left \lbrace j_1 ^a (x_3) ,
 j_0 ^c (x_1) j_0 ^d (x_2)  \right \rbrace   _{\mathrm{SSM-PCM}}
\Big] \nonumber\\
& 
= \int \limits _{-L_3} ^{L_4} \diff x_1 \diff x_2 \, \epsilon(x_2-x_1) 
\int \limits _{-L_5} ^{L_6} \diff x_3 \Big[
j_0 ^c (x_1) \Big( 
\mathbf{f}^a {}_{dc}  \left( k^{db} (x_3) - G^{db} \right) \nonumber\\ 
& \hspace*{58mm}
-\mathbf{f} ^b {}_{dc}  \left( k^{da} (x_3) - G^{da} \right) \Big) 
\partial_{x_2} \, \delta(x_2 - x_3 ) \Big] \, ,
\end{align}
where we defined $\epsilon(x_2-x_1) = \theta(x_2-x_1)  - \theta(x_1-x_2)$ in the last line. Integrating by parts gives the boundary term
\begin{align}
& B^{ab} = \Big( \mathbf{f} ^a {}_{dc}  \left( ( k^{db} (L_4) - G^{db})
\theta(L_6-L_4) + ( k^{db} (-L_3) - G^{db} ) \theta(L_5-L_3)  \right) \nonumber\\
&- \mathbf{f} ^b {}_{dc}  \left(( k^{da} (L_4) - G^{da}) 
\theta(L_6-L_4)  + ( k^{da} (-L_3) - G^{da}) \theta(L_5-L_3) 
 \right) \Big) \chargeY^{(0)\, c} \, .
\end{align}
Again, the result shows the difference between the boundary terms arising for the principal chiral and the symmetric space model. In the case of a principal chiral model, the boundary terms are not relevant for the Serre relations due to the Jacobi identity $\mathbf{f} _b {} ^{[cd}  \mathbf{f} ^{a]b} {}_e =0$. The situation is different for a symmetric space model, where the above result shows that generically we have
\begin{align}
\mathbf{f}_b {} ^{[cd}  B^{a]b} \neq 0 \, ,
\end{align} 
such that the Serre relations \eqref{Serre} are violated by the boundary terms. We must hence require $L_4 > L_6$ and $L_5 > L_3$ in order for the Yangian algebra to hold true. In this case, the boundary terms are absent and we only have the bulk term, which takes the following form after taking $L_i$ to infinity in the appropriate order: 
\begin{align}
\left \lbrace \chargeY^{(1)\, a} , \chargeY^{(1)\, b} \right \rbrace _{\mathrm{SSM-PCM}} 
&= -2 \int \diff x \, j_0 ^c (x) \left( 
	- 2 \mathbf{f}  ^{ab} {} _c  + \mathbf{f} _c {} ^{ad} k_d {} ^b (x) 
	-  \mathbf{f}_c {} ^{bd} k_d {} ^a (x) \right) \nn \\
&= 2 \, \mathbf{f} ^{ab} {} _c \, \chargeY ^{(0)\, c}  .
\end{align}
Here, we have used the relation \eqref{kidentity} to eliminate the terms involving $k$. Hence, the difference to the result for the principal chiral model is given by a term which drops out of the Serre relations. We can thus conclude that the Serre relations for symmetric space models are satisfied if a particular ordering prescription is chosen for taking the boundaries of the integration domains to infinity. The Yangian relations for a symmetric space model can hence be understood to fix a limit-ambiguity in the definition of the charges. The situation is thus slightly different from the principal chiral model, where the order of $L_4,L_6$ and $L_3,L_5$ is not relevant to establish the Serre relation.%


\paragraph{Master charges.}

The conserved charges associated with the master symmetry are compositions of the Yangian charges, see e.g.\ table \ref{tab:overview}. Hence, the respective algebra relations are inherited from the Yangian algebra. For instance, the level-0 Yangian charge $\chargeY$ and the level-1 master charge $\chargeC ^{(1)} = \tr \big( \chargeY \, \chargeY ^{(1)} \big)$ given in \eqref{eq:lev1master}, commute due to \eqref{eq:Yangcomms}:
\begin{align}
\{\chargeY_a,\chargeC^{(1)}\}=
\chargeY^b\{\chargeY_a,\chargeY^{(1)}_b\}
+
\{\chargeY_a,\chargeY^b\}\chargeY^{(1)}_b=0.
\end{align}

\chapter{Minimal Surfaces in AdS}
\label{chap:MinSurf}

We now turn to the discussion of minimal surfaces in Anti-de Sitter spaces of various dimensions, which all arise as submanifolds of 
$\AdS_5$. Since $\AdS_N$ is a symmetric space, we can apply the symmetries discussed in the previous chapter. We begin by showing that a given symmetry of the area functional also leaves the renormalized area appearing in the strong-coupling description of the Maldacena--Wilson loop invariant. The symmetries of the area functional discussed in the last chapter are thus also symmetries of the Maldacena--Wilson loop. 
In order to gain further insight into these symmetries, we study their action on the boundary curves directly. As a prerequisite for this, we discuss the coset parametrization for $\AdS_N$ in detail in section \ref{sec:AdSCoset}.

We then turn to the study of large master symmetry transformations in section \ref{sec:LargeTrans}. The calculation of the master symmetry transformation requires the knowledge of the minimal surface and the analysis is thus limited to cases, where a minimal surface solution is known. We discuss the four-cusp solution in detail, which is of particular interest, since it describes the four-gluon scattering amplitude at strong coupling. 
Generic minimal surfaces can be treated numerically and we recall some of the numerical results found in reference \cite{Klose:2016qfv}. 

The variations of the boundary curve can again be studied analytically and we derive the variations of a generic boundary curve under the master and level-1 Yangian-like symmetry in section \ref{sec:InfSymm}. We make contact with the results of reference \cite{Muller:2013rta}, where the vanishing of the level-1 charge $Q^{(1)}$ was used to derive the Yangian symmetry of the Maldacena--Wilson loop at strong coupling.  

\section{Symmetry and Renormalization}
\label{sec:Renorm}
We have seen in section \ref{sec:MWL} that the Maldacena--Wilson loop at strong coupling is described by the renormalized area of the minimal surface ending on the respective boundary curve. It is then natural to ask whether also the renormalized area is invariant under the master symmetry. In fact, we show that any symmetry of the area functional and the equations of motion is also a symmetry of the renormalized minimal area $A_{\mathrm{ren}}(\gamma)$. The proof given below is an adaptation of the proof of the conformal invariance of the renormalized minimal area which is briefly%
\footnote{I would like to thank Harald Dorn for sharing his notes on that proof with me.} 
described in reference \cite{Dorn:2015bfa}.   

An important tool in the argument is the known expansion of the minimal surface around the conformal boundary, which was first derived in reference \cite{Polyakov:2000ti,Polyakov:2000jg}. The expansion can be derived from the equations of motion and Virasoro constraints by treating the boundary value problem as an initial value problem in the $\tau$-coordinate and expanding around the boundary located at $\tau=0$. The boundary conditions are given by
\begin{align}
X^\mu ( \tau = 0 , \sigma ) &= x^\mu (\sigma) \, , &
y ( \tau = 0 , \sigma ) &= 0 \, .
\end{align}
Since the equations of motion are second-order differential equations, they are under-determined as an initial value problem. Hence, we do not expect that the equations of motion allow to fix all coefficients of the $\tau$-expansion. The expansion would be determined completely, if one provided the conjugate momentum 
$\delta A / \delta x^\mu(\sigma)$. For the minimal surface problem, the conjugate momenta are determined from the requirement that the minimal surface closes, which cannot be studied in the expansion around the conformal boundary. The conjugate momentum would usually be given by the first coefficient in the $\tau$-expansion and all higher coefficients would be determined in terms of it. The situation is different for minimal surfaces in $\AdS$ due to the divergence of the metric on the conformal boundary, which forces the minimal surface to leave the boundary perpendicularly. 

For the determination of the Polyakov--Rychkov expansion we fix conformal gauge, such that the area functional takes the form
\begin{align}
A =  \frac{1}{2} \int \diff \tau \wedge \diff \sigma \, 
	 \frac{\partial_i X^\mu \partial_i X^\mu + \partial_i y \, \partial_i y}{y^2} \, .
\label{AreaFunctional}	 
\end{align} 
The equations of motion are then given by
\begin{align}
\partial ^2  X^\mu - \frac{2 \, \partial_i X^\mu \, \partial_i y }{y} &=0 \, , &
\partial^2 y + \frac{ \left( \partial X \right)^2 - \left( \partial y \right) ^2}{y} 
&=0 \, , 
\label{AdSEOM}
\end{align}
and we note the Virasoro constraints 
\begin{align}
\left( \partial_\tau X \right) \left( \partial_\sigma X \right) 
+ \partial_\tau y  \, \partial_\sigma y &= 0 \, , &
\left( \partial_\tau X \right)^2 + \left( \partial_\tau y \right)^2
- \left( \partial_\sigma X \right)^2 - \left( \partial_\sigma y \right)^2 &= 0 \, .
\label{Virasoro}
\end{align}
With these equations at hand, we turn to the determination of the $\tau$-expansion. We fix the following convention for the $\tau$-expansion of a generic function $f(\tau, \sigma)$:
\begin{align}
f(\tau, \sigma) = \sum \limits _{n=-m} ^\infty f_{(n)} (\sigma) \, \tau ^n \, .
\label{def:expansion}
\end{align}
Here, we have allowed for a Laurent-expansion, which appears e.g.\ for the Noether current. The parametrization of the minimal surface is not divergent in the limit $\tau \to 0$ and from the equations of motion we find the coefficients
\begin{align}
X_{(1)} ^\mu (\sigma) &= 0 \, , & 
X_{(2)} ^\mu (\sigma) &= \frac{1}{2} \, \dot{x}^2(\sigma) \, \partial_\sigma 
	\left( \frac{\dot{x}^\mu(\sigma)}{\dot{x}^2(\sigma)} \right)  , \\
y_{(1)} (\sigma) &= \lvert \dot{x} \rvert \, , &
y_{(2)} (\sigma) &= 0 \, .
\label{PR1}
\end{align}
The third-order coefficients are not fixed by the equations of motion. From our above discussion we thus expect that the third-order coefficient of $X^\mu$ is related to the functional derivative of the minimal area. We hence consider a variation $\delta x^\mu(\sigma)$ of the boundary curve. The variation of the boundary curve induces a variation $\left(\delta X^\mu , \delta y \right)$ of the parametrization of the minimal surface. Let us then compute the variation of the renormalized minimal area,
\begin{align}
A_\mathrm{ren}(\gamma) = \lim \limits _{\varepsilon \to 0} \left \lbrace A(\gamma) \big \vert_{y \geq \varepsilon} - \frac{L(\gamma)}{\varepsilon} \right \rbrace \, , 
\label{def:Aren}
\end{align}
which is regulated by demanding $y \geq \varepsilon$, or equivalently $\tau \geq \tau_0(\s)$, where $\tau_0(\s)$ is defined by $y(\tau_0(\s),\s) =\varepsilon$, which we can rewrite as 
\begin{align}
\tau_0(\sigma) = \frac{\varepsilon}{\lvert \dot{x} (\s) \rvert} + \O ( \varepsilon^3) , 
\end{align}
employing the coefficients of $y$ derived above. Since we are varying around a minimal surface solution, we may employ that $(X^\mu,y)$ satisfy the equations of motion and hence the variation is given by a boundary term,
\begin{align*}
\delta A \big \vert _{y \geq \varepsilon} = 
	\int \limits _0 ^{2 \pi} \diff \sigma \hspace*{-2mm} 
	\int \limits _{\tau_0(\sigma)} ^c  \hspace*{-2mm} \diff \tau 
	\, \partial_i \, \frac{ \partial_i X^\mu \, \delta X_\mu + \partial_i y \, \delta y}{y^2}
	= \frac{1}{\varepsilon^2} \int \limits _0 ^{2 \pi} \diff \sigma 
	\left[ \tau_0^\prime (\sigma) \,  \partial_\sigma X^\mu \delta X_\mu 
	-  \partial_\tau X^\mu \delta X_\mu \right] .
\end{align*}
Here, we used that $\delta y(\tau_0(\s),\s)=0$ due to the definition of $\tau_0$ and employed the periodicity of the solutions in $\sigma$. Inserting the results \eqn{PR1} we then find
\begin{align*}
\delta A \big\vert _{y \geq \varepsilon} = \frac{\delta L(\gamma)}{\varepsilon} 
	- \int \limits _0 ^{2 \pi}
	\diff \sigma \, \frac{3 X_{(3)}^\mu}{\dx^2} \, \delta x_\mu \, ,
\end{align*}
from which we read off that
\begin{align}
X_{(3)} ^\mu (\s) = - \frac{\dx^2}{3} \, \dfrac{\delta A_\mathrm{ren}(\gamma) }{\delta x_\mu (\sigma) } \, . \label{funcder}
\end{align}
To fix the higher coefficients in the $\tau$-expansion, it is convenient to restrict the parametrization of the boundary curve to satisfy $\dx ^2 \equiv 1$. The residual reparametrization invariance in conformal gauge is sufficient to do so. We have worked with a general parametrization so far to avoid possible problems in calculating the variation above. In the text below, we will indicate working with an arc-length parametrization by using boundaries $[0 , L(\gamma)]$ for the variable $\sigma$, whereas we use the boundaries $[0 , 2 \pi]$ for generic parametrizations.
\begin{figure}
\centering
\includegraphics[width=125mm]{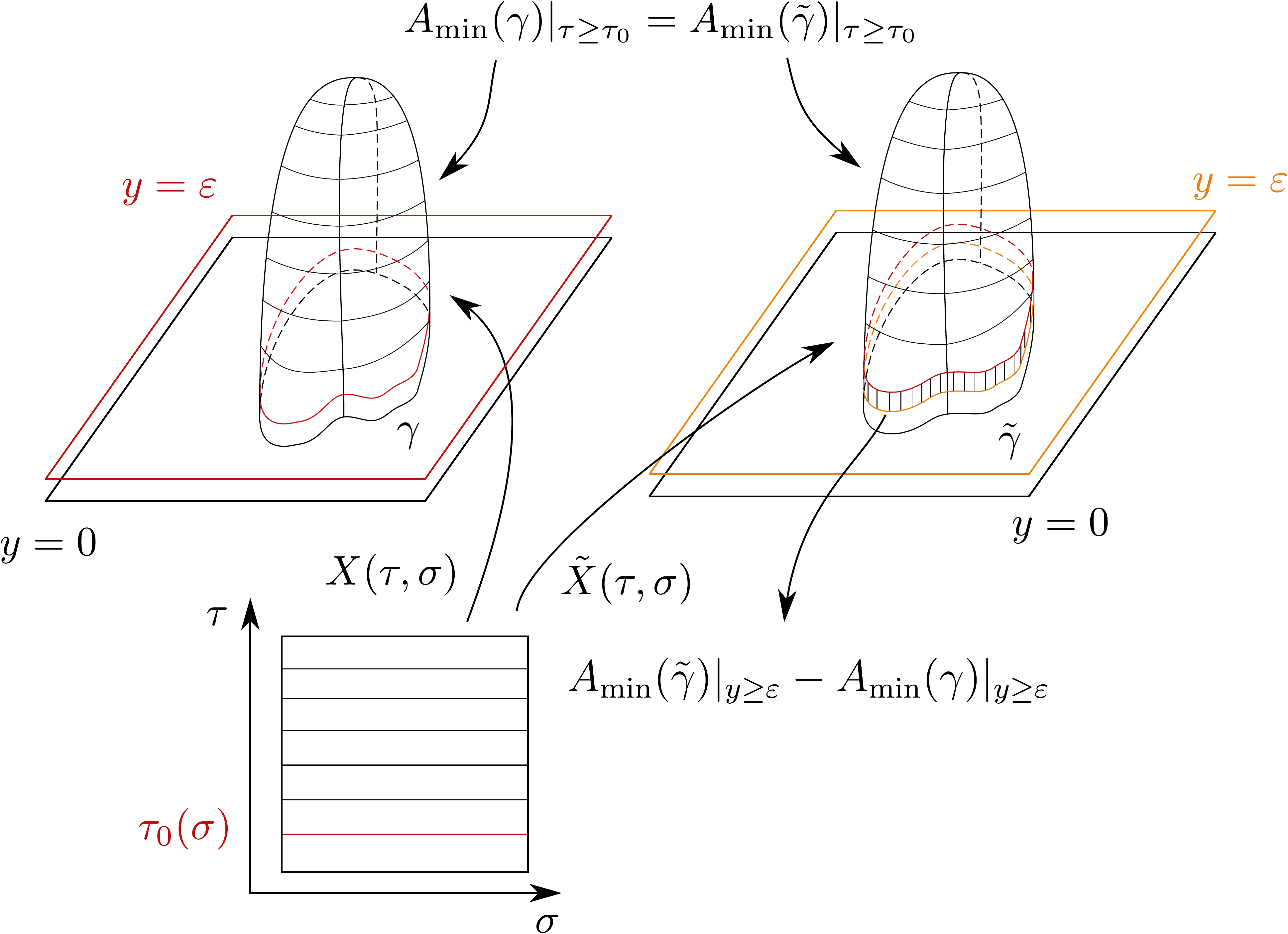}
\caption{Comparison of the minimal surfaces associated to the boundary curves $\gamma$ and $\tilde{\gamma}$.}
\label{fig:renormalization}
\end{figure}  

The third-order coefficient of $y(\tau , \sigma)$ can be obtained from the Virasoro constraints, which gives
\begin{align}
y_{(3)} (\sigma) = - \frac{\ddot{x}(\sigma) ^2}{3} \, .
\end{align}
The fourth-order coefficient of $X^\mu(\tau , \sigma)$ can also be obtained from the equations of motion, but we will not need it for our analysis. In summary, we have found the following expansion for the minimal surface solutions:
\begin{align}
X^\mu \left( \tau , \sigma \right) &= x^\mu (\sigma)
+ \frac{\tau^2}{2} \, \dot{x}^2(\sigma) \, \partial_\sigma \left( \frac{\dot{x}^\mu(\sigma)}{\dot{x}^2(\sigma)} \right) 
- \frac{\tau^3}{3} \, \dot{x}^2(\sigma) \, \frac{\delta A_{\mathrm{ren}}(\gamma)}{\delta x_\mu (\sigma)} 
+ \mathcal{O}\left(\tau^4 \right)  ,\label{eqn:expansion_boundary} \\
y \left( \tau , \sigma \right) &= \tau \, \lvert \dot{x}(\sigma) \rvert + \O (\tau^3) \, .
\label{eqn:expand_bound}
\end{align}
If we choose an arc-length parametrization, the above expansion simplifies to
\begin{align}
X^\mu \left( \tau , \sigma \right) &= x^\mu (\sigma)
+ \frac{\tau^2}{2} \, \ddot{x}^\mu (\sigma)
- \frac{\tau^3}{3} \, \frac{\delta A_{\mathrm{ren}}(\gamma)}{\delta x_\mu (\sigma)} 
+ \mathcal{O}\left(\tau^4 \right)  ,\label{eqn:expansion_boundary_arc} \\
y \left( \tau , \sigma \right) &= \tau \,  
	- \frac{\tau^3}{3} \, \ddot{x}(\sigma) ^2 + \O (\tau^4) \, .
\label{eqn:expand_bound_arc}
\end{align}
With the above expansion established, we can now turn to the question whether a generic symmetry of the area functional leads to a symmetry of the renormalized minimal area. Since we are considering a symmetry transformation, the transformed surface $\lbrace \tilde{X}^\mu (\tau , \sigma ) , \tilde{y}(\tau , \sigma ) \rbrace$ ending on the transformed boundary contour $\tilde \gamma$ is also a solution of the equations of motion in conformal gauge and thus we have
\begin{align*}
\tilde{X}^\mu ( \tau , \sigma) &= \tilde{x}^\mu (\sigma ) + \mathcal{O}(\tau^2) \, , &
 \tilde{y} ( \tau , \sigma) &= \tau \lvert \dot{\tilde{x}}(\sigma) \rvert + \mathcal{O}(\tau^3) \, .
\end{align*}
We note now that the cut-off of the original surface at $y= \varepsilon$ is not necessarily mapped to the cut-off of the transformed surface, which is situated at $\tilde{y} = \varepsilon$, see figure \ref{fig:renormalization}. In parameter space, the cut-off of the transformed surface corresponds to $\tilde{\tau}_0(\sigma)$, where we have again defined
$ \tilde{y} \left( \tilde{\tau}_0 (\sigma) , \sigma \right) = \varepsilon$. Using now that we are discussing a symmetry of the area functional, the transformed surface covers the same area in a given parameter region, i.e.\ we have
\begin{align*}
A_{\mathrm{min}}(\gamma) \big \vert _{\tau \geq \tau_0 (\sigma)} = A_{\mathrm{min}}(\tilde{\gamma}) \big \vert _{\tau \geq \tau_0 (\sigma)} \, .
\end{align*}
Correspondingly, the difference between the minimal areas for the original curve $\gamma$ and the transformed curve $\tilde{\gamma}$ is given by integrating over the surface between the two cut-offs situated at $\tau_0 (\sigma)$ and $\tilde{\tau}_0 (\sigma)$,  
\begin{align}
& A_{\mathrm{min}}(\gamma) \big \vert _{y \geq \varepsilon} 
	- A_{\mathrm{min}}(\tilde{\gamma}) \big \vert _{\tilde{y} \geq \varepsilon} 
	=  \frac{1}{2} \int \limits _{0} ^{2 \pi} \diff \sigma 
	\int \limits _{\tau_0 (\sigma)} ^{\tilde{\tau}_0 (\sigma)} \diff \tau \, 
	\frac{\partial_i \tilde{X}^\mu \partial_i \tilde{X}^\mu 
	+\partial_i \tilde{y} \partial_i \tilde{y}}{\tilde{y}^2} \nn \\
	& \qquad = \int \limits _{0} ^{2 \pi} \diff \sigma 
	\int \limits _{\tau_0 (\sigma)} ^{\tilde{\tau}_0 (\sigma)} \diff \tau \, 
	\left( \frac{1}{\tau^2} + \mathcal{O} ({\tau^0}) \right) \nn \\
	& \qquad =   \int \limits _{0} ^{2 \pi} \diff \sigma \, 
	\frac{\lvert \dot{x}(\sigma )\rvert }{\varepsilon} 
	- \int \limits _{0} ^{2 \pi} \diff \sigma \, 
	\frac{ \lvert \dot{\tilde{x}}(\sigma ) \rvert}{\varepsilon} + \mathcal{O}(\varepsilon) 
	= \frac{L(\gamma)}{\varepsilon} - \frac{L(\tilde{\gamma})}{\varepsilon} 
	+ \mathcal{O}(\varepsilon) \, .
\end{align}
Inserting the definition \eqref{def:Aren} of the renormalized minimal area, we thus find that it is invariant under the map $\gamma \mapsto \tilde{\gamma}$ induced by any symmetry of the model.

\section{Coset Space Construction}
\label{sec:AdSCoset}
In order to transfer the symmetries discussed in chapter 
\ref{chap:SSM}, we describe $\AdS_d$ as a coset space and introduce a particular set of coset representative, which will be convenient later on. We begin with the space $\AdS_3$, which is special since it can be identified with the Lie group $\grp{SL}(2, \mathbb{R})$. Generically, the coset spaces G/H cannot be equipped with a group structure. We then turn to the coset space construction for the higher-dimensional AdS-spaces, considering both Euclidean and Lorentzian signature. The isometry groups are given by
$\grp{SO}(2,d-1)$ or $\grp{SO}(1,d)$, respectively, and we have the coset spaces
\begin{align}
\AdS_d & \simeq \grp{SO}(2,d-1)/\grp{SO}(1,d-1) \, , &
\EAdS_d & \simeq \grp{SO}(1,d)/\grp{SO}(d) \, .
\end{align}
Due to the gauge freedom, there are different ways to describe the above spaces by choosing a set of coset representatives on G. In order to read off the corresponding point on 
$\MM = \grp{G} / \grp{H}$ from an element of G without first determining to which coset representative it is related by a gauge transformation, one can employ the gauge-invariant quantity given by
\begin{align}
\mathbb{G} =  \sigma(g) \, g^{-1} \, .
\label{G:gauge_invariant}
\end{align}
The gauge invariance follows directly from the fact that $\sigma$ acts as the identity on H. Note also that $\mathbb{G}$ is related to the Noether current (which is also gauge-invariant) by
\begin{align}
j = \mathbb{G}^{-1} \diff \mathbb{G} \, .
\end{align}

\subsection{The Group Manifold for \texorpdfstring{$\AdS_3$}{AdS3}}
The identification of $\AdS_3$ with the Lie group 
$\grp{SL}(2,\mathbb{R})$ is easiest to describe in embedding coordinates, where we map $\mathbb{R}^{(2,2)}$ to 
$\grp{GL}(2,\mathbb{R})$ by
\begin{align}
\left( Z_{-1} , Z_0 , Z_1 , Z_2 \right)
\; \; \mapsto \; \; 
\begin{pmatrix}
Z_{-1} + Z_2 & Z_1 + Z_0 \\
Z_1 - Z_0 & Z_{-1} - Z_2 
\end{pmatrix} 
= g (Z) \, .
\end{align}
With this identification, we have
\begin{align}
\det (g(Z)) = 1 \quad \Leftrightarrow \quad
-Z_{-1}^2 - Z_0 ^2 + Z_1^2 + Z_2 ^2 = -1 \, , 
\end{align}
such that $\AdS_3$ is identified with $\grp{SL}(2,\mathbb{R})$. It is then easy to see that left and right multiplication by elements of $\grp{SL}(2, \mathbb{R})$ give isometries of 
$\AdS_3$. As we have seen in section \ref{sec:AdS/CFT}, the embedding coordinates for $\AdS_3$ are related to Poincar{\'e} coordinates by
\begin{align}
Z_0 &= \frac{X_0}{y} \, , & 
Z_1 &= \frac{X_1}{y} \, , & 
Z_{-1} + Z_2 &= \frac{1}{y} \, , &
Z_{-1} - Z_2 &= \frac{y^2 + X_1 ^2 - X_0 ^2}{y} \, ,
\end{align}
and hence the identification of $\AdS_3$ with $\grp{SL}(2,\mathbb{R})$ takes the following form in Poincar{\'e} coordinates:
\begin{align}
g(X,y) = \frac{1}{y}
	\begin{pmatrix}
		1 & X_1 + X_0 \\
		X_1 - X_0 & y^2 + X_1^2 - X_0^2
	\end{pmatrix} \, .
\label{eqn:gPCM}
\end{align}
We can then convince ourselves that the trace on $\grp{SL}(2,\mathbb{R})$ reproduces the $\AdS$-metric. With 
$U_i = g^{-1} \partial_i g$, we find
\begin{align}
\tr \left( U_i \, U_j \right)
	= 2 \, \frac{\partial_i X_1 \, \partial_j X_1
	- \partial_i X_0 \, \partial_j X_0 + 
	\partial_i y \, \partial_j y}{y^2} \, . 
\end{align}
We can hence describe string theory in $\AdS_3$ as the principal chiral model on $\grp{SL}(2, \mathbb{R})$, 
\begin{align}
S = - \frac{T}{4}\int \tr \left( U \wedge \ast U \right) \, .
\end{align}
This is of course equivalent to the description as the symmetric space model on 
$\grp{SL}(2,\mathbb{R}) \times \grp{SL}(2,\mathbb{R}) / \grp{SL}(2,\mathbb{R})$. The description as a principal chiral model is more convenient in our case due to the absence of the three gauge degrees of freedom. We have seen in section \ref{sec:PCM} that the master symmetry can be formulated also for these models.

\subsection{The Coset Space for \texorpdfstring{$\AdS_d$}{AdSd}}
For the higher-dimensional $\AdS$-spaces, we employ a coset construction based on the respective isometry groups. We have discussed our conventions for the fundamental representations of the isometry groups $\grp{SO}(2,d-1)$ and $\grp{SO}(1,d)$ in section \ref{sec:Int_Symm} and we recall the commutation relations 
\begin{align}
\left[M_{\mu \nu} , M_{\rho \sigma} \right] = \eta_{\mu \rho} M_{\nu \sigma} 
	- \eta_{\mu \sigma} M_{\nu \rho} + \eta_{\nu \sigma} M_{\mu \rho} 
	- \eta_{\nu \rho} M_{\mu \sigma} \, , 
\end{align}
as well as 
\begin{align}
\left[ D, P_{\mu} \right] &=  P_{\mu} \, ,&  
\left[ M_{\mu \nu}, P_{\lambda} \right] &= \eta_{\mu \lambda} P_{\nu}  
	- \eta_{\nu \lambda} P_{\mu}   \, , & 
\left[ P_{\mu}, K_{\nu} \right] &=  2 \eta_{\mu \nu} \, D  - 2 M_{\mu \nu} \, , \nn \\
\left[ D, K_{\mu} \right] &= - K_{\mu}\, ,&  
\left[ M_{\mu \nu}, K_{\lambda} \right] &= \eta_{\mu \lambda} K_{\nu} 
	-  \eta_{\nu \lambda} K_{\mu}   \, .   \label{conf_algebra}
\end{align}
For the trace metric, we recall the expressions 
\begin{align}
\tr \left( M_{\mu \nu} M_{\rho \sigma} \right) &= 
	2 \, \eta_{\mu \sigma} \, \eta_{\nu \rho} - 2\, \eta_{\mu \rho} \, \eta_{\nu \sigma} \, , &
\tr \left(P_\mu \, K_\nu \right) &= 4 \, \eta_{\mu \nu} \, , & 
\tr \left( D \, D \right) &= 2 \, ,	
\label{eqn:Metric}
\end{align}
and the $\mathbb{Z}_2$ grading of the algebra led to the decomposition
\begin{align}
\alg{h} &= \mathrm{span} \left \lbrace M_{\mu \nu} , P_\mu - K_\mu \right \rbrace  \, , &
\alg{m} &= \mathrm{span} \left \lbrace P_\mu + K_\mu , D \right \rbrace \, .
\end{align}
The Lie algebra $\alg{h}$ of the gauge group is isomorphic to 
$\alg{so}(d)$ or $\alg{so}(1,d-1)$, respectively. 
A convenient choice of coset representatives is given by
\begin{align}
g(X,y) = e^{X \cdot P} \, y^D    \quad \Rightarrow \quad U = g^{-1} \, \diff g = \frac{\diff X^\mu}{y} \, P_\mu + \frac{\diff y}{y} \, D \, , \label{eqn:UXy}
\end{align}
and for the projections of the Maurer--Cartan form, we note
\begin{align}
A &=  \frac{\diff X^\mu}{2y} \left( P_\mu - K_\mu \right) \, ,  &
a &= \frac{\diff X^\mu}{2y} \left( P_\mu + K_\mu \right)+ \frac{\diff y}{y} \, D \, .
\end{align}
The metric of the coset space is obtained from the group metric introduced above and the projection $a$ of the Maurer--Cartan form as
\begin{align}
\tr \left( a_i \, a_j \right) = 2 \, 
	\frac{ \eta^{\mu \nu}  \partial_i X_\mu  \, \partial_j X_\nu 
	+ \partial _i y \, \partial_j y }{y^2} \, .
\end{align}
This shows that the parametrization $g(X,y)$ provides Poincar{\'{e}} coordinates for $\mathrm{\EAdS}_N$ or $\AdS_N$, respectively. Correspondingly, we may describe the string action in these coordinates by
\begin{align}
S = - \frac{T}{4} \int \tr \left( a \wedge \ast a \right) \, .
\end{align}
In order to get acquainted with our above choice of coset representatives, we derive the coordinate expressions for the
G-Symmetry of the model. The transformations are described by left-multiplication of the coset representatives by a constant $L \in \grp{G}$,
\begin{align}
g \left( X , y   \right) \mapsto L \cdot g \left( X, y  \right) = g \left( X ^\prime , y  ^\prime  \right) \cdot R \, .
\end{align}
Here, we need to allow for a general gauge transformation $R \in \grp{H}$. For an infinitesimal transformation, we replace $L$ by the generators $T_a$ of $\grp{G}$, which implies the variation
\begin{align}
\partial_\mu g \left( X , y  \right) \, \delta _a X^\mu 
+ \partial_y g \left( X , y  \right) \, \delta _a y 
= T_a \cdot g \left(X , y\right) 
- g \left( X , y\right) \cdot h_a \, ,
\end{align}
where $h_a \in \alg{h}$ is an element of the gauge Lie algebra. A more convenient expression in order to read off the variations $(\delta_a X^\mu , \delta_a y)$ is given by the components of the Maurer--Cartan form,
\begin{align}
U_\mu \delta _a X^\mu + U_y \delta_a y = g \left( X, y \right)^{-1} \cdot T_a \cdot g \left( X, y \right) - h_a \, .
\end{align}
Inserting equation \eqref{eqn:UXy}, we thus have
\begin{align}
\delta_a X^\mu \, P_\mu + \delta_a y \, D = y \left( g \left( X , y \right)^{-1} \cdot T_a \cdot g \left( X, y \right) - h_a \right) \, .
\end{align}
As the right hand side of this equation only contains the generators $P_\mu$ and $D$, the gauge transformation $h_a$ can be determined from the terms proportional to $K_\mu$ and $M_{\mu \nu}$ in $g^{-1} T_a g$, which gives
\begin{align*}
h_a = \frac{1}{4} \tr \left( g^{-1} T_a \, g \, P^\mu \right) \left(K_\mu - P_\mu \right) 
- \frac{1}{4} \tr \left( g^{-1} T_a \, g \, M^{\mu \nu} \right) M_{\mu \nu} \, .
\end{align*}
Using the expressions \eqref{eqn:Metric} for the group metric we then have
\begin{align}
\delta_a X^\mu &= \frac{y}{4} \left( \tr \left(  g^{-1} T_a g \,  K^\mu \right)  
+  \tr \left( g^{-1} T_a g \, P^\mu \right) \right) \, , &
\delta_a y &= \frac{y}{2} \tr \left( g^{-1} T_a g \, D \right) \, .
\label{eqn:varcoord}
\end{align}
We are particularly interested in the variation of the coordinates $x^\mu$ at the conformal boundary $y=0$. In order to take the boundary limit 
$y \to 0$, we employ the definition of $g(X,y)$ as given in \eqref{eqn:UXy} and compute the conjugation of any generator with $y^D$. This can be done by noting that our choice of basis is such that the commutation with $D$ is diagonal,
\begin{align}
\left[ D , T_a \right] = \Delta \left( T_a \right) \, T_a \, ,
\end{align}
which implies that
\begin{align}
y^D \, T_a \, y^{-D} = y^{\Delta \left( T_a \right)} \, T_a \, .
\end{align}
The above relation follows from a simple application of Hadamard's lemma. With the $n$-fold commutator defined by
$[A, B]_{(n)} = [A , [A, B]_{(n-1)}]$ and $[A,B]_{(0)} = B$, we have
\begin{align*}
y^D \, T_a \, y^{-D} 
&= e^{\ln (y) D} \, T_a \, e^{-\ln(y) D} 
= \sum \limits _{n=0} ^\infty \frac{1}{n!}
	\big[ \ln (y) D , T_a \big]_{(n)} \\
&= \sum \limits _{n=0} ^\infty 
	\frac{\left( \ln (y) \Delta \left( T_a \right) \right)^n}
	{n!} \, T_a 
= y^{\Delta \left( T_a \right)} \, T_a \, .
\end{align*}
Using the cyclicity of the trace in \eqref{eqn:varcoord} as well as $\Delta(K^\mu) =-1$, $\Delta(P^\mu) =1$ and $\Delta(D) =0$, we thus have 
\begin{align*}
\delta _a y \overset{y \to 0}{\longrightarrow} 0 \, , 
\end{align*}
which ensures that $\mathrm{AdS}$-isometries map the conformal boundary to itself and the variation of the boundary coordinates is given by
\begin{align}
\delta_a x^\mu = \frac{1}{4} \tr \left(  e^{-x \cdot P} T_a \, e^{x \cdot P}   K^\mu \right) 
	= \xi ^\mu _a (x) \, . 
\label{Def:ConfKilling}
\end{align}
Here, $\xi^\mu _a(x)$ form a basis of conformal Killing vectors of the flat boundary space, as expected. Concretely, we find the expressions 
\begin{align}
\xi^\mu _a (x) = \left \lbrace  \delta ^\mu _\nu , \,
	 x_\nu \delta ^\mu _\rho - x_\rho \delta ^\mu _\nu , \, 
	 x^\mu , \,
	x^2 \delta ^\mu _\nu - 2 x^\mu x_\nu \right \rbrace 
\end{align}
for the generators
\begin{align}
T_a = \left \lbrace P_\nu , \,
	M_{\nu \rho} , 	\,
	D , \, 
	K_\nu \right \rbrace . 
\end{align}
We note that the conformal Killing vectors $\xi^\mu _a(x)$ satisfy the algebraic identities
\begin{align}
\left \lbrace \xi _a , \xi _b \right \rbrace ^\mu 
	= \xi ^\nu _a \partial_\nu \, \xi ^\mu _b 
	- \xi ^\nu _b \partial_\nu \, \xi ^\mu _a 
	= f_{ba} {} ^c \, \xi ^\mu _c 
	= \mathbf{f} _{ab} {} ^c \, \xi ^\mu _c \, .
\end{align}
Here, as before, $f_{ab} {} ^c$ denote the structure constants of the underlying Lie group 
$\grp{SO}(1, N)$ or $\grp{SO}(2, N-1)$ in the basis of the generators $T_a$, 
\begin{align}
\left[ T_a , T_b \right] = f_{ab} {} ^c \, T_c \, ,
\end{align}
and $\mathbf{f} _{ab} {} ^c$ denote the structure constants of the symmetry variations. Moreover, the vectors $\xi^\mu _a(x)$ satisfy the conformal Killing equation
\begin{align}
\partial^\mu \xi^\nu _a +  \partial^\nu \xi^\mu _a = 
	\half \left( \partial_\rho \, \xi^\rho _a \right) \eta ^{\mu \nu} \, .
\end{align} 

\section{Large Symmetry Transformations}

We now turn to the discussion of large symmetry transformations for the circular and four-cusp solution. 

\label{sec:LargeTrans}
\subsection{Analytical Deformations}
\paragraph{The circular solution.}
The simplest minimal surface in Euclidean Anti-de Sitter space is the minimal surface for the boundary curve being a circle, which we have already discussed in section \ref{sec:MWL}. We recall that the minimal surface can be parametrized (in conformal gauge) by
\begin{align}
X_1 (\tau , \sigma) &= \frac{\cos \sigma}{\cosh \tau} \, , &
X_2 (\tau , \sigma) &= \frac{\sin \sigma}{\cosh \tau} \, , & 
y (\tau , \sigma) &= \tanh \tau \, . &  
\label{sol:circle} 
\end{align} 
The master symmetry deformation of this minimal surface was found in reference \cite{Dekel:2015bla} to be given by an AdS-isometry. This finding can be explained by the fact that the circle is a contour, for which the mapping between the boundary curve and the area of the associated minimal surface reaches a minimum. In order to see this, note that we have already calculated the variation of the renormalized minimal area for generic variations of the boundary curve in equation \eqref{funcder}, where we found that
\begin{align}
\delta A_{\mathrm{ren}} (\gamma) =  - \int \limits _0 ^{2 \pi}
	\diff \sigma \, \frac{3 X_{(3)}^\mu}{\dx^2} \, \delta x_\mu \, .  
\end{align}    
Expanding the above minimal surface solution around the boundary $\tau = 0$, we see that the variation around the circle vanishes, 
\begin{align}
\delta A_{\mathrm{ren}} \left(  \gamma = 
	\raisebox{-.9mm}{\includegraphics[height=2.2ex]{circle.pdf}}   
	\right) = 0 \, .
\end{align}
This suggests that the master symmetry deformation --- which leaves the renormalized area invariant --- should not lead to a different minimal surface, i.e.\ one that is not related to the original surface by an AdS-isometry and a reparametrization. The above finding is not quite sufficient to prove that there are no minimal surfaces with the same area as the circular one, since we cannot exclude the possibility of degenerate minimal boundary curves which are not related by conformal transformations. 

\paragraph{The four-cusp solution.}
Another example of an analytically-known minimal surface is the four-cusp solution discussed in reference \cite{Alday:2007hr}. The four-cusp solution is of particular interest here since it describes the four-gluon scattering amplitude at strong coupling. 

For any minimal surface ending on a light-like polygon, it is an interesting question whether the master symmetry again yields a light-like polygonal boundary curve and thus relates gluon scattering amplitudes at strong coupling. Since the minimal surfaces for polygons with more than four edges are not known analytically, the four-cusp solution is a natural starting point to address this question. One should point out, however, that conformal symmetry is very restrictive for light-like polygons with less than six edges, since no conformal cross ratios can be formed. This implies that all light-like polygons with four edges are related by conformal transformations. Hence, if the master symmetry maps the minimal surfaces for
light-like polygons to such, it will be equivalent to an AdS-isometry in the case of the four-cusp solution. 

Before we turn to the discussion of the master symmetry, let us briefly discuss the four-cusp solution. Interestingly, it is quite similar to the circular minimal surface in Euclidean 
$\AdS_3$. Defining $r^2 = X_1 ^2 + X_2 ^2$, the solution for the circular Wilson loop can be written as
\begin{align}
y(r) = \sqrt{R^2 - r^2} \, ,    
\end{align}    
for a circle with radius $R$, which we have set to one before. It is plausible to expect that the minimal surface ending on the hyperbola in Minkowski space can be obtained from the solution above. If we formally adapt the above solution, we obtain the parametrization
\begin{align}
X_0 (\sigma , r ) &= r \cosh \sigma \, , &
X_1 (\sigma , r ) &= r \sinh \sigma \, , &
y (\sigma , r ) &= \sqrt{r^2 - R^2} \, , 
\label{sol:Hyperbola}
\end{align}
ending on the hyperbola defined by $X_0^2 - X_1 ^2 = R^2$. Let us check that the above parametrization gives a solutions of the equations of motion. Assuming that $y$ is a function only of $r$, we have the area functional
\begin{align}
A = \int \diff r \, \diff \sigma \, 
	\frac{r \, \sqrt{y^\prime(r) ^2  - 1} }{y(r)^2} \, , 
\label{Area:Hyperbola}	
\end{align}
where we have assumed that $y'(r)^2 \geq 1$ as above, such that the induced signature on the surface is Euclidean. We then find the equations of motion 
\begin{align}
\partial_r \left( \frac{r \, y^\prime (r) }{ y(r)^2 \, \sqrt{y^\prime(r) ^2  - 1} } \right)
	+ \frac{2r \, \sqrt{y^\prime(r) ^2  - 1} }{y(r)^3} = 0 \, ,
\end{align}
and it is easy to check that the parametrization \eqref{sol:Hyperbola} indeed solves the equations of motion. One might now expect that the solution for the boundary curve being the light-cone centered at the origin, i.e.\ the single-cusp solution, can be obtained from the solution \eqref{sol:Hyperbola} by taking the limit $R \to 0$. Note however, that the induced metric becomes degenerate in this limit, such that the area vanishes. 

If we then consider surfaces with $y(r) = \alpha r$, with $\alpha > 1$ to ensure Euclidean signature, we find that the integrand%
\footnote{Naturally, the area is divergent since it corresponds to the divergent cusped Wilson loop.}
in equation \eqref{Area:Hyperbola} becomes
\begin{align}
\frac{\sqrt{\a^2-1}}{\a^2} \, \frac{1}{r} \, , 
\end{align}
and hence reaches zero in the limits $\alpha \to 1$ and $\alpha \to \infty$ and becomes extremal for $\alpha = \sqrt{2}$. This observation was made in reference \cite{Kruczenski:2002fb}, where the single-cusp solution
\begin{align}
X_0 (\sigma , r ) &= r \cosh \sigma \, , &
X_1 (\sigma , r ) &= r \sinh \sigma \, , &
y (\sigma , r ) &= \sqrt{2} \, r \, , 
\label{sol:Cusp}
\end{align}
was found. A parametrization solving the equations of motion in conformal gauge is given by
\begin{align}
X_0 &= e^\tau \, \cosh(\sigma) \, , &
X_1 &= e^\tau \, \sinh(\sigma) \, , &
y &= \sqrt{2} \, e^\tau \, .
\label{sol:SingleCusp}
\end{align}  
It was then noted in reference \cite{Alday:2007hr} that the single-cusp solution is related to the four-cusp solution by an AdS-isometry. In order to discuss the relation, it is convenient to use embedding coordinates $(Z_{-1}, Z_0 , Z_1 , Z_2, Z_3 , Z_4 )$ for $\AdS_5$, for which the AdS-isometries are given by $\grp{SO}(2,4)$-matrices. The embedding coordinates satisfy the constraint
\begin{align}
-  Z_{-1}  ^2 -  Z_0  ^2 
	+  Z_1 ^2 +  Z_2 ^2 +  Z_3 ^2 +  Z_4 ^2 = -1 \, . 
\label{AdS5:def}	 
\end{align}  
We translate the equations describing the minimal surface to embedding coordinates. We recall that they are related to the Poincar{\'e} coordinates by the relations
\begin{align}
Z_\mu &= \frac{X_\mu}{y} \, , &
Z_{-1} + Z_4 &= \frac{1}{y} \, , &
Z_{-1} -  Z_4 &= \frac{X^\mu X_\mu + y^2}{y} \, .
\end{align}
It is then easy to see that the equations 
\begin{align}
y^2 &= 2 \left( X_0^2 - X_1 ^2 \right) \, , &
X_2 &= 0 \, , & 
X_3 &= 0 \, , 
\label{eqn:SingleCusp}
\end{align}
describing the single-cusp solution in Poincar{\'e} coordinates translate to the equations
\begin{align}
Z_4 ^2 - Z_1 ^2 &= Z_{-1}^2 - Z_0^2 \, , & 
Z_2 &= 0 \, , &
Z_3 &= 0 \, . 
\end{align}
If we now apply the AdS-isometry 
\begin{align*}
Z_{-1} ^\prime &= \frac{1}{\sqrt{2}} \left(Z_0 - Z_{-1} \right) \, , &
Z_0 ^\prime &= \frac{1}{\sqrt{2}} \left(Z_0 + Z_{-1} \right) \, , &
Z_1 ^\prime &= \frac{1}{\sqrt{2}} \left(Z_1 + Z_4 \right) \, , \\
Z_2 ^\prime &= \frac{1}{\sqrt{2}} \left(Z_1 - Z_4 \right) \, , &
Z_3 ^\prime &= Z_3 \, , &
Z_4 ^\prime &= Z_2 \, ,
\end{align*}
the transformed surface satisfies the equations
\begin{align}
Z^\prime_1 Z ^\prime_2 &= Z_{-1} ^\prime Z_0 ^\prime \, , & 
Z_3 ^\prime &= 0 \, , &
Z_4 ^\prime &= 0 \, .  
\end{align}
In terms of Poincar{\'e} coordinates, these equations are in turn expressed as
\begin{align}
X_0 ^\prime &= X_1^\prime X_2^\prime \, , &
y^{\prime \, 2} &= \left( 1-X_1 ^{\prime \, 2} \right) \left( 1  - X_2 ^{\prime \, 2} \right) \, , &
X_3 ^\prime &= 0 \, . 
\end{align}
In this form it is easy to see that the boundary curve is described by the four light-like edges
\begin{align}
\big( X_1 &= \pm 1 \, , \quad X_0 = \pm X_2 \big) \, , &
\big( X_2 &= \pm 1 \, , \quad X_0 = \pm X_1 \big) \, .
\end{align}
The four-cusp solution is hence equivalent to the single-cusp solution, and we can use either one to study the master symmetry, since it commutes with the G-symmetries.

\paragraph{Master symmetry for the single-cusp solution.}
Since the parametrization is slightly simpler, we work with the single-cusp solution 
\eqref{sol:SingleCusp}. In finding the master symmetry deformation, we use the identification of 
$\AdS_3$ with $\grp{SL}(2,\mathbb{R})$ discussed in section \ref{sec:AdSCoset}. Thus we employ the expressions for the master symmetry of principal chiral models derived in section \ref{sec:PCM}. We recall that the master symmetry transformation is given by
\begin{align}
g_u = \chi ^l _u \, g \, \chi^{r \, -1} _u 
\end{align}
where $\chi ^{r/l}_u $ are defined by the auxiliary linear problems
\begin{align}
\diff \chi ^r_u  &= \chi _u ^r \cdot L^r _{p \, u} \, , & \diff \chi _u ^l &= \chi_u ^l \cdot L^l _{p \, u} \, .
\label{SingleCusp:auxiliary:problem}
\end{align}
The left- and right Lax connection $L_u ^{r/l}$ appearing above are deformations of the left- and right Maurer--Cartan currents 
$U^r = g^{-1} \diff g$ and $U^l = - \diff g g^{-1}$, which are both flat and conserved. Explicitly, we have the expressions 
\begin{align}
L_{p\, u} ^{r/l} = \frac{u^2}{1+u^2} \,  U_p ^{r/l} + \frac{u}{1+u^2} \, \ast U_p ^{r/l} \, .
\end{align}

\begin{figure}
\centering
\subcaptionbox{Boundary Curve}
[.52\linewidth]{\includegraphics[width=.40\linewidth]{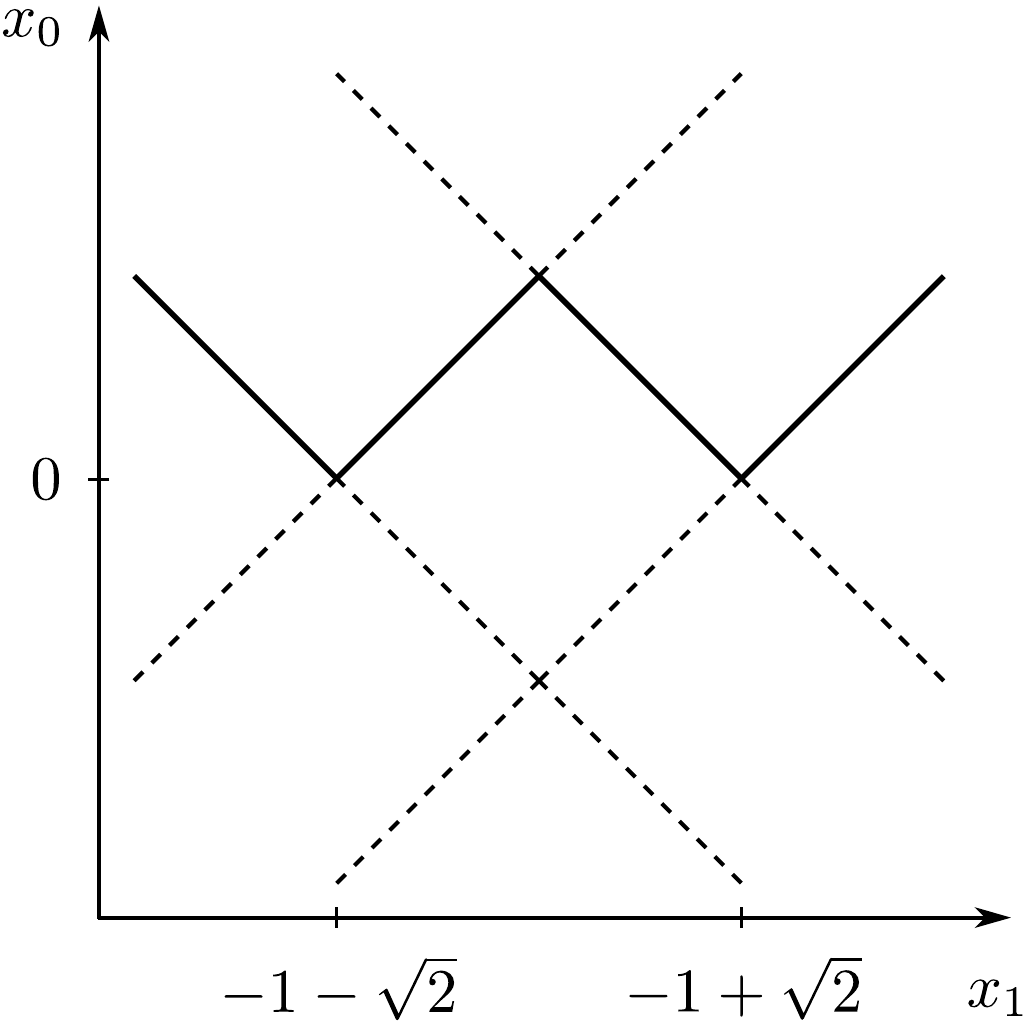}}
\subcaptionbox{Minimal Surface} 
[.42\linewidth]{\includegraphics[width=.30\linewidth]{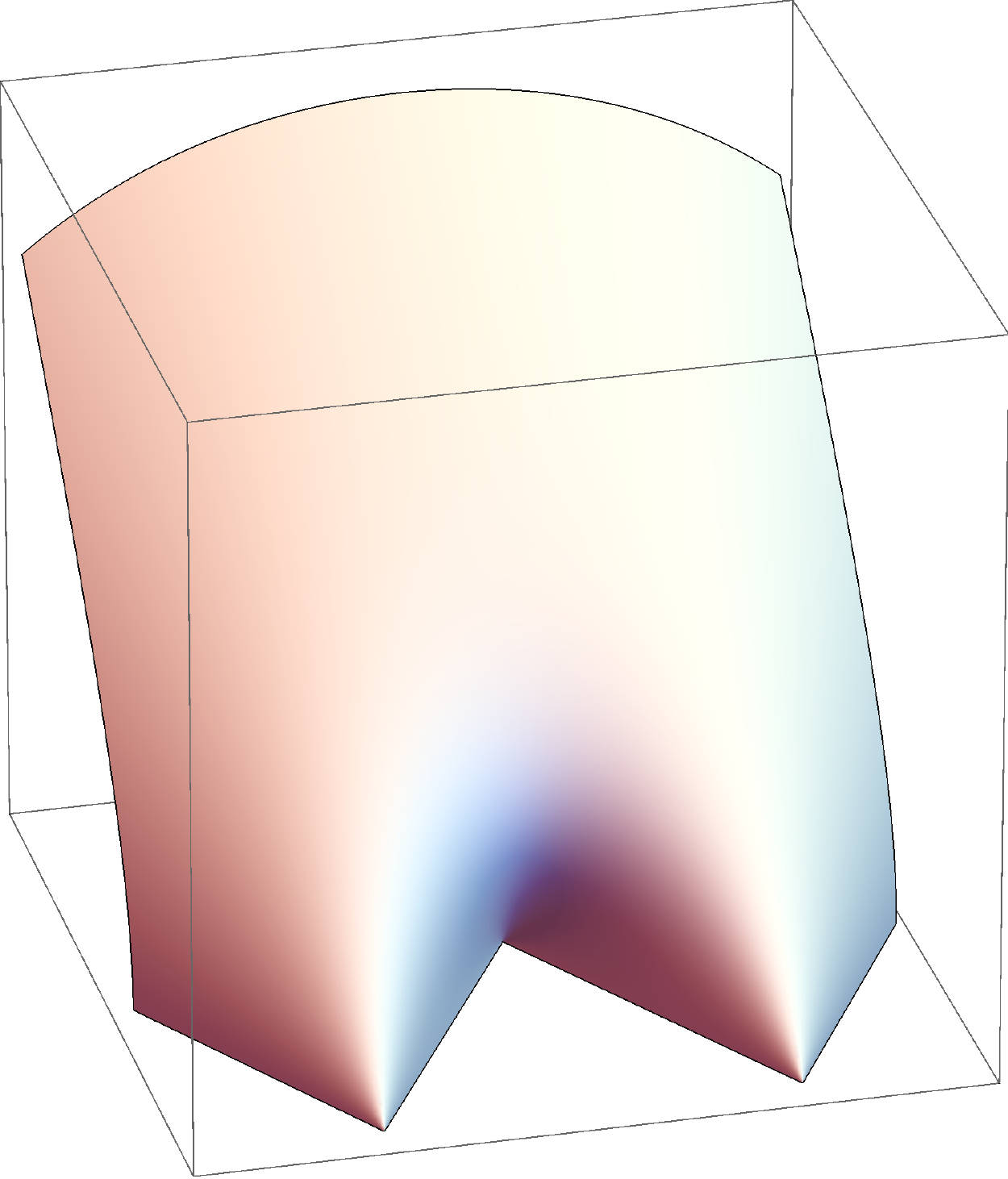} 
\vspace*{5mm}}
\caption{Display of the deformed boundary curve and minimal surface for $u=1$. The dashed lines in the picture of the boundary curve mark the parts of the solution of equation 
(\ref{eqn:Curve-equation}), which are not reached by the parametrization (\ref{eqn:def_sol}) from $y>0$.
}
\label{fig:SingleCuspDeform}
\end{figure}

With $g(X,y)$ as defined in equation \eqref{eqn:gPCM} and the single-cusp solution \eqref{sol:SingleCusp}, we have the parametrization
\begin{align}
g(\tau , \sigma) = \frac{1}{\sqrt{2}}  \begin{pmatrix}
e^{-\tau} & e^\sigma \\
- e^{-\sigma} & e^\tau 
\end{pmatrix} ,
\end{align}
and for the Maurer--Cartan current 
$U^r = g^{-1} \diff g$, we have
\begin{align}
U^r _\tau &= \frac{1}{2} \begin{pmatrix}
-1 & - e^{\tau + \sigma} \\
- e^{-\tau - \sigma} & 1 
\end{pmatrix}  , &
U^r _\sigma &= \frac{1}{2} \begin{pmatrix}
-1 & e^{\tau + \sigma} \\
e^{-\tau - \sigma} & 1 
\end{pmatrix} . 
\end{align} 
With $\ast U^r = U^r_\sigma \, \diff \tau - U^r_\tau \, \diff \sigma$, the Lax connection 
\begin{align*}
L^r = \frac{u}{1+u^2} \left( \ast U^r + u \, U^r \right) 
\end{align*}
has the components
\begin{align}
L^r _\tau &= \frac{u}{1+u^2} \left( U_\sigma ^r + u \,U_\tau ^r \right)
 , &
L^r _\sigma &= \frac{u}{1+u^2} \left(- U_\tau ^r + u \,U_\sigma ^r \right) .
\end{align}
Even though we are only dealing with $(2 \times 2)$ matrices and the functions appearing in the Lax connections are fairly simple, the auxiliary linear problems \eqref{SingleCusp:auxiliary:problem} are non-trivial to solve. Note however that 
\begin{align*}
U^r _\sigma + U^r _\tau
	&= \begin{pmatrix}
	-1 & 0 \\ 0 & 1
	\end{pmatrix} , &
U^r _\sigma - U^r _\tau
	&= \begin{pmatrix}
	0 & e^{\tau + \sigma} \\ e^{-\tau - \sigma} & 0
	\end{pmatrix} , &	
\end{align*}
and moreover that 
\begin{align*}
(1-u) L_\tau ^r + (1+u) L_\sigma ^r 
	&= u \left( U_\sigma - U_\tau \right) ,  \\
-(1+u) L_\tau ^r + (1-u) L_\sigma ^r 
	&=- u \left( U_\sigma + U_\tau \right) . 
\end{align*}
We can thus simplify the differential equations by changing coordinates according to 
\begin{align}
\begin{pmatrix}
	\tau \\ \sigma
\end{pmatrix} &= 
\begin{pmatrix}
1-u & - (1+u) \\ 1+u & 1-u
\end{pmatrix}
\begin{pmatrix}
x \\ y
\end{pmatrix} ,
\end{align}
such that we have
\begin{align*}
\partial_x &= (1-u) \partial_\tau + (1+u) \partial_\sigma \, , &
\partial_y &= - (1+u) \partial_\tau + (1-u) \partial_\sigma \, ,
\end{align*}
and correspondingly the auxiliary linear problem takes the form
\begin{align}
\partial_x \, \chi_u ^r &= \chi_u ^r 
	\begin{pmatrix}
		0 & u \, e ^{2x-2uy} \\
		u \, e^{-2x+2uy} & 0
	\end{pmatrix} , &
\partial_y \, \chi_u ^r &= \chi_u ^r 
	\begin{pmatrix}
		u & 0 \\
		0 & -u
	\end{pmatrix} .
\end{align}
Imposing the initial condition $\chi ^r _u (0,0)= \unit$, we then find the solution 
\begin{align*}
\chi_u ^r (x,y) = \frac{1}{\alpha}
	\begin{pmatrix}
	e^{uy-x} \left( \alpha \cosh(\alpha x ) + \sinh(\alpha x) \right)
	& u \, e^{x - uy} \sinh(\alpha x)  \\
	u \, e^{uy - x} \sinh(\alpha x) & 
	e^{x-uy} \left( \alpha \cosh(\alpha x ) - \sinh(\alpha x) \right)
	\end{pmatrix} ,
\end{align*}
where we have abbreviated $\alpha = \sqrt{1+u^2}$. 
In order to find the deformed solution, we also need to calculate the function $\chi_u ^l$, which is defined by the auxiliary problem for the Lax connection $L_u ^l$. The solution can however be constructed from the above solution from $\chi_u ^r (x,y)$. In order to see this, note that from 
$g U^r g^{-1} = - U^l$ we find
\begin{align*}
g \left( L_u ^r - U ^r \right) g^{-1}
&= g \left( \frac{-1}{1+u^2} \, U^r 
	+ \frac{u}{1+u^2} \, \ast U^r \right) 
= \frac{1}{1+u^2} \, U^l - \frac{u}{1+u^2} \, \ast U^l 
= L_{-1/u} ^l \, .
\end{align*}
We  can thus identify (see appendix \ref{app:Hodge})
\begin{align}
\chi ^l _u = g_0 \, \chi^r _{-1/u} \, g^{-1} \, . 
\end{align}
The deformed solution $g_u$ is thus given by 
\begin{align}
g_u = g_0 \, \chi^r _{-1/u} \, \chi^{r \, -1} _{u} \, ,
\end{align}
and we can read off the coordinates of the deformed solution from the general form \eqref{eqn:gPCM} of $g(X,y)$. In this way, we find
\begin{align}
X_{0 , u} (\tau , \sigma) &=
	\frac{\left(1 + u^2\right) 
	\cosh (\smi_u)}
	{\left(1 + u^2\right) \cosh (\spl_u)
	-\sqrt{1 + u^2} (u \sinh (\smi_u)
	+\sinh(\spl_u))} \, , \nn
	\\
X_{1 , u} (\tau , \sigma) &=
	\frac{\sqrt{1 + u^2} (\sinh (\smi_u) 
	- u \sinh (\spl_u))}
	{\left(1 + u^2\right) \cosh (\spl_u)
	- \sqrt{1 + u^2} (u \sinh (\smi_u) +\sinh (\spl_u))} \, , 
	\label{eqn:def_sol} \\
y_u(\tau, \sigma) &=
	\frac{\sqrt{2} \left(1 + u^2\right)}
	{\left(1 + u^2\right) \cosh (\spl_u)
	- \sqrt{1 + u^2} (u \sinh (\smi_u)+\sinh(\spl_u))} \, , \nn 
\end{align}
where we have introduced the abbreviations
\begin{align*}
\spl_u &= \frac{\tau + u \sigma}{\sqrt{1 + u^2}} \, , &
\smi_u &= \frac{\sigma - u \tau}{\sqrt{1 + u^2}} \, .
\end{align*}
The deformed solution given above indeed arises from the single-cusp solution \eqref{sol:SingleCusp} by the composition of an AdS-isometry and a reparametrization, but this is difficult to see in the form above. A more convenient way to describe the deformed surface is to use an equation such as equation \eqref{eqn:SingleCusp}, which does not refer to a specific parametrization. It is not difficult to see that the deformed solution given above satisfies the equation
\begin{align}
\left(u \left(X_1^2 - X_0^2 + y^2 -1 \right)+ 2 X_1 \right)^2 
-2 \left(u^2+1\right) \left(2 X_0^2 - y^2 \right) = 0 \, .
\label{eqn:Surf-equation}
\end{align}
While the parametrization of the deformed solution satisfies the above equation, it does not cover the whole space of solutions. However, while technically cumbersome, it is not difficult to find the branches of the space of solutions of the above equation, which are covered by the parametrization \eqref{eqn:def_sol}. We note that the deformed surface also covers negative values of $y$. This is not surprising since AdS-isometries do not commonly map the two regions covered by Poincar{\'e} coordinates to themselves. The deformed minimal surface for $y>0$ and $u=1$ is depicted in figure \ref{fig:SingleCuspDeform}.    
Another advantage of working with an equation such as \eqref{eqn:def_sol} is that the boundary limit $y \to 0 $ can be obtained easily. For the boundary curve we thus find the equation
\begin{align}
\left(u \left(x_1^2 - x_0^2 -1\right)+ 2 x_1 \right)^2
-4 \left(u^2+1\right) x_0^2 = 0 \, , 
\label{eqn:Curve-equation}
\end{align}
which describes the union of two light-cones centered at $x_0 = 0$ and
\begin{align}
x_1^+ &= \frac{-1 + \sqrt{1+u^2}}{u} \, ,  &
x_1^- &= \frac{-1 - \sqrt{1+u^2}}{u} \, ,
\end{align}
respectively. The center of the first light-cone moves from 0 to 1 as $u$ increases from 0 to 
$\infty$, whereas the center of the second moves from $-\infty$ to $-1$. Again, not the whole of the boundary curve described here is covered by the parametrization of the deformed surface, see figure \ref{fig:SingleCuspDeform}. 

It was pointed out above that the deformed surface is related to the original one by an AdS-isometry, i.e.\ the boundary curves are related by a conformal transformation, which is indicated by the transformed boundary curve being light-like everywhere. Equation \eqref{eqn:Surf-equation} in principle allows to find the AdS-isometry relating the two surfaces, but it is simpler to study this question infinitesimally. For an infinitesimal transformation, any variation tangential to the surface can be accounted for by an infinitesimal reparametrization. Hence, we only need to consider the part of the variation, which is parallel to the normal vector
\begin{align}
\left( X_0 ^{\mathrm{n}}(\tau, \sigma) , 
	X_1 ^{\mathrm{n}} (\tau, \sigma) , 
	y ^{\mathrm{n}} (\tau, \sigma) \right)
= \left( \sqrt{2} e^\tau \cosh (\sigma) , 
	\sqrt{2} e^\tau \cosh (\sigma) , e^\tau \right) .	
\label{eqn:normal_vec}	 
\end{align}
We compare this piece of the variation to a generic AdS-isometry, which can be parametrized by
\begin{align}
\delta X_0 &= a_0 + \omega X_1 + s X_0 + c_0 \left( X_1 ^2 - X_0 ^2 \right) 
	-2 X_0 \left( c_1 X_1 - c_0 X_0 \right) , \nn \\
\delta X_1 &= a_1 + \omega X_0 + s X_1 + c_1 \left( X_1 ^2 - X_0 ^2 \right) 
	-2 X_1 \left( c_1 X_1 - c_0 X_0 \right) , \\
\delta y &= s y - 2 y \left( c_1 X_1 - c_0 X_0 \right) . \nn	 
\end{align}
Here, the parameters $\left( a_\mu , \omega , s , c_\mu \right)$ correspond to generators 
$\left( P_\mu , M_{01} , D , K_\mu \right)$, respectively. Comparing the projection of the master symmetry variation
\begin{align}
\master \left( X_0 , X_1 , y \right)
	= \frac{\diff}{\diff u} \left( X_{0,u} , X_{1,u} , y_u \right) \big \vert _{u=0} 
\end{align}
onto the normal vector \eqref{eqn:normal_vec} shows that it can be accounted for by an AdS-isometry with all of the above parameters set to zero, except
\begin{align}
a_1 = c_1 = \frac{1}{2} \, .
\end{align}
We have hence found that the minimal surface described by the parametrization \eqref{eqn:def_sol} is related to the single-cusp solution by an AdS-isometry. 

\subsection{Numerical Deformations}
The examples discussed above could lead to the impression that the master symmetry transformation is typically given by a combination of a G-symmetry and a reparametrization. This is however only the case for the particularly symmetric boundary curves, for which minimal surface solutions are known. More complicated boundary curves have been studied in references \cite{Ishizeki:2011bf} and \cite{Dekel:2015bla} and there one observes deformations, which cannot be related to conformal transformations of the boundary curves.
\begin{figure}
\centering
\includegraphics[width=75mm]{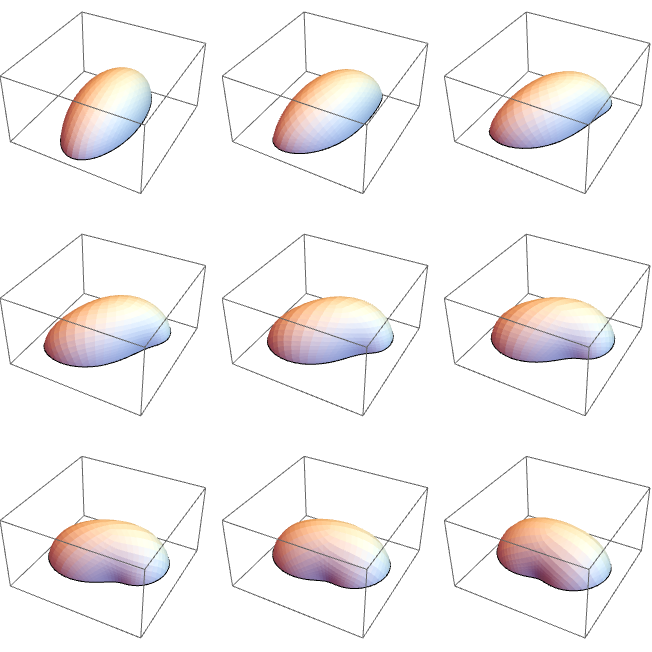}
\caption{Minimal surfaces arising from the master symmetry deformation of the minimal surface for an elliptical boundary curve. The values of the spectral parameter $\theta$ range from $0$ to $\pi$ in uniform steps. For values ranging from $\pi$ to $2\pi$, the deformation continues until the original shape is reached again. This figure has been reproduced from reference \cite{Klose:2016qfv}.}
\label{fig:ellipse-family}
\end{figure}

\begin{figure}
\centering
\includegraphics[width=75mm]{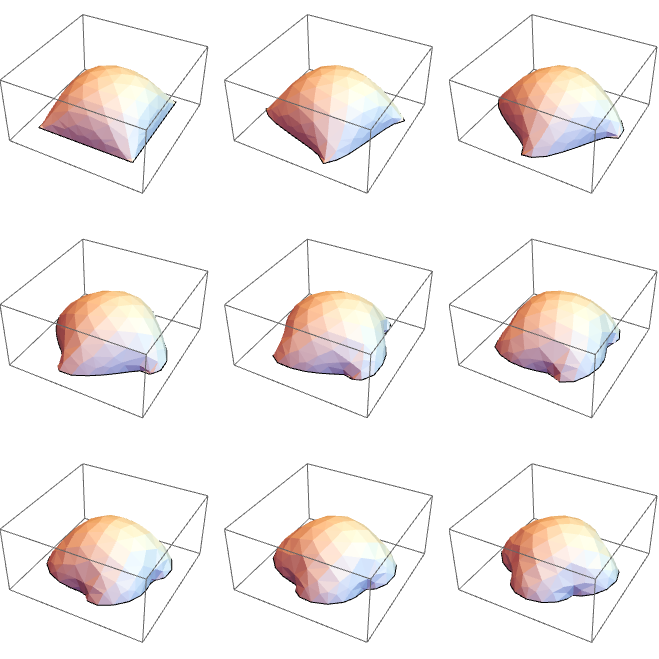}
\caption{Minimal surfaces arising from the master symmetry deformation of the minimal surface for the boundary curve being a square. The values of the spectral parameter $\theta$ range from $0$ to $\pi$ in uniform steps. For values ranging from $\pi$ to $2\pi$, the deformation continues until the original shape is reached again. This figure has been reproduced from reference \cite{Klose:2016qfv}.}
\label{fig:square-family}
\end{figure}

Not many minimal surfaces in Euclidean or Lorentzian Anti-de Sitter space are known analytically and for generic boundary curves we need to rely on numerical results in order to approximate the deformations. For the Euclidean case, a numerical approach to the calculation of the master symmetry deformation of minimal surfaces has been developed and described in reference \cite{Klose:2016qfv} and the numerical results obtained therein%
\footnote{While an author of this paper, the author of the present thesis has participated little in the work that lead to the algorithm described here.}
are shown in figures \ref{fig:ellipse-family} and \ref{fig:square-family}. 

In short, the algorithm that created the images referenced above works as follows: The original boundary curve is provided to the program \textit{Surface Evolver} \cite{brakke1992} in parametric form along with the AdS metric and an initial (crude) approximation of the minimal surface. The program then returns an approximate discrete minimal surface, from which a discrete approximation of the Maurer--Cartan current can be calculated. After finding a parametrization satisfying conformal gauge, the Maurer--Cartan current $U$ is deformed into the Lax connection $L_u$ and the deformed surface is approximated by integrating the defining relation \eqref{eqn:def-deformed-sol}, 
\begin{align*}
g_u^{-1} \diff g_u &= L_u \, , &
g_u (z_0) &= g (z_0) \, ,   
\end{align*}   
where the initial condition is taken at a point far away from the boundary, which has proven to give better numerical results. The coordinates of the deformed minimal surface can the be read off the group elements $g_u$ by employing the form \eqref{G:gauge_invariant}.

\section{Infinitesimal Symmetry Transformations}
\label{sec:InfSymm}
In the last section we have derived the master symmetry transformations for certain boundary curves, for which the minimal surfaces were known either analytically or numerically. For generic boundary curves, one could hope to bypass the minimal surface problem and find the transformed boundary curve directly from the original one, e.g.\ by finding a differential equation for the transformed boundary curve. Unfortunately, this seems not to be possible. We thus consider the infinitesimal transformations of generic boundary curves, focusing on the master and level-1 Yangian-like symmetry. 
Here, we consider two approaches. We begin by calculating the variation of the boundary curves from the variations discussed in section \ref{sec:IntCompl} using the Polyakov--Rychkov expansion. The variations obtained in this way include the third-order coefficient of the expansion, which cannot be fixed from the expansion around the boundary. This shows that calculating the transformation of the boundary curve indeed requires finding a minimal surface solution. 

Since we consider the variations on shell, the variation of the area is in general given by a boundary term, which is absent for the master symmetry itself but generically non-vanishing. The identity arising from equating the action of the variation on the minimal area and the corresponding boundary term is comprised in the Noether charge associated to the respective variation. The vanishing of these charges encodes the global information that the minimal surface closes and they thus contain information which is not captured in the expansion around the boundary curve. We evaluate the level-1 Yangian charge explicitly in order to obtain the corresponding Ward identity at strong coupling. This approach was already discussed in reference \cite{Muller:2013rta}, where the level-1 Yangian Ward identity for the Maldacena--Wilson loop at strong coupling was first derived. The result appears in new light due to the underpinning of the associated variation and the calculation of the variation of the boundary curve. The re-derivation presented here also serves the purpose of preparing the similar calculation for the Wilson loop in superspace, which we consider in chapter \ref{chap:SurfaceSuperspace}. The further interpretation of the symmetries is postponed to chapter \ref{chap:CarryOver}, where we discuss the possible continuations to weak or arbitrary coupling. 

\subsection{Variations of the boundary curves}
We first consider the variation of the boundary curve under the bilocal symmetries discussed in section \ref{sec:IntCompl}, i.e.\ for the level-0 master and the level-1 Yangian symmetry.

\paragraph{Level-0 master symmetry.}
We have already discussed the G-symmetry variations of generic boundary curves in section \ref{sec:AdSCoset}. This discussion can be generalized to arbitrary variations $\delta g = \eta g$ by replacing $T_a$ by $\eta$ in the variations \eqref{eqn:varcoord}, 
\begin{align}
\delta X^\mu &= \frac{y}{4} \left( \tr \left(  g^{-1} \eta g \,  K^\mu \right)  
+ \tr \left( g^{-1} \eta g \, P^\mu \right) \right) \, , &
\delta y &= \frac{y}{2} \tr \left( g^{-1} \eta g \, D \right) \, .
\label{eqn:varcoordgeneral}
\end{align}
Making use of these relations, we now turn to the master variation, which is given by
\begin{align}
\delC g &= \chi ^{(0)} \cdot g \, , & \chi ^{(0)} &= \int \ast j \, .
\end{align}
As we are aiming for the variation of the boundary curve, we only need to compute $ \chi ^{(0)} $ in the vicinity of the boundary. More precisely, we calculate $ \chi ^{(0)} $ at $y = \eps$ in an expansion in $\eps$. The expansion \eqref{eqn:expansion_boundary} of the minimal surface solution close to the conformal boundary is sufficient to do so. For the Noether current $j= -2 g a g^{-1}$, we note
\begin{align}
j = - 2 \, e^{XP} \left( 
	\frac{\diff X^\mu}{2 y^2} (K_\mu + y^2 P_\mu) 
	+ \frac{\diff y}{y} D \right) e^{-XP} \, , 
\end{align}
and by inserting the Polyakov--Rychkov expansion \eqref{eqn:expansion_boundary}, we find that $j_\sigma$ takes the form
\begin{align}
j_\sigma = \frac{1}{\tau^2} \, j_{\sigma (-2)} + j_{\sigma (0)} +
	\O(\tau) \, .
\label{jsigma_exp}
\end{align}
By using current conservation, we can then fix all components of 
$j_\tau$ except for $j_{\tau (0)}$, 
\begin{align}
j_\tau &= \frac{1}{\tau} \, \partial_\sigma j_{\sigma \, (-2)}
	+ j_{\tau \, (0)} - \tau \, \partial_\sigma j_{\sigma (0)}
 	+ \mathcal{O}(\tau^2) \, .
\label{jtau_exp}
\end{align}
In the discussion below we will need the coefficients $j_{\sigma \, (-2)}$ and $j_{\tau \, (0)}$. For the first coefficient, we note that
\begin{align}
j_{\sigma \, (-2)} = - e^{xP} \left(
	\frac{\dx_\mu}{\dx^2} K^\mu \right) e^{xP} 
	= - 4 \, \frac{\dot{x}_\mu}{\dot{x}^2} 
	\, \hat{\xi}^\mu (x) \, .
\end{align}
Here, the quantity $\hat{\xi}^\mu (x)$ is related to the conformal Killing vectors \eqref{Def:ConfKilling} by
\begin{align}
\hat{\xi}^\mu (x) = \xi ^\mu _a (x) \, T^a = \frac{1}{4} \, e^{x P} K^\mu e^{-x P} \, . 
\end{align}
For the computation of the coefficient 
$j_{\tau \, (0)}$, we note that the $\tau^{-1}$-coefficient in
\begin{align*}
\frac{\partial_\tau X^\mu}{2 y^2} (K_\mu + y^2 P_\mu) 
	+ \frac{\partial_\tau y}{y} D
\end{align*}
does not get mixed with the $\tau^0$-coefficient by the conjugation with $e^{XP}$, since the expansion of the $X$-coordinates is given by $X = x + \O(\tau^2)$. Hence we find the expression
\begin{align}
j_{\tau \, (0)} &= 4 \, \frac{\delta A_{\mathrm{ren}}(\gamma)}
	{\delta x^\mu}  \, \hat{\xi}^\mu (x) \, .
\label{eqn:jtau0}	
\end{align}
With these coefficients calculated, we turn to the calculation of $ \chi ^{(0)} $ for $y= \eps$ which corresponds to the point $(\tau_0 (\sigma), \sigma )$ in parameter space. The definition of  $ \chi ^{(0)} $ requires to choose some starting point on the worldsheet, which we take to be $( \tau = c , \sigma = 0 )$. Since $ \chi ^{(0)} $ is path-independent, we may use any path connecting the points $( c , 0 )$ and $(\tau_0 (\sigma), \sigma )$. We choose the composed path $\gamma = \gamma _1 \ast \gamma_2 $, where $\gamma_1$ connects $(c,0)$ to $(\tau_0 (\sigma), 0 )$ along $\sigma = 0$ and $\gamma_2$ connects $(\tau_0 (\sigma), 0 )$  to $(\tau_0 (\sigma), \sigma )$ along the $\sigma$-direction. We find
\begin{align*}
\int _{\gamma_1} \ast j &=
\int \limits _c ^{\tau_0 (\sigma)} \diff \tau \, j_\sigma ( \tau , 0 ) 
= - \frac{1}{\tau_0 (\sigma)} j_{\sigma \, (-2)} (0)  + \zeta 
+ \mathcal{O}(\eps) \, , \\
\int _{\gamma_2} \ast j 
&= \int \limits _0 ^\sigma \diff \sigma ^\prime 
\left( - j_\tau (\tau_0 (\sigma) , \sigma ^\prime)  \right) 
= \frac{1}{\tau_0 (\sigma)} \left( j_{\sigma \, (-2)} (0) - j_{\sigma \, (-2)} (\sigma) \right) 
- \int \limits _0 ^\sigma \diff \sigma ^\prime j_{\tau \, (0)} (\sigma  ^\prime) \, .
\end{align*}
Here, $\zeta$ is some $\sigma$-independent element of $\alg{g}$ and we may neglect the conformal transformation which it parametrizes. Combining the two results we find
\begin{align}
\chi ^{(0)} (\tau_0 (\sigma) , \sigma ) 
	&= \frac{4}{\eps} \frac{\dot{x}(\sigma)_\mu}
	{\lvert \dot{x}(\sigma) \rvert} \, 
	\hat{\xi}^\mu (x(\sigma)) + \zeta  
	- 4 \int \limits _0 ^\sigma \diff \sigma ^\prime \, 
	\frac{\delta A_{\mathrm{ren}}(\gamma)}
	{\delta x^\mu(\sigma ^\prime) } \,  
	\hat{\xi}^\mu (x(\sigma ^\prime)) + \mathcal{O}(\eps) \, .
\label{eqn:chi_expl}
\end{align}
We can then determine the variation of the coordinates. Let us first convince ourselves that the master symmetry does not move the boundary into the bulk, i.e.\ we have $\master y \to 0$ as $y$ approaches $0$. Making use of equation \eqref{eqn:varcoordgeneral} with $\eta=\chi^{(0)}$, we find
\begin{align}
\master y &= \frac{y}{2} \tr \left( g^{-1} \chi^{(0)} g D \right) = 
\frac{2\, \dot{x}(\sigma)_\mu}{ \lvert \dot{x}(\sigma) \rvert} 
\tr \left( y^{-D} e^{-X \cdot P} \hat{\xi}^\mu e^{X \cdot P} y^D D \right) 
+ \mathcal{O}(y) \nn \\
&=  \frac{2\, \dot{x}(\sigma)_\mu}{\lvert \dot{x}(\sigma) \rvert} 
\tr \left( K^\mu D \right) + \mathcal{O}(y) = \mathcal{O}(y) \, ,
\end{align}
which shows that the master symmetry maps the conformal boundary to itself. For the variation of the $X$-coordinates we find 
\begin{align}
\delC X^\mu &= \frac{1}{4} \tr \left( e^{-X \cdot P } \chi^{(0)} e^{X \cdot P } K^\mu \right) 
+ \mathcal{O}(y) =  \tr \left( \chi^{(0)} \hat{\xi}^\mu(x) \right) + \mathcal{O}(y)  \nn \\
&= \frac{1}{4y} \,  \frac{\dot{x}(\sigma)_\nu}{\lvert \dot{x}(\sigma) \rvert} \, \tr \left( K^\nu K^\mu \right) + \delta _{\zeta} X^\mu \nn \\
& \; -  4 \, G^{ab}  \int \limits _0 ^\sigma \diff \sigma ^\prime 
	\, \frac{\delta A_{\mathrm{ren}}(\gamma)}
	{\delta x^\nu(\sigma ^\prime) } \, 
	\xi^\nu _a (x(\sigma ^\prime)) \, \xi^\mu _b( x (\sigma) )
	+ \mathcal{O}(y) \, .
\end{align}
As the first term vanishes due to $\tr \left( K^\nu K^\mu \right)=0$, we can safely take the limit $y \to 0$. We neglect the conformal variation parametrized by $\zeta$ as it is independent of the point along the contour and depends on our choice of a starting point on the minimal surface. Then we obtain
\begin{align}
\master x^\mu (\sigma) = - 4 \, G^{ab}  \int \limits _0 ^\sigma \diff \sigma ^\prime \, \frac{\delta A_{\mathrm{ren}}(\gamma)}{\delta x^\nu(\sigma ^\prime) } \,  \xi^\nu _a (x(\sigma ^\prime)) \, \xi^\mu _b (x(\sigma)) \, .
\label{eqn:deltaCx}
\end{align}
The appearance of the third-order coefficient of the expansion 
\eqref{eqn:expansion_boundary} indicates that it is indeed necessary to compute the minimal surface solution in order to determine the master transformation of the boundary curve as it was done in references \cite{Kruczenski:2013bsa,Kruczenski:2014bla,Huang:2016atz,Dekel:2015bla}.

From the considerations of chapter \ref{chap:SSM} and the proof given  in section \ref{sec:Renorm} it is clear that the above variation is a symmetry of the minimal area. In fact, this is easy to see directly from the variation and without referring to the formalism introduced before, 
\begin{align}
\master A_{\mathrm{ren}}(\gamma) &= \int \limits _0 ^{2 \pi} \diff \sigma 
	\frac{\delta A_{\mathrm{ren}}(\gamma)}{\delta x^\mu(\sigma) } \, \master x^\mu (\sigma) \nn \\
&= - 4 \, G^{ab} \int \limits _0 ^{2 \pi} \diff \sigma \int \limits _0 ^{\sigma} \diff \sigma^\prime \,
	\frac{\delta A_{\mathrm{ren}}(\gamma)}{\delta x^\mu(\sigma) } \xi^\mu _b (x(\sigma))
	\frac{\delta A_{\mathrm{ren}}(\gamma)}{\delta x^\nu(\sigma^\prime) } \xi^\nu _a (x(\sigma^\prime)) 
\nn \\
&= -  2 \, G^{ab} \, \delta_b \left( A_{\mathrm{ren}}(\gamma) \right)
	\delta_a \left( A_{\mathrm{ren}}(\gamma) \right)
= 0 \, .
\label{deltaAsymm}
\end{align}
It is intriguing that the invariance of the minimal area under the master symmetry, which can be employed to construct all nonlocal conserved charges and their associated symmetry transformations, follows without referring to integrability. This corresponds to the finding that the conserved charge associated with the master symmetry itself is the Casimir of the G-symmetry charges. 

\paragraph{Level-1 Yangian-like symmetry.}

The procedure described above for the master symmetry variation can in principle be carried out for any of the variations described in section \ref{sec:IntCompl}, although that would require to extend the expansion \eqref{eqn:expand_bound} to higher orders. Here, we study the level-1 Yangian-like variation $\delta_\epsilon ^{(1)}$. 

Noting that the master variation is given by $\master g = G^{bc} \chi^{(0)}_b \, T_c g$, whereas a level-1 Yangian variation is given by
\begin{align}
\delta _a ^{(1)} g = \left[ T_a , \chi^{(0)} \right] g = f_a {} ^{bc} \chi^{(0)}_b \, T_c g \, ,
\end{align}
suggests that the variation of the boundary curve can be obtained from \eqref{eqn:deltaCx} by replacing $G^{bc}$ by  
$f_a {} ^{bc}$ to obtain 
\begin{align}
\delta _a ^{(1)} x^\mu (\sigma) = - 4 \, f_a {} ^{bc}   \int \limits _0 ^\sigma \diff \sigma ^\prime \, \frac{\delta A_{\mathrm{ren}}(\gamma)}{\delta x^\nu(\sigma ^\prime) } \,  \xi^\nu _b (x(\sigma ^\prime)) \, \xi^\mu _c (x(\sigma)) \, .
\label{eqn:deltaax}
\end{align}
However, we still need to discuss the divergent terms contained in the expression given for $\chi^{(0)}$ in equation \eqref{eqn:chi_expl}, which did not contribute to the master variation $\master$. For the variation of the original boundary curve, we find the additional term
\begin{align}
\frac{\dot{x}_\nu}{\lvert \dot{x} \rvert \eps} 
\tr \left( \left[ T_a \, , \, \hat{\xi}^\nu (x) \right] \hat{\xi}^\mu (x) \right) 
= \frac{\dot{x}_\nu}{\lvert \dot{x} \rvert \eps} 
\tr \left( \left[\hat{\xi}^\nu (x) \, , \, \hat{\xi}^\mu (x) \right] T_a \right) = 0 \, ,
\end{align}
which shows that our expectation \eqref{eqn:deltaax} is indeed correct. For the variation of $y$, however, we have
\begin{align}
\delta_a ^{(1)} y &= \frac{2\,\dot{x}_\mu}{\lvert \dot{x} \rvert} 
\tr \left( e^{-x \cdot P} \left[ T_a \, , \, \hat{\xi}^\mu (x) \right] e^{x \cdot P} D \right) \nn \\
& = \frac{2\, \dot{x}_\mu}{\lvert \dot{x} \rvert} 
\tr \left( \left[ e^{-x \cdot P}  \hat{\xi}^\mu (x)  e^{x \cdot P} \, , \, D  \right]  e^{-x \cdot P}  T_a  e^{x \cdot P} \right) 
= \frac{2 \, \dot{x}_\mu}{\lvert \dot{x} \rvert} \, \xi ^\mu _a (x) \, .
\end{align}
We thus see that the level-1 Yangian-like variation $\delta_a ^{(1)}$ generically shifts the boundary curve into the bulk. This behaviour is accompanied by a divergent boundary term arising from the application of $\delta_a ^{(1)}$ to the minimal area $A_{\mathrm{ren}}(\gamma)$. Let us also note that for certain choices of the boundary curve $\gamma$ and $\epsilon$, the boundary curve is not shifted into the bulk. The simplest example corresponds to $\epsilon = D$ and the boundary curve being a circle. In this case we find 
\begin{align*}
\delta_D ^{(1)} y = \frac{2 \, \dot{x}_\mu x^\mu }{\lvert \dot{x} \rvert} + \mathcal{O}(y)
= \mathcal{O}(y) \, .
\end{align*}

\subsection{Evaluation of Conserved Charges}
In order to obtain the Ward identity for the level-1 Yangian-like variations, we evaluate the conserved charge $Q^{(1)}$ on the minimal surface by employing the Polyakov--Rychkov expansion \eqref{eqn:expansion_boundary}. Let us begin by considering the conserved charge associated to the conformal symmetry, 
\begin{align}
Q^{(0)} = - \int \limits _0 ^{2 \pi} \diff \sigma \, j_\tau \, .
\end{align}
If we consider the Laurent expansion of the above charge in $\tau$, all of the coefficients expect $Q^{(0)} _{(0)}$ vanish due to the equations of motion, and hence we have
\begin{align}
Q^{(0)} = Q^{(0)} _{(0)} = - \int \limits _0 ^{2 \pi} \diff \sigma \, j_{\tau \, (0)} 
	= - 4 \int \limits _0 ^{2 \pi} \diff \sigma \, 
	\frac{\delta A_{\mathrm{ren}}(\gamma)} {\delta x^\mu (\sigma)}  
	\, \hat{\xi}^\mu (x(\sigma)) \, ,
\end{align}
where we have inserted the expression \eqref{eqn:jtau0} for $j_{\tau \, (0)}$. The vanishing of the charge thus encodes the conformal symmetry of the minimal surface problem, 
\begin{align}
\int \limits _0 ^{2 \pi} \diff \sigma \, 
	\frac{\delta A_{\mathrm{ren}}(\gamma)} {\delta x^\mu (\sigma)}  
	\, \delta_a x^\mu (\sigma) 
	= 0 \, . 
\end{align}  
We then turn to the level-1 Yangian charge
\begin{align}
Q^{(1)} = \frac{1}{2} \, \int \limits _{0} ^L \diff \sigma_1 \, \diff \sigma_2 \, 
	\varepsilon \left( \sigma_2 - \sigma_1 \right) 
	\left[ j_\tau (\sigma_1) , j_\tau (\sigma_2) \right]
	+ 2 \int \limits _{0} ^L \diff \sigma \, j_\sigma (\sigma) \, .
\end{align}
Here, we have introduced the notation $\varepsilon(\sigma) = \theta(\sigma) - \theta(-\sigma)$ and restricted to a parametrization by arc-length, which will simplify the calculations below. As for the level-0 charge, we focus on the $\tau^0$-coefficient, since all other coefficients vanish automatically due to the equations of motion. Making use of the expansion of $j_\tau$ as given in equation \eqref{jtau_exp}, we find the $\tau^0$-coefficient of $Q^{(1)}$ to be
\begin{align}
Q^{(1)} = \frac{1}{2} \, \int \limits _{0} ^L \diff \sigma_1 \, \diff \sigma_2 \, 
	\varepsilon \left( \sigma_2 - \sigma_1 \right) 
	\left[ j_{\tau \, (0)} (\sigma_1) , j_{\tau \, (0)} (\sigma_2) \right] + 
	Q^{(1)} _{\text{local}} \, , 
\end{align}
where the local term includes contributions that arise from the first term of $Q^{1}$ after integration by parts, 
\begin{align}
Q^{(1)} _{\text{local}}  &= \int \limits _{0} ^L \diff \sigma \big( 2 j_{\sigma \, (0)} 
	- \left[ j_\sigma , \partial_\sigma  j_{\sigma} \right] _{(-2)} \big) . 
\end{align}	
We can further simplify the calculation of the local term by using
$j_\sigma = - 2 g a_\sigma g^{-1}$ to get 
\begin{align}
2 j_{\sigma \, (0)} - \left[ j_\sigma , \partial_\sigma  j_{\sigma} \right] _{(-2)}
	= - 4 \left \lbrace g \left( a_\sigma + \tau ^2 \left[ a_\sigma , \partial_\sigma a_\sigma 
	+ \left[ U_\sigma , a_\sigma \right] \right] \right) g^{-1} \right \rbrace _{(0)} .
\end{align}
Here, it is preferable to do the conjugation with $e^{XP}$ last. Noting that
\begin{align*}
y^D U_\sigma y^{-D} &= \dot{X}^\mu P_\mu + \frac{\dot{y}}{y} \, D \, ,  &
y^D a_\sigma y^{-D} &= \frac{\dot{X}^\mu}{2y^2} \left( K_\mu + y^2 P_\mu \right) 
	+ \frac{\dot{y}}{y} \, D \, , 
\end{align*}
we find the intermediate result 
\begin{align*}
y^D \left( a_\sigma + \tau ^2 \left[ a_\sigma , \partial_\sigma a_\sigma 
	+ \left[ U_\sigma , a_\sigma \right] \right] \right) y^{-D} 
	= \frac{2 \dx ^\mu  + \tau^2 \left( \dx^\mu \ddx^2 + \dddot{x} ^\mu \right)}
	{2 \tau^2} K_\mu - \dx^\mu \ddx^\nu M_{\mu \nu} + \O (\tau) . 
\end{align*}
Here, we have made use of the arc-length parametrization, which e.g.\ allows to conclude that
\begin{align*}
\dot{y} &= \O (\tau^3) \, ,&
\dx \cdot \ddx &= 0 \, , &
\dx \cdot \dddot{x} &= - \ddx^2 \, . 
\end{align*}
The conjugation with $e^{XP}$ is now simplified by the finding 
\begin{align*}
\dx^\mu X_{(2)}^\nu \left[ P_\nu , K_\mu \right] - \dx^\mu \ddx^\nu M_{\mu \nu} = 0 \, ,
\end{align*}
which implies that 
\begin{align*}
\left \lbrace e^{X P} \left( \frac{\dx^\mu}{\tau ^2} K_\mu 
	- \dx^\mu \ddx^\nu M_{\mu \nu} \right) e^{- X P} \right \rbrace _{(0)} = 0 \, .
\label{bosoniczero}	
\end{align*}
We have hence found the local term to be
\begin{align}
Q^{(1)} _{\text{local}}  &= -2 \int \limits _{0} ^L \diff \sigma \,
	e^{xP} \left( \dx_\mu \ddx^2 + \dddot{x} _\mu \right) K^\mu e^{-xP} 
	= - 8 \int \limits _{0} ^L \diff \sigma \left( \dx_\mu \ddx^2 + \dddot{x} _\mu \right) 
	\hat{\xi}^\mu (x) \, . 
\end{align}
The full level-1 Yangian charge is hence given by
\begin{align}
Q^{(1)} &= 8 \int \limits _{0} ^L \diff \sigma_1 \, \diff \sigma_2 \, 
	\varepsilon_{2 1}
	\left[ \hat{\xi}^\mu _1, \hat{\xi}^\nu _2 \right] 
	\frac{\delta A_{\text{ren}}}{\delta x_1 ^\mu} \,
	\frac{\delta A_{\text{ren}}}{\delta x_2 ^\nu} 
	- 8 \int \limits _{0} ^L \diff \sigma 
	\left( \dx_\mu \ddx^2 + \dddot{x} _\mu \right) \hat{\xi}^\mu (x) \, ,
\end{align} 
where we have abbreviated $\varepsilon_{2 1} = \varepsilon(\sigma_2 - \sigma_1)$ and 
$\hat{\xi}^\mu _i = \hat{\xi}^\mu (x(\sigma_i))$. The vanishing of the level-1 charge hence encodes the identity
\begin{align}
\mathbf{f} \indices{_a ^{cb}} \, \int \limits _{0} ^L \diff \sigma_1 \, \diff \sigma_2 \, 
	\varepsilon_{21} \,  \xi ^\mu _{1 b} \, 
	\frac{\delta A_\mathrm{ren}}{\delta x_1^\mu  } \, 
	\xi ^\nu _{2 c} \, 
	\frac{\delta A_\mathrm{ren}}{\delta x_2^\nu } 
	- \int \limits _0 ^L \diff \sigma \, \xi ^\mu _a
	\left( \dx_\mu \, \ddot{x}^2 + \dddot{x}_\mu \right) = 0 \, . 
\label{Q1identity}	
\end{align}
We can interpret the above identity as the result of acting with the variation \eqref{eqn:deltaax} of the boundary curve on $A_\mathrm{ren}(\gamma)$ to obtain the local term as a boundary term. Note that the boundary term obtained in this way is not simply the boundary term arising from applying the variation $\delta_a ^{(1)}$ to the area functional, since we are not considering exactly this variation but rather only the variation \eqref{eqn:deltaax} of the boundary curve with $y=0$ fixed.  
A different interpretation of the identity \eqref{Q1identity} is that it arises from the application of the Yangian level-1 generator
\begin{align}
\J_a ^{(1)} = \mathbf{f} \indices{_a ^{cb}} \, \int \limits _{0} ^L 
	\diff \sigma_1 \, \diff \sigma_2 \, 
	\varepsilon_{21} \,  \xi ^\mu _{1 b} \,  \xi ^\nu _{2 c} \,  
	\frac{\delta^2}{\delta x^\mu _1 \delta x^\nu _2} 
	- \frac{\la}{4 \pi ^2} \int \limits _0 ^L \diff \sigma \, \xi^\mu _a  
	\left( \dx_\mu \, \ddot{x}^2 + \dddot{x}_\mu \right) 
\label{J1Bos}	
\end{align}
to the expectation value of the Maldacena--Wilson loop at strong coupling, 
\begin{align}
\left \langle W(\gamma) \right \rangle =
	\exp \left( - \ft{\sqrt{\la}}{2 \pi} A_\mathrm{ren}(\gamma) \right) .
\end{align}
This interpretation has been proposed in reference \cite{Muller:2013rta} and we discuss it in more detail in the next chapter. 

\section{Minimal Surfaces in \texorpdfstring{$\mathrm{S}^5$}{S5}}
\label{sec:S5}

So far, we have exclusively considered minimal surfaces in $\AdS_5$. For Maldacena--Wilson loops with non-constant sphere vectors $n^I$, however, the strong-coupling description also includes a minimal surface in $\mathrm{S}^5$ bounded by the curve $n^I (\s)$. Since $\mathrm{S}^5$ is a symmetric space as well, we can again employ the symmetries discussed in chapter \ref{chap:SSM}. In fact, since the $\mathrm{S}^5$-metric does not diverge upon approaching the boundary curve, the discussion is significantly simplified.  

As a preparation for our discussion of minimal surfaces in superspace in chapter \ref{chap:SurfaceSuperspace}, we describe the sphere $\mathrm{S}^5$ by using a coset construction based on $\grp{SU}(4)$-matrices. The Lie algebra $\alg{su}(4)$ is decomposed into a gauge and coset part as 
\begin{align}
\begin{aligned}
\alg{h} &= \mathrm{span} \left \lbrace \gamma^{r s} = 
	\quarter \, \left[ \gamma ^r \, , \, \gamma ^s \right]  \, , 
	\gamma ^{r 6} = \quarter \, \left[ \gamma ^r \, , \, \gamma ^5 \right] \, : 
	r,s \in   \left \lbrace 1 \, , \ldots , 4 \right \rbrace \right \rbrace 
	\simeq \mathfrak{so}(5) \, , \\
\alg {m} &= \mathrm{span} \left \lbrace \gamma^{r 5} = \ihalf \gamma^r \, , 
	\gamma ^{5 6} = - \ihalf \gamma^5 \, : 
	r \in   \left \lbrace 1 \, , \ldots , 4 \right \rbrace \right \rbrace .	
\end{aligned}	
\end{align}
Here, the matrices $\gamma^{IJ}= - \gamma^{JI}$ are constructed from the Dirac matrices $\lbrace \gamma^1 , \ldots , \gamma^5 \rbrace$, which are introduced in appendix \ref{app:u224}. They satisfy the Clifford algebra 
\begin{align}
\left \lbrace \gamma ^r  , \gamma ^s \right \rbrace &= 2 \delta ^{rs} \, \unit \, , &
\left \lbrace \gamma ^r  , \gamma ^5 \right \rbrace &= 0 \, , &
\left \lbrace \gamma ^5  , \gamma ^5 \right \rbrace &= 2 \, \unit \, . 
\end{align} 
A suitable choice of coset representatives can be adapted from reference \cite{Arutyunov:2009ga}, 
\begin{align}
m(\phi , z ) = \exp{\left( \ihalf \phi \gamma^5 \right)} \, 
	\left(1 + z^2 \right)^{-1/2} \left( \unit + i\, z^r \gamma^r \right)  \,. 
\label{eqn:su4coset}
\end{align} 
For the Maurer--Cartan form $U = m^{-1} \diff m = A + a$, we find using basic gamma-matrix manipulations
\begin{align}
A &= \frac{2 \, z^r \diff z^s}{1+z^2} \, \gamma^{rs} 
	+ \frac{2 \, z^r \diff \phi }{1+z^2} \, \gamma^{r6} \, , &
a&=  - \frac{1-z^2}{1+z^2} \, \diff \phi \, \gamma ^{56} 
	+ \frac{2 \, \diff z^r}{1+z^2} \, \gamma ^{r5} \, . 
\end{align}
For the metric of the coset space, we note
\begin{align}
\diff s^2 = - \tr \left( a \, a \right) 
	= \frac{4 \, \diff z ^2}{(1+z^2)^2} 
	+ \left( \frac{1-z^2}{1+z^2} \right)^2 \, \diff \phi ^2  
	= \diff N^I \, \diff N^I \, .
\end{align}
Here, we are using the trace metric with a factor of $-1$, since this factor will arise from the supertrace when we are considering the supercoset based on $\grp{SU}(2,2 \vert 4)$. The coordinates $\left(\phi , z^r \right)$ are related to the embedding coordinates of the sphere by
\begin{align}
N^r &= \frac{2\, z^r}{1+z^2} \, , &
N^5 + i N^6 &= \frac{1-z^2}{1+z^2} \, e^{i \phi} \,.
\end{align}
For $z \leq 1$, we get a one-to-one map between the coordinates. We are particularly interested in the Noether current $j = -2 m a m^{-1}$, for which we find the expression
\begin{align}
j &= 2 \left( \frac{1-z^2}{1+z^2} \right) ^2 \diff \phi \, \gamma ^{56} 
	+ 4 \, \frac{\cos \phi \, (1-z^2) z^r \, \diff \phi 
		- \sin \phi \left[
		(1-z^2) \diff z^r + 2 (z \diff z ) z^r \right]}
		{(1+z^2)^2}	\, \gamma ^{r6} \nn \\
	& - 4 \, \frac{\cos \phi \left[
		(1-z^2) \diff z^r + 2 (z \diff z ) z^r \right]
		+ \sin \phi (1-z^2) z^r \diff \phi }
		{(1+z^2)^2}	\, \gamma ^{r6}	
		+ \frac{8 z^r \diff z^s}{(1+z^2)^2} \, \gamma^{rs} \nn \\
 &= 2 \, N^I \, \diff N^J \, \gamma ^{IJ} \, . 
\label{S5Noether} 	
\end{align} 
We note that the Noether current is not divergent when we approach the boundary curve. This simplifies the calculation of the $\tau^0$-coefficient of the conserved charges. In fact, we only need to find the first coefficient in the $\tau$-expansion of the parametrization of the minimal surface, 
\begin{align}
N^I (\tau = 0 , \s ) & = n^I(\s) \, , &  
N^I (\tau , \s ) &= n^I(\s)+ \tau \, N^I_{(1)}(\s) + \O ( \tau^2 ) \, .
\end{align}
This coefficient can be identified with the variation of the minimal area, for which we note the functional%
\footnote{One should add a Lagrange multiplier term when working with embedding coordinates in order to enforce the constraint $N^2 = 1$. This term will not be relevant for our further considerations and so it is left out here.}
\begin{align}
A[N,h] = - \frac{1}{2} \, \int \tr \left( a \wedge \ast a \right) 
	= \frac{1}{2} \, \int \diff \tau \wedge \diff \sigma \, 
		h^{ij} \, \partial_i N^I \, \partial_j N^J  
\end{align} 
In order to determine the coefficient $N_{(1)}^I$, we consider a variation $\delta n^I$ of the boundary curve, which induces a variation $\delta N^I$ of the parametrization of the minimal area. Using that the parametrization of the minimal area satisfies the equations of motion, one only picks up a boundary term in computing the variation of the area and thence (we use conformal gauge)
\begin{align}
\delta A_{\mathrm{min}} =  - \int \diff \s \, N_{(1)} ^I \, \delta n^I \, .
\end{align}
Due to the use of embedding coordinates $N^I$ we have $n^I \delta n^I = 0$ and we conclude that
\begin{align*}
\dfrac{\delta A_{\mathrm{min}} }{\delta n^I (\s)} = - N_{(1)} ^I (\s) + \alpha(\s) n^I(\s)\,.
\end{align*}
The coefficient $\alpha(s)$ is determined from the condition $n^I N_{(1)}^I = 0$ and we find
\begin{align}
N_{(1)}^I (\s) = - \frac{\delta A_{\mathrm{min}} }{\delta n^I (\s)} 
	+ \left( n^J(\s) \, \frac{\delta A_{\mathrm{min}} }{\delta n^J (\s)} \right) n^I(\s) \,.
\end{align} 
With this information, we can go on to evaluate the conserved charges $Q^{(0)}$ and 
$Q^{(1)}$. For the components of the Noether current, we have 
\begin{align}
j_{\tau \, (0)} &= n^I (\sigma) \, \frac{\delta A_{\mathrm{min}} }{\delta n^J (\s)} 
	\, \gamma ^{IJ} \, , &
j_{\sigma \, (0)} &= n^I (\sigma) \, \dot{n}^J(\s)  
	\, \gamma ^{IJ} \, . 	
\end{align}
and the vanishing of the conserved charges thus encodes the condition
\begin{align}
\int \diff \s \left( n^I(\s) \, \dfrac{\delta A_{\mathrm{min}} }{\delta n^J (\s)} 
	- n^J(\s) \, \dfrac{\delta A_{\mathrm{min}} }{\delta n^I (\s)} \right) = 0 \, , 
\end{align}
corresponding to the $\grp{SO}(6)$-invariance of the surface, as well as the non-local condition
\begin{align}
&\int \diff \s_1 \, \diff \s_2 \, \epsilon_{21} \bigg(
	n_1 ^K \, \frac{\delta A_{\mathrm{min}}}{\delta n_2 ^K}
	\left( n_2 ^I \, \frac{\delta A_{\mathrm{min}}}{\delta n_1 ^J}
		- n_2 ^J \, \frac{\delta A_{\mathrm{min}}}{\delta n_1 ^I} \right)	 
	+ n_1 ^K n_2^K \, \frac{\delta A_{\mathrm{min}}}{\delta n_1 ^I} \, 
		\frac{\delta A_{\mathrm{min}}}{\delta n_2 ^J} \nn \\
	& \quad + n_1 ^I n_2^J \, \frac{\delta A_{\mathrm{min}}}{\delta n_1 ^K} \, 
		\frac{\delta A_{\mathrm{min}}}{\delta n_2 ^K}	
	\bigg) 
	=   \int \diff \s \left( n^I \dot{n}^J - n^J \dot{n}^I \right) \, .
\end{align}

\chapter{Away from Strong Coupling}
\label{chap:CarryOver}

It is natural to ask if and how the symmetries discussed in the last chapter can be extended to any value of the coupling constant $\lambda$. Even if a proof of the symmetry property for any value of the coupling constant is out of reach, establishing the symmetry for both large and small values of the coupling constant would show that it is not merely a specialty of either approximation. Below, we discuss different options to transfer the symmetries of minimal surfaces in $\AdS_5$ to arbitrary values of the coupling constant $\lambda$ and indicate which of these are successful. 

\section{Transferring Variations}
The most direct approach to extend the strong-coupling symmetries to smaller values of the coupling constant is to use the same transformations or variations at any value. One could then calculate a variation or large transformation for a specific boundary curve and check whether the one-loop expectation value \eqref{W1loop} of the Maldacena--Wilson loop over this curve is invariant. This approach was discussed by Dekel in reference \cite{Dekel:2015bla} for the spectral-parameter deformation introduced in reference \cite{Ishizeki:2011bf,Kruczenski:2013bsa}, which is the realization of the large master symmetry transformation in Euclidean $\AdS_3$. Building on the wavy expansion studied in references \cite{Polyakov:2000ti,Semenoff:2004qr}, he considered elliptical boundary curves as perturbations of the circle and computed the deformations of these curves to high orders. 
The weak-coupling analysis showed that the invariance observed at strong coupling is broken beyond a certain order in the expansion in the waviness. 

This finding shows that the master variation \eqref{eqn:deltaCx} is generically not a symmetry of the Maldacena--Wilson loop, 
\begin{align*}
\master \left \langle   W(\gamma) \right \rangle = 
-  4 \, G^{ab} \int \limits _0 ^{2 \pi} \diff \sigma \int \limits _0 ^{\sigma} \diff \sigma^\prime \,
	\frac{\delta \left \langle   W(\gamma) \right \rangle}{\delta x^\mu(\sigma) } \xi^\mu _b (x(\sigma))
	\frac{\delta A_{\mathrm{ren}}(\gamma)}{\delta x^\nu(\sigma^\prime) } \xi^\nu _a (x(\sigma^\prime)) 
\neq 0 \ .
\end{align*}
The variation can, however, be adapted in such a way that it becomes a symmetry. By the same reasoning as in \eqref{deltaAsymm} the variation defined as
\begin{align}
\master _{(\lambda)} x^\mu (\sigma) = - 4\,  G^{ab}  \int \limits _0 ^\sigma \diff \sigma ^\prime \, 
\frac{\delta \ln \left \langle   W(\gamma) \right \rangle}{\delta x^\nu(\sigma ^\prime) } \,  \xi^\nu _a (x(\sigma ^\prime)) \, \xi^\mu _b (x(\sigma)) 
\end{align}
constitutes a symmetry of the Maldacena--Wilson loop,
\begin{align}
\master _{(\lambda)} \left \langle   W(\gamma) \right \rangle = 
-  \frac{2}{\left \langle   W(\gamma) \right \rangle } \, G^{ab} 
	\, \delta_b  \left \langle   W(\gamma) \right \rangle 
	\delta_a \left \langle   W(\gamma) \right \rangle 
	= 0 \, .
\end{align}
While the derivation of the invariance is trivial, the action of the variation on different curves should be similarly non-trivial as in the numerical examples of the last chapter. One can adapt the level-1 Yangian-like symmetry variation \eqref{eqn:deltaax} similarly to obtain the variation
\begin{align}
\delta _{(\lambda) a} ^{(1)} x^\mu (\sigma) = - 4 \, f_a {} ^{bc} 
	\int \limits _0 ^\sigma \diff \sigma ^\prime \, 
	\frac{\delta \ln \left \langle   W(\gamma) \right \rangle}
	{\delta x^\nu(\sigma ^\prime) } \,  \xi^\nu _b (x(\sigma ^\prime)) 
	\, \xi^\mu _c (x(\sigma)) \, .
\label{eqn:deltaaxadapted}
\end{align}
Given that the strong-coupling limit of this variation gives a boundary term when acting on the area of the minimal surface, one would not expect the above variation to give a symmetry, but only that its action on the expectation value of the Maldacena--Wilson loop might be local as well. Unfortunately, such a behavior is difficult to see and could not be established at weak coupling for the one-loop expectation value \eqref{W1loop}.  

\section{Transferring Generators}
A different approach to extend the symmetries found at strong coupling is to carry over their generators rather than the variations; this approach was discussed in reference 
\cite{Muller:2013rta}. The conformal symmetry can (at any value of $\lambda$) be written as the invariance under the action of the level-0 generators
\begin{align}
\J_a ^{(0)} = \int \limits \diff \sigma \, \xi ^\mu _a (x(\sigma)) 
	\frac{\delta}{\delta x^\mu (\sigma)} \, .
\end{align}
The master symmetry becomes trivial in this approach, since it corresponds to the Casimir of the level-0 generators, $G^{a b} \J_a ^{(0)} \, \J_b ^{(0)}$. We have seen above that the vanishing of the level-1 Yangian charge $\mathcal{Q}^{(1)}$ can be rewritten as the invariance of the Maldacena--Wilson loop under the level-1 Yangian generator
\begin{align}
\J_a ^{(1)} = \mathbf{f} \indices{_a ^{cb}} \, 
	\int \limits _{0} ^L \diff \sigma_1 \, \diff \sigma_2 \, 
	\varepsilon_{21} \,  \xi ^\mu _{1 b} \,  \xi ^\nu _{2 c} \,  
	\frac{\delta^2}{\delta x^\mu _1 \delta x^\nu _2} 
	- \frac{\la}{4 \pi ^2} \int \limits _0 ^L \diff \sigma \, \xi^\mu _a  
	\left( \dx_\mu \, \ddot{x}^2 + \dddot{x}_\mu \right) \, ,
\label{gen:Level-1}	
\end{align}
which we decompose into a bi-local and a local part, 
$\J_a ^{(1)} = \J_{a , \mathrm{bi-lo}} ^{(1)} + \lambda \J_{a , \mathrm{lo}} ^{(1)}$. The application of this generator to the strong-coupling limit of the expectation value of the Maldacena--Wilson loop, 
\begin{align}
\left \langle W(\gamma) \right \rangle =
	\exp \left( - \ft{\sqrt{\la}}{2 \pi} A_\mathrm{ren}(\gamma) \right)  ,
\label{WStrong}	
\end{align}
gives the identity \eqref{Q1identity}. Here, we only consider the leading term in the expansion in $1/ \sqrt{\lambda}$ and neglect the double-derivative term appearing at the sub-leading order. The bilocal part of this generator shows the typical structure of a level-one Yangian symmetry generator as it is known from two-dimensional integrable field theories or scattering amplitudes, cf.\ the discussion in section \ref{sec:Int_Symm}. The Poisson algebra of the associated conserved charges suggests that these generators satisfy a Yangian algebra and it was indeed shown in references \cite{Dolan:2003uh,Dolan:2004ps} that generators of the form $\J_{a , \mathrm{bi-lo}} ^{(1)}$ satisfy the commutation relations of the Yangian algebra,
\begin{align}
\left[ \J^{(0)} _a \, , \,  \J^{(1)} _b \right] = \mathbf{f} \indices{_{ab}^c} \, \J^{(1)} _c \, .
\label{Yangian_Comm1}  
\end{align}
The Serre relations were discussed for the special case of the underlying Lie Group being 
$\grp{SU}(N)$ in reference \cite{Dolan:2004ps}. At least the commutation relation above can be transferred also to the local term. This follows from noting that
\begin{align*}
\J_{a , \mathrm{bi-lo}} ^{(1)} \left \langle W(\gamma) \right \rangle = \big( 
	\lambda \, \J_{a , \mathrm{lo}} ^{(1)} + \O \big( \sqrt{\lambda} \big) \big)
	\left \langle W(\gamma) \right \rangle 
\end{align*}
implies 
\begin{align*}
\big[ \J_a ^{(0)} , \J_{b , \mathrm{bi-lo}} ^{(1)} \big] 
	\left \langle W(\gamma) \right \rangle =
	\big( 
	\lambda \, \big[ \J_a ^{(0)} , \J_{b , \mathrm{lo}} ^{(1)} \big] 
	+ \O \big( \sqrt{\lambda} \big) \big)
	\left \langle W(\gamma) \right \rangle \, , 
\end{align*}
such that the commutation relation \eqref{Yangian_Comm1} transfers to the local part. We check this property explicitly in appendix~\ref{app:transf}, since it provides a strong consistency check of the calculation performed in section \ref{sec:InfSymm}. 
Even though we cannot apply the above argument to transfer the validity of the Serre relation for the bi-local pieces to the full generators, there is strong evidence that the above level-1 generators satisfy the commutation relations of the Yangian algebra
$\mathrm{Y}[ \alg{so}(2,4)]$. 

\begin{figure}
\centering
\includegraphics[width=60mm]{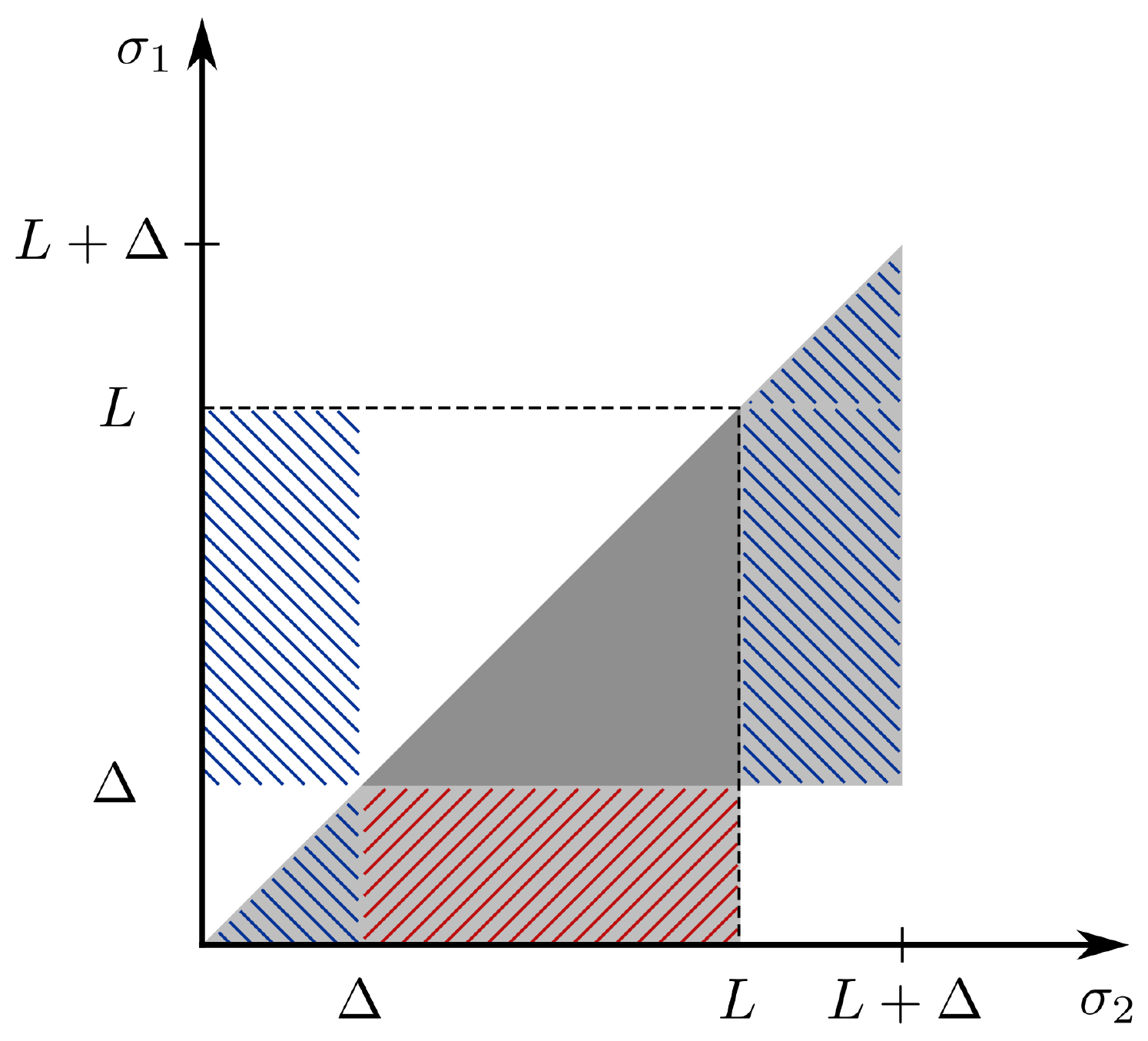}
\caption[Schematic drawing of the parameter regions.]
{Schematic drawing of the parameter regions $\sigma_2 > \sigma_1$ for the generators $\J_a ^{(1)}$ 
and $\tilde{\J}_a ^{(1)}$. The difference of the two generators is obtained as the difference 
that arises from integrating over the regions  
\includegraphics[height=1.8ex]{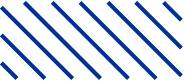}
and 
\includegraphics[height=1.8ex]{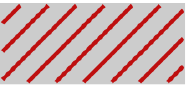}$\,.$
}
\label{fig:parameter}
\end{figure}

There is, however, an algebraic problem that prevents this generator from becoming a symmetry at weak coupling. We have noted before that all of the non-local variations depend on the choice of the point $z_0$ on the worldsheet where the initial condition $\chi_u (z_0)= \unit$ is imposed.
Changing this point amounts to the transformation $\chi _u \to L \cdot \chi_u$, see section \ref{sec:Master}. For the higher Yangian-like variations this implies
\begin{align*}
\delta_\eps ^{(u)} \; \; \to \; \; \delta_{\tilde{\eps}} ^{(u)} 
\qquad \text{with} \qquad \tilde{\eps} = L^{-1} \, \eps \, L \, . 
\end{align*}
For the generator $\J_a^{(1)}$, however, the base-point dependence is different. For explicitness, let us consider a curve $\gamma$ parametrized by $x:[0,L] \to \mathbb{R}^{(1,3)}$. Instead of $x(0)$ we could equally well choose a different starting point $x(\Delta)$ and obtain a different level-1 generator $\tilde{\J}_a^{(1)}$. The difference of the parameter regions for the bi-local integrals of the two generators is depicted in figure \ref{fig:parameter}. For the difference between the two level-1 generators, we find  
\begin{align}
\tilde{\J}_a ^{(1)} - \J_a ^{(1)} &= 2
	\left( \int _\Delta ^L  \diff \sigma _1 \int  _0 ^\Delta \diff \sigma _2
	- \int  _0 ^\Delta \diff \sigma _1 \int  _\Delta ^L \diff \sigma _2 \right)
	\mathbf{f} \indices{_a ^{cb}} \, \jay_b (\sigma_1) \, \jay_c (\sigma_2)  \nn \\
	&= 2 \, \mathbf{f} \indices{_a ^{cb}}  
	\left( \J_b ^{(0)} \, \jay_c ^\Delta  - \jay_b ^\Delta \, \J_c ^{(0)} \right) \, .
\label{calc:cyclic}	
\end{align}
Here, we have introduced the abbreviations 
\begin{align}
\jay_a ^\Delta &:= \int \limits _0 ^\Delta \diff \sigma \, \jay_a  (\sigma) \, , &
\jay_a  (\sigma) &= \xi ^\mu _a (x(\sigma)) \, \frac{\delta}{\delta x^\mu (\sigma)} \, .
\end{align}
If we act on an object that is invariant under the level-0 generators, all terms that contain the generator $\J_a ^{(0)}$ at the right-most position can be neglected. Then, we get
\begin{align}
\tilde{\J}_a ^{(1)} - \J_a ^{(1)} & \simeq 2 \, \mathbf{f} \indices{_a ^{cb}}  \, 
	\mathbf{f} \indices{_{bc} ^d}  \, \jay_d ^\Delta 
	= 4  \mathfrak{c} \, \jay_a ^\Delta \, .
\end{align}
The index structure appears for any Lie algebra, the numerical factor of $\mathfrak{c}$ is called the dual Coxeter number and depends on the underlying algebra; it accounts for the difference between the Killing metric and the trace metric which we are using to raise and lower the group indices. For the conformal algebra $\alg{so}(2,4)$, we note that $\alg{c}=2$. 

Since the starting points for the generators $\J_a ^{(1)}$ are chosen arbitrarily, both $\tilde{\J}_a ^{(1)}$ and $\J_a ^{(1)}$ should be symmetries in order to have level-1 Yangian symmetry. Hence, since $\Delta$ is arbitrary, the functional derivative $\delta/\delta x^\mu(\sigma)$ at any point on the loop would have to annihilate the result, which can clearly not be the case. Note that the base-point dependence does not contradict the strong-coupling result, since the difference between the two level-1 generators gives a sub-leading result when applied to \eqref{WStrong}.

We have thus seen that the invariance of the Maldacena--Wilson loop under the above level-1 generators, which we showed for asymptotically large $\la$, cannot extend in an expansion in $1/\sqrt{\la}$. This matches well with the finding that the generators \eqref{gen:Level-1} do not annihilate the one-loop expectation value of the Maldacena--Wilson loop, which was obtained in reference \cite{Muller:2013rta} by explicit calculation. It is, however, possible to obtain a cyclic generator of the form \eqref{gen:Level-1} by adding an appropriate `local' term. This was noted in reference \cite{Chicherin:2017cns} in the study of loop amplitudes in the bi-scalar field theory introduced in reference \cite{Gurdogan:2015csr}.

In order to find the appropriate local term, it is simpler to consider the level-1 Yangian generators for the discrete multi-site space discussed in section \ref{sec:Int_Symm}. We consider the generators
\begin{align}
\J_a ^{(1)} &= \mathbf{f} \indices{_a ^{cb}} 
	\sum \limits _{i < k} ^N \, \J_{b,i} \, \J_{c,k} 
	+ \sum \limits _{i=1} ^N v_i \, \J_{a,i} 
	= \J_{a, \mathrm{bi-lo}} ^{(1)} + \J_{a, \mathrm{lo}} ^{(1)} \, .
\end{align}
Here, we have added a generic local term $\J_{a, \mathrm{lo}} ^{(1)}$ with coefficients $v_i$. A non-constant choice for these coefficients would render the local term non-cyclic as well and the idea is to choose them in such a way that the sum with the bi-local piece 
$\J_{a, \mathrm{bi-lo}} ^{(1)}$ becomes cyclic under generic shifts of the starting point up to level-0 generators. Transferring our above result for the continuous case, we see that the bi-local piece transforms as 
\begin{align}
\J_{a, \mathrm{bi-lo}} ^{(1)} \big \vert _{m+1, N+m} - 
	\J_{a, \mathrm{bi-lo}} ^{(1)} \big \vert _{1, N} 
	\simeq 4  \alg{c} \sum \limits _{i=1} ^m \J_{a,i} \, , 
\end{align}
under a shift of the starting point from 1 to $m$. Here, we have adapted the notation of reference \cite{Chicherin:2017cns} to make the starting-point dependence explicit and again left out terms containing level-0 generators at the right-most position. For the local part of the level-1 generator, we find the transformation
\begin{align}
\J_{a, \mathrm{lo}} ^{(1)} \big \vert _{m+1, N+m} - 
	\J_{a, \mathrm{lo}} ^{(1)} \big \vert _{1, N} 
	&= \sum \limits _{i=m+1} ^{N+m} \, v_{i-m} \J_{a,i} 
	- \sum \limits _{i=1} ^{N} \, v_{i} \J_{a,i} \nn \\
	&= \sum \limits _{i=1} ^m \left( v_{i-m+N} - v_i \right) \J_{a,i}
	+ \sum \limits _{i=m+1} ^{N} \left( v_{i-m} - v_i \right) \J_{a,i} \, .
\end{align}   
For the full level-1 generator we thus have
\begin{align}
\J_{a} ^{(1)} \big \vert _{m+1, N+m} - 
	\J_{a} ^{(1)} \big \vert _{1, N} 
	= \sum \limits _{i=1} ^m \left( v_{i-m+N} - v_i + 4 \, \alg{c} \right) \J_{a,i}
	+ \sum \limits _{i=m+1} ^{N} \left( v_{i-m} - v_i \right) \J_{a,i} \, ,
\end{align}
and we aim to fix the coefficients $v_i$ by demanding that
\begin{align}
\J_{a} ^{(1)} \big \vert _{m+1, N+m} - 
	\J_{a} ^{(1)} \big \vert _{1, N} = b_m \, \J_a ^{(0)} \, .
\end{align}
Note that we allow the proportionality constant to depend on the length of the shift. We thus have the conditions
\begin{align}
v_{i-m+N} - v_i + 4 \, \alg{c} &= b_m \, , \qquad \forall i \in 
	\lbrace 1 , \ldots , m \rbrace \, , \\
v_{i-m} - v_i &= b_m \, , \qquad \forall i \in 
	\lbrace m+1 , \ldots , N \rbrace \, .	
\end{align}
It is then easy to see that the coefficients $v_i$ must be linear, $v_i = ri$, and solving for $r$, we find $v_i = - 4 \alg{c}i /N$. We have thus found that the level-1 generators 
\begin{align}
\J_a ^{(1)} &= \mathbf{f} \indices{_a ^{cb}} 
	\sum \limits _{i < k} ^N \, \J_{b,i} \, \J_{c,k} 
	- \frac{4 \alg{c}}{N} \sum \limits _{i=1} ^N i \, \J_{a,i} 
\end{align}
are indeed cyclic. 

In transferring this finding to the continuous generators needed for the discussion of the Maldacena--Wilson loop, we need to consider the requirement of reparametrization invariance as well. Note that the addition of a term of the form
\begin{align*}
- \frac{4 \alg{c}}{L(\gamma)} \int \diff \sigma \, \sigma \, \jay_a(\sigma)   
\end{align*}  
is not sensible, since it does not give a reparametrization invariant expression when applied to one. We can obtain a reparametrization invariant generator by replacing the above term by the (no longer local) piece 
\begin{align}
- \frac{4 \alg{c}}{L(\gamma)} \int \diff \sigma _1 \, \diff \sigma_2 \, \theta_{21} \, 
	\lvert \dx_1 \rvert \, \jay_a (\sigma_2) \, ,
\end{align}
which reproduces the initial guess for an arc-length parametrization. By going through the steps of the calculation \eqref{calc:cyclic}, it is then easy to see that the completed generator
\begin{align}
\J_a ^{(1)} &= 
	\int \diff \sigma_1 \, \diff \sigma_2 \, \theta_{21}
	\left( 2 \mathbf{f} \indices{_a ^{cb}} \, \jay_b (\sigma_1) \, \jay_c (\sigma_2) 
		- \frac{4 \alg{c}}{L} \, \lvert \dx_1 \rvert \, \jay_a (\sigma_2) \right)
	+ \J_{a, \mathrm{lo}} ^{(1)}  	
\label{Gen:Level-1_cyclic}	
\end{align}
is indeed cyclic up to level-0 generators. Here, $\J_{a, \mathrm{lo}} ^{(1)}$ denotes a generic local piece of the form appearing in equation \eqref{gen:Level-1}. We note however, that the above level-1 generator no longer satisfies the Yangian algebra relation
\begin{align}
\big[ \J_a ^{(0)} \, , \,  \J_b ^{(1)} \big] &= 
	\mathbf{f} \indices{_{ab} ^c} \, \J_c ^{(1)} \, ,
\end{align}
since reparametrization invariance required to make the counterpart of the coefficients $v_i$ curve-dependent. A short calculation reveals that the commutator with the special conformal level-0 generators is no longer of the form above. This indicates that the generator \eqref{Gen:Level-1_cyclic} does not provide a symmetry of the Maldacena--Wilson loop. Indeed, it seems not to be possible to find a local piece $\J_{a, \mathrm{lo}} ^{(1)}$ such that the generator  
\eqref{Gen:Level-1_cyclic} becomes a symmetry at weak coupling. We thus have to conclude that, while an intriguing possibility in general, the restoration of the cyclicity of the level-1 Yangian generator via the addition of an appropriate local term does not lead to a Yangian invariance of the Maldacena--Wilson loop. 

In the following chapters, the problem of having cyclic generators will be circumvented in another way, which is also crucial for the Yangian symmetry of super-amplitudes \cite{Drummond:2009fd}. For the superalgebra $\alg{psu}(2,2 \vert 4)$, the Killing metric and hence the dual Coxeter number 
$\alg{c}$ vanishes, such that the level-1 generators for the Yangian over the superconformal algebra $\alg{psu}(2,2 \vert 4)$ are automatically cyclic.  

In order to obtain a Yangian symmetry over the superconformal algebra, one needs to consider a supersymmetric extension of the Maldacena--Wilson loop into a Wilson loop in superspace, for which the underlying level-0 symmetry is given by $\alg{psu}(2,2 \vert 4)$. It was observed in reference \cite{Muller:2013rta} that the Wilson loop in superspace can be constructed from the requirement of supersymmetry and based on this insight the first two orders in an expansion in 
Gra{\ss}mann degrees were constructed. This was sufficient to show the Yangian invariance of the one-loop expectation value at the zero order in the Gra{\ss}mann expansion. 

The construction of the Wilson loop in superspace was subsequently completed in reference \cite{Beisert:2015jxa} and the full level-1 Yangian invariance at the one-loop order was shown in reference \cite{Beisert:2015uda}. In parallel, the strong-coupling description of the Wilson loop in superspace as well as its Yangian invariance were worked out in reference \cite{Munkler:2015gja}. The extension considered at strong coupling is also based on the requirement of supersymmetry and thus lifts the minimal surfaces in
$\AdS_5 \times \mathrm{S}^5$ to minimal surfaces of the superstring in the coset superspace
\begin{align}
\frac{\grp{PSU}(2,2 \vert 4)}{\grp{SO}(1,4) \times \grp{SO}(5)} \, ,
\end{align}
in which type IIB superstring theory in $\AdS_5 \times \mathrm{S}^5$ is constructed
\cite{Metsaev:1998it}. The discussion of these minimal surfaces and the consequences of the classical integrability \cite{Bena:2003wd} of type IIB superstring theory in $\AdS_5 \times \mathrm{S}^5$ are the subject we discuss in the remainder of this thesis. 
We begin by studying type IIB superstring theory in $\AdS_5 \times \mathrm{S}^5$ and its classical integrability in chapter \ref{chap:Semi-SSM}. Here, we also discuss the master symmetry appearing in this model. We then go on to discuss minimal surfaces ending on the superconformal boundary of the above coset space and transfer the symmetries of the superstring theory to these configurations in chapter \ref{chap:SurfaceSuperspace}. 

\chapter{Semisymmetric Space Models}
\label{chap:Semi-SSM}

As a prerequisite for our discussion of superspace Wilson loops at strong coupling, we discuss semisymmetric space models, which are the supersymmetric generalizations of the symmetric space models discussed in chapter \ref{chap:SSM}. In addition to reviewing the action and integrability following references \cite{Arutyunov:2009ga,Alday:2005gi,Zarembo:2010sg}, we introduce the master symmetry and discuss the so-called bonus symmetries. These are higher-level recurrences of charges, which are not conserved at the level zero.   

\section{The String Action}

The action for the type IIB Green-Schwarz superstring in $\AdS_5 \times \mathrm{S}^5$ was first constructed in reference \cite{Metsaev:1998it} employing a coset construction similar to the one of reference \cite{Henneaux:1984mh}, which discussed the Green-Schwarz superstring in flat space. For a detailed description, the reader is referred to the review \cite{Arutyunov:2009ga}, which we follow below. 

We consider a semisymmetric space $\grp{G}/\grp{H}$, which means that the Lie superalgebra $\alg{g}$ allows for a $\mathbb{Z}_4$-grading,
\begin{align}
\alg{g} &= \alg{g} ^{(0)} \oplus \alg{g} ^{(1)} 
	\oplus \alg{g} ^{(2)} \oplus \alg{g} ^{(3)} \, , & 
\left[ \alg{g} ^{(k)} , \alg{g} ^{(l)} \right] \subset
	\alg{g} ^{(k+l \, \mathrm{mod} 4)} \, .	
\end{align}
The decomposition is based on an automorphism $\Omega$, for which
\begin{align}
\alg{g} ^{(0)} = \left \lbrace X \in \alg{psu}(2,2 \vert 4) 
	: \Omega(X) = X \right \rbrace = \alg{h} \, ,
\end{align}
and the other subspaces correspond to the eigenvalues $i^k$. The case we are interested in here is the supergroup $\grp{PSU}(2,2 \vert 4)$, for which we have introduced the above structure in section \ref{sec:Int_Symm}, but the discussion extends to other cases as well. For 
$\grp{PSU}(2,2 \vert 4)$, the bosonic subgroup is given by 
$\grp{SU}(2,2) \times \grp{SU}(4)$ and corresponds to the isometry group 
$\grp{SO}(2,4) \times \grp{SO}(6)$ of $\AdS_5 \times \mathrm{S}^5$. We discuss the fundamental representation of the superconformal algebra in more detail in appendix \ref{app:u224} and the discussion there shows that the Lie algebra $\alg{g}^{(0)}$ of the gauge group is given by 
$\alg{so}(1,4) \oplus \alg{so}(5)$, such that the supercoset indeed describes a superspace with    
$\AdS_5 \times \mathrm{S}^5$ as the bosonic base. 

Below, we will actually consider the field $g$ to take values in the supergroup 
$\grp{SU}(2,2\vert4)$, such that we can employ the matrix representation discussed in section \ref{sec:Int_Symm} and appendix \ref{app:u224}. We will see below that the model nonetheless corresponds to the supercoset space
\begin{align}
\frac{\grp{PSU}(2,2 \vert 4)}{\grp{SO}(1,4) \times \grp{SO}(5)} \, . 
\end{align}
In order to construct the superstring action, we consider the Maurer--Cartan form
\begin{align}
U = g^{-1} \diff g = \A{0} + \A{1} + \A{2} + \A{3} \, ,
\end{align}
which we again split into its graded components. The action is then given by
\begin{align}
S = - \frac{T}{2} \, \int \tr \left( 
	\A{2} \wedge \ast \A{2} + i \, \tilde{\kappa} \, \A{1} \wedge \A{3} \right) .
\label{Superstring_Action}	
\end{align}
We note that we are considering a Euclidean worldsheet metric, which implies the appearance of the factor $i$ in front of the fermionic term, corresponding to the Wick-rotation of the term 
$\eps^{i j}$ \cite{Zarembo:2010sg}.
It is important to stress that we are not considering a Wick rotation in the target space, but consider minimal surfaces for which the induced metric and hence also the worldsheet metric has Euclidean signature. 
The action becomes real after Wick rotating to a Minkowskian worldsheet metric, for which the reality constraints imposed on $\alg{g}$ imply that the action is real. The relative factor $\tilde{\kappa}$ between the bosonic and fermionic contributions in the action is fixed by requiring kappa symmetry or integrability to satisfy 
$\tilde{\kappa}^2 = 1$ and we will set $\tilde{\kappa}=1$ in the following. 

We note that the three-form appearing in the non-local description of the Wess--Zumino term is exact in the case at hand \cite{Berkovits:1999zq}, such that it can be reduced to the two-dimensional integral appearing in the action \eqref{Superstring_Action}.  

Since the Wess--Zumino term does not couple to the worldsheet metric, the Virasoro constraints only involve the bosonic component $\A{2}$ of the Maurer--Cartan form and are analogous to the ones for symmetric space models, 
\begin{align}
\str \big( \A{2}_i \, \A{2} _j \big) 
	- \half \, h_{ij} \, h^{kl} \,  
	\str \big( \A{2}_k \, \A{2} _l \big) = 0 \, .
\label{Virasoro_SUSY}	
\end{align}

Let us now discuss the gauge symmetries of the above action. As already for the symmetric space models, gauge transformations are given by $g \mapsto g R$, where $R(\tau, \s)$ is a function taking values in the gauge group H. For these, the Maurer--Cartan form transforms as
\begin{align}
\A{0} \; \; & \mapsto R^{-1} \A{0} R + R^{-1} \diff R \, , &
\A{i} \; \; & \mapsto R^{-1} \A{i} R \, , &
\end{align}
such that the action of semisymmetric space models is invariant. In the case of $g$ taking values in $\grp{SU}(2,2 \vert 4)$, there is another gauge symmetry, which is given by the transformation%
\footnote{Note that the generator $C$ given in appendix \ref{app:u224} is defined in such a way that $iC \in \alg{su}(2,2 \vert 4)$. }
\begin{align}
g \; \; \mapsto \; \; g \, e^{i \beta C} = e^{i \beta C} g \, ,
\end{align}
where $\beta = \beta(\tau , \sigma)$ is local. We note that, since $iC \in \alg{g}^{(2)}$, this transformation differs from the gauge transformations described above. The Maurer--Cartan form transforms as 
\begin{align}
\A{2} \; \; \mapsto \; \; \A{2} + i \diff \beta \, C \, , 
\end{align}
and the invariance of the action follows from the degeneracy of the metric, 
\begin{align}
\str \left( T_a \, C \right) = 0 \qquad \forall \; T_a \in \alg{su}(2,2 \vert 4) \, . 
\end{align} 
Gauge-fixing the above gauge symmetry effectively reduces the model to the supercoset space with 
$\grp{G}=\grp{PSU}(2,2 \vert 4)$. 

Indeed, the global symmetry of the model is given by $\grp{PSU}(2,2 \vert 4)$, since the left-multiplication with the exponential of $C$ corresponds to a gauge transformation. The invariance under these transformations then follows as before from the invariance of the Maurer--Cartan form under the transformations $g \mapsto L g$.

Importantly, the above action has another gauge symmetry which is known as kappa symmetry. For the flat-space superstring, kappa symmetry has been observed in reference \cite{Green:1983wt} and accounts can be found in textbooks on superstring theory such as \cite{Green:1987sp}. The kappa symmetry of the Green-Schwarz superstring generalizes the local fermionic symmetries observed for superparticle actions \cite{deAzcarraga:1982dhu,Siegel:1983hh}, which are also called kappa symmetries. The appearance of kappa symmetry corresponds to the fact that half of the fermionic degrees of freedom decouple from the superstring theory equations of motion.   

For the discussion and proof of kappa symmetry in the coset description employed here, we refer the reader to reference \cite{Arutyunov:2009ga}, which in turn follows references \cite{McArthur:1999dy,Roiban:2000yy}. Here, we will be satisfied with just stating the kappa symmetry transformations. On the coset representatives, a kappa symmetry transformation acts by the multiplication of a local fermionic element from the right, 
\begin{align}
g \; \; & \mapsto g \cdot \exp{\kappa} \, , & 
\delta_\kappa \, U &= \mathrm{d}\kappa + \left[U \, , \, \kappa\right]  \,. 
\end{align}
Here, $\kappa$ takes values in the fermionic part of the superalgebra, 
$\kappa = \kappa^{(1)+(3)}$. The kappa symmetry symmetry transformation also includes a variation of the worldsheet metric, cf.\ reference \cite{Arutyunov:2009ga}. The transformations constitute a local gauge symmetry of the action provided that the supermatrix $\kappa$ is given by
\begin{align}
\kappa^{(1)} &= A_{i , -} ^{(2)} \, \mathcal{K} _+ ^{(1),i} 
	+ \mathcal{K} _+ ^{(1),i} \, A_{i , -} ^{(2)} \, , &
\kappa^{(3)} & = A_{i , +} ^{(2)} \, \mathcal{K}_- ^{(3),i} 
	+ \mathcal{K} _- ^{(3),i} \, A_{i , +} ^{(2)} \, . 
\label{Kappa_Transf}	
\end{align}
Here, we employ the projections 
$P_{\pm} ^{i j} = \half \left( \gamma^{ij} \pm i \, \eps^{i j} \right)$ on the worldsheet and denote
\begin{align}	
V_{\pm} ^i &= P_{\pm} ^{i j} \, V_j 
	= \half \left( \gamma^{ij} \pm i \, \eps^{i j} \right) V_j \, . 
\end{align}
Moreover, $\mathcal{K}$ denotes an arbitrary fermionic one form taking values in 
$\mathfrak{psu}(2,2\vert4)$ and $\mathcal{K}^{(1)}$, $\mathcal{K}^{(3)}$ are its projections in the algebra. Again, the appearance of the factor of $i$ is related to working with a Euclidean world-sheet metric. It implies that the kappa symmetry variations in this case typically violate the reality constraint imposed on the algebra elements. This is not the case if one works with a Minkowskian signature for the worldsheet metric. 

We then turn to the derivation of the equations of motion. For a generic variation $\delta g$, the Maurer--Cartan form transforms as
$\delta U = \diff \left( g^{-1} \delta g \right) + \left[ U , g^{-1} \delta g \right]$ and hence the variation of the action is given by
\begin{align*}
\delta S = - T \, \int \tr \left[ \left( 
	\diff \left( g^{-1} \delta g \right) + \left[ U , g^{-1} \delta g \right] \right)
	\wedge \left( \ast \A{2} - \ihalf A^{(1)-(3)} \right) \right] . 
\end{align*}
In this expression we denote 
\begin{align}
\ast \Lambda = \ast \A{2} - \ihalf A^{(1)-(3)} \, ,
\label{def:Lambda}
\end{align}
and employ \eqref{TraceTrick} to get
\begin{align}
\delta S = T \, \int \tr \left[ 
	\left( \diff \ast \Lambda + U \wedge \ast \Lambda + \ast \Lambda \wedge U \right)
	g^{-1} \delta g  \right] .
\end{align}
If we regard the above combination to take values in $\alg{su}(2,2 \vert 4)$, the equations of motion are given by 
\begin{align}
\diff \ast \Lambda + U \wedge \ast \Lambda + \ast \Lambda \wedge U = i \alpha \, C \, ,
\label{EOM_SUSY_C}
\end{align}
due to the degeneracy of the metric. Here, $\alpha$ is a generic two-form taking values in 
$\mathbb{R}$. Over $\alg{psu}(2,2 \vert 4)$, we have the equations of motion
\begin{align}
\diff \ast \Lambda + U \wedge \ast \Lambda + \ast \Lambda \wedge U = 0 \, .
\label{EOM_SUSY}
\end{align}
We will work with the $\alg{psu}(2,2 \vert 4)$-case in our further discussion of the integrability of the model, but return to the case of $\alg{su}(2,2 \vert 4)$ in section \ref{sec:Bonus}.

Let us now derive the Noether current associated to the G-symmetry of the model. For the variation 
$\delta_\epsilon g = \epsilon g $ we note that
$ \delta_\epsilon U = g^{-1} \diff \epsilon g$, such that
\begin{align}
\delta S = - T \, \int \tr \left[ 
	g^{-1} \diff \epsilon g \wedge 
	\left( \ast \A{2} - \ihalf A^{(1)-(3)} \right) \right] .
\end{align}
The associated Noether current is hence given by
\begin{align}
j = - 2 g \left( \A{2} +  \ihalf \ast A^{(1)-(3)} \right) g^{-1} 
	= - 2 g \Lambda  \, g^{-1} \, .
\end{align}
It is simple to check that the above current is indeed conserved, 
\begin{align}
\diff \ast j = - 2 \diff \left[ g \ast \Lambda  \, g^{-1} \right]
	= - 2 g \left( \diff \ast \Lambda + U \wedge \ast \Lambda + \ast \Lambda \wedge U \right) g^{-1}
	= 0 \, .
\end{align}
In contrast to the bosonic case, however, the Noether current is not flat, although a kappa-symmetry gauge exists in which it becomes flat \cite{Hatsuda:2004it,Hatsuda:2011mt}. However, we prefer not to exhaust our kappa symmetry gauge freedom in order to reach a flat Noether current, since we may employ a different approach than the BIZZ construction to obtain an infinite tower of conserved charges.

\section{Integrability}
The integrability of semisymmetric space models was shown in reference \cite{Bena:2003wd} by constructing a Lax connection as a deformation of the Maurer--Cartan form $U$. With the Lax connection at hand, we can employ the methods explained in chapter \ref{chap:SSM} to construct an infinite tower of conserved charges.

Let us first construct the Lax connection in our conventions. In order to do this, it is helpful to use the projections of the flatness condition 
$\diff U + U \wedge U = 0$ and the equations of motion \eqref{EOM_SUSY} on the various graded components of the algebra. For the flatness condition, we note the projections
\begin{align}
\begin{aligned}
\diff \A{0} + \A{0} \wedge \A{0} + \A{2} \wedge \A{2}
	+ \commwedge{\A{1}}{\A{3}} &= 0 \, , \\
\diff \A{2} + \A{1} \wedge \A{1} + \A{3} \wedge \A{3}
	+ \commwedge{\A{0}}{\A{2}} &= 0 \, , \\
\diff \A{1} + \commwedge{\A{0}}{\A{1}} + \commwedge{\A{2}}{\A{3}} &= 0 \, , \\
\diff \A{3} + \commwedge{\A{0}}{\A{3}} + \commwedge{\A{2}}{\A{1}} &= 0 \, .	
\end{aligned} 
\label{flatness_proj}
\end{align} 
Here, we have introduced the notation
\begin{align}
\commwedge{A}{B} = A \wedge B + B \wedge A \, ,
\end{align}
to abbreviate the frequently-occurring combinations $ A \wedge B + B \wedge A$, which give rise to a commutator of the components. In the following, we will sometimes find it useful to work with the quantities 
\begin{align}
\AAp &= \A{1} + \A{3} \, , & 
\AAm &= \A{1} - \A{3} \, , 
\end{align}
rather than $\A{1}$ and $\A{3}$ and we note that the flatness conditions of the latter can be rewritten as
\begin{align}
\begin{aligned}
\diff \AAp + \commwedge{\AAp}{\A{0}} + \commwedge{\AAp}{\A{2}} = 0 \, , \\
\diff \AAm + \commwedge{\AAm}{\A{0}} - \commwedge{\AAm}{\A{2}} = 0 \, .
\end{aligned}
\label{flatness_ferm}
\end{align}
The two equations above are of course equivalent, since their projections on the components $\alg{g}^{(1)}$ and $\alg{g}^{(3)}$ are the same. 

For the equations of motion, we note that the projection on $\alg{g}^{(0)}$ vanishes. For the other projections, we find using the flatness condition \eqref{flatness_proj}
\begin{align}
\diff \ast \A{2} + \commwedge{\A{0}}{\ast \A{2}} 
	- i \A{1} \wedge \A{1} + i \A{3} \wedge \A{3}    
	&= 0 \, , \nn \\ 
\commwedge{\ast \A{2}}{\A{3}} + i \commwedge{\A{2}}{\A{3}} &= 0 \, , \\	
\commwedge{\ast \A{2}}{\A{1}} - i \commwedge{\A{2}}{\A{1}} &= 0 \, . \nn
\end{align}
We can again rewrite the last two equations to find
\begin{align}
\diff \ast \A{2} + \commwedge{\A{0}}{\ast \A{2}} 
	- \ihalf \commwedge{\AAp}{\AAm} &= 0 \, , \nn \\ 
\commwedge{\ast \A{2}}{\AAp} - i \, \commwedge{\A{2}}{\AAm} &= 0 \, , \\
\commwedge{\ast \A{2}}{\AAm} - i \, \commwedge{\A{2}}{\AAp} &= 0 \, . \nn
\label{EOM_Susy_2}
\end{align}
Again, we note that either of the above equations is equivalent to the two fermionic projections of the equations of motion, since both take values in $\alg{g}^{(1)}$ and $\alg{g}^{3}$. 

In order to construct a Lax connection $L_u$, we follow references 
\cite{Bena:2003wd,Arutyunov:2009ga} and consider the ansatz
\begin{align}
L_u = \A{0} + \alpha_1 \, \A{2} + \alpha_2 \, \ast \A{2} 
	+ \alpha_3 \, \AAm + \alpha_4 \, \AAp \, .
\end{align}
We note that the above Lax connection has to reduce to the Lax connection 
\eqref{eqn:definition-Lax} of symmetric space models when we set the fermionic components zero, and so we have $\alpha_1 ^2 + \alpha_2 ^2 = 1$, or using the parametrization employed in chapter \ref{chap:SSM}, 
\begin{align}
\alpha_1 &= \frac{1-u^2}{1+u^2} \, , &
\alpha_2 &= - \frac{2u}{1+u^2} \, .
\label{bos_param}
\end{align}
We the consider the projections of the flatness condition
\begin{align}
\diff L_u + L_u \wedge L_u = 0
\end{align}
on the graded components of the algebra in order to fix the coefficients $\alpha_3$ and $\alpha_4$. For the projection on $\alg{g}^{(0)}$ we get
\begin{align}
\diff \A{0} + \A{0} \wedge \A{0} + \A{2} \wedge \A{2} 
	+ \left( \alpha_4 ^2 - \alpha_3 ^2 \right) \commwedge{\A{1}}{\A{3}} = 0 \, ,
\end{align}
and comparing with the flatness condition \eqref{flatness_proj} for $\A{0}$ gives the condition
\begin{align}
\alpha_4 ^2 - \alpha _3 ^2 = 1 \, . 
\end{align}
Making use of the flatness condition for $\A{2}$, we find the projection on the component 
$\alg{g}^{(2)}$ to be given by
\begin{align}
\alpha_2 \big( \diff \ast \A{2}  + \commwedge{\A{0}}{\A{2}} \big) 
	& + \left(\a _3 ^2 + \a_4 ^2 + 2 \a_3 \a_4 - \a_1 \right) 
	\A{1} \wedge \A{1} \nn \\
	& + \left(\a _3 ^2 + \a_4 ^2 - 2 \a_3 \a_4 - \a_1 \right) 
	\A{3} \wedge \A{3} 
	= 0 \, .
\end{align}
The comparison with the equations of motion \eqref{EOM_SUSY} then gives the relations
\begin{align}
\a_3 ^2 + \a_4 ^2 + 2 \a_3 \a_4 - \a_1 &= - i \a_2 \, , &
\a_3 ^2 + \a_4 ^2 - 2 \a_3 \a_4 - \a_1 &=  i \a_2 \, , &
\end{align}
or equivalently
\begin{align}
\a_3 \, \a_4 &= - \ihalf \, \a_2 \, , & 
\a_3 ^2 + \a_4 ^2 - \a_1 &= 0 \, .
\end{align}
Given the form \eqref{bos_param} of the coefficients of the bosonic components, we can already solve for the coefficients $\a_3$ and $\a_4$ for which we obtain
\begin{align}
\a_3 &= \frac{i u }{\sqrt{1+u^2}} \, , & 
\a_4 &= \frac{1}{\sqrt{1+u^2}} \, .
\end{align}
It remains to check that with these coefficients also the fermionic parts of the flatness condition vanish. Using the flatness conditions \eqref{flatness_ferm}, we find
\begin{align}
\diff L_u + L_u \wedge L_u 
	&= \a_2 \a_4 \commwedge{\ast \A{2}}{\AAp} 
	+ \a_3 (\a_1 + 1) \commwedge{\A{2}}{\AAm} \nn \\
	& + \a_2 \a_3 \commwedge{\ast \A{2}}{\AAm} 
	+ \a_4 (\a_1 - 1) \commwedge{\A{2}}{\AAp} \, . 
\label{Lax_flat}	
\end{align}
Using the relations $(\a_1 + 1 ) = 2 \a_4 ^2$ and $(\a_1 - 1 ) = 2 \a_3 ^2$ as well as 
$\a_3 \a_4 = - \ihalf \a_2$, we get
\begin{align}
i \a_3 (\a_1 + 1) &= \a_2 \a_4 \, , &
i \a_3 (\a_1 - 1) &= \a_2 \a_3 \, ,    
\end{align}
such that $\diff L_u + L_u \wedge L_u$ indeed vanishes due to the equations of motion 
\eqref{EOM_Susy_2}. In summary, we have found the Lax connection of semisymmetric space models to be given by
\begin{align}
L_u = \A{0} + \frac{1-u^2}{1+u^2} \A{2} - \frac{2u}{1+u^2} \! \ast \! \A{2} 
	+ \frac{iu}{\sqrt{1+u^2}} \, \AAm + \frac{1}{\sqrt{1+u^2}} \, \AAp  .
\label{Lax_SUSY}	
\end{align}
Employing the angular parametrization, 
\begin{align}
e^{i \theta} = \frac{1-iu}{1+iu} \, ,
\end{align}
we have the Lax connection
\begin{align}
L_\theta = \A{0} + \cos \theta \, \A{2} + \sin \theta \, \ast \! \A{2} 
	- i \sin \ft{\theta}{2} \, \AAm + \cos \ft{\theta}{2} \, \AAp  .
\label{Lax_SUSY_2}	
\end{align}
For the construction of the conserved charges, we can then proceed as before. We transform the Lax connection to obtain the flat connection 
\begin{align}
\ell _u &= g(L_u - U) g^{-1} \nn \\
	&= - \frac{2u^2}{1+u^2} \, a^{(2)} - \frac{2u}{1+u^2} \, \ast \! a^{(2)} 
	+ \frac{iu}{\sqrt{1+u^2}} \, a^{(1)-(3)}  
	+ \frac{1- \sqrt{1+u^2} }{\sqrt{1+u^2}} \, a^{(1)+(3)} \, . 
\end{align}
Here, we have introduced the abbreviations
\begin{align}
a^{(k)} = g \, \A{k} \, g^{-1}
\label{small_a}
\end{align}
for the frequently-occurring conjugations of the projections of the Maurer--Cartan form with 
$g$. This form of the Lax connection is convenient to derive an infinite tower of conserved charges. 

We can also apply it to convince ourselves that the Noether current of semisymmetric space models is generically not flat. If we express the above Lax connection in terms of the Noether current, we obtain
\begin{align}
\ell_u &= \frac{u}{1+u^2} \left( \ast j + u \, j \right)
	+ \frac{i\, u^2}{1+u^2} \, \ast \! a^{(1)-(3)} \nn \\
	&+ \frac{iu ( \sqrt{1+u^2} -1 ) }{\sqrt{1+u^2}} 
	\, a^{(1)-(3)} 
	+ \frac{1-\sqrt{1+u^2}}{\sqrt{1+u^2}} \, a^{(1)+(3)} \, . 
\end{align}
Expanding the above expression in $u$, 
\begin{align}
\begin{aligned}
\ell_u &= u \ast j + u^2 \left( j + i \ast a^{(1)-(3)} 
	-\half a^{(1)+(3)} \right) + \O(u^3) \\
	&= u \ast j - 2 u^2 \left( a^{(2)}
	+ \quarter a^{(1)+(3)} \right) + \O(u^3) \, ,
\end{aligned}	
\end{align}
we find that the flatness condition 
$\diff \ell_u + \ell_u \wedge \ell_u$ gives current conservation as well as the relation
\begin{align}
\diff j + j \wedge j = \half \, \diff a^{(1)-(3)} 
	+ i \, \diff \ast a^{(1)-(3)} \, ,
\label{jnotflat}	
\end{align}
showing that the Noether current is not flat, as claimed above. 

For the construction of the conserved charges, we again consider the auxiliary linear problem
\begin{align}
\diff \chi_u &= \chi_u \, \ell_u \, , &
\chi_u(z_0) &= \unit \, .
\end{align}
or the associated monodromy 
\begin{align}
t_u = \prexp \left( \int _\gamma \ell_u \right) .
\end{align}
Here, we again consider the curve $\gamma$ to cover the whole period of $\sigma$ at constant $\tau$. The $\tau$-dependence of the monodromy $t_u$ is then captured by the evolution equation 
\begin{align}
 \partial_\tau t_u (\tau)  = \left[t_u (\tau) \, , \, \ell_{u , \tau} (\tau , 0) \right] \, .
\end{align}
This equation implies that the matrices $t_u$ at different values of $\tau$ are related by a similarity transformation, such that all eigenvalues are conserved charges. In the case of a minimal surface, we can contract the curve $\gamma$ to a point, at which we then find
$t_u = \unit$, which extends to all other values of $\tau$ by the similarity transformations.  

The expansion of the monodromy gives the conserved charges (we restrict to conformal gauge)
\begin{align}
Q^{(0)} &= \int \diff \s  \, j_\tau  \, , \\
\tilde{Q}^{(1)} &= \int \diff \s_1 \, \diff \s_2 \, \theta_{21} \, 
	j_\tau (\s_1) \, j_\tau(\s_2) 
	- 2 \int \diff \s \left( a_\s ^{(2)} + \quarter a_\s ^{(1)+(3)} \right) 
	\, , \ldots  
\end{align}

\section{Master Symmetry}
The master symmetry discussed for symmetric space models in reference \cite{Klose:2016uur} was extended shortly afterwards to the pure spinor description of the superstring in $\AdS_5 \times \mathrm{S}^5$ \cite{Chandia:2016ueo}. Indeed, the generalization to semisymmetric space models is straight-forward, as we shall see below.  

As before, we introduce a deformation of the field $g$ by demanding that its associated Maurer--Cartan form be the Lax connection $L_u$, 
\begin{align}
g_u ^{-1} \diff g_u &= L_u \, ,  & 
g_u (z_0) &= g(z_0) \, .
\label{Master_SUSY}
\end{align} 
Again, we note that due to the use of a Euclidean world-sheet metric, there is no value of the spectral parameter except the undeformed case $u=0$, for which the Lax connection is real. The deformations $g_u$ will thus generically violate the reality constraints for $\grp{SU}(2,2 \vert 4)$ in the Euclidean case. 

We need to show that the above deformation leaves the action, the equations of motion and the Virasoro constraints invariant as in the bosonic case. The invariance of the Virasoro constraints \eqref{Virasoro_SUSY} can be transferred directly, since they only involve the component $\A{2}$ of the Maurer--Cartan form and the bosonic parts of the Lax connection are deformed in the same way as for the symmetric space models. 

By the same reasoning, we need only consider the fermionic part of the superstring action to show its invariance. Here, we note that the fermionic components of the Lax connection are given by 
\begin{align}
L_u ^{(1)} &= \frac{1+iu}{\sqrt{1+u^2}}\, \A{1} \, , &
L_u ^{(3)} &= \frac{1-iu}{\sqrt{1+u^2}}\, \A{3} \, , 
\end{align}
such that 
\begin{align}
\str \left( L_u ^{(1)} \wedge L_u ^{(3)} \right) 
	= \frac{(1+iu)(1-iu)}{(1+u^2)} \str 
	\left( \A{1} \wedge \A{3} \right) 
	= \str \left( \A{1} \wedge \A{3} \right) .
\end{align}
This establishes the invariance of the superstring action \eqref{Superstring_Action} under the master symmetry transformation \eqref{Master_SUSY}. 

The invariance of the equations of motion is similarly direct. We note that upon deforming the Maurer--Cartan form $U$ into the Lax connection $L_u$, the quantity $\Lambda$ introduced above is deformed to 
\begin{align}
\ast \Lambda_u &= \ast L_u ^{(2)} - \ihalf L_u ^{(1)-(3)} \nn \\
	&= \frac{2u}{1+u^2} \, \A{2} + \frac{1-u^2}{1+u^2} \, \ast \A{2} 
		- \frac{i}{2 \sqrt{1+u^2}} \, \AAm
		+ \frac{u}{2 \sqrt{1+u^2}} \, \AAp \, .
\end{align}
We then need to check that the equations of motion are still satisfied, 
\begin{align}
\diff \ast \Lambda_u + \commwedge{L_u}{\ast \Lambda_u} = 0 \, . 
\label{EOM_deformed}
\end{align}
In order to see this, it is convenient to split the above combination into two terms, 
\begin{align*}
\diff \ast \Lambda_u + \commwedge{L_u}{\ast \Lambda_u} = \diff \ast \Lambda_u 
	+ \commwedge{\A{0}}{\ast \Lambda_u}
	+ \commwedge{L_u - \A{0}}{\ast \Lambda_u} \, .
\end{align*}
For the first term, we find
\begin{align}
& \diff \ast \Lambda_u + \commwedge{\A{0}}{\ast \Lambda_u} = 
	\frac{2u}{1+u^2} \left( \diff \A{2} + \commwedge{\A{0}}{\A{2}} \right) \nn \\
	 &+ \frac{1-u^2}{1+u^2} \left( \diff \ast \A{2} + \commwedge{\A{0}}{\ast \A{2}} \right) 
	 - \frac{i}{2 \sqrt{1+u^2}} \left( \diff \AAm + \commwedge{\A{0}}{\AAm} \right) \nn \\
	 &+ \frac{u}{2 \sqrt{1+u^2}}  \left( \diff \AAp + \commwedge{\A{0}}{\AAp} \right) .
\end{align}
For the other term, we find using the fermionic parts of the equations of motion \eqref{EOM_Susy_2},  
\begin{align}
\commwedge{L_u - \A{0}}{\! \ast \Lambda_u} &= 
	\frac{2u}{1+u^2} \left( \A{1} \! \wedge \A{1} + \A{3} \! \wedge \A{3} \right)
	- \frac{i(1-u^2)}{(1+u^2)} \commwedge{\AAm\!}{\AAp} \nn \\
	&+ \frac{i}{2 \sqrt{1+u^2}} \commwedge{\A{2}}{\AAm} 
	+ \frac{u}{2 \sqrt{1+u^2}} \commwedge{\A{2}}{\AAp}  \, .
\end{align}
We can then employ the flatness conditions \eqref{flatness_proj} and \eqref{flatness_ferm} as well as the bosonic part of the equations of motion \eqref{EOM_SUSY} in order to see that the two contributions cancel each other. The master symmetry transformation thus indeed preserves the equations of motion. 

We can then carry over many of the results derived in section \ref{sec:Master} and section \ref{sec:IntCompl}. For example, we may rewrite the master symmetry transformation as 
\begin{align}
g_u &= \chi_u \, g \, , &
\diff \chi_u &= \chi_u \ell_u \, , &
\chi_u (z_0) &= \unit \, , 
\end{align}
and the master symmetry variation is again identified as
\begin{align}
\master \, g &= \chi^{(0)} \, g \, , &
\diff \chi^{(0)} &= \ast j \, , & 
\chi^{(0)} (z_0) &= 0 \, .
\end{align}
Moreover we note that the master symmetry transformation again commutes with the underlying 
G-symmetry and that changing the point where the initial condition is imposed, corresponds to a G-symmetry. In order to carry over the property that the concatenation of two master symmetries gives a master symmetry with different spectral parameter, we need to consider the deformation of the Lax connection $(L_{\theta_1} )_  {\theta_2}$. We can show by direct calculation that
\begin{align*}
(L_{\theta_1} )_  {\theta_2} = L_{\theta_1 + \theta_2} \, . 
\end{align*}
In fact, this can be seen from noting that it holds for the bosonic components of the Lax connection and that the coefficients of its fermionic components are completely determined by the requirement of flatness. We thus note that
\begin{align}
\left( M_{\theta_1} \circ M_{\theta_2} \right) (g) = M_{\theta_1 + \theta_2} (g) \, ,
\end{align}
with $M_{\theta} (g) = g_\theta$. By the same reasoning as the one presented around equation  
\eqref{eqn:delchi}, we then find the master variations of $g_u$ and $\chi_u$ to be given by
\begin{align}
\master g_u &= (1+u^2) \frac{\diff}{\diff u} \, g_u \, , &
\master \chi_u &= \left(1 + u^2 \right) \frac{\diff}{\diff u} \chi_u 
	- \chi_u \cdot \chi^{(0)} \, .
\end{align}

We can also apply the master symmetry to obtain a one-parameter family of conserved currents, which are the Noether currents associated to the deformed solution $g_u$, 
\begin{align}
j_u = - 2 g_u \, \Lambda_u \, g_u ^{-1} \, 
	&= \chi_u \bigg( \frac{2(u^2-1)}{1+u^2} \, a^{(2)} + \frac{4u}{1+u^2} \, \ast  a^{(2)} \nn \\
	& \quad  - \frac{i}{\sqrt{1+u^2}} \, \ast a^{(1)-(3)} 
	+ \frac{u}{\sqrt{1+u^2}} \, \ast a^{(1)+(3)} \bigg) \chi_u ^{-1} \, .
\end{align}
The relation between the associated conserved charges
\begin{align}
Q_u = \int \ast j_u 
\end{align}
and the charges obtained from the expansion of the monodromy is the same as for the symmetric space models. Again, we note that the master symmetry variation acts as a raising operator on the charges obtained from the expansion of the above one-parameter family, 
\begin{align}
Q_\theta &= \sum \limits _{n=0} ^{\infty}  \frac{\theta^n}{n!} \, Q ^{(n)} \, , & 
\master\, Q^{(n)} &= Q^{(n+1)} \, .
\end{align}
Last, we note that the conserved charge associated to the master symmetry itself is again the Casimir of the G-symmetry charges.

For the symmetric space models, we have discussed an analogous procedure to infer a one-parameter family of symmetry variations by using the master symmetry, which was called the integrable completion of the underlying symmetry variations. This structure is more difficult to transfer to semisymmetric space models and remains to be studied. 

\section{Bonus Symmetries}
\label{sec:Bonus}

Apart from the Yangian $Y[\mathfrak{psu}(2,2\vert4)]$ symmetries \cite{Drummond:2009fd,Beisert:2010gn}, the S-matrix of $\Nfour$ supersymmetric Yang--Mills theory has a so-called bonus symmetry \cite{Beisert:2011pn}, the level-1 recurrence of the hypercharge generator, which is itself not a symmetry. A similar situation has been observed before the discovery of the bonus symmetry for the $\mathrm{AdS}_5 \times \mathrm{S}^5$ worldsheet S-matrix, which is Yangian invariant \cite{Arutyunov:2006yd,Beisert:2007ds} and also exhibits an additional symmetry at level-1 \cite{Matsumoto:2007rh,deLeeuw:2012jf,Beisert:2014hya}, the so-called secret symmetry.

In the classical string theory, the bonus symmetry corresponds to the part proportional to $C$ of the level-1 Yangian charge. Such charges have been constructed in the pure spinor formalism in all odd levels of the Yangian \cite{Berkovits:2011kn}. Below, we construct conserved charges of this kind in \emph{all higher levels} of the Yangian. The presentation differs slightly from the one given in reference \cite{Munkler:2015xqa} as it draws some inspiration from the master symmetry discussed above.  

In order to construct these conserved charges, we enforce no constraint on the gauge symmetry associated to the central charge $C$, and hence we have the equations of motion 
\eqref{EOM_SUSY_C},
\begin{align}
\diff \ast \Lambda + U \wedge \ast \Lambda + \ast \Lambda \wedge U = i \, \alpha \, C \, .
\label{EOM_SUSY_C2}
\end{align}
This implies that the $C$-part of the Noether current $j$ is no longer conserved, 
\begin{align}
\diff \ast j &= - 2 g \left( \diff \ast \Lambda 
	+ \ast \Lambda \wedge U + U \wedge \ast \Lambda \right) g^{-1}
	= - 2i \, \alpha \, C \, ,
\end{align}
and, correspondingly, neither is the $C$-part of the associated conserved current $Q^{(0)}$. For the conserved current $Q^{(1)}$, however, a direct calculation shows that also the $C$-part of this charge is conserved. The charge is given by 
\begin{align}
Q^{(1)} = \int \diff \s_1 \, \diff \s_2 \, \theta_{21} \, 
	\left[ j_\tau (\s_1) , j_\tau (\s_2) \right] 
	- 4 \int \diff \sigma \left( a_\s ^{(2)} + \quarter a_\s ^{(1)+(3)} \right) \, , 
\end{align}
and its conservation follows from noting that, while the current conservation condition is altered as above, the condition \eqref{jnotflat} describing the deviation of the Noether current from the flatness condition is not altered by the inclusion of the central charge. We may rewrite this condition as 
\begin{align}
\partial_\tau \, a_\s ^{(2)} - \partial_\s \, a_\tau ^{(2)}
	+ \quarter \left(\partial_\tau \, a_\s ^{(1)+(3)} 
		- \partial_\s \, a_\tau ^{(1)+(3)} \right)
	= \half \, \left[j_\tau , j_\s \right] \, , 	
\end{align}
and together with the altered current conservation condition, it is easy to show that 
$Q^{(1)}$ is a conserved charge. Here, the appearance of the $C$-term is not relevant, since it only appears inside commutators. 

In order to generalize the above finding to higher-level charges, we consider the monodromy 
$t_u$. We note that the Lax connection is no longer flat given the equations of motion 
\eqref{EOM_SUSY_C2} and hence the auxiliary linear problem is no longer well-defined. The evolution equation for the monodromy thus no longer follows from the auxiliary linear problem and so we need to take a different approach. 

Concretely, we note that the Lax connection \eqref{Lax_SUSY} satisfies the condition 
\begin{align}
\diff L_u + L_u \wedge L_u = - \frac{2iu}{1+u^2} \, \alpha \, C \, . 
\end{align}
In order to see this, note that the Maurer--Cartan form is still flat, 
$\diff U + U \wedge U = 0$, and since $iC \in \alg{g}^{(2)}$, we only need to modify the equation involving $\diff \ast \A{2}$. The above finding transfers to the transformed Lax connection $\ell_u$, for which we have 
\begin{align}
\diff \ell_u + \ell_u \wedge \ell_u = - \frac{2iu}{1+u^2} \, \alpha \, C \, ,  
\end{align}
or in coordinates,
\begin{align}
\partial_\tau \, \ell_{u, \s} - \partial_\s \, \ell_{u , \tau}
	+ \left[ \ell_{u , \tau} , \ell_{u, \s} \right] 
	&= - \frac{2i \, u}{1+u^2} \, \alpha(\tau , \s) \, C \, .
\end{align}
Here, we have set $\alpha = \alpha(\tau , \s ) \, \diff \tau \wedge \diff \s$. For the $\tau$-dependence of the monodromy $t_u$, we then find
\begin{align}
\partial_\tau \, t_u &= \partial_\tau \, \prexp \bigg( 
	\int _0 ^{2 \pi} \diff \s \, \ell_{u, \s} \bigg) \nn \\
	&= \int _0 ^{2 \pi} \diff \s \, 
	\prexp \bigg( \int _0 ^{\s} \diff \s^\prime \, \ell_{u, \s} \bigg)
	\partial_\tau \, \ell_{u , \s} 	\,
	\prexp \bigg( \int _\s ^{2 \pi} \diff \s^\prime \, \ell_{u, \s} \bigg) \nn \\
	&=  \int _0 ^{2 \pi} \diff \s \, \partial_\s \bigg[
	\prexp \bigg( \int _0 ^{\s} \diff \s^\prime \, \ell_{u, \s} \bigg)
	\ell_{u, \tau} 	\,
	\prexp \bigg( \int _\s ^{2 \pi} \diff \s^\prime \, \ell_{u, \s} \bigg) \bigg] \nn \\
	& \qquad - \frac{2i\, u}{1+u^2} \, 
	\int \diff \s \, \alpha (\tau , \sigma) \, C \, 
	\prexp \bigg( \int _0 ^{\s} \diff \s^\prime \, \ell_{u, \s} \bigg)
	\prexp \bigg( \int _\s ^{2 \pi} \diff \s^\prime \, \ell_{u, \s} \bigg) \nn \\
	&= \left[t_u , \ell_{u, \tau} (\tau , 0) \right] 
	- \frac{2i\, u}{1+u^2} \, \tilde{\alpha}(\tau) \, C \, t_u \, .
\end{align}
Here, we have abbreviated
\begin{align*}
\tilde{\alpha}(\tau) = \int \diff \sigma \, \alpha (\tau , \s) 
\end{align*}
in the last step. We can use the above evolution equation to show that all expansion coefficients of the monodromy are proportional to the central charge. We have seen this explicitly for the coefficient $t^{(1)} = Q^{(0)}$ above and it follows by induction for the higher coefficients: For the $\tau$-dependence of the $n$-th order coefficient, we find
\begin{align}
\partial_\tau \, t^{(n)} = 
	\sum \limits _{m=1} ^{n-1} \left[ t^{(m)} , \ell ^{\,(n-m)} _{\tau} (\tau, 0) \right] 
	- i \, \tilde{\a} (\tau) \sum \limits _{m=0} ^{n-1} 
		c_m \, C \, t^{(m)} \, . 
\end{align}
By the induction hypothesis and noting that $C^2 =C$, we thus have 
$\partial_\tau \, t^{(n)} \propto C$. At this point, we make use of the fact that we are dealing with a minimal surface%
\footnote{The results presented here extend to the field theory case (i.e.\ imposing boundary conditions at spatial infinity), which is discussed in reference \cite{Berkovits:2011kn}. The conservation condition for the quasi-momenta extracted from the monodromy in the case of closed strings remains to be studied for the central extension considered here. 
}
with disk topology, which allows us to conclude that $ t^{(n)} \propto C$ as well. With this finding established, we can simplify the evolution equation to 
\begin{align}
\partial_\tau \, t_u &= - \frac{2i\, u}{1+u^2} \, \tilde{\alpha}(\tau) \, C \, t_u \, .
\label{Evolution_C} 
\end{align} 

We note now that the conserved level-1 charge discussed above was obtained from the application of the master symmetry to the G-symmetry charge rather than the expansion of the monodromy, from which it differs by adding a multiple of $(Q^{(0)} )^2$. We have established the connection between these charges for symmetric space models in equation \eqref{eqn:chiandjt} and building on this finding, it seems sensible to consider the combination
\begin{align*}
f_u = (1+u^2) \, \dot{t}_u \, t_u ^{-1} \, .
\end{align*}
Making use of equation \eqref{Evolution_C}, we find
\begin{align}
\partial_\tau \, f_u &= 
	- 2i \, u \, \tilde{\alpha}(\tau) \, \left[C , f_u \right] 
	- 2i \, (1+u^2) \, \partial_u \left( \frac{u}{1+u^2} \right) \tilde{\alpha}(\tau) \, C \nn \\
	&= \frac{1-u^2}{1+u^2} \; \partial_\tau \, t^{(1)} \, , 
\end{align}
where we have inserted the first order of equation \eqref{Evolution_C} in order to replace 
$\tilde{\alpha}(\tau)$ by $\partial_\tau \, t^{(1)}$, which captures the non-conservation of the 
$C$-term of the G-symmetry charge. We have thus found that also the $C$-part of the quantity
\begin{align}
\tilde{f}_u = (1+u^2) \, \dot{t}_u \, t_u ^{-1} 
	- \frac{1-u^2}{1+u^2} \; t^{(1)}  
\end{align}
is conserved. 

\chapter{Minimal Surfaces in Superspace}
\label{chap:SurfaceSuperspace}

We now turn to the supersymmetric generalization of the strong-coupling description of the Maldacena--Wilson loop, which is obtained by replacing the minimal area in 
$\AdS_5 \times \mathrm{S}^5$ by the area of a minimal surface in the supercoset space 
\begin{align*}
\frac{\grp{SU}(2,2 \vert 4)}{\grp{SO}(1,4) \times \grp{SO}(5)} \, ,  
\end{align*}
which we again renormalize appropriately. We thus describe the expectation value of the super Maldacena--Wilson loop at strong coupling by
\begin{align}
\left \langle \mathcal{W}(\gamma) \right \rangle 
	= e^{ -\frac{\sqrt{\la}}{2 \pi} \mathcal{A}_\mathrm{ren}(\gamma) } \, , 
\label{strongsuperWL}
\end{align}
where the area functional $\mathcal{A}$ is based on the superstring action discussed in the last chapter. Appropriate boundary conditions for this minimal surface have been suggested in reference \cite{Ooguri:2000ps} based on a set of generalized Poincar{\'e} coordinates for the supercoset space. Following their approach, we assume the boundary curve to take values in a $\Nfour$ non-chiral superspace, i.e.\ we have a parametrization
\begin{align}
\left( x^\mu (\s) , \theta \indices{_\a ^A} (\s) , \btheta_{A \da} (\s) , n^I(\s) \right) .
\end{align}
We will then see that the  divergence of the minimal area is proportional to the super-length of the curve, i.e.\ we have
\begin{align}
\mathcal{A}_\mathrm{min}(\gamma) \big \vert _{y \geq \varepsilon} 
	&= \frac{\mathcal{L}(\gamma)}{\varepsilon} + \mathcal{A}_\mathrm{ren}(\gamma) \, , & 
\mathcal{L}(\gamma) &= \int \diff \s \lvert \pi (\s) \rvert \, .
\end{align}
Here, $\pi ^\mu = \dx ^\mu + i \tr \big( \dot{\btheta} \s ^\mu \theta - \btheta \s ^\mu \dot{\theta} \big)$ describes the supermomentum of a superparticle moving along the respective contour in the boundary superspace and we have regulated the minimal area by imposing a cut-off $\varepsilon$ in the coordinate $y$ of $\AdS_5$.

With the description of the minimal surfaces in superspace understood, the Yangian symmetry at strong coupling can be established in the same way as in chapter \ref{chap:MinSurf} for the minimal surfaces in $\AdS_5$. This requires to determine the first coefficients of the expansion of the minimal surface around the conformal boundary in order to evaluate the conserved charges derived in the last chapter. 

\section{Poincar{\'e} Coordinates for the Supercoset}
\label{sec:Coordinates}

We begin by introducing a generalization of Poincar{\'e} coordinates for the supercoset space by employing the coset representatives introduced in reference \cite{Ooguri:2000ps}, 
\begin{align}
g(X,N,y,\Theta, \vartheta) = e^{ X \cdot P } \, 
	e^{\Theta _\a {} ^A \, Q_A {}^\a + \bTheta_{A \da} \, \Qb^{\da A} } \, 
	e^{\vart _A {} ^\a \, S_\a {} ^A + \bvart^{\da A} \, \Sb_{A \da} } 
	\, M(N) \, y^D \,. 
\label{cosetrep}
\end{align}
Here $(X,y)$ and $N$ are bosonic coordinates parametrizing the $\AdS_5$ and the $\mathrm{S}^5$ part, respectively. The 32 fermionic degrees of freedom are parametrized by the Gra{\ss}mann odd coordinates $\Theta, \bTheta, \vart$ and $\bvart$. The spherical part of the coset space is parametrized in the same way as in section \ref{sec:S5}, but written as a $\left(4 \vert 4 \right)$ supermatrix, 
\begin{align}
M(N) = \begin{pmatrix}
\unit_4 & 0 \\ 0 & m(N) 
\end{pmatrix} .
\end{align}
This specific choice of coset parametrization is not at all arbitrary. A crucial aspect is that all exponents have definite weight and that they are ordered by these weights. Moreover, the $y$-coordinate, which vanishes on the conformal boundary, is associated to the dilatation generator and
appears at the right-most position. This form is important for the discussion of the superconformal boundary as we shall see below. 

The Maurer--Cartan form for the above choice of coset representatives has been partially derived in reference \cite{Ooguri:2000ps} in so far as it was required for their discussion of the superconformal boundary. In order to determine the expansion of the minimal surface into the bulk space, we need the complete expression and so we derive it below. The calculation is lengthy but straightforward. Abbreviating
\begin{align}
\Omega := \Theta _\a {} ^A \, Q_A {}^\a + \bTheta_{A \da} \, \Qb^{\da A}  \, , \qquad \eta := \vart _A {} ^\a \, S_\a {} ^A + \bvart^{\da A} \, \Sb_{A \da} \, , 
\end{align}  
we need to compute
\begin{align*}
g^{-1} \diff g &= y^{-D} \, M^{-1} \, e^{-\eta} 
	\left( \diff X \cdot P + e^{-\Omega} \diff e^ {\Omega} \right) e^\eta \, M \, y^D \\
	& \quad + y^{-D} \, M^{-1} \left( e^{-\eta}  \diff e^\eta \right) M \, y^D 
	+ M^{-1} \diff M + \frac{\diff y}{y} D \,.
\end{align*}  
We note that 
\begin{align*}
e^{-\Omega} \diff e^ {\Omega} = \diff \Omega 
	+ i \left(\Theta _\a {} ^A \, \diff \bTheta _{A \da} 
		- \diff \Theta _\a {} ^A \, \bTheta _{A \da} \right) P^{\da \a} \, ,
\end{align*}
and a similar formula holds for $e^{-\eta}  \diff e^\eta$. Hence, we only have to calculate the remaining conjugations. Most of them can be carried out by using Hadamard's lemma, 
\begin{align*}
e^{A} \, B \, e^{-A} &= \sum \limits _{n=0} ^\infty \frac{1}{n!} \, 
	\left[ A   ,  B \right]_{(n)} \, , & 
\left[ A  ,  B \right]_{(n)} &= \big[A  ,  \left[ A  ,  B \right]_{(n-1)} \big] \, , &
\left[ A  ,  B \right]_{(0)} &= B \,.
\end{align*}
Here it is crucial that the exponents $X\cdot P$, $\Omega$ and $\eta$ have a definite non-zero weight, such that the above expansion breaks off after at most four orders. This follows from noting that the weights are additive,   
\begin{align*}
[D , [A , B ] ] &= \left( \Delta _A + \Delta_B \right) [A , B ]  \, ,
\end{align*}
and the fact that for $\alg{su}(2,2\vert 4)$ we only have half-integer weights ranging from $-1$ to $1$. In order to do the conjugations with $M(N)$, we consider the supermatrices explicitly.  Using the definitions given in appendix \ref{app:u224}, we note
\begin{align}
M^{-1} \, \left( \Theta  _\a {}^A \, Q _A {} ^\a \right) M &= 
	\begin{pmatrix} \unit_2 & 0 & 0 \\ 0 & \unit_2 & 0 \\ 0 & 0 & m^{-1} \end{pmatrix}
	\begin{pmatrix} 0 & 0 & 2 \, \Theta \\ 0 & 0 & 0 \\ 0 & 0 & 0 \end{pmatrix} 
	\begin{pmatrix} \unit_2 & 0 & 0 \\ 0 & \unit_2 & 0 \\ 0 & 0 & m \end{pmatrix} \nn \\
&=  \begin{pmatrix} 0 & 0 & 2 \, \Theta \, m \\ 0 & 0 & 0 \\ 0 & 0 & 0 \end{pmatrix} 
	= \left( \Theta  _\a {}^B m \indices{_B ^A} \right) Q _A {} ^\a 
	= \left( \Theta m \right)_\a {}^A  \,  Q _A {} ^\a \, .
\end{align}
Similarly, one finds:
\begin{align}
M^{-1}  \left( \bvart^{\da A} \, \Sb_{A \da} \right) M & = 
	\left(\bvart m  \right)^{\da A}  \Sb_{A \da} \, , & 
	M^{-1} \left( \bTheta_{A \da} \, \Qb^{\da A}  \right) M &= 
	\left( m^{-1}  \bTheta \right)_{A \da}  \Qb^{A \da} \, ,  \\
M^{-1} \left( \vart _A {} ^\a \, S_\a {} ^A  \right) M &= 
	\left( m^{-1}  \vart \right) _A {} ^\a  S_\a {} ^A \, , & 
	M^{-1} \left(  \Lambda \indices{_A ^B} R \indices{ ^A _B} \right) M &= 
	\left( m^{-1} \Lambda  m \right) \indices{_A ^B}  R \indices{ ^A _B} \,. \nn
\end{align}
The conjugations with $y^D$ again follow from the weights of the generators. For 
$[D,T^\Delta] = \Delta T^\Delta$, we get
\begin{align}
y^{-D} \, T^\Delta \, y^D = y^{-\Delta} \, T^\Delta \,.
\end{align}
Carrying out all conjugations in this way, we arrive at 
\begin{align}
U &= g^{-1} \, \diff g = - \frac{r_{\a \da} }{2 y} \, P^{\da \a} + \frac{\sigma}{y} D 
	+ \frac{1}{\sqrt{y}} \left( \varepsilon _\a {}^A \, Q_A {} ^\a 
		+ \widebar{\varepsilon}_{A \da} \, \Qb^{\da A} \right) 
	+ \la \indices{_\a ^\b} \, M \indices{_\b ^\a} 
	+ \bar{\la} \indices{^\db _\da} \, \bar{M} \indices{ ^\da _\db}  \nn \\
& + \Lambda \indices{_A ^B} R \indices{^A _B} + \gamma \, C 
	+ \sqrt{y} \left( \chi_A {}^\a \, S_\a {} ^A + \bar{\chi}^{\da A} \, \Sb_{A \da} \right) 
	- \frac{y}{2}\, \kappa ^{\da \a}   \, K_{\a \da} + M^{-1} \diff M 
\label{eqn:MaurerCartanForm}
\end{align} 
Here, we defined the coefficients
\begin{align}
\begin{aligned}
r_{\a \da} &=  \diff X_{\a \da} + 2 i \left( \diff \Theta  \, \bTheta  
	- \Theta \, \diff \bTheta \right) {} _{\a \da} \, , \\
r_\mu &=  \diff X_{\mu} - i \, \tr \left( \diff \bTheta \s_\mu \Theta  
		- \bTheta \s_\mu \diff  \Theta \right) \, , \\
\sigma &= \diff y + 2 y\,  \tr \left( \diff \bTheta \, \bvart 
	+ \vart\, \diff \Theta \right) \, , \\
\gamma &=  \tr \left( \vart \left(2 \diff\Theta + \zeta \right) 
	- \left(2 \diff\bTheta + \widebar{\zeta} \right) \bvart \right) , \\
\la \indices{_\a ^\b} &= - i \, \left[ \left(2 \diff \Theta + \zeta \right) 
	\vart \right] \indices{_\a ^\b} \, , 
\end{aligned}
&&
\begin{aligned}
\zeta _\a {} ^A &= i \, r_{\a \da} \, \widebar{\vartheta}^{\da A} \, , \\
\widebar{\zeta}_{A \da} &= - i \, \vartheta _A {} ^\a \, r_{\a \da} \, , \\
\varepsilon _\a {}^A &= \left[ \left( \diff \Theta + \zeta \right) m \right] _\a {} ^A \, , \\
\widebar{\varepsilon}_{A \da} &= \left[ m^{-1} \left( \diff \bTheta + \widebar{\zeta} \, \right)
	\right] _{A \da} \, , \\
\widebar{\la} \indices{^\db _\da} &= i \, \left[ \bvart 
	\left(2 \diff \bTheta +  \bar{\zeta} \right) \right] \indices{^\db _\da} \, .	
\end{aligned}
\end{align}
The remaining terms are given by
\begin{equation}
\begin{aligned}
\Lambda _A {} ^B &=
	\left[ m^{-1} \left(
		\vart \left(\diff\Theta + \half \zeta\right)
		- \left(\diff\bTheta + \half \bar{\zeta} \right) \bvart
	\right) m \right]_A {} ^B \, , \\
\chi _A {}^\a &=
	\left[ m^{-1} \left[ \left(
			4 \vart \left(\diff\Theta + \thirrd \zeta \right)
			- 2 \left(\diff\bTheta + \thirrd \bar{\zeta} \right) \bvart
		\right) \vart
		+ \diff\vart
	\right] \right] _A {}^\a \, , \\
\widebar{\chi} ^{\da A} &=
	\left[ \left[
		\diff\bvart
		+ \bvart \left(
			4 \left(\diff\bTheta + \thirrd \bar{\zeta} \right) \bvart
			- 2 \vart \left(\diff\Theta + \thirrd \zeta \right)
		\right)
	\right] m \right] ^{\da A} \, , \\
\kappa ^{\da \a} &=
	\left[ - 8 i \, \bvart \left[
		\left( \diff\bTheta + \quarter \bar{\zeta} \right) \bvart
		- \vart \left( \diff\Theta + \quarter \zeta \right)
	\right] \vart
	- 2 i \left( \diff\bvart \, \vart - \bvart \, \diff\vart \right) \right] ^{\da \a} \, .
\end{aligned}
\end{equation}
Here and in the further calculations, we have assigned the following canonical index positions:
\begin{alignat}{3}
&{\Theta_\a}^A \, , \bTheta_{A \da}  & \quad &\text{for variables conjugate to} \quad {Q_A}^\a \, , \, \Qb^{\da A} \, , \nn \\
&{\vart_A}^\a \, , \bvart^{\da A} & \quad &\text{for variables conjugate to} \quad {S_\a}^A \, , \, \Sb_{A \da} \, \,. \nn
\end{alignat}
The raising or lowering of a four-dimensional spinor index is always indicated explicitly. This notation allows us to write the index contractions in terms of matrix products as it is done above, see also appendix \ref{app:Spinor}. For our further calculations we note the $\mathbb{Z}_4$-decomposition of
$U= g^{-1} \diff g$ using the formulas given in appendix \ref{app:u224} for the graded components of the generators, 
\begin{align}
A^{(0)} &= - \frac{r_{\a \da} - y^2 \, \kappa_{\a \da}}{4y} 
	\left( P^{\da \a} - K^{\da \a} \right) + \la \indices{_\a ^\b} \, M \indices{_\b ^\a} \nn \\
	& \quad + \widebar{\la}  \indices{^\db _\da} \, \widebar{M} \indices{ ^\da _\db} 
	+  \left(\Lambda \indices{_A ^B} R \indices{^A _B}  + M^{-1} \diff M   \right)^{(0)} \, ,  \\
A^{(2)} &= - \frac{r_{\a \da} + y^2 \,  \kappa_{\a \da}}{4y} 
	\left( P^{\da \a} + K^{\da \a} \right)  + \frac{\s}{y} D  + \gamma \, C  \nn \\
	& \quad + \left(\Lambda \indices{_A ^B} R \indices{^A _B}  + M^{-1} \diff M   \right)^{(2)} 
	\, , \label{A2} \\ 
A^{(1)+(3)} &=  \frac{1}{\sqrt{y}} \left( \varepsilon _\a {}^A \, Q_A {} ^\a 
	+ \widebar{\varepsilon}_{A \da} \,  \Qb^{\da A}  
	+ y \left(  \chi_A {}^\a \, S_\a {} ^A 
	+ \widebar{\chi}^{\da A} \, \Sb_{A \da} \right) \right) , \\
A^{(1)-(3)} &= \frac{i}{\sqrt{y}} \Big( 
	\varepsilon_\a {} ^A  \, K_{A B}  \, S^{\a B} 
	-  \widebar{\varepsilon}_{A \da} \,  K^{A B} \, \Sb^{\, \da} {} _B  \nn \\ 
	& \quad + y \left(  \chi _A {} ^\a \, K^{A B} \, Q _{B \a} 
	- \widebar{\chi}^{A \da} \, K_{A B} \, \Qb^B {} _\da \,  \right) \Big)  .
\end{align}
The matrix $K = \big( K^{A B} \big) = \big( K_{A B} \big)$ appears explicitly in the $\mathbb{Z}_4$ decomposition and we recall that it is given by
\begin{align}
K = \begin{pmatrix}
0 & -1 & 0 & 0 \\ 1 & 0 & 0 & 0 \\ 0 & 0 & 0 & -1 \\ 0 & 0 & 1 & 0
\end{pmatrix} \,.
\label{Kdef}
\end{align}

\section{The Superconformal Boundary}

In the case of the Maldacena--Wilson loop, the minimal surface is required to end on the respective curve on the conformal boundary of $\AdS_5$. We have seen in chapter \ref{chap:Basics} that the discussion of the conformal boundary is simplified in Poincar{\'e} coordinates, where it is simply given by, where we simply approach the Minkowski space at $y=0$ in the boundary limit. In order to discuss the boundary conditions for the minimal surfaces in superspace corresponding to the Wilson loop in superspace, we employ the generalized Poincar{\'e} coordinates for the supercoset introduced above. 

There are two important aspects that need to be considered in the construction of the conformal boundary of our superspace. The geometric relation between bulk and boundary space requires that the super-isometries of the bulk space should reduce to superconformal transformations on the conformal boundary space when taking the boundary limit. Moreover, we should impose the right number of boundary conditions on the bulk coordinates in order to determine a minimal surface. For the bosonic coordinates, the equations of motion are second order differential equations and we impose a boundary condition for all bosonic coordinates. For the fermionic coordinates, the equations of motion are first order differential equations and we will thus only impose boundary conditions on half of the fermionic coordinates.

Let us first consider $\AdS_5$ once more. As we have seen in chapter \ref{chap:MinSurf}, the coset parametrization
\begin{align}
g_1(X,y) = e^{X \cdot P} \, y^D
\end{align}
provides Poincar{\'e} coordinates on $\AdS_5 \simeq \grp{SO}(2,4) / \grp{SO}(1,4)$. In this case, the Lie algebra $\alg{so}(2,4)$ is split as
\begin{align}
\alg{so}(2,4) &= \alg{h} \oplus \alg{m} \, , &
\alg{h} &= \mathrm{span} \left \lbrace P_\mu - K_\mu , M_{\mu \nu} \right \rbrace \, , &
\alg{m} &= \mathrm{span} \left \lbrace P_\mu + K_\mu , D \right \rbrace \,.
\end{align}
We can similarly write the boundary Minkowski space as the coset space $\grp{SO}(2,4)/\grp{H}$, where H is the subgroup of $\grp{SO}(2,4)$ generated by the algebra 
$\alg{h} = \mathrm{span} \lbrace K_\mu , M_{\mu \nu}, D \rbrace$. Note that in this way, we are not employing the coset description of a symmetric space for the boundary space, and the splitting of the Lie algebra,
\begin{align}
\mathfrak{so}(2,4) = \mathrm{span} \lbrace K_\mu , M_{\mu \nu}, D \rbrace \oplus \mathrm{span} \lbrace P_\mu \rbrace \, ,
\end{align}
does not satisfy the algebra relations \eqref{eqn:algebra-relations}. 
The coset space is naturally parametrized by
\begin{align}
g_2(x) = e^{x \cdot P} \, .
\end{align}
For either coset, an isometry or conformal transformation is obtained from the left-multiplication with a generic group element $t = e^{\bm{t}}$,
\begin{align}
t \cdot g_1(X,y) &= g_1(X^\prime , y ^\prime) \cdot h(X,y) \, , &   
t \cdot g_2(x) &= g_2(x^\prime) \cdot h(x) \, .
\end{align} 
Here, $h = e^{\bm{h}}$ are compensating gauge transformations taking values in the respective gauge groups. Infinitesimally, we then have
\begin{equation}
\begin{aligned}
\delta g_1(X,y) &= \bm{t} \cdot g_1 - g_1 \cdot \bm{h} (X,y) 
	= \partial_\mu g_1 \, \delta_{\bm{t}} X^\mu 
		+ \partial_y g_1 \, \delta_{\bm{t}} y \, ,  \\
\delta g_2(x) &= \bm{t} \cdot g_2 - g_2 \cdot \bm{h} (x) 
	= \partial_\mu g_2 \, \delta_{\bm{t}} x^\mu  \, . 
\end{aligned}
\end{equation}
Consider now for example a transformation parametrized by $\bm{t}= \varepsilon^\mu K_\mu$. One computes easily that the coordinates of the two coset spaces transform as
\begin{equation}
\begin{aligned}
\delta _{\varepsilon \cdot K } X^\mu &= X^2 \, \varepsilon^\mu 
	- 2 \left( \varepsilon \cdot X \right) X^\mu + y^2 \, \varepsilon^\mu \, , &
	\delta _{\varepsilon \cdot K }  y &= -2y \left( \varepsilon \cdot X \right) \, ,  \\
\delta _{\varepsilon \cdot K }  x^\mu &= x^2 \, \varepsilon^\mu 
	- 2 \left( \varepsilon \cdot x \right) x^\mu \,.
\end{aligned}
\end{equation}
We recognize that the transformations indeed agree in the boundary limit $y \to 0$ if we identify the $X$ coordinates. Before turning to the superspace, we reformulate the criterion that the isometries of the bulk space reduce to conformal transformations on the boundary space for a more general situation.

Assume we have two coset spaces $\MM_1 = \grp{G} / \grp{H}_1$ and 
$\MM_2 = \grp{G} / \grp{H}_2$ with the same group G but different stability groups $\grp{H}_1$ and $\grp{H}_2$. Correspondingly, we have two decompositions of the Lie algebra $\mathfrak{g}$,
\begin{align}
\mathfrak{g} &= \mathfrak{h}_1 \oplus \mathfrak{m}_1 \, , & 
\mathfrak{g} &= \mathfrak{h}_2 \oplus \mathfrak{m}_2 \, .
\end{align}
We do not assume either of the two cosets to be a symmetric space and impose no further restrictions on the decomposition. Let us assume that the coset space $\MM_1$ represents the bulk space and the coset space $\MM_2$ the boundary space. Then we have 
$\mathrm{dim}(\grp{H}_1) < \mathrm{dim}(\grp{H}_2)$, but we will not assume that $\grp{H}_1$ is a subset of $\grp{H}_2$. Furthermore, we assume that we can find coset representatives 
$g_1(x^m, y^i)$ for $\MM_1$ and $g_2(x^m)$ for $\MM_2$, which satisfy the relation
\begin{align}
g_1(x, y) = g_2(x) \, h_2(x, y) \, ,
\label{eqn:ReprRelation}
\end{align}
where $h_2(x, y) \in H_2$. In particular, the Maurer--Cartan forms for the two parametrizations are related by
\begin{align}
U_1 = g_1^{-1} \diff g_1 = h_2 U_2 h_2 ^{-1} + h_2 ^{-1} \diff h_2 \,. 
\label{MC_Relation} 
\end{align}
Under the left-multiplication by some group element $t = e^{\bm{t}} \in \grp{G}$, the coset representatives transform as
\begin{align}
g(Z) \mapsto g(Z') = t g(Z) h(Z) \, , \qquad \delta g(Z) = \bm{t} g(Z) - g(Z) \bm{h}(Z) \,.
\label{gtransf}
\end{align}
In order to extract the coordinate variations $\delta Z^M$, we rewrite the above transformation as
\begin{align}
\delta Z^M U_M = \delta Z^M g(Z)^{-1} \partial_M g(Z) = g(Z)^{-1} \left (
	\bm{t} g(Z) - g(Z) \bm{h}(Z)\right) .
\label{eqn:mcCosetTrans}	
\end{align}
We are interested in the difference of the variations of the coordinates $x^m$ and begin by evaluating the above variation for the first coset space. For the coordinates of the bulk coset space, $Z^M = (x^m, y^i)$, we find using the relation \eqref{MC_Relation}
\begin{align}
\delta_1 Z^M U_{1, M} =
	\delta_1 x^m \, h_2^{-1} U_{2, m} h_2
	+ \delta_1 x^m \, h_2^{-1} \partial_m h_2
	+ \delta_1 y^i \, h_2^{-1} \partial_i h_2 \, ,
\end{align}
which we rearrange as
\begin{align}
\delta_1 x^m U_{2,m} =
	\delta_1 Z^M \, h_2 U_{1,M} h_2^{-1}
	- \delta_1 x^m \left ( \partial_m h_2 \right ) h_2^{-1}
	- \delta_1 y^i \left ( \partial_i h_2 \right ) h_2^{-1} \;.
\end{align}
We can further rewrite the above relation by plugging in \eqref{eqn:mcCosetTrans}, 
\begin{align}
\delta_1 x^m U_{2,m} &=
	h_2 \, g_1^{-1} \left (
		\bm{t} g_1 - g_1 \bm{h}_1
	\right)h_2^{-1}
	- \delta_1 x^m \left ( \partial_m h_2 \right ) h_2^{-1}
	- \delta_1 y^i \left ( \partial_i h_2 \right ) h_2^{-1} \nn \\
&= g_2^{-1} \bm{t} g_2 + h_2 \bm{h}_1 h_2^{-1}
	- \delta_1 x^m \left ( \partial_m h_2 \right ) h_2^{-1}
	- \delta_1 y^i \left ( \partial_i h_2 \right ) h_2^{-1} \, .
\end{align}
This form is suitable to compare with the variation obtained for the other coset. Evaluating the relation \eqref{eqn:mcCosetTrans} for the coset space $\MM_2$ gives
\begin{align}
\delta_2 x^m U_{2,m} = g_2^{-1} \bm{t} g_2 - \bm{h}_2 \, . 
\end{align}
For the difference between the variations of the coordinates we thus have
\begin{align}
\Delta \left (\delta x^m \right ) U_{2,m}
&= \left( \delta_1 x^m - \delta_2 x^m \right) U_{2,m} \nn \\
&= h_2 \bm{h}_1 h_2^{-1} + \bm{h}_2
	- \delta_1 x^m \left ( \partial_m h_2 \right ) h_2^{-1}
	- \delta_1 y^i \left ( \partial_i h_2 \right ) h_2^{-1} \;.
\end{align}
We can read off $\Delta \left (\delta x^m \right )$ by projecting the above relation on 
$\alg{m}_2$. Only the first term on the right-hand side of the above equation contains contributions in $\alg{m}_2$ and hence we have
\begin{align}
\Delta \left (\delta x^m \right ) U_{2,m} \big \vert _{\alg{m}_2} 
	= h_2 \bm{h}_1 h_2^{-1} \big \vert _{\alg{m}_2} \, .
\end{align}
We thus note the following criterion for the condition that the variations obtained from the two coset constructions agree in the boundary limit $y\to 0$:
\begin{align}
\Delta \left (\delta x^m \right ) 
	\xrightarrow{\scriptstyle y \to 0} 0 
	\qquad \Leftrightarrow \qquad 
	h_2 \bm{h}_1 h_2^{-1} \big \vert _{\alg{m}_2} 
	\xrightarrow{\scriptstyle y \to 0}  0 \,. 
\label{crit1}
\end{align}
To understand this formulation of the boundary criterion better, let us consider $\AdS_5$ once more. The Maurer--Cartan form for the bulk coset space is given by 
\begin{align}
U_1 = g_1 ^{-1} \diff g_1 = \frac{\diff X^\mu}{y} \, P_\mu + \frac{\diff y}{y} \, D \,.
\end{align}
For the generator $\bm{t} = \varepsilon \cdot K$, we thus have the variation
\begin{align}
\delta _{\varepsilon \cdot K} Z^M \, U_{1, M} &= 
	\frac{1}{y} \left( \delta_{\varepsilon \cdot K}  X^\mu \, P_\mu 
	+ \delta_{\varepsilon \cdot K} y \, D \right) 
	= y^{-D} e^{- X \cdot P} \left ( \varepsilon \cdot K \right ) e^{X \cdot P} y^D 
	- \bm{h}_{1}  \\
&= \frac{2 \left( \varepsilon \cdot X \right) X^\mu 
	- X^2 \, \varepsilon^\mu}{y} \, P_\mu 
	+ 2 \left( \varepsilon \cdot X \right) D + 2 \, \varepsilon^\mu x^\nu M_{\mu \nu} + y\,  \varepsilon^\mu K_\mu - \bm{h}_{1} \, . \nn
\end{align}
We read off that
\begin{align}
\bm{h}_{1} = 2 \, \varepsilon^\mu x^\nu M_{\mu \nu} + 
	y \, \varepsilon^\mu \left(K_\mu -P_\mu \right)
\end{align}
in order to cancel the contributions proportional to $K_\mu$ and $M_{\mu \nu}$. For the situation at hand, we have $h_1 = y^D$ and $\alg{m}_2 = \mathrm{span}\lbrace P_\mu \rbrace$ and hence we have
\begin{align}
h_2 \bm{h}_1 h_2^{-1} \big \vert _{\alg{m}_2} = 
	y^D \left(2 \, \varepsilon^\mu x^\nu M_{\mu \nu} 
	- y \, \varepsilon^\mu \left(K_\mu -P_\mu \right) \right) 
	y^{-D} \big \vert _{\alg{m}_2} = y^2 \varepsilon^\mu P_\mu \, ,
\end{align}
which indeed vanishes in the boundary limit $y \to 0$.

With these preparations, we turn to the discussion of the conformal boundary of the supercoset space 
\begin{align*}
\frac{\grp{PSU}(2,2\vert 4)}{\grp{SO}(4,1) \times \grp{SO}(5)} \, .
\end{align*}
As a candidate for the conformal boundary, we consider the coset $\grp{PSU}(2,2 \vert 4)/\grp{H}_2$, where $\grp{H}_2$ is the subgroup generated by the subalgebra
\begin{align}
\mathfrak{h}_2 = \mathrm{span} \left \lbrace M_{\mu \nu}, D ,  K_\mu , S_\a {} ^A , \Sb_{A \da} \right \rbrace \oplus \mathfrak{so}(5) \, .
\end{align}
A suitable coset representative is given by \cite{Ooguri:2000ps}, 
\begin{align}
g_2(x,\theta,N) = e^{ x \cdot P } \, 
	e^{\theta _\a {} ^A \, Q_A {}^\a + \btheta_{A \da} \, \Qb^{\da A} } \, M(N) \,.
\label{boundarycoset}
\end{align}
The boundary superspace has only half as many fermionic degrees of freedom as the bulk space, which is due to the fermionic part of the superstring equations of motion being first order differential equations. They thus require less boundary conditions in order to determine a minimal surface solution. We will see explicitly in section \ref{sec:SuperExp} that the respective boundary conditions determine the minimal surface in the bulk space.

Following reference \cite{Ooguri:2000ps}, we now apply the criterion \eqn{crit1} to show that the bulk isometries reduce to superconformal transformations on the boundary space. The relation between the coset representatives of bulk and boundary space is given by $g=g_2 h_2$, where
\begin{align}
h_2 = M(N)^{-1} \, e^{\vart _A {} ^\a \, S_\a {} ^A + \bvart^{\da A} \, \Sb_{A \da} } 
	\, M(N) \, y^D =  e^{ (m^{-1} \vart ) _A {} ^\a \, S_\a {} ^A 
	+ ( \bvart m ) ^{\da A} \, \Sb_{A \da} } \, y^D \,.
\end{align}
Consider now an isometry of the bulk space parametrized by $\bm{t} \in \mathfrak{psu}(2,2 \vert 4)$. The coset representative transforms according to \eqn{gtransf},
\begin{align}
g^{-1} \, \delta g &= g^{-1} \bm{t}\,  g - \bm{h}_1 \, , & 
\bm{h}_1 \in \mathfrak{h}_1 = \mathrm{span} \lbrace M_{\mu \nu} , P_\mu - K_\mu \rbrace 
	\oplus \mathfrak{so}(5) \, ,
\end{align}
and we want to show that $h_2 \bm{h}_1 h_2^{-1} \big \vert _{\mathfrak{m}_2} \xrightarrow{\scriptstyle y \to 0}  0$. For this purpose it is not necessary to actually compute $\bm{h}_1$. Rather, decompose $\bm{h}_1$ as
\begin{align}
\bm{h}_1 = \bm{h}_1 \big \vert _{\mathfrak{h}_2} + \bm{h}_1 ^\prime \, 
	\quad \Rightarrow \quad 
	h_2 \, \bm{h}_1 \, h_2^{-1} \big \vert _{\mathfrak{m}_2} 
	= h_2 \, \bm{h}_1 ^\prime \, h_2^{-1} \big \vert _{\mathfrak{m}_2} .
\end{align}
The comparison of $\alg{h}_1$ and $\alg{h}_2$ shows that $\bm{h}_1 ^\prime$ is proportional to $P_\mu$, and we only have to determine the $y$-dependence, since
\begin{align}
\bm{h}_1 ^\prime = y^\omega c^\mu P_\mu \, 
	\quad \Rightarrow \quad  
	h_2 \, \bm{h}_1 ^\prime \, h_2^{-1} = \mathcal{O }\big( y^{\omega +1} \big) \,.
\end{align}
It is thus sufficient to show that for 
\begin{align*}
\bm{h}_1 = y^\omega c^\mu \left( P_\mu - K_\mu \right)+ \ldots \, , 
\end{align*}
we have $\omega = 1$. In order to see this, we note that $g = g_+ \, g_-$ with
\begin{align}
g_+ &= e^{ X \cdot P } \, e^{\Theta _\a {} ^A \, Q_A {}^\a 
		+ \bTheta_{A \da} \, \Qb^{\da A} } \, , & 
	g_- &= \, e^{\vart _A {} ^\a \, S_\a {} ^A 
		+ \bvart^{\da A} \, \Sb_{A \da} } \, M(N) \, y^D \, ,
\end{align}
where $g_+$ only contains generators with positive weights, while $g_-$ only contains generators with non-positive weights. The group elements $g_+$ can be considered as a set of coset representatives for a superspace obtained by factoring out the subgroup $H_+$ with Lie algebra
\begin{align}
\mathfrak{h}_+ = \mathrm{span} \left \lbrace M_{\mu \nu}, D ,  K_\mu , S_\a {} ^A , \Sb_{A \da} \right \rbrace \oplus \mathfrak{su}(4) \,.
\end{align}
In particular, $\mathfrak{h}_+$ only contains generators of non-positive weights. The Maurer--Cartan form for the coset representative $g_+$ can be obtained from \eqn{eqn:MaurerCartanForm} by setting $\vart = 0$, $N$ constant and $y=1$. 
We compare the variation of $g$ with the variation of $g_+$, for which we note
$g_+ ^{-1} \bm{t} g_+ = g_+ ^{-1} \delta g_+ + \bm{h}_+$, where $\bm{h}_+ \in \mathfrak{h}_+$.
For the variation of $g$, we thus have
\begin{align}
\delta Z^M \, U_M = g^{-1} \, \delta g = g_- ^{-1} \left( g_+ ^{-1} \delta g_+ + \bm{h}_+ \right) g_- - \bm{h}_1
\end{align}
In the above formula $\bm{h}_1$ compensates for those terms of $g_- ^{-1} \left( g_+ ^{-1} \delta g_+ + \bm{h}_+ \right) g_-$ which may not be put into the form $\delta Z^M \, U_M$. In particular, there is no need to compensate a term proportional to $P_\mu$ as these terms are the same in $g^{-1} \, \delta g$ and $g_- ^{-1} \left( g_+ ^{-1} \delta g_+  \right) g_-$, which may be seen easily from the calculation of the Maurer--Cartan form shown in section \ref{sec:Coordinates}. It is thus clear that the term in $\bm{h}_1$ proportional to $(P_\mu - K_\mu)$ compensates for a term in $g_- ^{-1} \left( g_+ ^{-1} \delta g_+ + \bm{h}_+ \right) g_-$, which is proportional to $K_\mu$. This term, however, is of order $\mathcal{O}(y)$, since $K_\mu$ has weight $-1$ and we are doing the conjugation with $y^D$ last. We thus see that
\begin{align*}
\bm{h}_1 = y \, c^\mu \left( P_\mu - K_\mu \right)+ \ldots \, ,
\end{align*}
as we have claimed above. 

This shows that the space parametrized by \eqn{boundarycoset} may indeed be viewed as the superconformal boundary of the space $\grp{PSU}(2,2\vert4) / \left( \grp{SO}(4,1) \times \grp{SO}(5) \right)$. Correspondingly, we impose the following set of boundary conditions on the minimal surface that describes the superspace Wilson loop at strong coupling:
\begin{equation}
\begin{alignedat}{3}
X^\mu (\tau = 0 , \s) &= x^\mu (\s) \, , & \qquad \quad 
y(0, \s) &= 0 \, , & \qquad \quad 
N^I(0, \s) &= n^I(\s) \, ,   \\
\Theta _\a {} ^A (0, \s ) &= \theta _\a {} ^A (\s) \, , & \qquad \quad
\bTheta_{A \da} (0, \s ) &= \btheta _{A \da} (\s) \, .
\end{alignedat}
\end{equation}

\section{The Bulk Expansion}
\label{sec:SuperExp}

We have seen in chapter \ref{chap:MinSurf} that in order to derive the symmetries of the minimal surfaces from the classical integrability of the underlying model, we have to determine the first few coefficients of the expansion of the minimal surface around the boundary. As before, we will fix them from the equations of motion, the Virasoro constraints and the functional derivatives of the area of the minimal surfaces.

For the superstring however, half of the fermionic components decouple from the equations of motion and we only get a unique solution by fixing a kappa symmetry gauge. In our case, it is convenient to set half of the coefficients of the fermionic part of $U=g^{-1} \diff g$ to zero. Concretely, we fix the conditions
\begin{align}
\varepsilon _\a {} ^2 &= \varepsilon _\a {} ^4  = 0 \, , &
\widebar{\varepsilon}_{2 \da} &= \widebar{\varepsilon}_{4 \da} = 0 \, , &
\chi \indices{ _1 ^\a} &= \chi \indices{ _3 ^\a} = 0 \, , & 
\widebar{\chi}^{\da 1} &= \widebar{\chi}^{\da 3} = 0 \,.
\label{kappagauge}
\end{align}
Written out for the supermatrix $A^{(1)+(3)}$, the above gauge condition takes the form
\begin{align}
A^{(1)+(3)} = \left( \begin{array} {cccc|cccc}
0 & 0 & 0 & 0 & \bullet & 0 & \bullet & 0 \\
0 & 0 & 0 & 0 & \bullet & 0 & \bullet & 0 \\
0 & 0 & 0 & 0 & 0 & \bullet & 0 & \bullet \\
0 & 0 & 0 & 0 & 0 & \bullet & 0 & \bullet \\ \hline
0 & 0 & \bullet & \bullet & 0 & 0 & 0 & 0 \\
\bullet & \bullet & 0 & 0 & 0 & 0 & 0 & 0 \\
0 & 0 & \bullet & \bullet & 0 & 0 & 0 & 0 \\
\bullet & \bullet & 0 & 0 & 0 & 0 & 0 & 0 \\
\end{array} \right) \, .
\end{align}

It is typically difficult to verify that a certain kappa symmetry gauge can be reached by kappa symmetry transformations, since the transformation involves a solution of the equations of motion and due to the complications arising in multiplying two exponentials. In order to check whether it is at least plausible that the above kappa symmetry gauge can be reached, we consider the simple case of a straight-line boundary curve and work to linear order in Gra{\ss}mann variables. 

We recall that the kappa symmetry transformations are given by
\begin{align}
g \; \; & \mapsto \; \; g \cdot \exp{\kappa} \, , & 
U \; \; & \mapsto \; \; U + \diff \kappa + \left[ U , \kappa \right] \, , 
\end{align}
where we have left out the gauge transformations that ensure staying within the given class of coset representatives. The kappa symmetry parameter $\kappa = \kappa^{(1)+(3)}$ is given by equation \eqref{Kappa_Transf}, 
\begin{align}
\kappa^{(1)} &= A_{i , -} ^{(2)} \, \mathcal{K} _+ ^{(1),i} 
	+ \mathcal{K} _+ ^{(1),i} \, A_{i , -} ^{(2)} \, , &
\kappa^{(3)} & = A_{i , +} ^{(2)} \, \mathcal{K}_- ^{(3),i} 
	+ \mathcal{K} _- ^{(3),i} \, A_{i , +} ^{(2)} \, ,  
\end{align}
where we have used the projections $V_{\pm} ^i = P_{\pm} ^{i j} \, V_j 
= \half \left( \gamma^{ij} \pm i \, \eps^{i j} \right) V_j$. In conformal gauge, we can write this more conveniently as
\begin{align}
\kappa = \kappa ^{(1)+(3)} = 
	i \left \lbrace A_\s ^{(2)}  , \mathcal{K} ^{(1)+(3)} \right \rbrace 
	+ \left \lbrace A_\tau ^{(2)}  , \mathcal{K} ^{(1)-(3)} \right \rbrace \, .
\label{kappaspec}
\end{align}
Here, $\mathcal{K}$ is again a generic odd supermatrix. For the straight-line boundary curve, we have the boundary conditions
\begin{align}
X(0,\s) &= (0 , 0 , 0 , \s) \, , & 
y(0,\s) &= 0  \, , & 
N^I(0,\s) &= n^I \, ,
\end{align}
where $n^I$ is constant. Since we are working up to linear order in Gra{\ss}mann variables, the fermionic parts of the minimal surface solution will not be relevant for us. For the bosonic part, we note that the equations of motion are solved by
\begin{align}
X(\tau, \s) &= (0 , 0 , 0 , \s) \, , & 
y(\tau,\s) &= \tau \, , 
\label{bosonicsol}
\end{align}
and correspondingly, we obtain
\begin{align}
A_\s^{(2)} &= - \frac{\dot{X}_{\a \da} }{4 \tau} \left( P^{\da \a} +  K^{\da \a} \right) \, , & 
A_\tau^{(2)} &= \frac{1}{\tau} D \,.
\end{align}
We can then calculate the kappa symmetry parameter $\kappa$ from equation \eqref{kappaspec}. We parametrize the supermatrix $\mathcal{K}$ as 
\begin{align}
\mathcal{K} = a_\a {} ^A Q_A {} ^\a - \widebar{a}_{A \da} \Qb^{\da A} 
	+ b_A {} ^\a S_\a {} ^A - \widebar{b}^{\da A} \Sb_{A \da} \, . 
\end{align}
This signs above are chosen in such a way that demanding 
$a_\a {} ^A = (\widebar{a}_{A \da}) ^\ast$ and similarly for $b$ would ensure 
$\kappa \in \alg{su}(2,2 \vert 4)$ on a worldsheet with Minkowskian signature. In the Euclidean case, the reality constraint cannot be realized. 
The anti-commutators of supermatrices appearing in \eqn{kappaspec} can be related to commutators due to the particular form of the supermatrix generators given in appendix~\ref{app:u224}. We have the following identities for $\mathcal{P} \in \left \lbrace P_\mu , K_\mu, D \right \rbrace$:
\begin{align*}
\left \lbrace \mathcal{P} \, , \, Q \right \rbrace 
	&= \left[ \mathcal{P} \, , \, Q \right] \, , &
\left \lbrace \mathcal{P} \, , \, S \right \rbrace 
	&= - \left[ \mathcal{P} \, , \, S \right] \, , \\
\left \lbrace \mathcal{P} \, , \, \Sb \right \rbrace 
	&= \left[ \mathcal{P} \, , \, \Sb \right]\, , &
\left \lbrace \mathcal{P} \, , \, \Qb \right \rbrace 
	&= - \left[ \mathcal{P} \, , \, \Qb \right] \, .
\end{align*}
For the parameter $\kappa$, we then find
\begin{align}
\kappa &= \frac{1}{2 \tau} \Big[ 
	\left( i \dx_{\a \da} \,  \widebar{b}^{\da A} + K^{AB} \, b_{B \a} \right) Q _A {} ^\a 
	+ \left(-i b  _A {} ^\a \, \dx_{\a \da} -K_{AB} \, \widebar{b}  _\da {}^B \right) 
	\Qb^{\da A} \nn \\
& \qquad + \left( i  \widebar{a}_{A \da} \, \dx^{\da \a} + K_{AB} \, a^{\a B} \right) S _\a {} ^A 
	+ \left(-i \dx ^{\da \a} \,  a_\a {} ^A - K^{AB} \, \widebar{a}^\da {} _B  \right) 
	\Sb_{A \da} \Big] \nn \\
&= c_\a {} ^A Q_A {} ^\a - \widebar{c}_{A \da} \,  \Qb^{\da A} 
	+ d_A {} ^\a S_\a {} ^A - \widebar{d}^{\, \da A} \Sb_{A \da} \, .
\end{align}
The parameters $c,\widebar{c},d, \widebar{d}$ are related among each other by 
\begin{align}
\bar{c}_{A \da} &= i \, \eps_{\da \db} \, \dx ^{\db \da} \,  c_\a {} ^B  \, K_{BA} \, , & 
\bar{d}^{\, \da A} &= -i   \, K^{AB} \, d_B {} ^\a \, \dx_{\a \db} \, \eps^{\db \da} \,.
\end{align}
This means that fixing $c_1$ and $\bar{c_1}$ determines $c_2$ and $\bar{c}_2$, fixing $c_3$ and $\bar{c_3}$ determines $c_4$ and $\bar{c}_4$ and likewise for $d, \bar{d}$. In particular, we observe that the kappa symmetry transformation has half the degrees of freedom of a generic fermionic element, as expected for this supercoset. Moreover, we observe that we cannot enforce $\kappa \in \mathfrak{su}(2,2 \vert 4 )$ by constraining $\mathcal{K}$. One may e.g.\ enforce $(c_\a {} ^1)^\ast = \bar{c}_{1 \da}$ but that leads to $(c_\a {} ^2)^\ast = - \bar{c}_{2 \da}$. 
We go on to calculate the fermionic part of the transformed Maurer--Cartan form $U^\prime$ from the relation
\begin{align}
A^\prime {}^{(1)+(3)} = A^{(1)+(3)} + \diff \kappa + \left[ A^{(0)+(2)} \, , \, \kappa \right] \, , 
\end{align}
noting that
\begin{align}
A_{\tau}^{(0)+(2)} &= \frac{1}{\tau} \, D \, , & 
A_\s^{(0)+(2)} &= - \frac{\dx_{\a \da}}{2\tau} \, P^{\a \da} \, . 
\end{align}
Then we find the following transformations for the parameters $\varepsilon$ and $\chi$:
\begin{equation}
\begin{alignedat}{2}
\chi ^\prime _\tau {} _A {} ^\a &= \chi _\tau {} _A {} ^\a 
	+ \left(\partial_\tau + \sfrac{1}{2\tau} \right) d_A {} ^\a \, ,& \qquad 
\widebar \chi^\prime _\tau {} ^{\da A}  &= \bar \chi _\tau {} ^{\da A} 
	+ \left(\partial_\tau + \sfrac{1}{2\tau} \right) \widebar{d} ^{\, \da A} \, , \\
\chi ^\prime _\s {} _A {} ^\a &= \chi _\s {} _A {} ^\a 
	+ \partial_\s \, d_A {} ^\a \, , & \qquad 
\bar \chi^\prime _\s {} ^{\da A}  &= \bar \chi _\s{} ^{\da A} 
	+ \partial_\s \, \bar{d} ^{\, \da A} \, , \\
\varepsilon^\prime _\tau {} _\a {} ^A &= \varepsilon _\tau {} _\a {} ^A 
	+ \left(\partial_\tau -\sfrac{1}{2\tau} \right) c_\a {} ^A \, ,& \qquad 
\bar \varepsilon^\prime _\tau {} _{A \da} &= \bar \varepsilon _\tau {} _{A \da}
	+ \left(\partial_\tau -\sfrac{1}{2\tau} \right) \bar{c} _{A \da} \, , \\
\varepsilon^\prime _\s {} _\a {} ^A &= \varepsilon _\s {} _\a {} ^A 
	+ \partial_\s \, c_\a {} ^A + \sfrac{i}{2\tau} 
	\, \dx_{\a \da} \, \bar{d}^{\, \da A} \, ,& \qquad
\bar \varepsilon^\prime _\s {} _{A \da} &= \bar \varepsilon _\s {} _{A \da} 
	+ \partial_\s \, \bar{c} _{A \da}  - \sfrac{i}{2\tau} \, d_A {} ^\a \, \dx_{\a \da} \, . 
\end{alignedat}
\end{equation}
Due to the form of the relations between the parameters $c , \bar{c}, d , \bar{d}$ explained above, it is thus clear that we can indeed reach the kappa symmetry gauge \eqref{kappagauge}, 
\begin{align}
\varepsilon _\a {} ^2 &= \varepsilon _\a {} ^4  = 0 \, , \qquad \bar{\varepsilon}_{2 \da} = \bar{\varepsilon}_{4 \da} = 0 \, , \qquad \chi ^\a _1 = \chi^\a _3 = 0 \, , \qquad \bar{\chi}^{1 \da} = \bar{\chi}^{3 \da} = 0 \,.
\end{align}
While we have motivated that the above kappa symmetry gauge can be reached, we have seen that the fermionic coefficients, which are not set zero, no longer obey the reality constraint 
$( \varepsilon _\a {} ^A ) ^ \ast = \bar{\varepsilon}_{A \da}$.
This finding is related to working with the Wick rotated superstring action for a Euclidean worldsheet, which implies the appearance of a factor $i$ in the kappa symmetry variations. This subtlety does, however, not affect our further calculation, for which the reality constraint is irrelevant.

\subsection{Equations of Motion}

We begin by studying the equations of motion \eqref{EOM_SUSY_C}. We recall that in conformal gauge they are given by
\begin{align}
\delta^{ij} \left( \partial_i \, \A{2}_j + \left[ \A{0} _i , \A{2}_j \right] \right) 
	- \ihalf \, \epsilon^{ij} \left[ \AAm _i , \AAp_j \right] &= \alpha(\tau, \s) \, C \, , 
	\label{boseom}\\
\delta^{ij} \left[ \A{2}_i , \AAp_j \right] 
	+ i \, \epsilon^{ij} \left[ \A{2} _i , \AAm_j \right] &= 0 \, .	
	\label{fermeom}
\end{align}
Since the Maurer--Cartan form is of order $\O(\tau^{-1})$ when approaching $y=0$, the equations of motion begin at the order $\O(\tau^{-2})$ and we consider them up to the next-to-leading order $\O(\tau^{-1})$. For this order, not all components of the Maurer--Cartan form \eqref{eqn:MaurerCartanForm} are relevant. The R-symmetry components, for example, only appear in the order $\O(\tau^0)$ of the bosonic equations of motion and we can hence neglect them here. This is plausible since we did not need to consider the equations of motion in the discussion of the minimal surfaces in $\mathrm{S}^5$ in section \ref{sec:S5}.  
The relevant components of the Maurer--Cartan form for our discussion of the equations of motion are given by
\begin{align*}
A^{(0)} &= - \frac{r_{\a \da}}{4y} 
	\left( P^{\da \a} - K^{\da \a} \right) + \la \indices{_\a ^\b} \, M \indices{_\b ^\a} 
	+ \widebar{\la}  \indices{^\db _\da} \, \widebar{M} \indices{ ^\da _\db}  \, ,  \\
A^{(2)} &= - \frac{r_{\a \da}}{4y} 
	\left( P^{\da \a} + K^{\da \a} \right)  + \frac{\s}{y} D \, , \\ 
A^{(1)+(3)} &=  \frac{1}{\sqrt{y}} \left( \varepsilon _\a {}^A \, Q_A {} ^\a 
	+ \widebar{\varepsilon}_{A \da} \,  \Qb^{\da A}  
	+ y \left(  \chi_A {}^\a \, S_\a {} ^A 
	+ \widebar{\chi}^{\da A} \, \Sb_{A \da} \right) \right) \, ,\\
A^{(1)-(3)} &= \frac{i}{\sqrt{y}} \Big( 
	\varepsilon_\a {} ^A  \, K_{A B}  \, S^{\a B} 
	-  \widebar{\varepsilon}_{A \da} \,  K^{A B} \, \Sb^{\, \da} {} _B  
	+ y \left(  \chi _A {} ^\a \, K^{A B} \, Q _{B \a} 
	- \widebar{\chi}^{A \da} \, K_{A B} \, \Qb^B {} _\da \,  \right) \Big) \, .
\end{align*}
Here, we have also left out the $C$-part of the projection $\A{2}$, since the $C$-part of the equations of motion provides no additional information. We recall the coefficients 
\begin{align*}
\zeta _\a {} ^A &= i \, r_{\a \da} \, \widebar{\vartheta}^{\da A} \, , &
\varepsilon _\a {}^A &= \left[ \left( \diff \Theta + \zeta \right) m \right] _\a {} ^A \, , &
\la \indices{_\a ^\b} &= - i \, \left[ \left(2 \diff \Theta + \zeta \right) 
	\vart \right] \indices{_\a ^\b} \, ,
\\
\widebar{\zeta}_{A \da} &= - i \, \vartheta _A {} ^\a \, r_{\a \da} \, , &
\widebar{\varepsilon}_{A \da} &= \left[ m^{-1} \left( \diff \bTheta 
	+ \widebar{\zeta} \, \right) \right] _{A \da} \, , &
\widebar{\la} \indices{^\db _\da} &= i \, \left[ \bvart 
	\left(2 \diff \bTheta +  \bar{\zeta} \right) \right] \indices{^\db _\da} \, , 		
\end{align*}
as well as
\begin{align*}
r_{\a \da} &=  \diff X_{\a \da} + 2 i \left( \diff \Theta  \, \bTheta  
	- \Theta \, \diff \bTheta \right) {} _{\a \da} \, , \\
\sigma &= \diff y + 2 y\,  \tr \left( \diff \bTheta \, \bvart 
	+ \vart\, \diff \Theta \right) \, , \\	
\chi _A {}^\a &=
	\left[ m^{-1} \left[ \left(
			4 \vart \left(\diff\Theta + \thirrd \zeta \right)
			- 2 \left(\diff\bTheta + \thirrd \bar{\zeta} \right) \bvart
		\right) \vart
		+ \diff\vart
	\right] \right] _A {}^\a \, , \\
\widebar{\chi} ^{\da A} &=
	\left[ \left[
		\diff\bvart
		+ \bvart \left(
			4 \left(\diff\bTheta + \thirrd \bar{\zeta} \right) \bvart
			- 2 \vart \left(\diff\Theta + \thirrd \zeta \right)
		\right)
	\right] m \right] ^{\da A} \, .		
\end{align*}

We begin by considering the leading order $\mathcal{O}(\tau^{-2})$ of the bosonic equation \eqn{boseom}, from which we find that
\begin{align*}
\left( \left( \partial_\tau y \right) r_{\tau \, \mu} \left( P^\mu + K^\mu \right) + \left( \left( \partial_\tau y \right)^2 - r_{\tau}^2 - r_\s^2 \right) D \right) _{(0)} = 0 \, , \\
\Rightarrow r_{\tau \, (0)} ^\mu  = 0 \, , \qquad \qquad  (y_{(1)}) ^2 = \left( r_{\s \, (0)} \right) ^2 \,.
\end{align*}
Here, we are using the notation introduced in equation \eqref{def:expansion} for the coefficients of the $\tau$-expansion. We identify the leading term of $r_\s$ with the supermomentum
\begin{align}
\pi ^\mu = r_{\s \, (0)}^\mu = \dx ^\mu 
	+ i \tr \Big( \, \dbt \s ^\mu \theta - \bt \s ^\mu \dt \, \Big)
\end{align}
of a superparticle moving along the boundary curve. As $y$ should be positive and the boundary curve space-like we note that $y_{(1)} = \sqrt{\pi^2} = \lvert \pi \rvert$. We now restrict the parametrization of the curve to satisfy
\begin{align}
\lvert \pi \rvert = 1 \quad \Rightarrow \quad 
\pi \cdot \dot{\pi} = 0 \quad \Rightarrow \quad 
\dot{\pi}^2 + \pi \cdot \ddot{\pi} = 0 \,.
\end{align}
This is the super analogue of the arc-length condition we employed in section~\ref{sec:InfSymm}. Considering the fermionic equations \eqn{fermeom} at leading order in $\tau$ leads to the following set of equations:
\begin{align}
\varepsilon_{\tau \, (0)} {} _\a {} ^A + i \pi_{\a \da} \, \bar{\varepsilon} _{\tau \, (0)} {} _B{}^\da K^{BA} &= 0  \label{ferml1} \\
 \bar{\varepsilon}_{\tau \, (0) \, A \da} + i \, \varepsilon_{\tau \, (0)}{}^{\a B} \, \pi _{ \a \da} \, K_{BA} &=0 \label{ferml2}  \\
i \, \bar{\varepsilon}_{\s \, (0) A \da} \, \pi ^{\da \a } +  \varepsilon_{\s \, (0)} {}^{\a B} \, K_{BA} &=0  \label{ferml3} \\
i \, \pi ^{\da \a} {\varepsilon_{\s \, (0)} {} _\a } ^A  + \bar{\varepsilon}_{\s \, (0)} {} _B {}^\da \, K^{BA} &= 0 \label{ferml4}
\end{align}
The equations \eqn{ferml1} and \eqn{ferml2} as well as \eqn{ferml3} and \eqn{ferml4} are equivalent to each other as one would expect as they stem from the coefficients of $Q$ and $\Qb$ or $S$ and $\Sb$ in \eqn{fermeom}. Moreover, the real and imaginary parts of these equations are equivalent. Inserting for example \eqn{ferml1} into \eqn{ferml2}, we obtain
\begin{align}
\bar{\varepsilon}_{\tau \, (0) \, A \da} = \tilde{\kappa}^2 \, \bar{\varepsilon}_{\tau \, (0) \, C \db} \, \pi ^{\db \a} \, \pi_{\a \da} \, K^{C B} K_{B A} = \tilde{\kappa}^2 \bar{\varepsilon}_{\tau \, (0) \, A \da} 
\end{align}
Here, we have left the parameter $\tilde{\kappa}$ introduced in the string action \eqref{Superstring_Action} open in order to show that the equations are less constraining for $\tilde{\kappa}^2 = 1$ when one has kappa symmetry. In this case, a unique solution is only found due to the gauge-fixing condition \eqref{kappagauge} for the kappa symmetry. 
Consider for example equation \eqn{ferml1}. Due to the kappa symmetry gauge \eqn{kappagauge} either $\varepsilon_{\tau \, (0)} {} _\a {}^A$ or $\bar{\varepsilon} _{\tau \, (0)} {} _B {}^\da K^{BA}$ are vanishing for any given value of $A$. Proceeding in the same way for equation \eqn{ferml3} we conclude that
\begin{align}
\varepsilon_{\tau \, (0)} = \varepsilon_{\s \, (0)} &= 0 \, , & 
\widebar{\varepsilon}_{\tau \, (0)} =  \widebar{\varepsilon}_{\s \, (0)} &= 0 \,.
\end{align}
Note in particular that we do not employ reality conditions such as $\bar{\varepsilon}_{ A \da} = \left( \varepsilon _\a {} ^A \right)^\ast$ as they become problematic for the Wick rotated superstring action. Given that $r_{\tau \, (0)}=0$ and hence $\zeta_{\tau \, (0)}=0$, we conclude that
\begin{align}
\Theta_{(1)} &= 0 \, , &
\bTheta_{(1)} &= 0 \, , & 
X_{(1)} &= 0 \,. \label{X10}
\end{align}
Setting  $\varepsilon_{\s \, (0)} = 0 = \bar \varepsilon_{\s \, (0)}$ enforces that
\begin{align}
\vart _{(0)} {} _A {}^\a &=  i \,  \dbt _{A \da} \, \pi^{\da \a} \, , & 
\bvart _{(0)} {}^{\da A} &= - i \, \pi ^{\da \a} \, \dt _\a {}^A \,. \label{vart0}
\end{align}
It is thus indeed inconsistent to specify boundary conditions for the $\vart$ variables, as we have claimed before based on the order of the fermionic part of the equations of motion. 

We now turn to the next-to leading order in the bosonic equation \eqn{boseom}. Due to our above finding, we have $A^{(1) \pm (3)} = \mathcal{O}(\tau ^{1/2})$ and we can hence neglect the fermionic contributions also at the next-to leading order. Moreover, we note the following identities for the coefficients of the Cartan form:
\begin{align}
\sigma_i &= \partial _i y + \mathcal{O}(\tau^2) \, , & 
\la_{\tau \, \a} {} ^\b &= \mathcal{O}(\tau) \, , & 
\la_{\s \, \a} {} ^\a - \bar \la _\s {} ^\da {} _\da &= \mathcal{O}(\tau) \,.
\end{align}
Evaluating \eqn{boseom} then leads to the following equations:
\begin{align}
0&= \delta^{i j} \left(  \left( \partial_i y \right)  \left( \partial_j y \right)  
	- y \, \partial _i \partial _j y - r_i \cdot r_j \right) _{(1)} \, , \label{bosnl1} \\
0&= \delta ^{i j} \left( 2 \, \partial_i y \,   r_{j \, \a \da} 
	- y \, \partial_i \,  r_{j \, \a \da} 
	+ 2i y \, \la \indices{ _{i \,\a} ^\b} r_{j \, \b \da} 
	- 2i y \, r_{j \, \a \db} \, \bar{\la} \indices{_i ^\db _\da} 
	\right)_{(1)} \,. \label{bosnl2}
\end{align}
Making use of the relations \eqn{X10} and \eqn{vart0}, we find that
\begin{align*}
r_\tau ^2 &= \mathcal{O}(\tau^2) \, , & 
r_\s^2 &= \pi^2 + \mathcal{O}(\tau^2) \, , & 
\delta ^{i j} \Big( 2i y  \, \la \indices{ _{i \,\a} ^\b} 
	r_{j \, \b \da} - 2i y \, r_{j \, \a \db} \, 
	\bar{\la} \indices{_i ^\db _\da} \Big) &= \mathcal{O}(\tau^2) \, , 
\end{align*}
and we may thus conclude that
\begin{align}
y_{(2)} &= 0 \, , & \qquad 
r_{\tau \, (1)}^\mu &= \dot{\pi}^\mu 
	\quad \Rightarrow \quad 
	X_{(2)} ^\mu = \dot{\pi}^\mu + i \tr \big( 
	\bTheta_{(2)} \s^\mu \theta - \bt \s^\mu \Theta_{(2)} \big) \,.
\end{align}

Last, we consider the next-to leading order in the fermionic equation \eqn{fermeom}. We find the following conditions, this time leaving out equivalent conditions:
\begin{align}
 \, \widebar{\chi}_{\tau \, (0)} {} ^{A \da} 
 	- i \, \pi ^{\da \a} \varepsilon_{\s \, (1)} {} _\a {} ^A 
 	- \left( \widebar{\varepsilon}_{\s \, (1)} {} _B {} ^\da 
 		-i \, \pi ^{\da \a} \chi _{\tau \, (0) B \a} \right) K^{BA} &= 0 
 	\label{fermnl1} \, ,\\
\widebar{\varepsilon}_{\tau \, (1) A \da} 
	- i \, \chi_{\s \, (0)} {} _A {} ^\a \, \pi_{\a \da}  
	- \left( \widebar{\chi}_{\s \, (0)} {} _\da {}^B 
		- i \, \varepsilon_{\tau \, (1)} {}^{\a B} \, \pi_{\a \da} \right) K_{BA} &= 0 \,. 
\end{align}
Due to the kappa symmetry gauge \eqn{kappagauge}, we can decompose these equations into the conditions
\begin{align}
\widebar{\chi}_{\tau \, (0)} {} ^{A \da} 
	+ K^{A B} \, \widebar{\varepsilon}_{\s \, (1)} {} _B {}^\da &= 0 \, , &  
\varepsilon_{\s \, (1)} {} _\a {} ^A +   K^{A B} \,  \chi _{\tau \, (0) B \a} &=0 \, , \label{fermnl2} \\
\widebar{\varepsilon}_{\tau \, (1) A \da}  
	+  K_{AB} \, \widebar{\chi}_{\s \, (0)} {} _\da {} ^B &= 0 \, , & 
\chi_{\s \, (0)} {} _A {}^\a  
	+ K_{A B} \, \varepsilon_{\tau \, (1)} {}^{\a B} &= 0 \,.
\end{align}
The latter conditions allow to solve for $\Theta_{(2)}$ and $\bTheta_{(2)}$,
\begin{equation}
\begin{aligned}
\Theta_{(2)} {} _\a {} ^A &= - \dot{\pi}_{\a \da} \, \pi ^{\da \b} 
	\, \dt  _\b {} ^A + \widebar K^{\prime \, A B} 
	\left( 4 \, \big( \dbt \pi \dt \big)_B {} ^C  
	- i \,  \delta ^C _B \, \partial _\s \right) 
	\big( \pi _{\a \db} \, \dbt  _C {} ^\db \big) \label{theta2} \\
\bTheta _{(2) \, A \da} &= - \, \dbt _{A \db} \, \pi ^{\db \a} 
	\, \dot{\pi}_{\a \da} -  K^\prime _{A B} 
	\left( 4 \, \big( \dbt \pi \dt \big)_C {} ^B 
	+ i \,  \delta ^B _C \, \partial _\s \right) 
	\big( \dt ^{\b C} \, \pi_{\b \da} \big)
\end{aligned}
\end{equation}
Here, the matrices $\widebar K^{\prime  A B}$ and $K^\prime _{A B}$ are given by
\begin{align}
K^\prime _{A B} &= u _A {} ^C \, u _B {} ^D \, K_{CD} 
	= \left( u K u^T \right)_{AB} \, , &
\widebar K^{\prime  A B} = \big( \left(u^{-1} \right)^T K u^{-1} \big) ^{AB}
\end{align}
Due to equations \eqn{eqn:su4coset} and \eqn{Kdef}, this can be rewritten as
\begin{align}
\begin{aligned}
K^\prime _{AB} &= \left( n^5 \unit + n^6 \gamma^5 + i n^r \gamma ^r \right) 
	_A {} ^C \, K_{CB} \, , \\
\widebar K^{\prime  AB} &= K^{AC} 
	\left( n^5 \unit - n^6 \gamma^5 - i n^r \gamma ^r \right)_C {} ^B \,.
\end{aligned}
\end{align}
From equation \eqn{fermnl2} we find the following condition for $\vart_{(1)}$:
\begin{align}
\vart_{(1) \, A \a} &= i \, K^\prime _{AB} \, \pi_{\a \da} \, 
	\bvart _{(1)} {} ^{\da B} \, , & 
\bvart _{(1)} {} ^{\da A} &= -i \, \widebar{K}^{\prime A B} \, 
	\pi^{\da \a} \, \vart_{(1) \, B \a} \,.
\end{align}
The above results are lengthy and it can be cumbersome to use them in the calculations that follow. For many purposes however, it suffices to note that they can be written in the form
\begin{align}
A_\tau ^{(1)-(3)} &= -i \, A_\s ^{(1)+(3)} + \mathcal{O}( \tau^{3/2} ) \, , & 
A_\s ^{(1)-(3)} = i \, A_\tau ^{(1)+(3)} + \mathcal{O}( \tau^{3/2} ) \,. \label{fermeqnA}
\end{align}

\subsection{Virasoro constraints}

The coefficient $y_{(3)}$ can be determined from the Virasoro constraints \eqref{Virasoro_SUSY}, specifically from the condition
\begin{align}
\str \big( \A{2}_\tau \, \A{2} _\tau \big) - \str \big( \A{2}_\s \, \A{2} _\s \big) = 0 \, .
\label{Vir1} 	
\end{align}
We recall that $A^{(2)}$ is given by \eqn{A2},
\begin{align*}
A_i ^{(2)} &=  \frac{r_i ^\mu  + y^2 \,  \kappa_i ^\mu }{2y} \left( P_\mu + K_\mu \right)  
	+ \frac{\s_i}{y} D  + \gamma_i \, C  
	+ \left(\Lambda \indices{_{i A} ^B} R \indices{^A _B}  
	+ M^{-1} \partial_i M \right)^{(2)} \,.   
\end{align*} 
The relation required to determine $y_{(3)}$ appears at the order $\O(\tau^0)$, such that the R-symmetry terms are important. From section~\ref{sec:S5}, we recall that
\begin{align*}
\left\langle P^{(2)} \left( M^{-1} \partial_i M \right), \, P^{(2)} \left( M^{-1} \partial_j M   \right) \right\rangle = \partial _i N^I \, \partial _j N^I \, ,
\end{align*}
and noting that $\Lambda_\tau = \mathcal{O}(\tau)$, we find
\begin{align}
\str \big( \A{2}_\tau \, \A{2} _\tau \big) = 
	\frac{1}{y^2} \left(r_\tau ^2 + \s_\tau ^2 \right) + N_{(1)} ^2  + \mathcal{O}(\tau) \,.
\end{align}
The other term is more involved since $\Lambda _{\s \, (0)}$ does not vanish,
\begin{align}
\Lambda \indices{_{\s A} ^B} = i \big( u^{-1}  
	 \dbt \, \pi \, \dt \,  u \big)_A {} ^B  
	+ \mathcal{O}(\tau) = i \big( u^{-1} \Sigma_\s u \big)_A {} ^B 
	+ \mathcal{O}(\tau) \, .
\end{align}
Making use of the trace identity \eqref{TrId_RSymm} and the results obtained in
section~\ref{sec:S5}, we find
\begin{align}
& \str \big( \Lambda \indices{_{\s A} ^B}  R \indices{^A _B} \, 
	\Lambda \indices{_{\s C} ^D} P^{(2)} \left( R \indices{^C _D} \right) \big) 
	= \Lambda \indices{_{\s A} ^B} \, \Lambda \indices{_{\s C} ^D}  
	\left( 4 K_{D B} K^{A C} - 4 \delta ^A _D \, \delta ^C _B 
	+ 2 \delta ^A _B \, \delta ^C _D \right) \nn \\
& \quad = - 4 \, \widebar K^{\prime A C} \, K^\prime _{DB} \, 
	\big( \dbt \, \pi \, \dt \big)_A {} ^B 
	\big( \dbt \, \pi \, \dt \big)_C {} ^D 
	+ 4 \tr \big( \dbt \, \pi \dt \dbt \, \pi \dt \big) 
	- 2 \tr \big(\dbt \, \pi \dt \big)^2  \nn \\
&\quad = - 4 \tr \big( \widebar K^\prime \, \Sigma_s K^{\prime} \Sigma_s ^T \big) 
	+ 4 \tr \big( \dbt \, \pi \dt \dbt \, \pi \dt \big) 
	- 2 \tr \big(\dbt \, \pi \dt \big)^2  \, , 
\end{align}	
as well as	
\begin{align}	
\str \big( \Lambda \indices{_{\s A} ^B}  R \indices{^A _B} \,
	P^{(2)} \left( M^{-1} \partial_\s M \right) \big)  
	= 4 \Lambda \indices{_{\s A} ^B} \big( a _{\s \, \mathrm{S}^5} \big) \indices{_B ^A}   
	= - 4i \, n^I \dot{n}^J \, \tr \big( \dt \gamma^{IJ} \dbt \, \pi \big)\,. 
\end{align}
Combining these findings, we have
\begin{align}
\str \big( A_\s ^{(2)} \, A_\s ^{(2)} \big) &= \frac{r_\s ^2}{y^2} 
	+ 2 \, \pi \cdot \kappa_{s \, (0)} + \dot{n}^2 
	+ 4 \tr \big( \dbt \, \pi \dt \dbt \, \pi \dt \big) 
	- 2 \tr \big(\dbt \, \pi \dt \big)^2 \nn \\ 
& \quad - 4 \tr \big( \widebar K^\prime \, \Sigma_\s K^{\prime} \Sigma_\s ^T \big)  
	- 4i \, n^I \dot{n}^J \, \tr \big( \dt \gamma^{IJ} \dbt \, \pi \big) 
	+ \mathcal{O}(\tau) \,.
\end{align}
Due to the results obtained in the last section, we can express all coefficients appearing above in terms of the boundary data,  
\begin{align}
\begin{aligned}
r_\s^2 &= 1- \tau^2 \left( \dot{\pi}^2 
	- 2i \tr \big( \bTheta _{(2)} \pi \dt
	- \dbt \pi \Theta _{(2)} \big) \right) + \mathcal{O}(\tau^3) \, , \\ 
r_\tau ^2 &= \tau ^2 \dot{\pi}^2 + \mathcal{O}(\tau^3) \, , \\
\sigma _\tau ^2 &= 1 + \tau^2 \left( 2 y_{(3)} 
	- 4i \tr \big( \bTheta _{(2)} \pi \dt 
	- \dbt \pi \Theta _{(2)} \big) \right) + \mathcal{O}(\tau^3) \, , \\
\pi \cdot \kappa _ {\s \, (0)} &= 6 \tr \big( 
	\dbt \, \pi \dt \dbt \, \pi \dt \big) 
	-i \tr \big(\dbt  \pi  \ddot{\theta} - \ddot{\btheta}  \pi  \dt \big) 
	+  \mathcal{O}(\tau) \,. 
\end{aligned}	
\label{pikappa}
\end{align}
Inserting these results into the Virasoro constraint \eqn{Vir1} allows to solve for $y_{(3)}$,
\begin{align}
y_{(3)} & = -\dot{\pi}^2 + 3i \tr \big( \bTheta _{(2)} \pi \dt
	- \dbt \pi \Theta _{(2)} \big) 
	+ 8 \tr \big( \dbt \, \pi \dt \dbt \, \pi \dt \big) 
	-i \tr \big(\dbt  \pi  \ddot{\theta} - \ddot{\btheta} \pi \dt \big) \nn \\
& - \tr \big(\dbt \, \pi \dt\big)^2  
	- 2 \tr \big(\widebar K^\prime \, \Sigma_\s K^{\prime} \Sigma_\s ^T \big) 
	-2 i \, n^I \dot{n}^J \, \tr \big( \dt \gamma^{IJ} \dbt \, \pi \big) 
	+ \half \left(\dot{n}^2 - N_{(1)}^2 \right) . 
\end{align}

\subsection{Variation of the Minimal Area}
The solutions of the first few orders of the equations of motion allow to extract the divergence of the minimal area. Using that 
\begin{align*}
r_\s ^2 &= \pi ^2 + \O(\tau^2) \, , & 
r_\tau ^2 &= \O(\tau^2) \, , & 
y &= \lvert \pi \rvert \tau + \O(\tau^2) \, , \\
\s_\tau ^2 &= \pi ^2 + \O(\tau^2) \, , & 
\s_\s ^2 &= \O(\tau^2) \, ,  
\end{align*}
we find the regulated area of the minimal surface in superspace to be given by
\begin{align}
\mathcal{A}_{\mathrm{min}}(\gamma) \big \vert _{y \geq \varepsilon} &= 
	\frac{1}{2} \int \limits _0 ^{2 \pi} \diff \sigma \!
	\int \limits _{\tau_0 (\s)} ^c \diff \tau \left( 
	\str \left( \A{2}_\tau \A{2}_\tau \right) 
	+ \str \left( \A{2}_\s \A{2}_\s \right) \right) \nn \\
& = \int \limits _0 ^{2 \pi} \diff \sigma \!
	\int \limits _{\tau_0 (\s)} ^c \diff \tau \left( \frac{1}{\tau^2} +\O(\tau^0) \right)
	= \frac{\mathcal{L}(\gamma)}{\varepsilon} + \left( \mathrm{finite} \right) \, .
\end{align}
Here, we have defined $\tau_0(\s)$ as before by 
$y(\tau_0(\s), \s) = \varepsilon$ and introduced the length of the curve $\gamma$ in superspace, 
\begin{align}
\mathcal{L}(\gamma) = \int \diff \s \, \lvert \pi(\s) \rvert \,.
\end{align}
If we then define the renormalized area as before to be given by 
\begin{align}
\mathcal{A}_\mathrm{ren}(\gamma) &= 
	\lim \limits_{\varepsilon \to 0} \left \lbrace 
	\mathcal{A}_{\mathrm{min}}(\gamma) \big \vert _{y \geq \varepsilon} 
	- \frac{\mathcal{L}(\gamma)}{\varepsilon} \right \rbrace \, , 
\end{align}
the arguments given in section \ref{sec:Renorm} can be applied to infer that the invariance of the area functional under $\grp{PSU}(2,2 \vert 4)$-transformations carries over to the renormalized minimal area. 

With the renormalization prescription established, we turn to the discussion of the variational derivatives of the area. As before, they are related to the third-order coefficients of $X$ and $\Theta$. Using that the parametrization of the minimal surface satisfies the equations of motion, the variation only contains boundary terms and we have
\begin{align}
\delta \mathcal{A}_{\mathrm{min}}(\gamma) \big \vert _{y \geq \varepsilon} 
	&= \int \limits _a ^b \diff \s \! 
	\int \limits _{\tau_0(\s)} ^c \diff \tau \left \lbrace 
	\partial _\tau \str \big( g^{-1} \delta g \, \Lambda ^\tau \big) 
	+ \partial _\s \str \big(  g^{-1} \delta g  \, \Lambda ^\s \big)
	\right \rbrace  \, ,
\end{align}
with $\Lambda$ given by equation \eqref{def:Lambda}, 
\begin{align*}
\Lambda ^i = \delta^{ij} \, \A{2}_j - \ihalf \, \epsilon^{ij} \AAm _j \, .
\end{align*}
Since the expression $g^{-1} \delta g$ does not contain any derivatives which may be restricted by choosing a special parametrization, we can safely demand that the parametrization satisfy $\lvert \pi \rvert = 1$. Moreover, due to the kappa symmetry invariance of the action, we can restrict ourselves to the kappa symmetry gauge \eqn{kappagauge}. This allows us to apply the results derived so far, in particular we have $\tau_0(\s) = \varepsilon + \mathcal{O} (\varepsilon^3)$. Thus, using the periodicity of the parametrization in $\s$, we find that
\begin{align}
\delta \mathcal{A}_{\mathrm{min}}(\gamma) \big \vert _{y \geq \varepsilon}  
	= - \int \limits _0 ^L \diff \s 
	\str \big( g^{-1} \delta g \, \Lambda ^\tau \big) (\tau_0 (\s), \s)
	+ \mathcal{O}(\varepsilon) \,.
\end{align}
To compute the above result explicitly, we use the expression \eqn{eqn:MaurerCartanForm} for $g^{-1} \delta g$ and in particular that
\begin{align}
\delta X (\tau , \s) &= \delta x(\s) + \mathcal{O}(\tau^2) \, , & 
\delta \Theta (\tau , \s) &= \delta \theta (\s) + \mathcal{O}(\tau^2)  \, , 
\end{align}
since the first-order coefficients of $X$ and $\Theta$ vanish identically. Applying the trace-identities given in appendix~\ref{app:u224}, we find
\begin{align}
\delta \mathcal{A}_{\mathrm{min}}(\gamma) \big \vert _{y \geq \varepsilon} 
	= \frac{\delta \mathcal{L}(\gamma)}{\varepsilon} + ( \mathrm{finite} ) \, ,
\end{align}
and the finite term is given by
\begin{align}
\delta \mathcal{A}_{\mathrm{ren}}(\gamma) & = 
	\int \limits _0 ^L \diff \s \big \lbrace \delta x_\mu \, b^\mu 
	+ \delta \theta _\a {} ^A \left(4 \xi _A {} ^\a 
	- i \, \bt _{A \da} \, b^{\da \a} \right) \nn \\
& \qquad \quad + \delta \bt_{A \da} \left( 4 \bar{\xi}^{\da A} 
	- i \, b^{\da \a} \, \theta _\a {} ^A \right) 
	- \delta n^I  N_{(1)} ^I \big \rbrace 
\label{variation} 
\end{align}
Here, we defined
\begin{align}
b^\mu &= - \half r_{\tau \, (2)} ^\mu - \kappa _{\tau \, (0)} ^\mu 
	+ 2i \tr \big( \big( \bvart_{(0)} \vart_{(1)} - \bvart_{(1)} \vart_{(0)} 
	+ 2 \, n^I N_{(1)} ^J \,  \bvart_{(0)} \gamma^{IJ} \vart_{(0)} \big) 
	\bar{\sigma}^\mu \big) \, ,  \label{coeffder1} \\
\xi _A {} ^\a &= \vart_{(1)} {} _A {} ^\a 
	+  n^I N_{(1)} ^J \left( \gamma ^{IJ} \vart_{(0)} \right) _A {} ^\a  
	\, , \qquad 
	\bar{\xi}^{\da A} = - \bvart _{(1)} ^{\da A} 
	+ n^I N_{(1)} ^J \left( \bvart_{(0)} \gamma ^{IJ} \right)^{\da A} \,.  
\label{coeffder2} 
\end{align}
We read off the functional derivatives of the regulated  minimal area from \eqn{variation}:
\begin{align}
\begin{aligned}
b^\mu (\s) &= \frac{\delta \mathcal{A}_{\mathrm{ren}}(\gamma) }{\delta x_\mu (\s)} \, , \\
N_{(1)} ^I (\s) &= - \frac{\delta \mathcal{A}_{\mathrm{ren}(\gamma)}}
	{\delta n^I (\s)} 
	+ \left( n^J(\s) \frac{\delta \mathcal{A}_{\mathrm{ren}}(\gamma)}
	{\delta n^J (\s)} \right) 
	n^I (\s) \, , \\
\xi _A {}^\a (\s) &= \frac{1}{4} \left( 
	\frac{\delta \mathcal{A}_{\mathrm{ren}}(\gamma)}{\delta \theta _\a {} ^A(\s)} 
	+ i \, \bt _{A \da} (\s) \, \s_\mu ^{\da \a} \, 
	\frac{\delta \mathcal{A}_{\mathrm{ren}}(\gamma)}{\delta x_\mu (\s)} \right) \, , \\
\bar{\xi}^{\da A} (\s) &=  \frac{1}{4} \left( 
	\frac{\delta \mathcal{A}_{\mathrm{ren}}(\gamma)}{\delta \bt_{A \da}(\s)} 
	+ i\, \s_\mu ^{\da \a} \, \theta _\a {} ^A (\s) \, 
	\frac{\delta \mathcal{A}_{\mathrm{ren}}(\gamma)}{\delta x_\mu (\s)} \right) \, .
\end{aligned}
\label{varder}
\end{align}
The relations between the functional derivatives of the minimal area and the coordinates $X_{(3)}, \Theta_{(3)}$ and $\vart_{(1)}$ take a much more complicated form than for the minimal surface in $\AdS_5$. 
The important point for us, however, is to identify the above coefficients in the Noether current $j$, where they appear naturally. In order to do this, we do not need to find expressions for $X_{(3)}, \Theta_{(3)}$ and $\vart_{(1)}$ in terms of the above functional derivatives and hence we have made no attempt to derive such relations above. 

\section{Yangian Symmetry}
With the expansion of the minimal surface around a generic smooth boundary curve established, we turn to the evaluation of the conserved charges in order to derive the Yangian symmetry of the superspace Wilson loop at strong coupling. 

\subsection{Level Zero}
We begin by discussing the G-symmetry charge $Q^{(0)}$, which encodes the superconformal invariance of the area of the minimal surface. The invariance under large superconformal transformations was already established by generalizing the argument given in section \ref{sec:Renorm}. The derivation presented below provides an explicit representation of the superconformal generators in the boundary superspace and serves as a preparation for our discussion of the Yangian invariance. 

We are thus interested in the $\tau^0$-component of the Noether current $j= - 2 g \Lambda g^{-1}$. Making use of relation \eqref{fermeqnA}, we have
\begin{align}
j_{\tau \, (0)} = -2 \left \lbrace g  \Lambda _\tau g^{-1} \right \rbrace _{(0)} 
	&= - 2 \left \lbrace g \left(\A{2}_{\tau} - \ihalf A_\s ^{(1)-(3)} 
		\right) g^{-1} \right \rbrace _{(0)}  \nn \\
	&= -2 \left \lbrace g \left(\A{2}_{\tau} + \half 
		A_\tau ^{(1)+(3)} \right) g^{-1} \right \rbrace _{(0)} \,.
\end{align}
For the last step, notice that the terms within the brackets agree up to terms of order 
$\O(\tau ^{3/2})$ and the conjugation with $g$ lowers the order of a fermionic term at most by
$\sqrt{\tau}$. The difference is hence not relevant for the $\tau^0$-component. 
The conjugations can be done in the same way as in the calculation of the Maurer--Cartan form explained in section \ref{sec:Coordinates}, and we arrive at 
\begin{align}
j_{\tau \, (0)} &= \Big \lbrace \sfrac{1}{\tau} \, e^{X \cdot P + \Omega} 
	\left( - \dot{\pi}^\mu \, K_\mu - 2 D 
	- \left(\vart _{(0)}{} _A {} ^\a  \, S_\a {} ^A 
	+ \bvart_{(0)}^{\, \da A} \, \Sb_{A \da} \right) \right)
	e^{-X \cdot P - \Omega} \Big \rbrace _{(0)} \nn \\
& + 2 \Big \lbrace  \, e^{X \cdot P + \Omega} 
	\left( \half b^\mu \, K_\mu - \xi _A {}^\a \, S_\a {} ^A 
	+ \widebar{\xi}^{\da A} \, \Sb_{A \da} + n^I N_{(1)} ^J \, \Gamma ^{I J}  
	\right)e^{-X \cdot P - \Omega} \Big \rbrace _{(0)} \, .
\end{align}
The first term above vanishes, since the expansions of the coordinates $X$ and $\Theta$ are of the form $X = x + \mathcal{O}(\tau^2)$ and $\Theta = \theta + \mathcal{O}(\tau^2)$ and it remains to compute
\begin{align}
j_{\tau \, (0)} = \hspace*{-.5mm} 2 \hspace*{.5mm}  e^{x \cdot P +  \theta Q + \bt \Qb } \!
	\left( \half b^\mu \, K_\mu - \xi _A {}^\a \, S_\a {} ^A 
	+ \widebar{\xi}^{\da A} \, \Sb_{A \da} + n^I N_{(1)} ^J \, \Gamma ^{I J}  
	\right) \! e^{-x \cdot P - \theta Q - \bt \Qb } \,. 
\label{Jtau0}
\end{align}
The coefficients $b$, $\xi$ and $N_{(1)}$ defined in \eqn{coeffder1} and \eqn{coeffder2} are exactly those that were identified with the functional derivatives of the minimal area in the last section. We can thus write the resulting expression in the form
\begin{align}
j_{\tau \, (0)} (\s) = 2 \, \jay_a (\s) \left( \mathcal{A}_\mathrm{ren} (\gamma) \right) 
	\, T^a \, .
\label{jtau0}	
\end{align}
Here, $T^a = G^{ab} \, T_b$ span the dual basis to the generators defined in appendix \ref{app:u224}. Note that we use the inverse of the metric $G_{ab}$ on $\alg{u}(2,2 \vert 4)$ to raise and lower the group indices. The densities $\jay_a (\s)$ denote functional derivative operators, which act on the minimal area $\mathcal{A}_\mathrm{ren} (\gamma)$ above. Their form can be read off from the relation 
\begin{align}
\jay_a (\s) \left( \mathcal{A}_\mathrm{ren} (\gamma) \right) 
	= \half \, \str \big( j_{\tau \, (0)} (\s) \, T_a \big) \,.
\label{defjay}	
\end{align}
The functional derivative operators $\jay_a (\s)$ are given explicitly in appendix \ref{app:densities}. We note that the superspace representation obtained in this way also includes the generator
\begin{align}
b(\s) = \frac{1}{2} \left( \theta \indices{_a ^A} (\s)
	\, \frac{\delta}{\delta \theta \indices{_a ^A} (\s)} 
	- \bt_{A \da} (\s) \, \, \frac{\delta}{\delta \bt_{A \da} (\s) }  \right) , 
\end{align} 
which does not give rise to a symmetry of the minimal area, since it corresponds to the $C$-part of $Q^{(0)}$, which is not conserved. Moreover, we also read off the central charge generator $c(\s) = 0$ from the $B$-part of the conserved charge, which vanishes since the model is constructed on $\grp{SU}(2,2 \vert 4)$. The algebra of the generators is compatible with setting 
$c(\s)=0$ and is given by
\begin{align}
\big[ \jay_a (\s) ,  \jay_b (\s^\prime) \big \rbrace 
	= \mathbf{f} \indices{_{ab} ^c} \, \jay_c (\s) \, \delta( \s - \s^\prime ) \, .
\end{align}
Again, the structure constants $\mathbf{f} \indices{_{ab} ^c}$ are related to the structure constants $f \indices{_{ab} ^c}$ of the supermatrices $T_a$ by 
\begin{align}
\mathbf{f} \indices{_{ab} ^c} = f \indices{_{ba} ^c} 
	=  - \left( -1 \right) ^{\lvert a \rvert \lvert b \rvert}  f \indices{_{ab} ^c} \, . 
\end{align}
We have thus seen that the vanishing of the level-0 charge $\mathcal{Q}^{(0)}$ can be rewritten as the invariance of the superspace Wilson loop under the level-0 Yangian generators
\begin{align}
\J_a^{(0)} = \int \diff \s \; \jay_a (\s) \, .
\label{Ja0}
\end{align}

\subsection{Level One}
The level-1 Yangian generators follow from evaluating the level-1 Yangian charge
\begin{align}
Q^{(1)} = \frac{1}{2} \, \int \diff \s_1 \, \diff \s_2 \, \varepsilon_{21} \, 
	\left[ j_\tau (\s_1) , j_\tau (\s_2) \right] 
	- 4 \int \diff \sigma \left( a_\s ^{(2)} + \quarter a_\s ^{(1)+(3)} \right) \, .
\end{align}
Here, we have employed the abbreviation $a^{(k)} = g \A{k} g^{-1}$ introduced in equation \eqref{small_a} and we recall that $\varepsilon_{21}$ denotes 
\begin{align}
\varepsilon (\s_2 - \s_1 ) = \theta(\s_2 - \s_1 ) - \theta(\s_2 - \s_1 ) \, .
\end{align}
Again, we are only interested in the $\tau^0$-component and note that its calculation can be simplified by using that the $\tau$-component of the Noether current can be written in the form given in \eqref{jtau_exp}, 
\begin{align}
j_\tau &= \frac{1}{\tau} \, \partial_\sigma j_{\sigma \, (-2)}
	+ j_{\tau \, (0)} - \tau \, \partial_\sigma j_{\sigma (0)}
 	+ \mathcal{O}(\tau^2) \, ,
\end{align}
due to current conservation. The $\tau^0$-component is hence given by
\begin{align}
Q^{(1)} _{(0)} &= \frac{1}{2} \, \int \diff \s_1 \, \diff \s_2 \,
	\varepsilon_{21} \, 
	\left[ j_{\tau \, (0)} (\s_1) , j_{\tau \, (0)}  (\s_2) \right] \nn \\
	& \quad - \int \diff \s \big( 4 a_{\s \, (0)}  ^{(2)} + a_{\s \, (0)}  ^{(1)+(3)} 
	+ \left[ j_\s , \partial_\s j_\s \right] _{(-2)} \big)
	= Q^{(1)} _{\mathrm{bi-lo}} + Q^{(1)} _{\mathrm{lo}} \, .
\end{align} 
Let us first evaluate the bi-local part. Using the expression \eqref{jtau0} derived for 
$j_{\tau \, (0)}$ above, we find 
\begin{align}
\mathcal{Q}^{(1)} _{\mathrm{bi-lo}} &= 2 \, 
	f \indices{^{bc} _a}  \int \diff \s_1 \, \diff \s_2 \, 		
	\varepsilon_{21} \, \jay_b(\s_1) 
	\left( \mathcal{A}_\mathrm{ren} (\gamma) \right) 
	j_c(\s_2) \left( \mathcal{A}_\mathrm{ren} (\gamma) \right) T^a \nn \\
&= 2 \, \mathbf{f} \indices{^{cb} _a}  
	\int \diff \s_1 \, \diff \s_2 \, \varepsilon_{21} \, 
	\jay_b(\s_1) \left( \mathcal{A}_\mathrm{ren} (\gamma) \right)  
	\jay_c(\s_2) \left( \mathcal{A}_\mathrm{ren} (\gamma) \right)  
	T^a \, . 
\label{non-local}
\end{align}
In the last step, we have switched to the structure constants 
$\mathbf{f} \indices{_{ab} ^c}$ of the functional derivative operators. The index raising and lowering is then done via the metric 
$\mathbf{G}^{ab}$ obtained for these generators. In appendix \ref{app:densities}, we find that it is related to the components of the metric for the corresponding supermatrices by
\begin{align}
\mathbf{G}^{ab} = G^{ba} = (-1) ^{\lvert a \rvert} G^{ab} \, .
\end{align}
Correspondingly, we have
\begin{align*}
f \indices{^{bc} _a} &= G^{bd} \, G^{ce} f\indices{ _{de} ^g} \, G_{ga} 
	=  \left(-1 \right)^{\lvert b \rvert 
		+ \lvert c \rvert  + \lvert a \rvert} \mathbf{G}^{bd} \, 
		\mathbf{G}^{ce} \mathbf{f} \indices{ _{ed} ^g} \, \mathbf{G}_{ga}
	= \left(-1 \right)^{\lvert b \rvert 
		+ \lvert c \rvert  + \lvert a \rvert} 
		\mathbf{f} \indices{^{bc} _a} 
	= \mathbf{f} \indices{^{bc} _a} \, .	
\end{align*}
Here, we have used that $\left( \lvert b \rvert + \lvert c \rvert  + \lvert a \rvert \right) \in \lbrace 0, 2 \rbrace$ for non-vanishing 
$\mathbf{f} \indices{^{bc} _a}$ due to the $\mathbb{Z}_2$ grading of the Lie super algebra.
The bi-local part found above thus shows the typical structure of a level-1 Yangian generator, as expected.  

It remains to compute the local term, for which we note the expression
\begin{align}
Q^{(1)} _{\mathrm{lo}} &= -4 \, \int \diff \s \left \lbrace 
	g \left( \A{2} _\s + \quarter \AAp_\s 
		+ \tau^2 \left[ \Lambda_\s , \partial_\s 
			\Lambda_\s + [U_\s , \Lambda_\s ] \right] \right) g^{-1}
	\right \rbrace _{(0)} \, .
\label{Q1lo}		
\end{align}
While we have pulled out the conjugations with $g$ in order to make use of the cancellations between the two terms at an earlier stage of the calculation, it turns out to be convenient to already discuss the conjugations with $M y^D$ when considering these terms individually, in particular because after the conjugation with $y^D$ the orders in $\tau$ will not get lowered and we may already discard terms that are of order $\mathcal{O}(\tau)$. We recall the expressions found for the components of the Maurer--Cartan form in section \ref{sec:Coordinates}, 
\begin{align*}
A^{(2)} &= - \frac{r_{\a \da} + y^2 \,  \kappa_{\a \da}}{4y} 
	\left( P^{\da \a} + K^{\da \a} \right)  + \frac{\s}{y} D  + \gamma \, C  
	+ \left(\Lambda \indices{_A ^B} R \indices{^A _B} + M^{-1} \diff M  \right)^{(2)} 
	\, , \label{A2n} \\ 
A^{(1)+(3)} &=  \frac{1}{\sqrt{y}} \left( \varepsilon _\a {}^A \, Q_A {} ^\a 
	+ \widebar{\varepsilon}_{A \da} \,  \Qb^{\da A}  
	+ y \left(  \chi_A {}^\a \, S_\a {} ^A 
	+ \widebar{\chi}^{\da A} \, \Sb_{A \da} \right) \right) \, ,
\end{align*}
as well as the coefficients
\begin{align}
\sigma &= \diff y + 2 y\,  \tr \left( \diff \bTheta \, \bvart 
	+ \vart\, \diff \Theta \right) \, , \nn \\
\gamma &=  \tr \left( \vart \left(2 \diff\Theta + \zeta \right) 
	- \left(2 \diff\bTheta + \widebar{\zeta} \right) \bvart \right) , \\
\Lambda _A {} ^B &=
	\left[ m^{-1} \left(
		\vart \left(\diff\Theta + \half \zeta\right)
		- \left(\diff\bTheta + \half \bar{\zeta} \right) \bvart
	\right) m \right]_A {} ^B	 \, . \nn
\end{align}
Inserting the solutions \eqref{vart0} for the components $\vart_{(0)}$ and $\bvart_{(0)}$, we find
\begin{align}
\begin{aligned}
\sigma _\s &= 2y \tr \big( \dbt \, \bvart_{(0)} + \vart_{(0)} \theta \big) 
	+ \O(\tau^2)  = 2 i \, \tau \tr \big( \dbt \pi \dt - \dbt \pi \dt \big) 
	+ \O(\tau^2)  =  \O(\tau^2)  \, , \\
\gamma _\s &= 2i \tr \big(  \dbt \pi \dt \big) + \O(\tau) \, .	
\end{aligned}
\end{align}
We note moreover that in the above expression for $\A{2}$, the projections are left implicit for the R-symmetry part. The part stemming from the $M^{-1} \diff M$-term can be inferred from the result \eqref{S5Noether} for the Noether current in the coset description of $\mathrm{S}^5$ discussed in section \ref{sec:S5}. For the other term, we make use of the explicit form of the projections given in appendix~\ref{app:u224} and note that 
\begin{align}
\Lambda _{\s \, A} {} ^B P^{(2)} \left(R \indices{^A _B} \right) = 
	\ft{1}{4} \left( \Lambda _{\s \, A} {} ^B  - K^{B C} 
	\Lambda _{\s \, C} {} ^D  K_{DA} \right) R \indices{^A _B} \, .
\end{align}
Using that $\varepsilon_{\s (0)} = 0$, we thus find
\begin{align}
& M y^D \left( A_\s ^{(2)} + \quarter A_\s ^{(1)+(3)} \right) y^{-D} M^{-1} 
	= - \frac{r_{\s \, \a \da} + y^2 \kappa _{\s \, \a \da}}{4y^2} \, 
	\left( K^{\da \a} + y^2 P^{\da \a} \right) 
	- \quarter \, \rho_A {} ^B  R \indices{^A _B} \nn \\ 
& \qquad + 2i \tr \big(  \dbt \pi \dt \big) \, C
	+ \quarter \left( \left(m \chi_\s \right)_A {} ^\a S_\a {} ^A 
	+ \left( \widebar{\chi}_\s m^{-1} \right)^{\da A} \Sb_{A \da} \right) 
	+ \mathcal{O}(\tau) \,. 
\end{align}
Here, the coefficient of the R-symmetry part is given by
\begin{align}
\rho_A {} ^B =   n^I \dot{n} ^J \, \left(\gamma ^{IJ} \right) _A {} ^B 
	-  i \big( \dbt  & \pi \dt \big)_A {} ^B  
	+ i \widebar K^{\prime \, BC}  \big( \dbt \pi 
		\dt \big)_C {} ^D  K^\prime _{DA} 
	+ \ihalf \delta ^B _A  \tr \big( \dbt  \pi \dt  \big)  
	\,. \label{rhoAB}
\end{align}
Next, we consider the commutator term in \eqref{Q1lo} and compute 
\begin{align}
\left(\ast \right) = \big[ M y^D \left( \tau^2 \Lambda _\s \right) y^{-D} M^{-1}  ,
	M y^D \left( \partial_\s \Lambda _\s + 
	\left[ U_\s  ,  \Lambda _\s \right] \right) y^{-D} M^{-1} \big] \,.
\end{align}
In this expression we may replace $\Lambda _\s = A_\s^{(2)} + \ihalf A_{\tau} ^{(1)-(3)} = A_\s^{(2)} + \half A_\s ^{(1)+(3)} + \mathcal{O}\left( \tau ^{3/2} \right)$ as the unwanted terms are at least of order $\tau$. Consider then the term on the left-hand side. As we shall see shortly, the right-hand side of the commutator is of order $\mathcal{O}(\tau^{-2} )$ and we can hence neglect all terms that are of order $\mathcal{O}(\tau^3 )$ in the computation of the expression on the left-hand side. This leads to finding
\begin{align}
M y^D \left( \tau^2 \Lambda _\s \right) y^{-D} M^{-1} & = 
	- \frac{\tau^2}{4y^2} \left( r_{\s \, \a \da} + y^2 \kappa _{\s \, \a \da} \right) 
	\left( K^{\da \a} + y^2 P^{\da \a} \right) 
	- \frac{\tau^2}{4} \, \rho_A {} ^B  R \indices{^A _B} \nn \\
&  + \frac{\tau^2}{2} \left( \left( m \chi_\s \right)_A {} ^\a S_\a {} ^A 
	+ \left( \bar{\chi}_\s m^{-1} \right)^{\da A} \Sb_{A \da} \right) 
	+ \left( \mathrm{negl.} \right) \,.
\end{align}
Here, the bracket $\left( \mathrm{negl.} \right)$ denotes terms, which can be neglected. 
Note in particular that only the term proportional to $K^{\da \a}$ is of order $\mathcal{O}(\tau^0)$, while all other terms are of order $\mathcal{O}(\tau^2)$. We need thus only compute the right-hand side of the above commutator up to $\mathcal{O}(\tau^{-2})$ for the generators that commute with $K^{\da \a}$ and to $\mathcal{O}(\tau^0)$ for those that don't. With these simplifications we have
\begin{align}
 M y^D \left( \partial_\s \Lambda _\s + 
 	\left[ U_\s , \Lambda _\s \right] \right) y^{-D} M^{-1} 
 	= - \frac{\dot{\pi}_{\a \da}}{4 \tau^2} \left( K^{\da \a} + \tau^2 P^{\da \a} \right) 
 	+ \frac{r_\s^2}{y^2} D + \left( \mathrm{negl.} \right) ,
\end{align}
and note that $r_\s^2 = 1 + \mathcal{O}(\tau)^2$. Combining these results, we find
\begin{align}
\left(\ast \right) &= - \frac{\tau^2 \, r_\s ^2}{4 y^4} 
	\left( r_{\s \, \a \da} + y^2 \kappa _{\s \, \a \da} \right) 
	\left( K^{\da \a} - y^2 P^{\da \a} \right) 
	- \pi^\mu \dot{\pi}^\nu M_{\mu \nu} \nn \\
& \qquad + \quarter  \left( \left(m \chi_\s \right)_A {} ^\a S_\a {} ^A 
	+ \left( \widebar{\chi}_\s m^{-1} \right)^{\da A} \Sb_{A \da} \right) 
	+ \mathcal{O}(\tau) \,.
\end{align}
Accordingly, we have
\begin{align}
& \quad M y^D \left( A_\s ^{(2)} + \quarter A_\s ^{(1)+(3)} 
	+ \tau^2 \big[ \Lambda _\s ,  \partial_\s \Lambda _\s 
	+ \left[ U_\s , \Lambda _\s \right] \big] \right) y^{-D} M^{-1} = \nn \\
&= - \frac{ y^2 + \tau ^2 r_\s ^2}{4  y^4}  
	\left( r_\s ^{\da \a} + y^2 \kappa_\s ^{\da \a} \right) K_{\a \da} 
	- \frac{y^2 - \tau ^2 r_\s ^2}{4 y^2}  \, r_\s ^{\da \a} \, P_{\a \da} 
	- \quarter \,  \rho_A {} ^B  R \indices{^A _B} 
	- \pi ^\mu \dot{\pi}^\nu M_{\mu \nu} \nn \\
& \quad + \half \left( \left(m \chi_\s \right)_A {} ^\a S_\a {} ^A 
	+ \left( \widebar{\chi}_\s m^{-1} \right)^{\da A} \Sb_{A \da} \right) 
	+ 2i \tr \big(  \dbt \pi \dt \big) \, C
	+ \mathcal{O}(\tau) \nn \\
&= - \frac{\pi^{\da \a}}{2 \tau^2} \, K_{\a \da} 
	- \quarter \left( r_{\s \, (2)}^{\da \a} + 2 \kappa _{\s \, (0)} ^{\da \a} 
	+ \left( \pi \cdot r_{\s \, (2)} - 2 y_{(3)} \right) \pi^{\da \a} \right) \, K_{\a \da} 
	\nn \\
& \quad - \quarter \,  \rho_A {} ^B  R \indices{^A _B} 
	- \pi ^\mu \dot{\pi}^\nu M_{\mu \nu}  
	+ 2i \tr \big(  \dbt \pi \dt \big) \, C \nn \\
& \quad - \half \big( 3 \, \dbt \pi  \dt \dbt  \pi  
	- i \, \partial_\s \big( \dbt \pi \big) \big)_A {} ^\a \, S_\a {} ^A 
	- \half \big( 3 \, \pi  \dt \dbt  \pi \dt 
	+ i \, \partial_\s \big( \pi \dt \big) \big)^{\da A} \, \Sb_{A \da}  
	+ \mathcal{O}(\tau) \,.
\end{align}
Rather conveniently, the terms proportional to $P_{\a \da}$ have cancelled out. This simplifies the computation of the conjugations with $e^\eta$, for which we find
\begin{align}
& e^{\eta} M y^D \! \left( A_\s ^{(2)} + \quarter A_\s ^{(1)+(3)} 
	+ \tau^2 \big[ \Lambda _\s ,  \partial_\s \Lambda _\s 
	+ \left[ U_\s , \Lambda _\s \right] \big] \right) y^{-D} M^{-1} e^{-\eta} 
	= \frac{\pi ^\mu}{\tau ^2} K_\mu - \pi ^\mu \dot{\pi}^\nu M_{\mu \nu} \nn \\
& + 2i \tr \big(  \dbt \pi \dt \big) \, C 
	+ \half ( N_{(1)} ^2 \, \pi^\mu + l^\mu ) K_\mu 
	- \quarter f_A {}^\a \, S_\a {}^A 
	+ \quarter \widebar{f}^{\da A} \, \Sb_{A \da} 
	- \quarter \,  \rho_A {} ^B  R \indices{^A _B}  \,. \label{etoeta}
\end{align}
Here, the coefficients $l^\mu, \, f_A {}^\a $ and $\widebar{f}^{\da A} $ are given by 
(cf.\ equations \eqref{theta2}, \eqn{pikappa} and \eqn{rhoAB})
\begin{equation}
\begin{aligned}
l^\mu &= r_{\s \, (2)} ^\mu + \left( \pi \cdot r_{\s \, (2)} - 2 y_{(3)} 
	- N_{(1)}^2 \right) \pi ^\mu 
	- 4i \tr \big( \bvart_{(0)} \, \rho \, \vart_{(0)} \, \bs ^\mu \big) \\
	& \quad - 3i \tr \big( \dt \dbt \big( \dot{\pi} \bs ^\mu \pi 
	-  \pi \bs ^\mu \dot{\pi}  \big) \big) \nn \\
&= \ddot{\pi}^\mu + \left( \dot{\pi}^2 - \dot{n}^2 \right) \pi ^\mu  
	- \left( 2 \pi \cdot \kappa_{\s \, (0)} - 4i \tr \big( \big( \Theta_{(2)} \dbt 
	- \dt \bTheta_{(2)} + \dt \rho \dbt \big) \pi \big) 
	\right) \pi ^\mu  \nn \\
& \quad + 4i \tr \big(\dt \rho \dbt \s ^\mu \big) 
	- 3i \tr \big( \dt \dbt \big( \dot{\pi} \bs ^\mu \pi 
	-  \pi \bs ^\mu \dot{\pi}  \big) \big)  \, ,\\
f _A {}^\a  &= \big[ 6  \, \dbt \pi  \dt \dbt  \pi  
	- 2i \, \ddot{\btheta} \, \pi  - 4i \, \dbt \, \dot{\pi} 
	- 4i \, \rho \dbt \pi \big] _A {}^\a   \, ,\\
\bar{f}^{\da A}  &= \big[  - 6 \, \pi  \dt \dbt \pi \dt
	- 2i \, \pi  \ddot{\theta} - 4i \, \dot{\pi} \dt 
	+ 4i \, \pi \dt \rho \big]^{\da A}   \,.
\end{aligned}
\end{equation}
We have chosen not to absorb the term involving $N_{(1)} ^2$ into the definition of $l^\mu$, since this term is the only one of the above terms that contains functional derivatives of the minimal area. 
The coefficients are introduced in such a way that the remaining conjugation with 
$e^{x \cdot P + \theta Q + \btheta Q}$ leads to an expression that resembles the one found for 
$j_{\tau \, (0)}$ in equation \eqref{Jtau0}, 
\begin{align}
j_{\tau \, (0)} &= 2 \, e^{x \cdot P + \theta Q + \btheta Q} 
	\Big( \half \, \partial^\mu  \, K_\mu 
	- \quarter \left( \partial_A {} ^\a 
		+ i \, \bt_{A \da} \, \partial ^{\da \a}  \right) \, S _\a {} ^A \nn \\
& + \quarter \left( \partial^{\da A} 
		+ i \, \partial^{\da \a} \, \theta_\a {} ^A  \right) \, \Sb_{A \da} 
	- \quarter \left( \gamma^{IJ} \right)_A {} ^B \, n^I \partial ^J  \, R ^A {} _B  \Big)
	( \mathcal{A}_{\mathrm{ren}}(\gamma) )
	e^{- x \cdot P - \theta Q - \btheta Q} \nn \\
&= 2 \, \jay_a (\s) \left( \mathcal{A}_\mathrm{ren} (\gamma) \right) \, T^a \nn \\
&= 2 \left( \, \jay_a ^\mu \, \partial_\mu + \jay_{a \a} {} ^A \, \partial_A {} ^\a 
	+ \jay_{a \, A \da} \, \partial ^{\da A} + \jay_{a A} {} ^B \,  
	\left( \gamma^{IJ} \right)_B {} ^A \, n^I \partial ^J \, \right) 
	 \left( \mathcal{A}_\mathrm{ren} (\gamma) \right) T^a \,.	
\end{align}
Here, we have used the abbreviations
\begin{align}
\partial^\mu &= \frac{\delta }{\delta x_\mu (\s)} \, , & 
\partial ^\a _A &= \frac{\delta }{\delta \theta _\a ^A(\s)} \, , & 
\widebar{\partial} ^{A \da} &=  \frac{\delta }{\delta \btheta_{A \da}(\s)} \, , & 
\partial ^I &= \frac{\delta }{\delta n^I (\s)} \, ,
\end{align}
for the functional derivatives appearing in the densities $\jay_a (\s)$. We can thus write a part of the local term as
\begin{align}
& e^{x \cdot P + \theta Q + \btheta \Qb} 
	\big( \half l^\mu K_\mu - \quarter f_A {}^\a \, S_\a {}^A 
	+ \quarter \widebar{f}^{\da A} \, \Sb_{A \da} 
	- \quarter \,  \rho_A {} ^B  R \indices{^A _B} \big) 
	e^{- x \cdot P - \theta Q - \btheta \Qb} \nn \\
&= \big( \jay_a ^\mu \, l_\mu 
	+ \jay_{a \a} {} ^A \left(f_A {} ^\a - i \, \bt_{A \da} \, l^{\da \a} \right) 
	+ \jay_{a \, A \da} \left(\bar{f}^{\da A} -i \, l^{\da \a} \theta_\a {} ^A \right) 
	+ \jay_{a A} {} ^B \,  \rho_B {} ^A \big) \, T^a	\, .
\end{align}
There is an additional term which is similar to the one which evaluated to zero in the bosonic calculation, see equation \eqref{bosoniczero}. Here, it does give a contribution which we denote by 
\begin{align}
\jay_a ^{(1)\, \prime} T^a \! &= \! \Big \lbrace e^{X \cdot P + \Theta Q + \bTheta \Qb} 
	\Big( \frac{\pi^\mu}{\tau^2} \, K_\mu  - \pi ^\mu \dot{\pi}^\nu M_{\mu \nu} 
	+ 2i  \tr \! \big(  \dbt \pi \dt \big)   C \Big) 
	e^{- X \cdot P - \Theta Q - \bTheta \Qb} \Big \rbrace_{\!(0)} .
\label{jayprime}	
\end{align}
Explicit expressions for the coefficients above are given in appendix \ref{app:densities}. Combining these terms with the ones discussed before, we obtain the level-one densities
\begin{align}
\jay_a ^{(1)} &= \jay_a ^\mu \, l_\mu 
	+ \jay_{a \a} {} ^A \! \left(f_A {} ^\a - i \, \bt_{A \da} \, l^{\da \a} \right) \!
	+ \jay_{a \, A \da} \! \left(\bar{f}^{\da A} -i \, l^{\da \a} \theta_\a {} ^A \right) \!
	+ \jay_{a A} {} ^B \,  \rho_B {} ^A + \jay_a ^{(1)\, \prime} .
\end{align}
The local part of the level-1 Yangian charge is thus given by
\begin{align}
Q^{(1)} _{\mathrm{lo}} &= - 4 \int \diff \s \left \lbrace 
	\jay_a ^{(1)} (\s) \, T^a + e^{x \cdot P + \theta Q + \btheta \Qb} 
	\left( \half N_{(1)}^2 \pi ^\mu K_\mu \right) 
	e^{- x \cdot P - \theta Q - \btheta \Qb} 
	\right \rbrace \nn \\
	&= - 4 \int \diff \s \Big \lbrace 
	\jay_a ^{(1)} (\s) \, T^a 
	+ \left( \jay_a ^ \mu \, \pi _\mu 
	- i \, \jay_{a \a} {} ^A \, \bt_{A \da} \, \pi^{\da \a} 
	- i \, \jay_{a \, A \da} \, \pi ^{\da \a} \, \theta _\a {} ^A \right) \nn \\
& \hspace*{23mm} \mathrm{P}^{IJ} \, 
	\partial^I  \left( \mathcal{A}_\mathrm{ren} (\gamma) \right) 
	\partial^J  \left( \mathcal{A}_\mathrm{ren} (\gamma) \right)   
	\Big \rbrace
	\, .
\end{align}
Here, we have introduced the projection $\mathrm{P}^{IJ} = \delta^{IJ} - n^I n^J$ onto the tangent space of $\mathrm{S}^5$ at the point $n^I$. The combination with the bi-local part \eqref{non-local} of $Q^{(1)}$ then gives the full level-1 Yangian charge
\begin{align}
Q^{(1)} _a &= 2 \, \mathbf{f} \indices{^{cb} _a}  
	\int \diff \s_1 \, \diff \s_2 \, \varepsilon_{21} \, 
	\jay_b(\s_1) \left( \mathcal{A} \right)  
	\jay_c(\s_2) \left( \mathcal{A} \right)  
	- 4 \int \diff \s \, \jay_a ^{(1)} (\s) 	\nn \\
&- 4 \int \diff \s  \left( \jay_a ^ \mu \, \pi _\mu 
	- i \, \jay_{a \a} {} ^A \, \bt_{A \da} \, \pi^{\da \a} 
	- i \, \jay_{a \, A \da} \, \pi ^{\da \a} \, \theta _\a {} ^A \right) 
	\mathrm{P}^{IJ} \, 	\partial^I  \left( \mathcal{A} \right) 
	\partial^J  \left( \mathcal{A} \right) . 	
\end{align}
The vanishing of the level-1 charge $\mathcal{Q}^{(1)}$ can be rewritten as the level-1 Yangian invariance of the Wilson loop in superspace, which is given by
\begin{align}
\left \langle \mathcal{W}(\gamma) \right \rangle 
	= e^{ -\frac{\sqrt{\la}}{2 \pi} \mathcal{A}_\mathrm{ren}(\gamma) } 
\end{align}
at strong coupling. The corresponding level-1 Yangian generators are given by
\begin{align}
\J_a ^{(1)} &= \mathbf{f} \indices{^{cb} _a} 
	\int \diff \s_1 \, \diff \s_2 \, \varepsilon_{21} \, 
	\jay_b(\s_1) \, \jay_c(\s_2) 
	- \frac{\lambda}{2 \pi ^2} \, \int \diff \s \, \jay_a ^{(1)}(\s) \nn \\
& - 2 \int \diff \s \left( \jay_a ^ \mu \, \pi _\mu 
	- i \, \jay_{a \a} {} ^A \, \bt_{A \da} \, \pi^{\da \a} 
	- i \, \jay_{a \, A \da} \, \pi ^{\da \a} \, \theta _\a {} ^A \right)
	\mathrm{P}^{IJ} \, \frac{\delta ^2}{\delta n^I \delta n^J} \, .
\end{align}
Concretely, applying the above generators to the super Wilson loop gives
\begin{align*}
\J_a^{(1)} \left \langle \mathcal{W}(\gamma) \right \rangle 
	= \frac{1}{8  \pi^2} \big( \la \, \mathcal{Q}_a ^{(1)} 
	+ \mathcal{O} \big(\sqrt{\la} \big) \big) 
	\left \langle \mathcal{W}(\gamma) \right \rangle \, .
\end{align*}
In order to further discuss the generators, we consider the level-1 momentum generator 
$\mathrm{P}^{(1) \, \mu}$. Explicitly, we have
\begin{align}
\mathrm{P}^{(1) \, \mu} &= 2 \int \diff \s_1 \, \diff \s_2 \, \varepsilon_{21} \, 
	\left[ \left( m^{\mu \nu} (\s_1) - d (\s_1 ) \eta^{\mu \nu} \right) p_\nu (\s_2)
	+ \ft{i}{4} q_A {} ^\a (\s_1) \bs^\mu _{\a \da} \widebar{q} ^{\, \da A} (\s_2) \right] \nn \\
& - 2 \int _0 ^L \diff \s \, \pi ^\mu \, \mathrm{P}^{IJ} 
	\frac{\delta^2 }{\delta n^I  \delta n^J }  
	\label{P1Ferm}  \\
& - \frac{\la}{2 \pi^2} \int _0 ^L \diff \s \, 
	\Big \lbrace \ddot{\pi}^\mu + \left( \dot{\pi}^2 - \dot{n}^2 \right) \pi ^\mu 
	+ 4i \tr \big(\dt \rho \dbt \s ^\mu \big) 
	- 3i \tr \big( \dt \dbt \big( \dot{\pi} \bs ^\mu \pi 
	-  \pi \bs ^\mu \dot{\pi}  \big) \big) \nn \\
& \qquad \qquad - \tr \big(12 \, \dbt \pi \dt \dbt \pi \dt 
	+ 2i \big(\ddot{\theta} \dbt -  \dt \ddot{\btheta} \big) \pi 
	- 4i \big(\Theta_{(2)} \dbt - \dt \bTheta_{(2)} 
	+ \dt \rho \dbt \big) \pi  \big) \pi ^\mu \Big \rbrace \, . \nn
\end{align}
The expressions for the variables $\Theta_{(2)}$ and $\rho$ can be found in equations \eqn{theta2} and \eqn{rhoAB}. Let us compare this generator to the level-1 momentum generator of the Yangian 
$\mathrm{Y}[\alg{so}(2,4)]$ found for the minimal surfaces in $\AdS_5$. From equation 
\eqref{J1Bos}, we infer that it is given by 
\begin{align}
\mathrm{P} ^{(1) \, \mu} _{\AdS} &=  \! \int \diff \s_1 \, \diff \s_2 \, \varepsilon_{21}  
	\left( m^{\mu \nu} (\s_1) - d (\s_1 ) \eta^{\mu \nu} \right) p_\nu (\s_2)
	- \frac{\lambda}{4 \pi ^2} \int _0 ^L \! \diff \s 
	\left( \dx^\mu \, \ddx^2 + \dddot{x}^\mu \right) \, .
\end{align}
The difference in the normalization of the two generators stems from the use of a different matrix representation%
\footnote{For the discussion in chapter \ref{chap:MinSurf}, it is preferable to employ a matrix representation that can easily be adapted to a different dimension of the space or signature of the metric. In the present discussion, we have employed a representation that fits well for the discussion of the superconformal algebra.}
for the conformal algebra in chapter \ref{chap:MinSurf} and the corresponding subalgebra of the superconformal algebra in the present chapter, cf.\ appendix \ref{app:u224}. This leads to a difference in the metric used to raise and lower the group indices (compare equations \eqref{eqn:Metric} and \eqref{Metric_Str}), which accounts for the different normalization. 

If we set the fermionic variables zero and take $n^I$ to be constant, the full generator will reduce to the AdS-generator above. This shows that the coefficient of the local term of the bosonic Yangian symmetry generator constructed in reference \cite{Muller:2013rta} is not altered by the inclusion of the fermionic degrees of freedom, which was considered to possibly lead a matching of the coefficients of the generators obtained at weak and strong coupling in reference \cite{Muller:2013rta}. In fact, this finding can be understood from noting that the full superspace calculation reduces to the AdS-calculation at every step, when the fermionic degrees of freedom are set to zero. 

However, if we do not restrict $n^I$ to be constant, the local term for the generator 
$\mathrm{P}^{(1) \, \mu}$ will be different from the one obtained from the AdS-calculation and contain a structurally new contribution involving two functional derivatives acting on the same point of the loop. The difference between the generators arises from employing the Virasoro constraints to determine the coefficient $y_{(3)}(\s)$. The Virasoro constraints include contributions from both Anti-de Sitter space and the sphere and hence the coefficient $y_{(3)}(\s)$ involves the terms 
$N_{(1)}^2$ and $\dot{n}^2$.  

It is interesting to compare the level-1 momentum generator obtained here to the one obtained for the weak coupling side in reference \cite{Beisert:2015uda}. The bi-local contributions follow from the same typical structure of a level-1 Yangian generator, but the local terms are derived differently. Recall that the comparison of the local terms at Gra{\ss}mann order zero showed that the structure of the local terms is the same while the relative factor between the bi-local and local contribution differs. Unfortunately, this structure seems not to prevail to the full superspace result as the comparison between the generator \eqref{P1Ferm} and the so-called remainder function for the respective generator in reference \cite{Beisert:2015uda} shows.  

In addition to the level-1 Yangian $\mathrm{Y}[\alg{psu}(2,2 \vert 4)]$-generators, the level-1 recurrence of the hypercharge generator $\mathrm{B}^{(0)}$ was observed to be a symmetry of the Wilson loop in superspace in the weak coupling analysis of reference \cite{Beisert:2015uda}. We have constructed corresponding conserved charges in all higher levels in chapter 
\ref{chap:Semi-SSM} and the evaluation of the $C$-part of the conserved charge $Q^{(1)}$ leads to the bonus symmetry generator%
\footnote{The result in reference \cite{Munkler:2015xqa}, where the strong-coupling bonus symmetry was shown, contains an error, which has been corrected here. }
\begin{align}
\mathrm{B}^{(1)} &= \frac{1}{4} \int \diff \s_1 \, \diff \s_2 \, \varepsilon_{21} 
	\left( q_A {}^\a (\s_1) \, s_\a {} ^A (\s_2) + 
		\widebar{q}^{\, \da A}(\s_1) \, \widebar{s}_{A \da}(\s_2) \right) \nn \\
&  + 2i \int \diff \s \, \tr \left[ \theta \bt \pi \right] \mathrm{P}^{IJ} 
	\frac{\delta^2}{\delta n^I \delta n^J }	\label{B1}  \\
& - \frac{\la}{2 \pi^2} \int \diff \s \Big \lbrace 
	3 \tr \left[ \big( \dt \bt - \theta \dbt \big) \pi \dt \dbt \pi 
	+ 2 \theta \bt \big(\dot{\pi} \dt \dbt \pi 
		- \pi \dt \dbt \dot{\pi} \big) \right] \nn \\
& \qquad + 2 i \tr \left[ \big( \theta \rho \dbt - \dt \rho \bt \big) \pi \right] 
	- i \tr \left[ \theta \btheta \pi \right] \left( \dot{n}^2 - \dot{\pi}^2 \right) 
	+ 8 \tr \left[\theta \btheta \eps \big( \dt \rho \dbt \big)^T \eps \right] \nn \\
& \qquad + i \tr \left[ \theta \btheta \pi \right] 
	\tr \left[\big( 12 \dt \dbt \pi \dt \dbt
		+ 2i \big( \dt \ddot{\bt} - \ddot{\theta} \dt \big) 
		+ 4i \big( \Theta_{(2)} \dbt - \dt \bTheta_{(2)} 
		+ \dt \rho \dbt \big) \big) \pi \right]  \Big \rbrace \nn \, .	
\end{align}
In the discussion in chapter \ref{chap:CarryOver}, we have noted that the Yangian level-1 generators over the superconformal algebra are cyclic due to the vanishing of the dual Coxeter number, which implies that the contraction 
$\mathbf{f} \indices{_a ^{cb}} \, \mathbf{f} \indices{_{bc} ^d}$ vanishes over 
$\alg{psu}(2,2\vert 4)$. Indeed, this argument ensures the cyclicity of the level-1 
$\mathrm{Y}[\alg{psu}(2,2\vert 4)]$-generators. For the above level-1 hypercharge generator, we note that the respective contraction in $\alg{u}(2,2 \vert 4)$ leads to the finding
\begin{align}
\mathrm{B}^{(1)} - \widetilde{\mathrm{B}} ^{(1)} \, \propto
	\int \limits _0 ^\Delta \diff \s \, c(\s) \, .
\end{align}
Here, the tilded generator again denotes the hypercharge generator with starting point $x(\Delta)$ rather than $x(0)$, as in the discussion of chapter \ref{chap:CarryOver}. The cyclicity then follows from the identical vanishing of $c(\s)$. 

We have seen in chapter \ref{chap:Semi-SSM} that the conserved charges of the model allow for the construction of the higher-level recurrences of the hypercharge generator in any higher level of the Yangian. While deriving these generators would likely require to extend the expansion of the minimal surface solution to higher orders in the $\tau$-expansion, the procedure described above for the level-1 generator should extend to these levels as well. In contrast to the Yangian generators in 
$\mathrm{Y}[\alg{psu}(2,2\vert 4)]$, the higher-level recurrences of the hypercharge generator cannot be obtained from commutators of the lower-level generators. Each of these infinitely many generators could thus describe an independent constraint, which the minimal surfaces or Wilson loop in superspace would have to obey.

\newpage

\clearpage 	
\chapter{Conclusion and Outlook} 
\label{chap:conclusion}

We have studied hidden symmetries in $\Nfour$ supersymmetric Yang--Mills theory or the corresponding string theory in $\AdS_5$. Here, we have focused on the Maldacena--Wilson loop or its strong-coupling description in terms of minimal surfaces. These are described by classical string theory and we employed the integrability of this theory in order to construct symmetries of the Maldacena--Wilson loop at strong coupling. One goal was to extend the results of reference \cite{Muller:2013rta} to the superspace Wilson loop and to establish the connection to the lambda deformations described by Kruczenski and collaborators \cite{Ishizeki:2011bf,Kruczenski:2013bsa}. 

The study of the symmetries of minimal surfaces in $\AdS_5$ can be generalized without creating additional complications to the same study in symmetric spaces and we have discussed the symmetries of symmetric space models in detail in chapter \ref{chap:SSM}. Here, we have in particular clarified the relation between the Yangian symmetry and the lambda deformation, which was shown to generalize to the master symmetry. This symmetry has an important role within the symmetry structure of these models: It acts as a raising operator on the infinite tower of conserved charges, thus generating the spectral parameter, and can be employed to construct all symmetry variations from the underlying global symmetry of the model. Additionally, we have calculated the symmetry algebra of the symmetry variations as well as the classical Poisson algebra of the conserved charges. The latter calculation closes a gap in the existing literature by extending the result of reference \cite{MacKay:1992he} to symmetric space models. 

With this foundation, we have turned to the specific case of minimal surfaces ending in the conformal boundary of $\AdS_5$, for which we showed that the renormalization of the area is compatible with the symmetries of the area functional. While we mostly considered smooth contours, we have taken a first step toward the study of the master symmetry for light-like polygons by constructing the master symmetry transformation of the four-cusp solution. For generic, smooth boundary curves, we have derived the variations arising from the master symmetry and the variations associated to the Yangian symmetry charges. Peculiarly, we found that the invariance under the first master symmetry variation is easily seen without referring to the underlying integrability of the model. This stands in a surprising contrast to its fundamental role in the symmetry structure of the model. 

The form of the variation suggests a natural way to extend the variation to a coupling-dependent variation, which is (somewhat trivially) a symmetry at any value of the coupling constant. This explains Dekel's finding \cite{Dekel:2015bla} that the transformation obtained at strong coupling does not constitute a symmetry at weak coupling, which is not related to whether or not we consider an extension into superspace in this case. Furthermore, we have discussed several attempts to transfer the strong-coupling Yangian symmetry to weak coupling. We noted that the key issue here is the cyclicity of the level-one Yangian generators. Since this problem arises when we read off the generator from the Yangian Ward identity at strong coupling, and is not present for the Yangian-like symmetry variations, it is natural to attempt to carry over these variations to weak coupling. 
A different attempt utilized the finding of reference \cite{Chicherin:2017cns} that the Yangian generators can be rendered cyclic by the addition of an appropriate local term. Unfortunately, both approaches are unsuccessful in establishing a symmetry at weak coupling and we conclude that a bosonic Yangian symmetry of the Maldacena--Wilson loop seems not to be present.   

Thus concluding the analysis of the hidden symmetries of the Maldacena--Wilson loop, we turned to the discussion of Wilson loops in superspace. Substantial evidence for the Yangian invariance of their one-loop expectation value had been provided in reference \cite{Muller:2013rta} and the Yangian invariance was indeed later established in reference \cite{Beisert:2015uda}.  
At strong coupling, following reference \cite{Ooguri:2000ps}, we describe the Wilson loop in superspace by imposing boundary conditions in the conformal boundary of the superspace appearing in the description of type IIB superstrings in $\AdS_5 \times \mathrm{S}^5$. This is the natural supersymmetric extension of the strong-coupling description of the Maldcena--Wilson loop. 

The symmetries of the theory describing the minimal surface were discussed in chapter \ref{chap:Semi-SSM}. In addition to reviewing the construction of an infinite tower of conserved charges following from the integrability of the model, we described the master symmetry for semisymmetric space models and showed that that the model contains an infinite tower of bonus symmetries. These had been shown to be present in all odd levels of the Yangian 
\cite{Berkovits:2011kn}. Here, we have seen that they can indeed be found in all levels except the level zero, which contains the underlying global symmetry. 

The Yangian symmetry of the superspace Wilson loop at strong coupling followed from extending the analysis of reference \cite{Muller:2013rta} to the full superspace appearing in the description of type IIB superstring theory in $\AdS_5 \times \mathrm{S}^5$. We constructed the expansion of the minimal surface around the boundary curve and derived the level-one Yangian as well as the level-one bonus symmetry generator from the evaluation of the corresponding conserved charges. Together with the weak-coupling analysis of reference \cite{Beisert:2015uda}, this completes the study of the Yangiain symmetry of the Wilson loop in superspace which is thus established both at weak and at strong coupling. The comparison with the generators derived there shows that the local term depends on the coupling constant in a non-trivial way. 

There are several interesting directions in which the work described in this thesis could be extended. A natural question concerns the relation between the master symmetry and dual (super-)conformal symmetry, which has recently been studied for minimal surfaces in 
$\AdS_3$ \cite{Dekel:2016oot}. The latter symmetry can be seen to originate from the self-T-duality of the (super-)string theory \cite{Berkovits:2008ic,Beisert:2008iq,Beisert:2009cs}. It would thus be interesting to work out the relation between the T-duality and the master symmetry. Moreover, it could be illuminating to discuss the relation between the master symmetry and the B{\"a}cklund transformations considered in the literature on integrable models, cf.\ e.g.\ reference \cite{Arutyunov:2005nk}. 

Another interesting direction concerns the role of the master symmetry within the symmetry structure of symmetric space models in the case of closed strings. In this thesis, we have focused on minimal surfaces and only pointed out the differences that should appear when one works with closed strings. In particular, rather than the whole monodromy, only its eigenvalues are conserved charges in this case. It is natural to expect that the master symmetry would similarly act as a raising operator also on these charges, but this remains to be studied. It would also be interesting to see whether the presence of the bonus symmetries leads to novel conserved charges in this case as well. 

We have described the master symmetry for symmetric space and principal chiral models and have seen in chapter \ref{chap:Semi-SSM} that it can also be formulated for semisymmetric space models. It is not clear, however, that a symmetry of this kind can be constructed in any integrable theory. To gain further insight into this question, it would be interesting to study other integrable models such as the (chiral) Gross--Neveu or the Landau--Lifshitz model. The latter is particularly interesting since it appears on both sides of the AdS/CFT correspondence \cite{Kruczenski:2003gt}. 
Other interesting cases include the $\eta$-deformation \cite{Delduc:2013qra} of the 
$\AdS_5 \times \mathrm{S}^5$-superstring theory or the deformations of simpler, bosonic theories. The work of reference \cite{Delduc:2016ihq} on q-deformed symmetries for deformed principal chiral models could be an interesting starting point for such an analysis. 

In the application of the symmetries established here to Maldacena--Wilson loops or Wilson loops in superspace, a pressing question is if and how these symmetries can be applied in calculations. In order to make progress in this direction, it would be important to understand the structure of the symmetry invariants in order to elaborate how the symmetries constrain the functional form of the vacuum expectation value of the operator. 

While the question of the functional dependence on the contour data is certainly difficult for generic, smooth contours, the constraints should be more transparent in the case of light-like polygons, where the result only depends on the cusp points or the conformal cross-ratios formed from them. With the analysis of the four-cusp solution, we have taken a first step toward studying the master symmetry in the case of light-like polygons. The analysis of the four-cusp case suggests that the master symmetry indeed maps light-like polygons to such. However, since all light-like polygons with four cusps are related to each other by a conformal transformation, the four-cusp case is too simple to provide an interesting result. The relevant configuration would be the case of six cusps, where there is a collection of conformally distinct polygons, which can be described in terms of three conformal cross ratios. Unfortunately, since a six-cusp solution is not known, we cannot directly calculate the master symmetry transformation in order to see whether it acts on the cross-ratios in an interesting way. However, the methods described in reference \cite{Alday:2009yn} or the mathematically related work of reference \cite{Gaiotto:2008cd} could help to make progress in this direction. 

\begin{appendix}  

\chapter{Spinor Conventions}
\label{app:Spinor}
In this appendix, we introduce our conventions for spinors in four, six and ten dimensions and provide several technical identities which are helpful for the manipulations of their indices. The spinor conventions in four dimensions are applied in the superspace calculations of chapter \ref{chap:SurfaceSuperspace}, the conventions for six and ten dimensions are mainly needed in the discussion of the dimensional reduction from  $\mathcal{N}\!=1$ supersymmetric Yang-Mills theory in ten dimensions to $\mathcal{N}\!=4$ supersymmetric Yang-Mills theory in four dimensions, which is also used in our discussion of the Maldacena--Wilson loop. We note that the exposition provided here is based on the discussion given in the author's master's thesis \cite{Master}, which in turn employed the conventions of references \cite{Belitsky:2003sh} and \cite{Groeger}. However, the conventions have to be adapted from a mostly minus to a mostly plus metric and they are thus provided here again.  

\section{Four-dimensional Minkowski Space}
\label{Conventions}
We consider Minkowski space with the mostly plus metric $\eta = \mathrm{diag}(-1,1,1,1)$ and write Dirac spinors as
\begin{align*}
\Psi = \begin{pmatrix} \Psi_\a \\ \tilde{\Psi} ^\da \end{pmatrix} \, .
\label{DiracSpinor}
\end{align*}
The spinor indices $\a$ and $\da$ are raised and lowered according to the rules
\begin{align}
\la^\a &= \varepsilon ^{\a \b} \la _\b \, , & 
\la _\a &= \la ^\b \varepsilon_{\b \a} \, , &
\bar \la _\da &= \varepsilon_{\da \db} \bar \la ^\db \, , &
\bar \la ^\da &= \bar \la _\db \varepsilon ^{\db \da} \, , 
\end{align}
and the epsilon tensors are given by
\begin{align}
\varepsilon^{1 2} = \varepsilon_{1 2} = 1 \, , \quad \varepsilon^{\dot{1} \dot{2}} = \varepsilon_{\dot{1} \dot{2}} = -1 \quad \Rightarrow \quad 
	\varepsilon ^{\a \b} \varepsilon _{\g \b} = \delta ^\a _\g \, , \quad \varepsilon ^{\da \db} \varepsilon _{\dg \db} = \delta ^\da _\dg \, .
\end{align}
Moreover, we note the following conventions for sigma matrices:
\begin{equation}
\begin{alignedat}{2}
\left( \sigma ^\mu \right) ^{\da \a} &= \left( \unit_2 , \vec{\s} \right) ^{\da \a} \, , & \qquad \left( \bs ^\mu \right) _{\a \da} &= \left( \unit_2, -\vec{\s} \right) _{\a \da} \, , \\
\left( \sigma ^{\mu \nu} \right) \indices{_\a ^\b} &= \ihalf \left( \bs ^\mu \s ^\nu - \bs ^\nu \s ^\mu \right) \indices{_\a ^\b} \, , & \qquad \left( \bs ^{\mu \nu} \right)\indices{^\da _\db} &= \ihalf \left( \s ^\mu \bs ^\nu - \s ^\nu \bs ^\mu \right) \indices{^\da _\db} \, .
\end{alignedat}
\end{equation}
These matrices satisfy the following identities:
\begin{equation}
\begin{alignedat}{2}
\bs ^\mu _{\a \dg} \, \s ^{\nu \, \dg \b} + \bs ^\nu _{\a \dg} \, \s ^{\mu \, \dg \b} &= - 2\, \eta^{\mu \nu} \, \delta ^\b _\a \, , \qquad \qquad &  \s ^{\mu \, \da \a} \, \bs _{\mu \, \b \db} &=- 2 \, \delta ^\a _\b \, \delta ^\da _\db \, ,\\
\s ^{\mu \, \da \g} \, \bs ^\nu _{\g \db} +  \s ^{\nu \, \da \g} \, \bs ^\mu _{\g \db}  &= - 2\, \eta^{\mu \nu} \, \delta ^\da _\db \, , & \bs ^\mu _{\a \da} \, \s ^{\nu \, \da \a} &= - 2 \, \eta^{\mu \nu} \, ,
\end{alignedat}
\end{equation}
as well as the trace-identities
\begin{equation}
\begin{aligned}
\half \, \Tr(\widebar \sigma^{\mu}\, \sigma^{\nu}\, \widebar \sigma^{\rho}\,\sigma^{\kappa}) &=
\eta^{\mu\nu}\, \eta^{\rho\kappa} + \eta^{\nu\rho}\, \eta^{\mu\kappa} -
\eta^{\mu\rho}\, \eta^{\nu\kappa} -i\,\epsilon^{\mu\nu\rho\kappa} \, , \\
\half \, \Tr(\sigma^{\mu}\, \widebar \sigma^{\nu}\, \sigma^{\rho}\, \widebar\sigma^{\kappa}) &=
\eta^{\mu\nu}\, \eta^{\rho\kappa} + \eta^{\nu\rho}\, \eta^{\mu\kappa} -
\eta^{\mu\rho}\, \eta^{\nu\kappa} +i\,\epsilon^{\mu\nu\rho\kappa} \,.
\end{aligned}
\end{equation}
From the identities given above, we can infer that the matrices 
\begin{align}
\gamma ^\mu = \begin{pmatrix} 0 & \widebar{\sigma}^\mu _{\a \db} \\ \sigma^{\mu \, \da \b} & 0 \end{pmatrix} \, , \qquad \gamma ^5 = i \gamma ^0 \gamma ^1 \gamma ^2 \gamma ^3 = \begin{pmatrix} \unit & 0 \\ 0 & - \unit \end{pmatrix} \, ,
\end{align}
provide a representation of the Clifford algebra in four dimensions, i.e.\ they satisfy
\begin{align}
\left \lbrace \gamma ^\mu , \gamma^\nu \right \rbrace = - 2 \eta ^{\mu \nu} \unit \, .
\end{align}
We moreover introduce the charge conjugation matrix $C_{(1,3)} = i \gamma ^2$, which satisfies the identities
\begin{align}
C_{(1,3)} ^\dag C_{(1,3)} &= \unit \, , &  
C_{(1,3)} ^\dag \gamma ^\mu C_{(1,3)} &= - \left( \gamma ^\mu \right) ^\ast . 
\label{DefC}
\end{align}
These identities ensure that the Majorana condition
\begin{align}
\Psi ^{(c)}  = C_{(1,3)} \Psi ^\ast = \Psi
\end{align}
is Lorentz-invariant. For the spinors given in equation \eqref{DiracSpinor}, the Majorana condition implies $\tilde{\Psi}_\da = (\Psi_\a )^\ast$. We note that in four dimensions, the Majorana condition cannot be enforced for left- or right-handed Weyl spinors, which take the form  
\begin{align*}
\begin{pmatrix} \Psi_\a \\ 0 \end{pmatrix} &= P_L \Psi = 
	\half \left( \unit + \gamma ^5 \right) \Psi \, , &
\begin{pmatrix} 0 \\ \tilde{\Psi} ^\da \end{pmatrix} &= P_R \Psi = 
	\half \left( \unit - \gamma ^5 \right) \Psi \, .
\end{align*}
In fact, the Weyl and Majorana conditions are known to be compatible only if the dimension of the underlying spacetime satisfies $d \equiv 2 \, (\mathrm{mod} \, 8)$. 

We assign bispinors to four-vectors and two-tensors by 
\begin{align}
\begin{aligned}
x^{\da \a} &= \s ^{\mu \, \da \a} \, x_\mu \, , \\
B \indices{_\a ^\b} &= B_{\mu \nu} \, \left( \sigma ^{\mu \nu} \right) \indices{_\a ^\b} \, ,
\end{aligned}
&&
\begin{aligned}
x_{\a \da} &= \bs ^\mu _{\a \da} \, x_\mu \, , \\
\bar{B}\indices{^\da _\db} &= B_{\mu \nu} \,  
	\left( \bs ^{\mu \nu} \right) \indices{^\da _\db} \, ,
\end{aligned}
\end{align}
and note the identities
\begin{align}
x^{\da \a} \, x_{\a \db} &= -x^2 \, \delta ^\da _\db \, , & 
x_{\a \da} \, x^{\da \b} &= -x^2 \delta ^\b _\a \, , &
x_{\a \da} \, y^{\da \a} &= - 2 \, x y \, .
\end{align}
In the calculations in chapter \ref{chap:SurfaceSuperspace}, we handle many fermionic variables with different indices and it is convenient to assign canonical index positions so that we can omit writing out the indices. The canonical index positions are given by:
\begin{alignat}{3}
&{\Theta_\a}^A \, , \bTheta_{A \da}  & \quad &\text{for variables conjugate to} \quad {Q_A}^\a \, , \, \Qb^{\da A} \, , \nn \\
&{\vart_A}^\a \, , \bvart^{\da A} & \quad &\text{for variables conjugate to} \quad {S_\a}^A \, , \, \Sb_{A \da} \, \,. \nn
\end{alignat}
Whenever a spinor index is raised or lowered into a different position, we spell out the indices explicitly. In a matrix product, the indices of bispinors are positioned accordingly. We provide the following examples for clarity:
\begin{align*}
\tr \big( \bTheta \, \bvart + \vart \Theta \big) &= 
	\bTheta_{A \da} \bvart^{\da A} + \vart_A {} ^\a \Theta_\a {} ^A \, , &  
\big[ \dbt \pi \dt \dbt  \pi \big] _A {} ^\a &=  \dbt _{A \db} \, \pi ^{\db \b} \,
	 \dt_\b {}^B \, \dbt_{B \dg} \, \pi^{\dg \a} \,.
\end{align*}

\section{Six-dimensional Euclidean Space}
We consider Euclidean space $\mathbb{R}^6$ with the metric $\delta_{IJ}$. The gamma matrices for the six-dimensional Euclidean space can be written as
\begin{align}
\hat{\gamma} ^I &= \begin{pmatrix} 0 & \Sigma^{I\, A B} \\ 
	\BSi ^I _{A B} & 0  \end{pmatrix} \, , &
\hat{\gamma} ^7 &= i \prod \limits _{I=1} ^6  \hat{\gamma}^I = 
	\begin{pmatrix} \unit_4 & 0 \\ 0 & - \unit_4  \end{pmatrix} .
\end{align} 
Here, $I$ runs from $1$ to $6$ while the upper or lower indices $A, B$ take values in $\left\lbrace 1, 2, 3, 4 \right\rbrace$. 
The sigma matrices are defined by
\begin{align}
(\Sigma^{1 \, A B}, \ldots , \Sigma^{6 \, A B} ) &= 
	(\eta_{1 \, A B}, \eta_{2 \, A B}, \eta_{3 \, A B}, i \weta_{1 \, A B}, 
	i \weta_{2 \, A B}, i \weta_{3 \, A B}) \, ,  \\
(\BSi^1 _{ A B}, \ldots , \BSi^6 _{ A B} ) &= 
	(\eta_{1 \, A B}, \eta_{2 \, A B}, \eta_{3 \, A B}, - i \weta_{1 \, A B}, 
	- i \weta_{2 \, A B},- i \weta_{3 \, A B}) \, , 
\end{align}
where $\eta_{i \, A B}$ and $\weta_{i \, A B}$ denote the 't Hooft symbols
\begin{align}
\eta_{i \, A B} :&= \epsilon_{i A B 4} + \delta_{i A} \delta_{4 B} - \delta_{i B} \delta_{4 A} \, ,  \qquad \widebar{\eta}_{i \, A B} := \epsilon_{i A B 4} - \delta_{i A} \delta_{4 B} + \delta_{i B} \delta_{4 A} \, .
\end{align}
Here, $\eps_{A B C D}$ denotes the four-dimensional epsilon tensor normalized by 
\begin{align*}
\eps_{1 2 3 4} = \eps^{1 2 3 4} = 1 \, . 
\end{align*}
Note that the sigma matrices are defined antisymmetric. The Clifford algebra relation 
$\hat{\gamma}^I \hat{\gamma}^J + \hat{\gamma}^J \hat{\gamma}^I = 
- 2 \delta ^{I J} \, \unit $ follows from the relations
\begin{align}
\begin{aligned}
\Sigma ^{I \, A B} \, \BSi ^J _{B C} + \Sigma ^{J \, A B} \, \BSi^I _{B C} &= 
	- 2 \, \delta^{I J} \, \delta ^A _C \, , \\ 
\BSi^I _{A B} \, \Sigma ^{J \, B C}  + \BSi^J _{A B} \,  \Sigma ^{I \, B C} &= 
	- 2 \, \delta^{I J} \, \delta ^C _A \, . 
\end{aligned}	
\label{6DClifford}
\end{align}
Proving the above involves a straightforward but lengthy calculation. 

Similar to the four-dimensional case, we assign a $(4 \times 4)$-matrix to a vector $X_I$ by the prescription
\begin{align}
X^{A B} &= \ft{1}{\sqrt{2}} \Sigma ^{I \, A B} X ^I \, , &
\widebar{X}_{A B} &= \ft{1}{\sqrt{2}} \BSi ^I _{A B} X^I \, .
\end{align}
These matrices are related by
\begin{align*}
\widebar{X}_{A B} &= \half \eps_{A B C D} \, X^{C D} \, , & 
X^{A B} &= \half \eps^{A B C D} \, \widebar{X}_{C D}  \, ,
\end{align*} 
and for the contraction of the $\grp{SU}(4)$ indices, we note
\begin{align}
X^{A B} \widebar{Y}_{A B} = \widebar{X}_{A B} Y^{A B} = 2 X^I Y^{I} \, .
\end{align} 

\section{Ten-dimensional Minkowski Space}
\label{sec:10dspinor}
We consider $\mathbb{R}^{(1,9)}$ with the metric 
$\eta^{m n} = \mathrm{diag} (-1 , 1 , \ldots , 1 )$. For the discussion of the dimensional reduction it is convenient to construct a representation of the ten-dimensional Clifford algebra from the four- and six-dimensional Dirac matrices introduced above. This can be achieved e.g.\ by defining
\begin{align*}
\Gamma ^m = \begin{cases} \unit_8 \otimes \gamma ^\mu & \text{for} \; M = \mu \in \lbrace 0 , 1 , 2 , 3 \rbrace  \\
\hat{\gamma}^I \otimes \gamma ^5 & \text{for} \; M = I + 3 \in \lbrace 4, 5, 6, 7, 8, 9 \rbrace  \end{cases} \: , \qquad \Gamma ^{11} = \hat{\gamma}^7 \otimes \gamma^5 \, .
\end{align*} 
Using the Clifford relations in four and six dimensions, one then shows that the above matrices satisfy the ten-dimensional Clifford algebra
$\lbrace \Gamma ^m \, ,  \, \Gamma ^n \rbrace = - 2 \, \eta ^{m n} \unit$. 
Similarly, we can write the spinors in $\mathbb{C}^{32} \simeq \mathbb{C}^8 \otimes \mathbb{C}^4$
in the form of a tensors product as
\begin{align}
\xi = \begin{pmatrix} \xi^A \\ \xi_A  \end{pmatrix} \otimes \begin{pmatrix} \xi_\a \\ \tilde{\xi}^\da  \end{pmatrix} \, . \label{10dspinor}
\end{align}
Note however, that this is not the most general form of an element of $\mathbb{C}^{32}$, as not every element of a tensor product space can be written as the tensor product of two vectors. To avoid writing linear combinations of vectors, we will write the spinors in the following form:
\begin{align*}
\xi = \left( \xi_\a ^A \, , \, \tilde{\xi}^{A\, \da} \, , \, \xi_{A \, \a} \, , \, \tilde{\xi}^\da _A \right) ^T \, .
\end{align*}
How to multiply the gamma matrices with these spinors can be read off from \eqn{10dspinor} and from the position of the indices, see reference \cite{Master} for explicit formulas.   

The ten-dimensional $\mathcal{N}\!=1$ supersymmetric Yang-Mills theory employs Majorana-Weyl spinors in ten dimensions. A left-handed Weyl spinor satisfies the condition
\begin{align*}
\Gamma ^{11} \xi = \left( \xi_\a ^A \, , \,- \tilde{\xi}^{A\, \da} \, , \,- \xi_{A \, \a} \, , \, \tilde{\xi}^\da _A \right) ^T = \xi \: ,
\end{align*}
and thus it has the form
\begin{align*}
\xi = \left( \xi_\a ^A \, , \, 0 \, , \, 0 \, , \, \tilde{\xi}^\da _A \right) ^T  .
\end{align*}
For the Majorana condition, we employ the charge conjugation matrix
\begin{align}
C_{1, \, 9} = \begin{pmatrix} 0 & \unit_4 \\ \unit_4 & 0  \end{pmatrix} 
	\otimes C_{1, \, 3} \, .  
\end{align}
A left-handed Majorana-Weyl spinor then satisfies the condition
\begin{align}
C_{1, \, 9}\, \xi ^\ast = \left( \eps _{\da \db}\, \tilde{\xi}^\db _A \, , \, 0 \, , \, 0\, , \, \eps^{\a \b} \, \xi^A _\b \right) ^T = \xi \, ,
\end{align}
which imposes the restrictions $( \tilde{\xi}_{A \, \da} ) ^\ast = \xi ^A _\a $ and 
$( \xi^{A \, \a} ) ^\ast = \tilde{\xi}^\da _A$. The 32 complex degrees of freedom of a Dirac spinor in ten dimensions are hence reduced to 16 real  degrees of freedom for a Majorana-Weyl spinor in ten dimensions. The conjugate spinor is as usual defined by $\widebar{\xi}= \xi^\dag \Gamma ^0$ and the relations needed for the dimensional reduction can be obtained from the explicit expressions given above for the Dirac matrices in four, six and ten dimensions. More details can be found in references \cite{Groeger,Master}. 

\chapter{The Fundamental Representation of
\texorpdfstring{$\alg{u}(2,2 \vert 4)$}{u(2,2;4)}} 
\label{app:u224}

In this appendix, we discuss the fundamental representation of the Lie superalgebra 
$\alg{u}(2,2 \vert 4)$. We introduce a specific set of generators and collect useful formulae for the calculations of chapters \ref{chap:Semi-SSM} and \ref{chap:SurfaceSuperspace}. 

Following the conventions established in reference \cite{Drummond:2009fd}, we choose the following basis elements for the Lie superalgebra $\alg{u}(2,2 \vert 4)$: 
\begin{align}
\left( \begin{array}{c c c} 
	0 & P_\mu & Q _A {} ^\a \\
	K_\mu & 0 & \widebar{S}_{A \da} \\ 
	S _\a {} ^A & \widebar{Q}^{\da A} & R \indices{^A _B}
\end{array} \right) = 
\left( \begin{array}{c c c} 
	0 & i \bs_\mu & 2 \, E \indices{^\a _A} \\
	i \s_\mu & 0 & 2 \, E_{\da A} \\ 
	- 2 \, E \indices{^A _\a} & -2 \, E^{A \da} 
	& 4 \, E \indices{^A _B} - \delta ^A _B \, \unit_4
\end{array} \right) \, .
\end{align}
This equation is to be read as
\begin{align}
P_\mu = \left( \begin{array}{c c c} 
	0 & i \bs_\mu & 0 \\
	0 & 0 & 0 \\ 
	0 & 0 & 0
\end{array} \right)
\end{align}
and similarly for the other generators. The notation $E \indices{^A _B}$ denotes a matrix with entry 1 in the position $(A,B)$ and all other entries vanishing. The remaining generators of  $\mathfrak{u}(2,2\vert4)$ are given by
\begin{align}
M_{\mu \nu}  &= -\frac{i}{2} \left( \begin{array}{c c c} 
	\s_{\mu \nu} & 0 & 0 \\
	0 & \bs_{\mu \nu}  & 0 \\ 
	0 & 0 & 0
	\end{array} \right) \, , &
C &=  + \frac{1}{2} \left( \begin{array}{c c} 
	\unit_4  & 0 \\ 
	0 & \unit_4  \\
	\end{array} \right) \, , \\	
D  &= + \frac{1}{2} \left( \begin{array}{c c c} 
	\unit_2  & 0 & 0 \\ 
	0 & -\unit_2 & 0 \\
	0 & 0 & 0 
	\end{array} \right) \, , &
B &=  - \frac{1}{2} \left( \begin{array}{c c} 
	0  & 0 \\ 
	0 & \unit_4  \\
	\end{array} \right)  \,.
\end{align}

We note that the fermionic generators do not satisfy the reality constraint. They are constructed in such a way that the linear combinations $\Theta _\a {} ^A Q_A {}^\a + \bTheta _{A \da} \bar{Q}^{\da A}$ and $\vart _A {} ^\a S_\a  {}^A  + \bvart ^{\da A} \Sb_{A \da}$ are elements of 
$\alg{u}(2,2 \vert 4)$, provided that 
\begin{align}
\bTheta _{A \da} &= \left( \Theta _\a {} ^A  \right) ^\ast \, , &
\bvart ^{\da A} = \left( \vart_A {}^\a \right) ^\ast \, ,
\end{align}
such that $\Theta$ and $\bTheta$ or $\vart$ and $\bvart$ form the components of a left- or right-handed ten-dimensional Majorana--Weyl spinor. The R-symmetry generators $R\indices{^A _B}$ do not satisfy the reality condition as well, only appropriate linear combinations do. These are found by constructing a representation of $\alg{su}(4)$, which we may e.g.\ obtain as in reference \cite{Arutyunov:2009ga} from the representation of the $\grp{SO}(5)$ Clifford algebra, which is given by
\begin{equation}
\begin{alignedat}{3}
\gamma ^1 &= \begin{pmatrix} 0 & -i \sigma^2 \\ i \sigma^2 & 0  \end{pmatrix} \, , & \qquad \gamma ^2 &= \begin{pmatrix} 0 & i \sigma^1 \\ -i \sigma^1 & 0  \end{pmatrix} \, , & \qquad \gamma ^3 &= \begin{pmatrix} 0 & \unit_2 \\ \unit_2 & 0  \end{pmatrix} \, , \\
\gamma ^4 &= \begin{pmatrix} 0 & -i \sigma^3 \\ i \sigma^3 & 0  \end{pmatrix} \, , & \qquad \gamma ^5 &= \begin{pmatrix} \unit_2 & 0 \\ 0 & - \unit_2  \end{pmatrix} \, .
\end{alignedat}
\end{equation}
These matrices satisfy the Clifford relation 
$\lbrace \gamma ^r ,  \gamma ^s \rbrace = 2 \, \delta^{r s} \, \unit_4$
and we employ them to construct a set of matrices $\gamma^{I J} = - \gamma^{J I}$, which form a basis of $\mathfrak{su}(4) \simeq \mathfrak{so}(6)$,
\begin{align}
\begin{aligned}
\gamma^{r s} &= \quarter \, \left[ \gamma ^r  ,  \gamma ^s \right] \, , \\
\gamma^{r 6} &= \quarter \, \left[ \gamma ^r  ,  \gamma ^5 \right] \, ,
\end{aligned}
&&
\begin{aligned}
\gamma^{r 5} &= \ihalf \gamma^r \, , \\
\gamma^{5 6} &= - \ihalf \gamma^5 \, ,
\end{aligned}
\end{align}
where $r,s$ take values in $\lbrace 1 \, , \ldots , 4 \rbrace$. For these matrices we note the commutation relations
\begin{align}
\left[ \gamma^{I J} \, , \, \gamma^{K L} \right] &= \delta ^{I L} \, \gamma^{J K} + \delta ^{J K} \, \gamma^{I L} - \delta ^{I K} \,  \gamma^{J L} - \delta ^{J L} \, \gamma^{I K} \, .
\end{align}
The $\mathfrak{su}(4)$ sub-algebra of the superconformal algebra is spanned by the matrices
\begin{align}
\Gamma ^{IJ} = \begin{pmatrix} 0 & 0 \\  0 & \gamma ^{I J} \end{pmatrix} \, ,
\end{align}
which are related to the generators $R\indices{^A _B}$ by 
$\Gamma^{IJ} = \quarter \, \left( \gamma^{IJ} \right){ } \indices{_A ^B} R \indices{^A _B} $.
The use of the generators $R\indices{^A _B}$ is advantageous when calculating commutation relations. 

We have picked the above generators in such a way that they satisfy the commutation relations
\begin{align*}
\left[ T_a , T_b \right] = \mathbf{f} \indices{_{ba} ^c} \, T_c 
	= f \indices{_{ab} ^c} \, T_c \, , 
\end{align*}
where $\mathbf{f} \indices{_{ab} ^c}$ denote the structure constants of the generators introduced in section \ref{sec:Int_Symm}, 
$\left[ t_a , t_b \right] = \mathbf{f} \indices{_{ab} ^c} \, t_c \, $. We collect the commutation relations here once more. 
The commutators with the generators $M$ and $R$ only depend on the set of indices and their position:
\begin{align}
\Big[ M \indices{_\a ^\b} \, , \, J_\g \Big] 
	&= 2i \, \delta ^\b _\g \, J_\a - i \delta ^\b _\a \, J_\g \, , & 
\Big[ M \indices{_\a ^\b} \, , \, J^\g \Big] 
	&= - 2i \, \delta ^\g _\a \, J^\b + i \delta ^\b _\a \, J^\g \, , \\
\Big[ \widebar{M} \, \indices{^\da _\db} \, , \, J^\dg \Big] 
	&= 2i \, \delta ^\dg _\db \, J^\da - i \delta ^\da _\db \, J^\dg \, , & 
\Big[ \widebar{M} \, \indices{^\da _\db} \, , \, J_\dg \Big] 
	&= - 2i \, \delta ^\da _\dg \, J_\db + i \delta ^\da _\db \, J_\dg \, ,\\
\Big[ R \indices{^A _B} \, , \, J^C \Big] 
	&= 4 \, \delta ^C _B \, J^A - \delta ^A _B \, J^C \, , & 
\Big[ R \indices{^A _B} \, , \, J_C \Big] 
	&= - 4 \, \delta ^A _C \, J_B + \delta ^A _B \, J_C \, .
\end{align}
The commutators with the dilatation $D$ and hypercharge generator $B$ are specified by a weight $\Delta(T_a)$ or a hypercharge $\mathrm{hyp}(T_a)$,
\begin{align*}
\left[ D, T_a \right] &= \Delta(T_a) \, T_a \, , & 
\left[ B, T_a \right] &= \mathrm{hyp}(T_a) \, T_a \, .
\end{align*}
The non-vanishing weights or hypercharges of the generators are given by
\begin{align}
\begin{aligned}
\Delta(P) &= 1 \, , \\
\Delta(K) &=-1 \, ,
\end{aligned}
&&
\begin{aligned}
\Delta(Q, \Qb) &= \half \, , \\
\mathrm{hyp}(Q,\Sb) &= \half \, ,
\end{aligned}
&&
\begin{aligned}
\Delta(S, \Sb) &= - \half  \, , \\
\mathrm{hyp}(\Qb,S) &=- \half \, ,
\end{aligned}
\end{align}
Moreover, we note the following commutation relations:
\begin{align}
\Big[ K_{\a \da} \, , \, Q_A {}^\b \Big] 
	&= -2i \, \delta^\b _\a \widebar{S}_{A \da} \, , & 
\Big[ K_{\a \da} \, , \, \widebar{Q}^{\db A} \Big] 
	&= + 2i \, \delta^\db _\da S _\a {}^A \, , \nn \\
\Big[ P^{\da \a} \, , \, S_\b {} ^A \Big] 
	&= + 2i \, \delta^\a _\b \widebar{Q}^{\da A} \, , & 
\Big[ P^{\da \a} \, , \, \widebar{S}_{A \db} \Big] 
	&= - 2i \, \delta^\da _\db Q_A {}^\a \,	, \\ 
\Big \lbrace Q _A {}^\a \, , \, \widebar{Q}^{\da B} \Big \rbrace 
	&= -2i \, \delta ^B _A P^{\da \a} \, , & 
\Big \lbrace S _\a {} ^A \, , \, \widebar{S}_{B \da} \Big \rbrace 
	&= -2i \, \delta ^A _B K_{\a \da} \, . \nn		
\end{align}
The remaining non-vanishing commutators are given by
\begin{equation}
\begin{aligned}
\Big[ P_\mu \, , \, K_\nu \Big] &= + 2 \eta _{\mu \nu} \, D - 2 M_{\mu \nu} \, , \\
\Big \lbrace Q _A {} ^\a \, , \, S_\b {}^B \Big \rbrace &= -2i \, \delta ^B _A \, M \indices{_\b ^\a} - \delta ^\a _\b \, R \indices{^B _A} - 2 \, \delta ^B _A \, \delta ^\a _\b \left(D + C \right) \, ,\\
\Big \lbrace \widebar{Q}^{\da A} \, , \, \widebar{S}_{B \db} \Big \rbrace &= -2i \, \delta ^A _B \, \widebar{M} \, \indices{^\da _\db} - \delta ^\da _\db \, R \indices{^A _B} + 2 \, \delta ^A _B \, \delta ^\da _\db \left(D - C \right) \,.
\end{aligned}
\end{equation}
The metric $G_{a b} = \str (T_a T_b)$ on the algebra has the following components:
\begin{align}
\begin{aligned}
\str ( P^{\da \a} \, K_{\b \db} )
	&= - 4 \, \delta ^\a _\b \, \delta ^\da _\db \, , \\
\str ( Q _A {} ^\a \, S_\b {} ^B ) 
	&= -4 \, \delta ^B _A \, \delta ^\a _\b \, , \\
\str ( \widebar{Q}^{\da A} \, \widebar{S}_{B \db} )
	&=  4 \, \delta ^A _B \, \delta ^\da _\db \, , 	\\
\str ( D \,  D ) &= 1 \, ,
\end{aligned}
&& 
\begin{aligned}
\str ( M \indices{_\a ^\b} , M \indices{_\g ^\eps} )
	&= - 4 \, \delta ^\b _\g \, \delta ^\eps _\a 
	+ 2\, \delta ^\b _\a \, \delta ^\eps _\g \, , \\
\str ( \widebar{M} \indices{^\da _\db} \, \widebar{M} \indices{^\dg _{\dot{\eps}}} ) 
	&= - 4 \, \delta ^\dg _\db \, \delta ^\da _{\dot{\eps}} 
	+ 2\, \delta ^\da _\db \, \delta ^\dg _{\dot{\eps}}	\, , \\ 
\str ( R \indices{^A _B} \, R \indices{^C _D} ) 
	&= - 16 \, \delta ^A _D \, \delta ^C _B + 4 \, \delta ^A _B \, \delta ^C _D	\, , \\
\str ( B \, C ) &= 1 \, .	
\end{aligned}	
\label{Metric_Str}
\end{align} 
All other entries are vanishing. We note that the metric $G_{a b}$ satisfies the symmetry property 
$G_{a b} = \left(-1 \right) ^{\lvert a \rvert} G_{b a}$, 
where $\lvert a \rvert = \mathrm{deg}(T_a)$ denotes the Gra{\ss}mann degree of a homogeneous basis element, i.e.\ $\lvert a \rvert = 0$ or 1 for an even or odd generator, respectively.
Moreover, we note the identity
\begin{align}
\str \big( P^{(2)} ( R \indices{^A _B} ) P^{(2)} ( R \indices{^C _D} ) \big) &= 
	\str \big( R \indices{^A _B}   P^{(2)} ( R \indices{^C _D} ) \big) \nn \\
	&= - 4 K_{B D} \, K^{A C} - 4 \delta ^A _D \, \delta ^C_B 
	+ 2 \delta ^A _B \, \delta ^C _D  \, ,
\label{TrId_RSymm}	
\end{align}
for the projections of the generators $R \indices{^A _B}$. 

Let us now work out the projections of the supermatrix generators onto the graded components. For a general supermatrix
\begin{align*}
N =  \begin{pmatrix}
m & \theta \\ 
\eta & n
\end{pmatrix} 
\end{align*}
these projections are given explicitly in reference \cite{Arutyunov:2009ga},
\begin{align*}
N^{(0)} \! &= \frac{1}{2} \! \begin{pmatrix}
	m - K m^t K^{-1} & 0 \\
	0 & n - K n^t K^{-1}
\end{pmatrix} \! , & 
N^{(2)} \! &= \frac{1}{2} \! \begin{pmatrix}
	m + K m^t K^{-1} & 0 \\
	0 & n + K n^t K^{-1}
\end{pmatrix} \! , \nn \\
N^{(1)} \! &= \frac{1}{2} \! \begin{pmatrix}
	0 & \theta - i K \eta^t K^{-1} \\
	\eta + i K \theta^t K^{-1}  & 0
\end{pmatrix} \! , & 
N^{(3)} \! &= \frac{1}{2} \! \begin{pmatrix}
0 & \theta + i K \eta^t K^{-1} \\ 
\eta - i K \theta^t K^{-1}  & 0
\end{pmatrix} \! .
\end{align*}
Making use of these identities we find the bosonic subspaces to be given by 
\begin{align}
\begin{aligned}
\mathfrak{g}^{(0)} &= \mathrm{span} 
	\left \lbrace  M_{\mu \nu} , \half \left(P_\mu - K_\mu \right)
	, \Gamma ^{r s} , \Gamma ^{r 6} \right \rbrace  \, , \\ 
\mathfrak{g}^{(2)} &= \mathrm{span} 
	\left \lbrace  C , D , \half \left(P_\mu + K_\mu \right) 
	, \Gamma ^{r 5} , \Gamma^{5 6} \right \rbrace \, .
\end{aligned}	
\end{align}
Here, we have used the property
\begin{align*}
K \left( \gamma ^a \right) ^{\mathrm{t}} = \gamma ^a K 
\end{align*}
to find the graded components of the $\alg{su}(4)$-subalgebra and we note that the matrices 
$\Gamma ^{r s}$ and $\Gamma ^{r 6}$ form a representation of $\alg{so}(5)$. 

For the fermionic generators we introduce the notation $A^{(1) \pm (3)} = A^{(1)} \pm A^{(3)}$ and note that $ ( Q , S , \widebar{Q} , \widebar{S} ) ^{(1)+(3)} 
	=  ( Q , S , \widebar{Q} , \widebar{S} )$ and
\begin{align}
\begin{aligned}
\big( Q \indices{_A ^\a} \big) ^{(1)-(3)} &= i \, K_{A B} \, 
	\epsilon^{\a \b} \, S \indices{_\b ^B} \, , \\
\left( S \indices{_\a ^A} \right) ^{(1)-(3)} &= i \, K^{A B}  \, 
	Q \indices{_B ^\b} \, \epsilon_{\b \a} \, ,
\end{aligned}
&&
\begin{aligned}
\left( \widebar{Q}^{\da A} \right) ^{(1)-(3)} &= - i \, K^{A B}  \, 
	\widebar{S}_{B \db} \, \epsilon^{\db \da}  \, , \\
\left( \widebar{S}_{A \da} \right) ^{(1)-(3)} &= - i \, K_{A B} 
	\, \epsilon_{\da \db}  \, \widebar{Q}^{\db B}  \,.
\end{aligned}	
\end{align}	

\chapter{Matrix-valued Differential Forms}
\label{app:Hodge}

We provide a brief overview for the use of $\alg{g}$-valued differential forms and derive a number of helpful formulas for the calculations with these forms. Since the worldsheet is two-dimensional, we are only considering zero-forms (functions), one-forms and two-forms. 

Two such forms can be combined by using the wedge product $\wedge$, which is an anti-symmetrized tensor product of differentials. However, the anti-symmetry of the wedge product on the basis one-forms,
\begin{align*}
\diff \sigma ^i \wedge \diff \sigma ^j = - \diff \sigma ^j \wedge \diff \sigma ^i \, ,
\end{align*}
does not carry over to general $\alg{g}$-valued one-forms, since the coefficients $\omega_i$ and $\rho_i$ do not commute in general, such that we get the relation 
\begin{align}
\omega \wedge \rho + \rho \wedge \omega = \left[ \omega_i , \rho _j \right] \, 
	\diff \sigma^i \wedge \diff \sigma^j \, . 
\end{align}
Applying the differential $\diff$ maps a $k$-form to a $(k+1)$-form and hence the differential of a two-form vanishes in our case. For a function $f$ or a one-form $\omega$ taking values in $\alg{g}$, we note
\begin{align*}
\diff f &= \partial_i f \, \diff \sigma^i \, , & 
\diff \omega &= \partial_j \, \omega_i \, \diff \sigma^j \wedge \diff \sigma ^i \, .  
\end{align*} 
In calculating the differential of a product, the order needs to be respected as well. For two functions $L$ and $R$ taking values in G, we have 
\begin{align*}
\diff \left( L \hspace*{.5mm} \omega  R \right) =
	\diff L \wedge \omega R + L \hspace*{.5mm} \diff \omega R 
	- L \hspace*{.5mm} \omega \wedge \diff R \, , 
\end{align*}
and in particular, for conjugations with $g$ we find using 
$\diff g^{-1} = - g^{-1} \diff g \, g^{-1}$ and
$U = g^{-1} \diff g$
\begin{align}
\diff ( g \omega g^{-1} ) = g \left( \diff \omega + 
	U \wedge \omega  + \omega \wedge U \right) g^{-1}.
\label{dconjg}	
\end{align}
A special situation occurs when the trace allows for cyclic shifts, but the order of the differential forms has to be respected as well. For this situation, we find the identity
\begin{align}
\tr \left( \left[ \omega , X \right] \wedge \rho \right) 
	= \tr \left( \left[ \rho_j , \omega_i \right] X \right) \diff \sigma^i \wedge \diff \sigma^j 
	= - \tr \left( \left( \omega \wedge \rho + \rho \wedge \omega \right) X \right) \, , 
\label{TraceTrick}	
\end{align}
where $X$ denotes a function taking values in $\alg{g}$.

In the index-free notation of differential forms, the worldsheet metric is incorporated in the Hodge star operator, which maps a $k$-form to a $(2-k)$-form, since we are working on the two-dimensional worldsheet. We introduce the Hodge star operator for a general signature of the worldsheet such that the reader can infer how our discussion of symmetric space models can be adapted to a Minkowskian worldsheet. The Hodge star operator acts on a one-form as
\begin{align}
\ast \, \diff \sigma^i = \sqrt{ \vert h \vert } \, 
	h^{i j} \, \epsilon_{j k} \, \diff \sigma ^k .
\end{align}
Here, we fix the convention $\eps_{\tau \sigma} = \eps_{1 2} = 1$ for the Levi-Civita symbol and denoted $h= \det(h_{ij})$. For zero- and two-forms, we note
\begin{align}
\ast 1 & = \sqrt{\vert h \vert} \, (\diff \tau \wedge \diff \sigma) \, , &
\ast (\diff \tau \wedge \diff \sigma) = \frac{\sqrt{\vert h \vert}}{h} \, .
\end{align}
Applying the Hodge star operator twice produces a sign depending on the rank of the form and the signature of the worldsheet. For a general $k$-form $r$, we have 
\begin{align}
\ast \ast r = (-1)^{k(2-k)} \, \sign(h) \, r \, .
\end{align}
This is easy to see for zero- or two-forms and follows for one-forms using the identity
\begin{align*}
\eps^{ik} \, h_{kl} \, \eps^{l j} = - h \, h^{i j} 
\end{align*}
for the worldsheet metric. For moving the Hodge star operator past the wedge product, we note the identity
\begin{align}
\omega \wedge \ast \rho &= - \ast \omega \wedge \rho \, , 
\end{align} 
where $\omega$ and $\rho$ are two general one-forms. The notation using differential forms and the Hodge star operator is related to index notation by
\begin{align*}
\omega \wedge \ast \rho & =  \left( \diff \tau \wedge \diff \sigma \right) 
	\sqrt{\vert h \vert} \, h^{i j} \omega_i \, \rho_j \, , &  
\diff \ast \omega &= \left( \diff \tau \wedge \diff \sigma \right) 
	\partial_i \big( \sqrt{\vert h \vert} h^{i j} \omega_j \big) \, . 	
\end{align*}
An important aspect in our discussion of symmetric space models is the presence of flat connections, i.e.\ one-forms $\omega$ satisfying 
\begin{align}
\diff \omega + \omega \wedge \omega = 0 \, .
\end{align}
The flatness of $\omega$ allows us to solve the differential equation
\begin{align}
\diff f = f \omega \, ,  
\label{aux:problem}
\end{align}
since we have
\begin{align}
0 = \diff ^2 f = \diff ( f \omega) = 
	f ( \diff \omega + \omega \wedge \omega ) = 0 \, .
\end{align}
The Poincar{\'e} lemma guarantees that a unique solution exists once an initial condition for $f$ is specified. Here, we apply the lemma on the worldsheet $\Sigma$, where we do not identify points related by the periodic boundary conditions. The lemma can then be applied since the worldsheet is star-shaped. 
With a solution to equation \eqref{aux:problem} given, we consider gauge transformation $f \mapsto f h$. The flat connection associated to $f$ then transforms as 
\begin{align}
\omega \; \; \mapsto \; \; \omega  ^\prime 
	=( f h )^{-1} \diff ( f h ) 
	= h^{-1} \omega h + h^{-1} \diff h 
\label{connection_transf}
\end{align}
The transformed connection is flat by construction which can easily be checked by direct calculation: With 
$r = \diff h \, h^{-1}$, we have $\diff r - r \wedge r =0$ and hence, using 
$ \omega ^\prime = h^{-1} ( \omega + r ) h$, 
\begin{align*}
\diff \omega ^\prime + \omega ^\prime \wedge \omega ^\prime
	&= h^{-1} \left( \diff \omega + \diff r 
	- r \wedge ( \omega + r ) - ( \omega + r ) \wedge r 
	+ ( \omega + r ) \wedge ( \omega + r ) \right) h = 0 \, . 
\end{align*} 
The flatness of the transformed connection does not depend on whether we are actually considering a gauge transformation, i.e.\ whether 
$h \in \grp{H}$, and the transformation \eqref{connection_transf} is often referred to as a gauge transformation of the connection $\omega$ also in this case.    

The differential equation \eqref{aux:problem} has a unique solution for any connection if we restrict to a curve 
$\gamma(s) = (\tau(s) , \sigma(s))$. Along the curve, we then have the component
\begin{align}
\omega _s = \omega_\tau \, \frac{\diff \tau}{\diff s}
	+ \omega_\sigma \, \frac{\diff \sigma}{\diff s} \, ,
\end{align}
and we impose the ordinary differential equation
\begin{align}
\partial_s f(\gamma(s)) = f(\gamma(s)) \, \omega_s (\gamma(s)) 
\, ,
\label{aux:problem:curve}
\end{align}
which has a unique solution given some initial value 
$f(\gamma(s_0))$. In the case of a flat connection, the existence of a solution to equation \eqref{aux:problem} implies that the solution to the above equation becomes path-independent, which is not the case for generic connections. We can obtain a formal solution by integrating equation \eqref{aux:problem:curve} to get the integral equation
\begin{align}
f(\gamma(s)) = f(\gamma(s_0)) + 
	\int \limits _{s_0} ^s \diff s_1 f(\gamma(s_1)) \,
	\omega_s (\gamma(s_1)) \, .  
\end{align}
If we read the above equation as a recursion for $f(\gamma(s))$, we obtain the formal solution
\begin{align}
f(\gamma(s)) = f(\gamma(s_0)) \cdot 
	\prexp \left( \int _{s_0} ^s \diff s_1 \,
	\omega_s (\gamma(s_1)) \right) \, .
\end{align}
The path-ordered exponential above is obtained from expanding the exponential and ordering the integrals by putting higher values of $s$ to the right, which we may express formally as
\begin{align}
\overrightarrow{\mathrm{P}} [ \omega_s(s_1) \, \omega_s(s_2) ] =
	\begin{cases}
		\omega_s(s_1) \, \omega_s(s_2) & \quad 
		\text{if} \quad s_1 < s_2 \, , \\
		\omega_s(s_2) \, \omega_s(s_1) & \quad 
		\text{if} \quad s_2 < s_1 \, .\\
	\end{cases}
\end{align}
It is now evident from equation \eqref{connection_transf} that the path-ordered exponential over the transformed connection is related to the one over the original connection by
\begin{align}
\prexp \left( \int _{s_0}  ^{s_1} \diff s \, \omega_s (\gamma(s)) \right) 
	= h(\gamma(s_0))^{-1} \prexp \left( \int _{s_0}  ^{s_1} \diff s \, \omega ^\prime _s (\gamma(s)) \right) h(\gamma(s_1)) \, .
\end{align}
This can be concluded from the fact that both sides of the equality solve the differential equation \eqref{aux:problem:curve} for $\omega ^\prime$ and reduce to 
$\unit$ at $s_1=s_0$. 

\chapter{Transformation of the Local Term}
\label{app:transf}

We prove that the local term
\begin{align}
\J_{a, \, \mathrm{lo}} ^{(1)} = - \frac{\la}{4 \pi ^2} 
	\int \limits _0 ^L \diff \s \, \xi^\mu _a (x) 
	\left( \dx_\mu \, \ddot{x}^2 + \dddot{x}_\mu \right) 
\label{local}
\end{align}
of the bosonic Yangian symmetry generators derived in section \ref{sec:InfSymm} indeed transforms as
\begin{align}
\left[ \J^{(0)} _a  , \J_{b , \,  \mathrm{lo}} ^{(1)} \right] 
	= \mathbf{f} \indices{_{ab}^c} \, \J_{c  , \, \mathrm{lo}} ^{(1)} \,.  
\label{level1transf}
\end{align}
Here, the coefficients $\xi^\mu _a (x)$ are the conformal Killing vectors introduced in equation \eqref{Def:ConfKilling}, for which we found the relations 
\begin{align}
\xi ^\nu _a \partial_\nu \, \xi ^\mu _b 
	- \xi ^\nu _b \partial_\nu \, \xi ^\mu _a 
	&= \mathbf{f} _{ab} {} ^c \, \xi ^\mu _c \, , & 
\partial^\mu \xi^\nu _a +  \partial^\nu \xi^\mu _a &= 
	\half \left( \partial_\rho \, \xi^\rho _a \right) \eta ^{\mu \nu} \, .	
\end{align}
The latter relation is known as the conformal Killing equation and for $d>2$, we may apply it to derive the additional relations 
\begin{align}
\quarter \left( \eta^{\mu \lambda} \, \partial^\nu + \eta ^{\nu \lambda} \, 
	\partial^\mu - \eta ^{\mu \nu} \, \partial^\lambda \right) 
	\left(\partial_\kappa \, \xi ^\kappa _a \right) 
	&= \partial ^\mu \partial^\nu \, \xi ^\lambda _a \, , &  
\partial_\mu \partial_\nu \partial_\lambda \, \xi ^\kappa _a &= 0 \,. 
\label{Killing}
\end{align}
Moreover, we have fixed the parametrization in equation \eqref{local} to satisfy $\dx^2 = 1$, which implies that $\dx \cdot \ddot{x} = 0$. As before, the use of such a parametrization is indicated by stating the boundaries $0$ and $L$ of the integration domain.
In order to derive the transformation behaviour \eqn{level1transf}, we need to rewrite the local term \eqref{local} as a reparametrization invariant curve integral, since the variations typically violate this constraint. 

It is easy to see that the expression given in \eqref{local} is not reparametrization invariant, since the second and third derivatives transform in the wrong way, for example we have
\begin{align}
\partial ^2 _\s  x ^\mu (\tilde{\s}(\s)) = 
	\tilde{\s}^\prime (\s) \ddx ^\mu (\tilde{\s}(\s)) 
	+ \tilde{\s}^{\prime \prime} (\s) \dx ^\mu (\tilde{\s}(\s)) \, .
\end{align}
Here, the double derivative of the reparametrization function $\tilde{\s}(\s)$ is not cancelled by the transformation of the integration measure and hence the above term does not lead to a reparametrization invariant integral. In order to address this problem, we divide derivatives of 
$x^\mu (\s)$ by $\lvert \dx \rvert$ before acting with the next derivative. In this way, higher derivatives of the reparametrization function do not appear. Based on this idea, we find the following expression for the local term \eqref{local},  
\begin{align}
\J_{a , \, \mathrm{lo}} ^{(1)} = - \frac{\la}{4 \pi ^2} 
	\int  \diff \s \, \xi^\mu _a (x) \, 
	\bigg[ \dx_\mu \bigg( \frac{1}{\lvert \dx \rvert} 
	\partial_\s \bigg( \frac{\dx_\mu}{\lvert \dx \rvert } \bigg) \bigg)^2  
	+ \partial_\s \bigg( \frac{1}{\lvert \dx \rvert }  
	\partial_\s \bigg( \frac{\dx_\mu}{\lvert \dx \rvert } \bigg) \bigg) \bigg] \, .
\end{align}
It is a simple exercise to check that the above expression is reparametrization invariant and reproduces \eqn{local} for an arc-length parametrization. We can thus compute the variation of 
$\J_{b , \, \mathrm{lo}} ^{(1)}$, 
\begin{align}
\delta \J_{b , \,  \mathrm{lo}} ^{(1)} &= - \frac{\la}{4 \pi ^2} 
	\int \limits _0 ^L \diff \s \Big \lbrace 
	\left( \partial _\rho \, \xi ^\mu _b \right) 
	\left( \dx_\mu \, \ddot{x}^2 + \dddot{x}_\mu \right) 
	- \partial_\s \left[ \xi ^\mu _b \left( \eta_{\mu \rho} \, \ddot{x}^2 
	- 4 \, \dx_\mu \dx_\rho \, \ddot{x}^2 \right) \right] \nn \\ 
& \qquad + 2 \, \partial_\s ^2 \left( \xi ^\mu _b \, \dx_\mu \ddx_\rho \right) 
	- \partial_\s \left[ \left( \partial_\s \, \xi ^\mu _b \right) \ddot{x}_\mu \dx_\rho 
	+ \left( \partial_\s ^2 \, \xi ^\mu _b \right) 
	\left( \eta_{\mu \rho} - \dx_\mu \dx_\rho \right) \right] 
	\Big \rbrace \delta x^\rho(s) \, .
\end{align}
Note that we have reverted back to an arc-length parametrization after calculating the variation. Using the above result, one finds
\begin{align}
\left[ \J^{(0)} _a  ,  \J_{b , \,  \mathrm{lo}} ^{(1)} \right] 
	&= \int \limits _0 ^L \diff \s \,  \xi ^\rho _a (x) \, 
	\frac{\delta \J_{b , \,  \mathrm{lo}} ^{(1)} }{\delta x^\rho(\s)} \nn \\
&= - \frac{\la}{4 \pi ^2} \int \limits _0 ^L \diff \s 
	\Big \lbrace \xi ^\rho _a \left( \partial _\rho \, \xi ^\mu _b \right) 
	\left( \dx_\mu \, \ddot{x}^2 + \dddot{x}_\mu \right) 
	+ \left( \partial_\s \, \xi ^\rho _a \right) \xi_{\rho b} \, \ddot{x}^2 
	+ \left( \partial_\s ^3 \, \xi ^\rho _a \right) \xi_{\rho b}  \nn \\
& \hspace*{-15mm} + \left( \partial_\s \, \xi ^\rho _a \right) 
	\big[ -4 \, \xi ^\mu _b \dx_\mu \dx_\rho \, \ddot{x}^2 
	- 2  \, \partial_\s  \left( \xi ^\mu _b \, \dx_\mu \ddx_\rho \right) 
	+ \left( \partial_\s \, \xi ^\mu _b \right) \ddot{x}_\mu \dx_\rho 
	- \left( \partial_\s ^2 \, \xi ^\mu _b \right) \, \dx_\mu \dx_\rho  
	\big] \Big \rbrace . 
\label{step1}
\end{align}
Due to the use of an arc-length parametrization and the identities \eqn{Killing}, one finds that
\begin{align*}
\partial _\s ^3 \, \xi ^\rho _a = \left( \partial ^\lambda \, \xi ^\rho _a \right) 
	\, \dddot{x}_\lambda + 3 \, \partial_\s 
	\left( \partial ^\lambda \, \xi ^\rho _a \right) \ddot{x}_\lambda 
	= -\left( \partial ^\rho \, \xi ^\lambda _a \right) \, \dddot{x}_\lambda 
	+ \half \left(\partial_\kappa \, \xi ^\kappa _a \right) \dddot{x}^\rho 
	+ 3 \, \partial_\s \left( \partial ^\lambda \, \xi ^\rho _a \right) \ddot{x}_\lambda \, .
\end{align*}
We can thus rewrite the first line of \eqn{step1} as 
\begin{align*}
\int \limits _0 ^L \diff \sigma \, 
	\left( \xi ^\rho _a  \partial _\rho \, \xi ^\mu _b 
	- \xi ^\rho _b  \partial _\rho \, \xi ^\mu _a  \right) 
	\left( \dx_\mu \, \ddot{x}^2 + \dddot{x}_\mu \right) 
	+ \half \, \xi ^\rho _b \big[ \left( \partial_\kappa \, \xi ^\kappa _a \right) 
	\left( \dx_\rho \, \ddot{x}^2 + \dddot{x}_\rho \right) 
	+ 6 \, \partial_\s \left( \partial ^\lambda \, \xi ^\rho _a \right) \ddot{x}_\lambda  \big].
\end{align*}
For the commutator \eqref{level1transf}, we thus find
\begin{align}
\left[ \J^{(0)} _a  , \J_{b , \,  \mathrm{lo}} ^{(1)} \right] 
	= \mathbf{f} \indices{_{ab}^c} \, \J_{c  , \, \mathrm{lo}} ^{(1)} 
	- \frac{\lambda}{4 \pi ^2} \, R_{ab}
\end{align}
and still need to show that
\begin{align}
R_{ab} = \int \limits _0 ^L \diff \sigma \Big \lbrace  
	\left( \partial_\s \, \xi ^\rho _a \right)
	&\big[ -4 \, \xi ^\mu _b \dx_\mu \dx_\rho \, \ddot{x}^2 
	- 2  \, \partial_\s  \left( \xi ^\mu _b \, \dx_\mu \ddx_\rho \right) 
	+ \left( \partial_\s \, \xi ^\mu _b \right) \ddot{x}_\mu \dx_\rho 
	- \left( \partial_\s ^2 \, \xi ^\mu _b \right) \, \dx_\mu \dx_\rho  \big] \nn \\
	+ \half \, \xi ^\mu _b \,  &\big[ \left( \partial_\kappa \, \xi ^\kappa _a \right) 
	\left( \dx_\mu \, \ddot{x}^2 + \dddot{x}_\mu \right) 
	+ 6 \, \partial_\s \left( \partial ^\lambda \, \xi  _{\mu a} \right) 
	\ddot{x}_\lambda  \big] \Big \rbrace = 0 \,. 
\label{step2}
\end{align}
We begin by rewriting the terms in the first line. Using the conformal Killing equation, we find e.g.\ for the first term 
\begin{align}
\int \limits _0 ^L \! \diff \s \left( \partial_\s \, \xi ^\rho _a \right) 
	\left( \xi ^\mu _b \dx_\mu \dx_\rho \, \ddot{x}^2 \right) 
	&= \int \limits _0 ^L \! \diff \s \left( \partial^\lambda \, \xi ^\rho _a \right) 
		\dx_\lambda \dx_\rho \, \xi ^\mu _b \dx_\mu \ddot{x}^2 
	= \frac{1}{4}  \int \limits _0 ^L \! \diff \s 
		\left( \partial_\kappa \, \xi ^\kappa _a \right) 
		\xi ^\mu _b \dx_\mu \, \ddot{x}^2 \, .	
\end{align}
We can treat two more terms analogously to find
\begin{align}
\int \limits _0 ^L \diff \s \left( \partial_\s \, \xi ^\rho _a \right) 
	\left(  \left( \partial_\s ^2 \, \xi ^\mu _b \right) 
	\, \dx_\mu \dx_\rho \right) 
	&= \frac{1}{4} \int \limits _0 ^L \diff \s 
	\left( \partial_\kappa \, \xi ^\kappa _a \right) 
	\left( \partial_\s ^2 \, \xi ^\mu _b \right) \, \dx_\mu \, , \nn \\
\int \limits _0 ^L \diff \s \left( \partial_\s \, \xi ^\rho _a \right) 
	\left( \partial_\s \, \xi ^\mu _b \right) \ddot{x}_\mu \dx_\rho 
	&= \frac{1}{4} \int \limits _0 ^L \diff \s 
	\left( \partial_\kappa \, \xi ^\kappa _a \right) 
	\left( \partial_\s \, \xi ^\mu _b \right) \ddot{x}_\mu \,.
\end{align}
The remaining term in the first line is more complicated to rearrange. After integrating by parts, we have
\begin{align}
\int \limits _0 ^L \diff \s \left( \partial_ \s ^2 \, \xi ^\rho _a \right)
	\ddx_\rho \, \xi ^\mu _b \dx_\mu 
	&= \int \limits _0 ^L \diff \s \left[
		\left( \partial ^\lambda \xi ^\rho _a \right) \ddx_\lambda \ddx_\rho
		+ \left( \partial^\kappa \partial ^\lambda \xi ^\rho _a \right) 
		\dx_\kappa \dx_\lambda \ddx_\rho \right] \xi ^\mu _b \dx_\mu \nn \\
	&= \frac{1}{4} \int \limits _0 ^L \diff \s \left[
		\left( \partial_\kappa \, \xi ^\kappa _a \right) \ddx ^2 
		- \left( \partial ^\rho \partial_\kappa \xi ^\kappa _a \right)
		\ddx_\rho \right] \, \xi ^\mu _b \dx_\mu \, .
\end{align}
We then note that, since 
$\partial_\s \left( \partial ^\rho \partial_\kappa \xi ^\kappa _a \right) = 0$, we can integrate by parts again to obtain 
\begin{align}
\int \limits _0 ^L \diff \s \left( \partial_\s ^2 \, \xi ^\rho _a \right) 
	\ddx_\rho \, \xi^\mu _b \dx_\mu 
	&= \frac{1}{4} \int \limits _0 ^L \diff \s 
	\left( \partial_\kappa \, \xi ^\kappa _a \right)  
	\left( \xi^\mu _b \, \dx_\mu \ddot{x}^2 
	- \partial _\s ^2 \left( \xi ^\mu _b \dx_\mu \right) \right) \,.
\label{step3}	
\end{align}
We have thus rewritten the first line of equation \eqref{step2} as
\begin{align}
- \frac{1}{4} \int \limits _0 ^L \diff \s \left( \partial_\kappa \, \xi ^\kappa _a \right) 
	\left( 2 \, \xi^\mu _b \left( \dx_\mu \, \ddot{x}^2 + \dddot{x}_\mu \right) 
		+ 3 \, \partial_\s \left( \left( \partial_\s \xi ^\mu _b \right) \dx_\mu \right) 
		\right) \, . 
\end{align}
We then note that using the same reasoning that led us to equation \eqref{step3}, we can rearrange
\begin{align}
\int \limits _0 ^L \diff \s \, \partial_\s \left( \partial ^\la \, \xi ^\mu _a \right) 
	\, \xi_{\mu b} \, \ddx_\la 
	&= \frac{1}{4} \int \limits _0 ^L \diff \s  \left( \partial_\kappa \, \xi ^\kappa _a \right) 
	\partial_\s \left( \left(\partial_\s  \xi^\mu _b \right) \, \dx_\mu \right) \, . 
\end{align}
Inserting the last two results into \eqn{step2}, one indeed finds $R_{ab} = 0$ which concludes the proof.

\chapter{Densities of the Yangian Generators}
\label{app:densities}

\section{Level Zero}
In this appendix we provide the differential generators $\jay_a (\s)$ obtained from equation \eqref{defjay}, 
\begin{align}
\jay_a (\s) \left( \mathcal{A}_\mathrm{ren} (\gamma) \right) 
	= \half \, \str \big( j_{\tau \, (0)} (\s) \, T_a \big) \, ,
\end{align}
which we write out explicitly in the form 
$p^\mu (\s) \left( \mathcal{A}_\mathrm{ren} (\gamma) \right)  
	=\half \, \str \big( j_{\tau \, (0)} (\s) \, P^\mu \big)$ and similarly for all other generators. We use the short-hand notation
\begin{align}
\partial^\mu &= \frac{\delta }{\delta x_\mu (\s)} \, , & 
\partial _A {} ^\a  &= \frac{\delta }{\delta \theta _\a {} ^A(\s)} \, , & 
\widebar{\partial} ^{\da A} &=  \frac{\delta }{\delta \btheta_{A \da}(\s)} \, , & 
\partial ^I &= \frac{\delta }{\delta n^I (\s)} \, ,
\end{align}
and note the generators
\begin{align}
\begin{aligned}
p^{\mu} &= \partial^{\mu}  \, , \\
q_A {} ^\a &= - \partial _A {} ^\a  + i \bt_{A \da} \, \partial ^{\da \a} \, , \\
m \indices {_\a ^\b} &= n \indices {_\a ^\b} 
	- \half \, \delta ^\b _\a \, n \indices {_\g ^\g} \, , \\
n \indices {_\a ^\b} &= -2i \, \theta _\a {} ^A \, \partial _A {} ^\b  
	+ i \, x_{\a \da} \, \partial ^{\da \b} \, , \\
b &= \half \left( \theta \, \partial _\theta - \bt \, \widebar{\partial} _{\theta} \right) \, ,	
\end{aligned}
&&
\begin{aligned}
d &= \half \left( \theta \, \partial _\theta + \bt \, \widebar{\partial} _{\theta} \right) 
	+ x \cdot \partial _x \, , \\
\widebar{q}^{\, \da A} &= - \widebar{\partial}^{\da A} 
	+ i \theta _\a {} ^A \, \partial ^{\da \a} \, , \\
\widebar{m} \indices{ ^\da _\db} &= \widebar{n} \indices{ ^\da _\db} 
	- \half \delta ^\da _\db \, \widebar{n} \indices{ ^\dg _\dg} \, , \\
\widebar{n} \indices{ ^\da _\db} &= 2i \, \bt_{A \db} \, \widebar{\partial}^{A \da} 
	- i \, x_{\a \db} \, \partial ^{\da \a}	\, , \\
c &= 0  \, . 	
\end{aligned}	
\label{levelzero1}
\end{align}
The remaining generators are given by
\begin{equation}
\begin{aligned}
 r \indices{^A _B} &= 4 \big( \big( \gamma ^{IJ} \big) _B {} ^A  \, n^I \partial ^J  
 	+ \bt _{B \da} \, \widebar{\partial} ^{A \da}  
 	- \theta _\a  {} ^A \, \partial _B {} ^\a  \big) 
 	- \delta ^A _B  \big(  \bt \, \widebar{\partial}_\theta  
 	- \theta \,  \partial _\theta \big) \, ,\\
s^A _\a &= i \, x^{-} _{\a \da} \, \widebar{\partial}^{\da A} 
	+ x^{+} _{\a \da} \, \theta _b {} ^A \, \partial ^{\da \b} 
	- 4 \, \theta _\a {} ^B \, \theta _\b {} ^A \, \partial _B {} ^\b  
	+ 4 \theta _\a {}^B  \, \big( \gamma ^{IJ} \big) _B {} ^A \, n^I \partial ^J \, , \\
\widebar{s}_{A \da} &= -i \, x^{+} _{\a \da} \, \partial _A {} ^\a 
	- x^{-} _{\a \da}  \, \bt_{A \db} \, \partial ^{\db \a} 
	- 4 \, \btheta _{A \db} \, \btheta _{B \da} \widebar{\partial}^{\db B} 
	+ 4 \big( \gamma ^{IJ} \big) _A {} ^B \, \bt _{B \da} \, n^I \partial ^J \, , \\
k_{\a \da} &= i \, x^+ _{\a \db} \, \widebar{n}^\db {} _\da 
	- i x^- _{\b \da} \, n _\a {} ^\b 
	- x^+_{\a \db} \, x^- _{\b \da} \, \partial ^{\db \b} 
	- 8i \left( \theta \, \gamma ^{IJ} \, \bt \right)_{\a \da} \, n^I \partial ^J   
\end{aligned}
\label{levelzero2} 
\end{equation}
Here, we introduced the chiral and anti-chiral coordinates
\begin{align}
x^+ _{\a \da} &= x_{\a \da} + 2i \, \theta _\a {}^A \, \bt _{A \da} \, , & 
x^- _{\a \da} &= x_{\a \da} - 2i \, \theta _\a {}^A \, \bt _{A \da} \, .
\end{align}  
These generators satisfy the commutation relations given for the superconformal algebra in section \ref{sec:Int_Symm}, i.e.\ we have 
\begin{align}
\Big[ \jay_a (\s)  ,  \jay_b (\s^\prime) \Big \rbrace 
	&= \mathbf{f} \indices{_{ab} ^c}  \, \delta(\s - \s^\prime) \, \jay_c (\s) \, .
\end{align}
As pointed out before, the structure constants above are related to the structure constants 
$f \indices{_{ab} ^c}$ of the supermatrix generators introduced in appendix \ref{app:u224} by the relation
\begin{align}
\mathbf{f} \indices{_{ab} ^c} = f \indices{_{ba} ^c} 
	= - (-1)^{ \lvert a \rvert \lvert b \rvert}  f \indices{_{ab} ^c} \, .
\label{f_rel}	
\end{align}
For the discussion of the level-1 Yangian generators, we also need the components of the metric on $\alg{u}(2,2 \vert 4)$ for the above generators. In order to obtain these, we construct a basis of matrix generators, which satisfy the above commutation relations, by manipulating the basis introduced in appendix \ref{app:u224}. Concretely, we have the assignment
\begin{align}
R( \jay_a ) = \mathbf{T}_a = 
	\begin{cases}
		T_a \quad & \text{if} \quad \Delta \in 
			\left \lbrace - 1 , - \half , 1 \right \rbrace \, , \\
		-T_a \quad & \text{if} \quad \Delta \in 
			\left \lbrace 0 ,\half  \right \rbrace \, .	 
	\end{cases}
\end{align}
In order to see that the basis thus introduced has the structure constants
\begin{align}
\left[ \mathbf{T}_a , \mathbf{T}_b \right] 
	= \mathbf{f} \indices{_{ab} ^c} \, \mathbf{T}_c \, ,
\end{align}
note that the weights are additive in a commutator and that the fermionic generators have half-integer weights, whereas the bosonic generators have integer weights. It is then easy to see that all except the odd-odd commutators change their sign, as equation \eqref{f_rel} demands. Noting moreover that the metric of two generators is only non-vanishing if the weights add up to zero, 
\begin{align}
G_{ab} \neq 0 \quad \Rightarrow \quad \Delta_a + \Delta_b = 0 \, , 
\end{align}
we observe moreover that the metric 
$\mathbf{G}_{ab} = \str \left( \mathbf{T}_a \, \mathbf{T}_b \right)$ is given by
\begin{align}
\mathbf{G}_{ab} = G_{ba} = \left(-1 \right)^{\lvert a \rvert} G_{ab} \, . 
\end{align}

\section{Level 1}

\label{app:level1}
We provide the parts of the level-1 densities defined in equation \eqref{jayprime}, 
\begin{align}
\jay_a ^{(1)\, \prime} T^a \! &= \! \Big \lbrace e^{X \cdot P + \Theta Q + \bTheta \Qb} 
	\Big( \frac{\pi^\mu}{\tau^2} \, K_\mu  - \pi ^\mu \dot{\pi}^\nu M_{\mu \nu} 
	+ 2i  \tr \! \big(  \dbt \pi \dt \big)   C \Big) 
	e^{- X \cdot P - \Theta Q - \bTheta \Qb} \Big \rbrace_{\!(0)} .
\end{align}
A direct calculation gives
\begin{align}
\left( p^{(1)\, \prime} \right)^{\mu} &= 0 \,  , &
\left( q^{(1)\, \prime} \right)_A {} ^\a &= 
	2i\,  \bTheta_{(2)\, A \da} \, \pi^{\da \a} \, , &
\left( \widebar{q}^{(1)\, \prime} \right)^{\da A} &= 
	2i \, \pi^{\da \a} \theta _{(2)} {} _\a {} ^A \, .
\end{align}
For the remaining generators, we have
\begin{align}
\begin{aligned}
d^{(1)\, \prime} &= i \tr \big( \bTheta_{(2)} \pi \theta - \bt \pi \Theta_{(2)} \big) \, , \\
b^{(1)\, \prime} &= 2i \tr \left( \bTheta_{(2)} \pi \theta 
	+ \btheta \pi \Theta _{(2)} + \dbt \pi \dt \right) \, , \\
\left( m^{(1)\, \prime} \right) \indices {_\a ^\b} 
	&= 4 \big( \theta \bTheta_{(2)}  \, \pi \big) _\a {} ^\b 
	- 2 \, \delta ^\b _\a \, \tr \big( \theta \bTheta_{(2)}  \, \pi \big) \, , \\
\left( \bar{m}^{(1)\, \prime} \right) \indices{ ^\da _\db} 
	&= - 4 \, \big( \pi \,  \Theta_{(2)} \bt \big) ^\da {} _\db 
	+ 2 \, \delta ^\da _\db \, \tr \big( \pi \,  \Theta_{(2)} \bt \big) \, ,\\
\left( r^{(1)\, \prime} \right) \indices{^A _B} 
	&= - 8i \big( \bt \pi \Theta_{(2)} + \bTheta_{(2)} \pi \theta  \big)_B {} ^A   
	+ 2i \, \delta ^A _B \, 
	\tr \big( \bt \pi \Theta_{(2)} +  \bTheta_{(2)}  \pi \theta \big)  \, , \\
\left( s^{(1)\, \prime} \right)_\a {} ^A 
	&= \left( 4 \left( \dot{\pi} - 2i \, \Theta_{(2)} \bt \right)  \, \pi \, \theta 
	+ 2 \, x^- \, \pi \Theta_{(2)} \right)_\a {} ^A \, , \\
\left( \bar{s}^{(1)\, \prime} \right)_{A \da} 
	&= - \left( 4 \, \bt \, \pi \left( \dot{\pi} + 2i \, \theta \bTheta_{(2)} \right)  
	+ 2 \, \bTheta_{(2)} \, \pi \, x^+ \right) _{A \da} \, , \\
\left(k^{(1)\, \prime} \right) _{\a \da} 
	&= - 4i \left(x^+ \, \pi \, \Theta_{(2)} \bt 
	- \theta \bTheta_{(2)} \, \pi \, x^- \right)_{\a \da} \, .
\end{aligned} 
\end{align}

\end{appendix}
 
\newpage
\bibliographystyle{nb}
\phantomsection
\addcontentsline{toc}{chapter}{Bibliography}
\bibliography{PhDThesis}

\begin{thebibliography}{100}
\ifx\href\asklfhas\newcommand{\href}[2]{#2}\fi
\ifx\arxivref\asklfhas\newcommand{\arxivref}[2]{\href{http://arxiv.org/abs/#1}{#2}}\fi
\ifx\doiref\asklfhas\newcommand{\doiref}[2]{\href{http://dx.doi.org/#1}{#2}}\fi
\raggedright
\small
\parskip 0pt

\bibitem{Muller:2013rta}
D.~M{\"u}ller, H.~M{\"u}nkler, J.~Plef\-ka, J.~Pollok and K.~Za\-rembo,
\textit{``{Yangian Symmetry of smooth Wilson Loops in $\mathcal{N} = $ 4 super
  Yang-Mills Theory}''},
\textsf{\doiref{10.1007/JHEP11(2013)081}{JHEP~1311,~081~(2013)}},
\texttt{\arxivref{1309.1676}{arxiv:1309.1676}}.

\bibitem{Munkler:2015gja}
H.~M{\"u}nkler and J.~Pollok,
\textit{``{Minimal surfaces of the ${{AdS}}_{5}\times {S}^{5}$ superstring and
  the symmetries of super Wilson loops at strong coupling}''},
\textsf{\doiref{10.1088/1751-8113/48/36/365402}{J.~Phys.~A48,~365402~(2015)}},
\texttt{\arxivref{1503.07553}{arxiv:1503.07553}}.

\bibitem{Munkler:2015xqa}
H.~M{\"u}nkler,
\textit{``{Bonus Symmetry for Super Wilson Loops}''},
\textsf{\doiref{10.1088/1751-8113/49/18/185401}{J.~Phys.~A49,~185401~(2016)}},
\texttt{\arxivref{1507.02474}{arxiv:1507.02474}}.

\bibitem{Klose:2016uur}
T.~Klose, F.~Loebbert and H.~M{\"u}nkler,
\textit{``{Master Symmetry for Holographic Wilson Loops}''},
\textsf{\doiref{10.1103/PhysRevD.94.066006}{Phys.~Rev.~D94,~066006~(2016)}},
\texttt{\arxivref{1606.04104}{arxiv:1606.04104}}.

\bibitem{Klose:2016qfv}
T.~Klose, F.~Loebbert and H.~M{\"u}nkler,
\textit{``{Nonlocal Symmetries, Spectral Parameter and Minimal Surfaces in
  AdS/CFT}''},
\textsf{\doiref{10.1016/j.nuclphysb.2017.01.008}{Nucl.~Phys.~B916,~320~(2017)}},
\texttt{\arxivref{1610.01161}{arxiv:1610.01161}}.

\bibitem{Dorn:2012cn}
H.~Dorn, H.~M{\"u}nkler and C.~Spielvogel,
\textit{``{Conformal geometry of null hexagons for Wilson loops and scattering
  amplitudes}''},
\textsf{\doiref{10.1134/S1063779614040066}{Phys.~Part.~Nucl.~45,~692~(2014)}},
\texttt{\arxivref{1211.5537}{arxiv:1211.5537}}.

\bibitem{Yang:1954ek}
C.-N.~Yang and R.~L.~Mills,
\textit{``{Conservation of Isotopic Spin and Isotopic Gauge Invariance}''},
\textsf{\doiref{10.1103/PhysRev.96.191}{Phys.~Rev.~96,~191~(1954)}}.

\bibitem{Aad:2012tfa}
ATLAS Collaboration, G.~Aad et~al.,
\textit{``{Observation of a new particle in the search for the Standard Model
  Higgs boson with the ATLAS detector at the LHC}''},
\textsf{\doiref{10.1016/j.physletb.2012.08.020}{Phys.~Lett.~B716,~1~(2012)}},
\texttt{\arxivref{1207.7214}{arxiv:1207.7214}}.

\bibitem{Chatrchyan:2012xdj}
CMS Collaboration, S.~Chatrchyan et~al.,
\textit{``{Observation of a new boson at a mass of 125 GeV with the CMS
  experiment at the LHC}''},
\textsf{\doiref{10.1016/j.physletb.2012.08.021}{Phys.~Lett.~B716,~30~(2012)}},
\texttt{\arxivref{1207.7235}{arxiv:1207.7235}}.

\bibitem{Aad:2015zhl}
ATLAS, CMS Collaboration, G.~Aad et~al.,
\textit{``{Combined Measurement of the Higgs Boson Mass in $pp$ Collisions at
  $\sqrt{s}=7$ and 8 TeV with the ATLAS and CMS Experiments}''},
\textsf{\doiref{10.1103/PhysRevLett.114.191803}{Phys.~Rev.~Lett.~114,~191803~(2015)}},
\texttt{\arxivref{1503.07589}{arxiv:1503.07589}}.

\bibitem{Brink:1976bc}
L.~Brink, J.~H.~Schwarz and J.~Scherk,
\textit{``{Supersymmetric Yang-Mills Theories}''},
\textsf{\doiref{10.1016/0550-3213(77)90328-5}{Nucl.~Phys.~B121,~77~(1977)}}.

\bibitem{Gliozzi:1976qd}
F.~Gliozzi, J.~Scherk and D.~I.~Olive,
\textit{``{Supersymmetry, Supergravity Theories and the Dual Spinor Model}''},
\textsf{\doiref{10.1016/0550-3213(77)90206-1}{Nucl.~Phys.~B122,~253~(1977)}}.

\bibitem{Maldacena:1997re}
J.~M.~Maldacena,
\textit{``{The Large N limit of superconformal field theories and
  supergravity}''},
\textsf{\doiref{10.1023/A:1026654312961}{Int.~J.~Theor.~Phys.~38,~1113~(1999)}},
\texttt{\arxivref{hep-th/9711200}{hep-th/9711200}},
[Adv. Theor. Math. Phys. 2, 231 (1998)].

\bibitem{Pestun:2016zxk}
V.~Pestun et~al.,
\textit{``{Localization techniques in quantum field theories}''},
\texttt{\arxivref{1608.02952}{arxiv:1608.02952}}.

\bibitem{Pestun:2007rz}
V.~Pestun,
\textit{``{Localization of gauge theory on a four-sphere and supersymmetric
  Wilson loops}''},
\textsf{\doiref{10.1007/s00220-012-1485-0}{Commun.~Math.~Phys.~313,~71~(2012)}},
\texttt{\arxivref{0712.2824}{arxiv:0712.2824}}.

\bibitem{Zarembo:2016bbk}
K.~Zarembo,
\textit{``{Localization and AdS/CFT Correspondence}''},
\texttt{\arxivref{1608.02963}{arxiv:1608.02963}}.

\bibitem{Minahan:2002ve}
J.~A.~Minahan and K.~Zarembo,
\textit{``{The Bethe ansatz for N=4 superYang-Mills}''},
\textsf{\doiref{10.1088/1126-6708/2003/03/013}{JHEP~0303,~013~(2003)}},
\texttt{\arxivref{hep-th/0212208}{hep-th/0212208}}.

\bibitem{Beisert:2003tq}
N.~Beisert, C.~Kristjansen and M.~Staudacher,
\textit{``{The Dilatation operator of conformal N=4 superYang-Mills theory}''},
\textsf{\doiref{10.1016/S0550-3213(03)00406-1}{Nucl.~Phys.~B664,~131~(2003)}},
\texttt{\arxivref{hep-th/0303060}{hep-th/0303060}}.

\bibitem{Beisert:2003yb}
N.~Beisert and M.~Staudacher,
\textit{``{The N=4 SYM integrable super spin chain}''},
\textsf{\doiref{10.1016/j.nuclphysb.2003.08.015}{Nucl.~Phys.~B670,~439~(2003)}},
\texttt{\arxivref{hep-th/0307042}{hep-th/0307042}}.

\bibitem{Beisert:2010jr}
N.~Beisert et~al.,
\textit{``{Review of AdS/CFT Integrability: An Overview}''},
\textsf{\doiref{10.1007/s11005-011-0529-2}{Lett.~Math.~Phys.~99,~3~(2012)}},
\texttt{\arxivref{1012.3982}{arxiv:1012.3982}}.

\bibitem{Dolan:2003uh}
L.~Dolan, C.~R.~Nappi and E.~Witten,
\textit{``{A Relation between approaches to integrability in superconformal
  Yang-Mills theory}''},
\textsf{\doiref{10.1088/1126-6708/2003/10/017}{JHEP~0310,~017~(2003)}},
\texttt{\arxivref{hep-th/0308089}{hep-th/0308089}}.

\bibitem{Drummond:2008vq}
J.~M.~Drummond, J.~Henn, G.~P.~Korchemsky and E.~Sokatchev,
\textit{``{Dual superconformal symmetry of scattering amplitudes in N=4
  super-Yang-Mills theory}''},
\textsf{\doiref{10.1016/j.nuclphysb.2009.11.022}{Nucl.~Phys.~B828,~317~(2010)}},
\texttt{\arxivref{0807.1095}{arxiv:0807.1095}}.

\bibitem{Drummond:2009fd}
J.~M.~Drummond, J.~M.~Henn and J.~Plefka,
\textit{``{Yangian symmetry of scattering amplitudes in N=4 super Yang-Mills
  theory}''},
\textsf{\doiref{10.1088/1126-6708/2009/05/046}{JHEP~0905,~046~(2009)}},
\texttt{\arxivref{0902.2987}{arxiv:0902.2987}}.

\bibitem{Beisert:2017pnr}
N.~Beisert, A.~Garus and M.~Rosso,
\textit{``{Yangian Symmetry and Integrability of Planar N=4 Supersymmetric
  Yang-Mills Theory}''},
\textsf{\doiref{10.1103/PhysRevLett.118.141603}{Phys.~Rev.~Lett.~118,~141603~(2017)}},
\texttt{\arxivref{1701.09162}{arxiv:1701.09162}}.

\bibitem{Maldacena:1998im}
J.~M.~Maldacena,
\textit{``{Wilson loops in large N field theories}''},
\textsf{\doiref{10.1103/PhysRevLett.80.4859}{Phys.~Rev.~Lett.~80,~4859~(1998)}},
\texttt{\arxivref{hep-th/9803002}{hep-th/9803002}}.

\bibitem{Rey:1998ik}
S.-J.~Rey and J.-T.~Yee,
\textit{``{Macroscopic strings as heavy quarks in large N gauge theory and
  anti-de Sitter supergravity}''},
\textsf{\doiref{10.1007/s100520100799}{Eur.~Phys.~J.~C22,~379~(2001)}},
\texttt{\arxivref{hep-th/9803001}{hep-th/9803001}}.

\bibitem{Drukker:1999zq}
N.~Drukker, D.~J.~Gross and H.~Ooguri,
\textit{``{Wilson loops and minimal surfaces}''},
\textsf{\doiref{10.1103/PhysRevD.60.125006}{Phys.~Rev.~D60,~125006~(1999)}},
\texttt{\arxivref{hep-th/9904191}{hep-th/9904191}}.

\bibitem{Alday:2007hr}
L.~F.~Alday and J.~M.~Maldacena,
\textit{``{Gluon scattering amplitudes at strong coupling}''},
\textsf{\doiref{10.1088/1126-6708/2007/06/064}{JHEP~0706,~064~(2007)}},
\texttt{\arxivref{0705.0303}{arxiv:0705.0303}}.

\bibitem{Brandhuber:2007yx}
A.~Brandhuber, P.~Heslop and G.~Travaglini,
\textit{``{MHV amplitudes in N=4 super Yang-Mills and Wilson loops}''},
\textsf{\doiref{10.1016/j.nuclphysb.2007.11.002}{Nucl.~Phys.~B794,~231~(2008)}},
\texttt{\arxivref{0707.1153}{arxiv:0707.1153}}.

\bibitem{Drummond:2008aq}
J.~M.~Drummond, J.~Henn, G.~P.~Korchemsky and E.~Sokatchev,
\textit{``{Hexagon Wilson loop = six-gluon MHV amplitude}''},
\textsf{\doiref{10.1016/j.nuclphysb.2009.02.015}{Nucl.~Phys.~B815,~142~(2009)}},
\texttt{\arxivref{0803.1466}{arxiv:0803.1466}}.

\bibitem{Master}
H.~M{\"u}nkler,
\textit{``{Yangian Symmetry of Maldacena--Wilson loops}''},
Master's Thesis, Humboldt-Universit{\"a}t zu Berlin (2013).

\bibitem{Ishizeki:2011bf}
R.~Ishizeki, M.~Kruczenski and S.~Ziama,
\textit{``{Notes on Euclidean Wilson loops and Riemann Theta functions}''},
\textsf{\doiref{10.1103/PhysRevD.85.106004}{Phys.~Rev.~D85,~106004~(2012)}},
\texttt{\arxivref{1104.3567}{arxiv:1104.3567}}.

\bibitem{Kruczenski:2013bsa}
M.~Kruczenski and S.~Ziama,
\textit{``{Wilson loops and Riemann theta functions II}''},
\textsf{\doiref{10.1007/JHEP05(2014)037}{JHEP~1405,~037~(2014)}},
\texttt{\arxivref{1311.4950}{arxiv:1311.4950}}.

\bibitem{Dekel:2015bla}
A.~Dekel,
\textit{``{Wilson Loops and Minimal Surfaces Beyond the Wavy Approximation}''},
\textsf{\doiref{10.1007/JHEP03(2015)085}{JHEP~1503,~085~(2015)}},
\texttt{\arxivref{1501.04202}{arxiv:1501.04202}}.

\bibitem{Beisert:2015jxa}
N.~Beisert, D.~M{\"u}ller, J.~Plefka and C.~Vergu,
\textit{``{Smooth Wilson loops in $ \mathcal{N}=4 $ non-chiral superspace}''},
\textsf{\doiref{10.1007/JHEP12(2015)140}{JHEP~1512,~140~(2015)}},
\texttt{\arxivref{1506.07047}{arxiv:1506.07047}}.

\bibitem{Beisert:2015uda}
N.~Beisert, D.~M{\"u}ller, J.~Plefka and C.~Vergu,
\textit{``{Integrability of smooth Wilson loops in $ \mathcal{N}=4 $
  superspace}''},
\textsf{\doiref{10.1007/JHEP12(2015)141}{JHEP~1512,~141~(2015)}},
\texttt{\arxivref{1509.05403}{arxiv:1509.05403}}.

\bibitem{Chicherin:2017cns}
D.~Chicherin, V.~Kazakov, F.~Loebbert, D.~M{\"u}ller and D.-l.~Zhong,
\textit{``{Yangian Symmetry for Bi-Scalar Loop Amplitudes}''},
\texttt{\arxivref{1704.01967}{arxiv:1704.01967}}.

\bibitem{Metsaev:1998it}
R.~R.~Metsaev and A.~A.~Tseytlin,
\textit{``{Type IIB superstring action in AdS(5) x S**5 background}''},
\textsf{\doiref{10.1016/S0550-3213(98)00570-7}{Nucl.~Phys.~B533,~109~(1998)}},
\texttt{\arxivref{hep-th/9805028}{hep-th/9805028}}.

\bibitem{Ooguri:2000ps}
H.~Ooguri, J.~Rahmfeld, H.~Robins and J.~Tannenhauser,
\textit{``{Holography in superspace}''},
\textsf{\doiref{10.1088/1126-6708/2000/07/045}{JHEP~0007,~045~(2000)}},
\texttt{\arxivref{hep-th/0007104}{hep-th/0007104}}.

\bibitem{Bena:2003wd}
I.~Bena, J.~Polchinski and R.~Roiban,
\textit{``{Hidden symmetries of the AdS(5) x S**5 superstring}''},
\textsf{\doiref{10.1103/PhysRevD.69.046002}{Phys.~Rev.~D69,~046002~(2004)}},
\texttt{\arxivref{hep-th/0305116}{hep-th/0305116}}.

\bibitem{Chandia:2016ueo}
O.~Chandia, W.~D.~Linch and B.~C.~Vallilo,
\textit{``{Master symmetry in the AdS$_{5} \times$ S$^{5}$ pure spinor
  string}''},
\textsf{\doiref{10.1007/JHEP01(2017)024}{JHEP~1701,~024~(2017)}},
\texttt{\arxivref{1607.00391}{arxiv:1607.00391}}.

\bibitem{Hawking:1973uf}
S.~W.~Hawking and G.~F.~R.~Ellis,
\textit{``{The Large Scale Structure of Space-Time}''},
Cambridge University Press (2011),
Cambridge.

\bibitem{book:Scho}
M.~Schottenloher,
\textit{``{A Mathematcal Introduction to Conformal Field Theory}''},
2$^\mathrm{nd}$ edition,
Springer (2008),
Berlin Heidelberg.

\bibitem{Dorn:Lecture_Notes}
H.~Dorn,
\textit{``{Konforme Invarianz in der Quantenfeldtheorie}''},
Lecture Notes, Humboldt-Universit{\"a}t zu Berlin (2009).

\bibitem{Penrose:1962ij}
R.~Penrose,
\textit{``{Asymptotic properties of fields and space-times}''},
\textsf{\doiref{10.1103/PhysRevLett.10.66}{Phys.~Rev.~Lett.~10,~66~(1963)}}.

\bibitem{Dirac:1936fq}
P.~A.~M.~Dirac,
\textit{``{Wave equations in conformal space}''},
\textsf{\doiref{10.2307/1968455}{Annals~Math.~37,~429~(1936)}}.

\bibitem{Weinberg}
S.~Weinberg,
\textit{``{The Quantum Theory of Fields, Volume III Supersymmetry}''},
Cambridge University Press (2000),
Cambridge.

\bibitem{Arutyunov:2009ga}
G.~Arutyunov and S.~Frolov,
\textit{``{Foundations of the $AdS_5 \times S^5$ Superstring. Part I}''},
\textsf{\doiref{10.1088/1751-8113/42/25/254003}{J.~Phys.~A42,~254003~(2009)}},
\texttt{\arxivref{0901.4937}{arxiv:0901.4937}}.

\bibitem{Beisert:2010kp}
N.~Beisert,
\textit{``{Review of AdS/CFT Integrability, Chapter VI.1: Superconformal
  Symmetry}''},
\textsf{\doiref{10.1007/s11005-011-0479-8}{Lett.~Math.~Phys.~99,~529~(2012)}},
\texttt{\arxivref{1012.4004}{arxiv:1012.4004}}.

\bibitem{Cornwell}
J.~F.~Cornwell,
\textit{``{Group Theory in Physics, Volume III Supersymmetries and
  Infinite-Dimensional Algebras}''},
Academic Press (1989),
London.

\bibitem{Drinfeld:1985rx}
V.~G.~Drinfeld,
\textit{``{Hopf algebras and the quantum Yang-Baxter equation}''},
\textsf{Sov.~Math.~Dokl.~32,~254~(1985)},
[Dokl. Akad. Nauk. Ser. Fiz. 283, 1060 (1985)].

\bibitem{Drinfeld:1986in}
V.~G.~Drinfeld,
\textit{``{Quantum groups}''},
\textsf{\doiref{10.1007/BF01247086}{J.~Sov.~Math.~41,~898~(1988)}},
[Translated from Zapiski Nauchnykh Seminarov Leningradskogo Otdeleniya
  Matematicheskogo Instituta im. V. A. Steklova AN SSSR, Vol. 155, pp. 18–49,
  1986.].

\bibitem{Bernard:1990jw}
D.~Bernard,
\textit{``{Hidden Yangians in 2-D massive current algebras}''},
\textsf{\doiref{10.1007/BF02099123}{Commun.~Math.~Phys.~137,~191~(1991)}}.

\bibitem{MacKay:1992he}
N.~J.~MacKay,
\textit{``{On the classical origins of Yangian symmetry in integrable field
  theory}''},
\textsf{\doiref{10.1016/0370-2693(92)90280-H}{Phys.~Lett.~B281,~90~(1992)}},
[Erratum: Phys. Lett. B308, 444 (1993)].

\bibitem{Bernard:1992ya}
D.~Bernard,
\textit{``{An Introduction to Yangian Symmetries}''},
\textsf{\doiref{10.1142/S0217979293003371}{Int.~J.~Mod.~Phys.~B7,~3517~(1993)}},
\texttt{\arxivref{hep-th/9211133}{hep-th/9211133}}.

\bibitem{Dolan:2004ps}
L.~Dolan, C.~R.~Nappi and E.~Witten,
\textit{``{Yangian symmetry in D = 4 superconformal Yang-Mills theory}''},
\texttt{\arxivref{hep-th/0401243}{hep-th/0401243}},
in: Proceedings, 3rd International Symposium on Quantum theory and symmetries
  (QTS3): Cincinnati, USA, September 10-14, 2003, p. 300 - 315, 2004.

\bibitem{Bargheer:2009qu}
T.~Bargheer, N.~Beisert, W.~Galleas, F.~Loebbert and T.~McLoughlin,
\textit{``{Exacting N=4 Superconformal Symmetry}''},
\textsf{\doiref{10.1088/1126-6708/2009/11/056}{JHEP~0911,~056~(2009)}},
\texttt{\arxivref{0905.3738}{arxiv:0905.3738}}.

\bibitem{Beisert:2010gn}
N.~Beisert, J.~Henn, T.~McLoughlin and J.~Plefka,
\textit{``{One-Loop Superconformal and Yangian Symmetries of Scattering
  Amplitudes in N=4 Super Yang-Mills}''},
\textsf{\doiref{10.1007/JHEP04(2010)085}{JHEP~1004,~085~(2010)}},
\texttt{\arxivref{1002.1733}{arxiv:1002.1733}}.

\bibitem{CaronHuot:2011kk}
S.~Caron-Huot and S.~He,
\textit{``{Jumpstarting the All-Loop S-Matrix of Planar N=4 Super
  Yang-Mills}''},
\textsf{\doiref{10.1007/JHEP07(2012)174}{JHEP~1207,~174~(2012)}},
\texttt{\arxivref{1112.1060}{arxiv:1112.1060}}.

\bibitem{Ferro:2011ph}
L.~Ferro,
\textit{``{Yangian Symmetry in N=4 super Yang-Mills}''},
\texttt{\arxivref{1107.1776}{arxiv:1107.1776}}.

\bibitem{Loebbert:2016cdm}
F.~Loebbert,
\textit{``{Lectures on Yangian Symmetry}''},
\textsf{\doiref{10.1088/1751-8113/49/32/323002}{J.~Phys.~A49,~323002~(2016)}},
\texttt{\arxivref{1606.02947}{arxiv:1606.02947}}.

\bibitem{MacKay:2004tc}
N.~J.~MacKay,
\textit{``{Introduction to Yangian symmetry in integrable field theory}''},
\textsf{\doiref{10.1142/S0217751X05022317}{Int.~J.~Mod.~Phys.~A20,~7189~(2005)}},
\texttt{\arxivref{hep-th/0409183}{hep-th/0409183}}.

\bibitem{Beisert:2010jq}
N.~Beisert,
\textit{``{On Yangian Symmetry in Planar N=4 SYM}''},
\textsf{\doiref{10.1142/9789814350198_0039}{Gribov-80~memorial~volume:~Quantum
  chromodynamics and beyond,~413~(2011)}},
\texttt{\arxivref{1004.5423}{arxiv:1004.5423}}.

\bibitem{Torrielli:2010kq}
A.~Torrielli,
\textit{``{Review of AdS/CFT Integrability, Chapter VI.2: Yangian Algebra}''},
\textsf{\doiref{10.1007/s11005-011-0491-z}{Lett.~Math.~Phys.~99,~547~(2012)}},
\texttt{\arxivref{1012.4005}{arxiv:1012.4005}}.

\bibitem{Torrielli:2011gg}
A.~Torrielli,
\textit{``{Yangians, S-matrices and AdS/CFT}''},
\textsf{\doiref{10.1088/1751-8113/44/26/263001}{J.~Phys.~A44,~263001~(2011)}},
\texttt{\arxivref{1104.2474}{arxiv:1104.2474}}.

\bibitem{Rocen}
A.~Roc{\'e}n,
\textit{``{Yangians and their representations}''},
PhD Thesis, University of York (2010).

\bibitem{Sohnius:1981sn}
M.~F.~Sohnius and P.~C.~West,
\textit{``{Conformal Invariance in N=4 Supersymmetric Yang-Mills Theory}''},
\textsf{\doiref{10.1016/0370-2693(81)90326-9}{Phys.~Lett.~B100,~245~(1981)}}.

\bibitem{Mandelstam:1982cb}
S.~Mandelstam,
\textit{``{Light Cone Superspace and the Ultraviolet Finiteness of the N=4
  Model}''},
\textsf{\doiref{10.1016/0550-3213(83)90179-7}{Nucl.~Phys.~B213,~149~(1983)}}.

\bibitem{Howe:1983sr}
P.~S.~Howe, K.~S.~Stelle and P.~K.~Townsend,
\textit{``{Miraculous Ultraviolet Cancellations in Supersymmetry Made
  Manifest}''},
\textsf{\doiref{10.1016/0550-3213(84)90528-5}{Nucl.~Phys.~B236,~125~(1984)}}.

\bibitem{Brink:1982pd}
L.~Brink, O.~Lindgren and B.~E.~W.~Nilsson,
\textit{``{N=4 Yang-Mills Theory on the Light Cone}''},
\textsf{\doiref{10.1016/0550-3213(83)90678-8}{Nucl.~Phys.~B212,~401~(1983)}}.

\bibitem{Brink:1982wv}
L.~Brink, O.~Lindgren and B.~E.~W.~Nilsson,
\textit{``{The Ultraviolet Finiteness of the N=4 Yang-Mills Theory}''},
\textsf{\doiref{10.1016/0370-2693(83)91210-8}{Phys.~Lett.~B123,~323~(1983)}}.

\bibitem{Sohnius:1985qm}
M.~F.~Sohnius,
\textit{``{Introducing Supersymmetry}''},
\textsf{\doiref{10.1016/0370-1573(85)90023-7}{Phys.~Rept.~128,~39~(1985)}}.

\bibitem{tHooft:1973alw}
G.~'t~Hooft,
\textit{``{A Planar Diagram Theory for Strong Interactions}''},
\textsf{\doiref{10.1016/0550-3213(74)90154-0}{Nucl.~Phys.~B72,~461~(1974)}}.

\bibitem{Gubser:1998bc}
S.~S.~Gubser, I.~R.~Klebanov and A.~M.~Polyakov,
\textit{``{Gauge theory correlators from noncritical string theory}''},
\textsf{\doiref{10.1016/S0370-2693(98)00377-3}{Phys.~Lett.~B428,~105~(1998)}},
\texttt{\arxivref{hep-th/9802109}{hep-th/9802109}}.

\bibitem{Witten:1998qj}
E.~Witten,
\textit{``{Anti-de Sitter space and holography}''},
\textsf{Adv.~Theor.~Math.~Phys.~2,~253~(1998)},
\texttt{\arxivref{hep-th/9802150}{hep-th/9802150}}.

\bibitem{Aharony:1999ti}
O.~Aharony, S.~S.~Gubser, J.~M.~Maldacena, H.~Ooguri and Y.~Oz,
\textit{``{Large N field theories, string theory and gravity}''},
\textsf{\doiref{10.1016/S0370-1573(99)00083-6}{Phys.~Rept.~323,~183~(2000)}},
\texttt{\arxivref{hep-th/9905111}{hep-th/9905111}}.

\bibitem{DHoker:2002nbb}
E.~D'Hoker and D.~Z.~Freedman,
\textit{``{Supersymmetric gauge theories and the AdS / CFT correspondence}''},
\texttt{\arxivref{hep-th/0201253}{hep-th/0201253}},
in: Strings, Branes and Extra Dimensions: TASI 2001: Proceedings, p. 3 - 158,
  2002.

\bibitem{Plefka:2005bk}
J.~Plefka,
\textit{``{Spinning strings and integrable spin chains in the AdS/CFT
  correspondence}''},
\textsf{\doiref{10.12942/lrr-2005-9}{Living~Rev.~Rel.~8,~9~(2005)}},
\texttt{\arxivref{hep-th/0507136}{hep-th/0507136}}.

\bibitem{Nastase:2015wjb}
H.~Nastase,
\textit{``{Introduction to the ADS/CFT Correspondence}''},
Cambridge University Press (2015),
Cambridge.

\bibitem{Polchinski:1995mt}
J.~Polchinski,
\textit{``{Dirichlet Branes and Ramond-Ramond charges}''},
\textsf{\doiref{10.1103/PhysRevLett.75.4724}{Phys.~Rev.~Lett.~75,~4724~(1995)}},
\texttt{\arxivref{hep-th/9510017}{hep-th/9510017}}.

\bibitem{Wilson:1974sk}
K.~G.~Wilson,
\textit{``{Confinement of Quarks}''},
\textsf{\doiref{10.1103/PhysRevD.10.2445}{Phys.~Rev.~D10,~2445~(1974)}}.

\bibitem{Dorn:1986dt}
H.~Dorn,
\textit{``{Renormalization of Path Ordered Phase Factors and Related Hadron
  Operators in Gauge Field Theories}''},
\textsf{\doiref{10.1002/prop.19860340104}{Fortsch.~Phys.~34,~11~(1986)}}.

\bibitem{Smit:2002ug}
J.~Smit,
\textit{``{Introduction to quantum fields on a lattice: A robust mate}''},
\textsf{Cambridge~Lect.~Notes~Phys.~15,~1~(2002)}.

\bibitem{Polyakov:1979ca}
A.~M.~Polyakov,
\textit{``{Gauge Fields as Rings of Glue}''},
\textsf{\doiref{10.1016/0550-3213(80)90507-6}{Nucl.~Phys.~B164,~171~(1980)}}.

\bibitem{Dotsenko:1980wb}
V.~S.~Dotsenko and S.~N.~Vergeles,
\textit{``{Renormalizability of Phase Factors in the Nonabelian Gauge
  Theory}''},
\textsf{\doiref{10.1016/0550-3213(80)90103-0}{Nucl.~Phys.~B169,~527~(1980)}}.

\bibitem{Brandt:1981kf}
R.~A.~Brandt, F.~Neri and M.-a.~Sato,
\textit{``{Renormalization of Loop Functions for All Loops}''},
\textsf{\doiref{10.1103/PhysRevD.24.879}{Phys.~Rev.~D24,~879~(1981)}}.

\bibitem{Erickson:2000af}
J.~K.~Erickson, G.~W.~Semenoff and K.~Zarembo,
\textit{``{Wilson loops in N=4 supersymmetric Yang-Mills theory}''},
\textsf{\doiref{10.1016/S0550-3213(00)00300-X}{Nucl.~Phys.~B582,~155~(2000)}},
\texttt{\arxivref{hep-th/0003055}{hep-th/0003055}}.

\bibitem{Frenkel:1984pz}
J.~Frenkel and J.~C.~Taylor,
\textit{``{Nonabelian Eikonal Exponentiation}''},
\textsf{\doiref{10.1016/0550-3213(84)90294-3}{Nucl.~Phys.~B246,~231~(1984)}}.

\bibitem{Korchemsky:1993uz}
G.~P.~Korchemsky and G.~Marchesini,
\textit{``{Resummation of large infrared corrections using Wilson loops}''},
\textsf{\doiref{10.1016/0370-2693(93)90015-A}{Phys.~Lett.~B313,~433~(1993)}}.

\bibitem{Magnea:2014vha}
L.~Magnea,
\textit{``{Progress on the infrared structure of multi-particle gauge theory
  amplitudes}''},
\textsf{PoS~LL2014,~073~(2014)},
\texttt{\arxivref{1408.0682}{arxiv:1408.0682}},
in: Proceedings, 12th DESY Workshop on Elementary Particle Physics: Loops and
  Legs in Quantum Field Theory (LL2014): Weimar, Germany, April 27 - May 2,
  2014, p. 073, 2014.

\bibitem{White:2015wha}
C.~D.~White,
\textit{``{An Introduction to Webs}''},
\textsf{\doiref{10.1088/0954-3899/43/3/033002}{J.~Phys.~G43,~033002~(2016)}},
\texttt{\arxivref{1507.02167}{arxiv:1507.02167}}.

\bibitem{Korchemsky:1987wg}
G.~P.~Korchemsky and A.~V.~Radyushkin,
\textit{``{Renormalization of the Wilson Loops Beyond the Leading Order}''},
\textsf{\doiref{10.1016/0550-3213(87)90277-X}{Nucl.~Phys.~B283,~342~(1987)}}.

\bibitem{Grozin:2015kna}
A.~Grozin, J.~M.~Henn, G.~P.~Korchemsky and P.~Marquard,
\textit{``{The three-loop cusp anomalous dimension in QCD and its
  supersymmetric extensions}''},
\textsf{\doiref{10.1007/JHEP01(2016)140}{JHEP~1601,~140~(2016)}},
\texttt{\arxivref{1510.07803}{arxiv:1510.07803}}.

\bibitem{Makeenko:2006ds}
Y.~Makeenko, P.~Olesen and G.~W.~Semenoff,
\textit{``{Cusped SYM Wilson loop at two loops and beyond}''},
\textsf{\doiref{10.1016/j.nuclphysb.2006.05.002}{Nucl.~Phys.~B748,~170~(2006)}},
\texttt{\arxivref{hep-th/0602100}{hep-th/0602100}}.

\bibitem{Correa:2012nk}
D.~Correa, J.~Henn, J.~Maldacena and A.~Sever,
\textit{``{The cusp anomalous dimension at three loops and beyond}''},
\textsf{\doiref{10.1007/JHEP05(2012)098}{JHEP~1205,~098~(2012)}},
\texttt{\arxivref{1203.1019}{arxiv:1203.1019}}.

\bibitem{Henn:2013wfa}
J.~M.~Henn and T.~Huber,
\textit{``{The four-loop cusp anomalous dimension in $ \mathcal{N} =$ 4 super
  Yang-Mills and analytic integration techniques for Wilson line integrals}''},
\textsf{\doiref{10.1007/JHEP09(2013)147}{JHEP~1309,~147~(2013)}},
\texttt{\arxivref{1304.6418}{arxiv:1304.6418}}.

\bibitem{Zarembo:2002an}
K.~Zarembo,
\textit{``{Supersymmetric Wilson loops}''},
\textsf{\doiref{10.1016/S0550-3213(02)00693-4}{Nucl.~Phys.~B643,~157~(2002)}},
\texttt{\arxivref{hep-th/0205160}{hep-th/0205160}}.

\bibitem{Drukker:2007dw}
N.~Drukker, S.~Giombi, R.~Ricci and D.~Trancanelli,
\textit{``{More supersymmetric Wilson loops}''},
\textsf{\doiref{10.1103/PhysRevD.76.107703}{Phys.~Rev.~D76,~107703~(2007)}},
\texttt{\arxivref{0704.2237}{arxiv:0704.2237}}.

\bibitem{Dymarsky:2009si}
A.~Dymarsky and V.~Pestun,
\textit{``{Supersymmetric Wilson loops in N=4 SYM and pure spinors}''},
\textsf{\doiref{10.1007/JHEP04(2010)115}{JHEP~1004,~115~(2010)}},
\texttt{\arxivref{0911.1841}{arxiv:0911.1841}}.

\bibitem{Cardinali:2012sy}
V.~Cardinali, L.~Griguolo and D.~Seminara,
\textit{``{Impure Aspects of Supersymmetric Wilson Loops}''},
\textsf{\doiref{10.1007/JHEP06(2012)167}{JHEP~1206,~167~(2012)}},
\texttt{\arxivref{1202.6393}{arxiv:1202.6393}}.

\bibitem{Berenstein:1998ij}
D.~E.~Berenstein, R.~Corrado, W.~Fischler and J.~M.~Maldacena,
\textit{``{The Operator product expansion for Wilson loops and surfaces in the
  large N limit}''},
\textsf{\doiref{10.1103/PhysRevD.59.105023}{Phys.~Rev.~D59,~105023~(1999)}},
\texttt{\arxivref{hep-th/9809188}{hep-th/9809188}}.

\bibitem{Siegel:1979wq}
W.~Siegel,
\textit{``{Supersymmetric Dimensional Regularization via Dimensional
  Reduction}''},
\textsf{\doiref{10.1016/0370-2693(79)90282-X}{Phys.~Lett.~B84,~193~(1979)}}.

\bibitem{Brezin:1977sv}
E.~Br{\'e}zin, C.~Itzykson, G.~Parisi and J.~B.~Zuber,
\textit{``{Planar Diagrams}''},
\textsf{\doiref{10.1007/BF01614153}{Commun.~Math.~Phys.~59,~35~(1978)}}.

\bibitem{FunctionAtlas}
K.~Oldham, J.~Myland and J.~Spanier,
\textit{``An Atlas of Functions''},
Springer (2009),
New York.

\bibitem{Drukker:2000rr}
N.~Drukker and D.~J.~Gross,
\textit{``{An Exact prediction of N=4 SUSYM theory for string theory}''},
\textsf{\doiref{10.1063/1.1372177}{J.~Math.~Phys.~42,~2896~(2001)}},
\texttt{\arxivref{hep-th/0010274}{hep-th/0010274}}.

\bibitem{Drukker:2000ep}
N.~Drukker, D.~J.~Gross and A.~A.~Tseytlin,
\textit{``{Green-Schwarz string in AdS(5) x S**5: Semiclassical partition
  function}''},
\textsf{\doiref{10.1088/1126-6708/2000/04/021}{JHEP~0004,~021~(2000)}},
\texttt{\arxivref{hep-th/0001204}{hep-th/0001204}}.

\bibitem{Kruczenski:2008zk}
M.~Kruczenski and A.~Tirziu,
\textit{``{Matching the circular Wilson loop with dual open string solution at
  1-loop in strong coupling}''},
\textsf{\doiref{10.1088/1126-6708/2008/05/064}{JHEP~0805,~064~(2008)}},
\texttt{\arxivref{0803.0315}{arxiv:0803.0315}}.

\bibitem{Vescovi:2016zzu}
E.~Vescovi,
\textit{``{Perturbative and non-perturbative approaches to string sigma-models
  in AdS/CFT}''},
PhD Thesis, Humboldt-Universit{\"a}t zu Berlin (2016).

\bibitem{Forini:2015bgo}
V.~Forini, V.~Giangreco M.~Puletti, L.~Griguolo, D.~Seminara and E.~Vescovi,
\textit{``{Precision calculation of 1/4-BPS Wilson loops in AdS$_5\times
  S^5$}''},
\textsf{\doiref{10.1007/JHEP02(2016)105}{JHEP~1602,~105~(2016)}},
\texttt{\arxivref{1512.00841}{arxiv:1512.00841}}.

\bibitem{Faraggi:2016ekd}
A.~Faraggi, L.~A.~Pando~Zayas, G.~A.~Silva and D.~Trancanelli,
\textit{``{Toward precision holography with supersymmetric Wilson loops}''},
\textsf{\doiref{10.1007/JHEP04(2016)053}{JHEP~1604,~053~(2016)}},
\texttt{\arxivref{1601.04708}{arxiv:1601.04708}}.

\bibitem{Forini:2017whz}
V.~Forini, A.~A.~Tseytlin and E.~Vescovi,
\textit{``{Perturbative computation of string one-loop corrections to Wilson
  loop minimal surfaces in AdS$_5 \times$ S$^5$}''},
\textsf{\doiref{10.1007/JHEP03(2017)003}{JHEP~1703,~003~(2017)}},
\texttt{\arxivref{1702.02164}{arxiv:1702.02164}}.

\bibitem{Alday:2008yw}
L.~F.~Alday and R.~Roiban,
\textit{``{Scattering Amplitudes, Wilson Loops and the String/Gauge Theory
  Correspondence}''},
\textsf{\doiref{10.1016/j.physrep.2008.08.002}{Phys.~Rept.~468,~153~(2008)}},
\texttt{\arxivref{0807.1889}{arxiv:0807.1889}}.

\bibitem{Henn:2009bd}
J.~M.~Henn,
\textit{``{Duality between Wilson loops and gluon amplitudes}''},
\textsf{\doiref{10.1002/prop.200900048}{Fortsch.~Phys.~57,~729~(2009)}},
\texttt{\arxivref{0903.0522}{arxiv:0903.0522}}.

\bibitem{Bern:2005iz}
Z.~Bern, L.~J.~Dixon and V.~A.~Smirnov,
\textit{``{Iteration of planar amplitudes in maximally supersymmetric
  Yang-Mills theory at three loops and beyond}''},
\textsf{\doiref{10.1103/PhysRevD.72.085001}{Phys.~Rev.~D72,~085001~(2005)}},
\texttt{\arxivref{hep-th/0505205}{hep-th/0505205}}.

\bibitem{Dixon:1996wi}
L.~J.~Dixon,
\textit{``{Calculating scattering amplitudes efficiently}''},
\texttt{\arxivref{hep-ph/9601359}{hep-ph/9601359}},
in: QCD and beyond. Proceedings, Theoretical Advanced Study Institute in
  Elementary Particle Physics, TASI-95, Boulder, USA, June 4 - 30, 1995, p. 539
  - 584, 1995.

\bibitem{Henn:2014yza}
J.~M.~Henn and J.~C.~Plefka,
\textit{``{Scattering Amplitudes in Gauge Theories}''},
\textsf{\doiref{10.1007/978-3-642-54022-6}{Lect.~Notes~Phys.~883,~pp.1~(2014)}}.

\bibitem{Drummond:2007cf}
J.~M.~Drummond, J.~Henn, G.~P.~Korchemsky and E.~Sokatchev,
\textit{``{On planar gluon amplitudes/Wilson loops duality}''},
\textsf{\doiref{10.1016/j.nuclphysb.2007.11.007}{Nucl.~Phys.~B795,~52~(2008)}},
\texttt{\arxivref{0709.2368}{arxiv:0709.2368}}.

\bibitem{Drummond:2006rz}
J.~M.~Drummond, J.~Henn, V.~A.~Smirnov and E.~Sokatchev,
\textit{``{Magic identities for conformal four-point integrals}''},
\textsf{\doiref{10.1088/1126-6708/2007/01/064}{JHEP~0701,~064~(2007)}},
\texttt{\arxivref{hep-th/0607160}{hep-th/0607160}}.

\bibitem{Drummond:2007aua}
J.~M.~Drummond, G.~P.~Korchemsky and E.~Sokatchev,
\textit{``{Conformal properties of four-gluon planar amplitudes and Wilson
  loops}''},
\textsf{\doiref{10.1016/j.nuclphysb.2007.11.041}{Nucl.~Phys.~B795,~385~(2008)}},
\texttt{\arxivref{0707.0243}{arxiv:0707.0243}}.

\bibitem{Drummond:2010km}
J.~M.~Drummond,
\textit{``{Review of AdS/CFT Integrability, Chapter V.2: Dual Superconformal
  Symmetry}''},
\textsf{\doiref{10.1007/s11005-011-0519-4}{Lett.~Math.~Phys.~99,~481~(2012)}},
\texttt{\arxivref{1012.4002}{arxiv:1012.4002}}.

\bibitem{CaronHuot:2010ek}
S.~Caron-Huot,
\textit{``{Notes on the scattering amplitude / Wilson loop duality}''},
\textsf{\doiref{10.1007/JHEP07(2011)058}{JHEP~1107,~058~(2011)}},
\texttt{\arxivref{1010.1167}{arxiv:1010.1167}}.

\bibitem{Mason:2010yk}
L.~J.~Mason and D.~Skinner,
\textit{``{The Complete Planar S-matrix of N=4 SYM as a Wilson Loop in Twistor
  Space}''},
\textsf{\doiref{10.1007/JHEP12(2010)018}{JHEP~1012,~018~(2010)}},
\texttt{\arxivref{1009.2225}{arxiv:1009.2225}}.

\bibitem{Belitsky:2011zm}
A.~V.~Belitsky, G.~P.~Korchemsky and E.~Sokatchev,
\textit{``{Are scattering amplitudes dual to super Wilson loops?}''},
\textsf{\doiref{10.1016/j.nuclphysb.2011.10.014}{Nucl.~Phys.~B855,~333~(2012)}},
\texttt{\arxivref{1103.3008}{arxiv:1103.3008}}.

\bibitem{Belitsky:2012nu}
A.~V.~Belitsky,
\textit{``{Conformal anomaly of super Wilson loop}''},
\textsf{\doiref{10.1016/j.nuclphysb.2012.04.022}{Nucl.~Phys.~B862,~430~(2012)}},
\texttt{\arxivref{1201.6073}{arxiv:1201.6073}}.

\bibitem{Beisert:2012xx}
N.~Beisert, S.~He, B.~U.~W.~Schwab and C.~Vergu,
\textit{``{Null Polygonal Wilson Loops in Full N=4 Superspace}''},
\textsf{\doiref{10.1088/1751-8113/45/26/265402}{J.~Phys.~A45,~265402~(2012)}},
\texttt{\arxivref{1203.1443}{arxiv:1203.1443}}.

\bibitem{Eichenherr:1979ci}
H.~Eichenherr and M.~Forger,
\textit{``{On the Dual Symmetry of the Nonlinear Sigma Models}''},
\textsf{\doiref{10.1016/0550-3213(79)90276-1}{Nucl.~Phys.~B155,~381~(1979)}}.

\bibitem{Cartan:1926}
{\'E}.~Cartan,
\textit{``{Sur une classe remarquable d'espaces de Riemann}''},
\textsf{Bull.~Soc.~Math.~France~54,~214~(1926)}.

\bibitem{Cartan:1927}
{\'E}.~Cartan,
\textit{``{Sur une classe remarquable d'espaces de Riemann. II}''},
\textsf{Bull.~Soc.~Math.~France~55,~114~(1927)}.

\bibitem{Helgason:2001}
S.~Helgason,
\textit{``Differential geometry, Lie groups and symmetric spaces''},
American Mathematical Society (2001),
Providence.

\bibitem{ONeill:1983}
B.~O'Neill,
\textit{``{Semi-Riemannian Geometry}''},
Academic Press, San Diego (1983).

\bibitem{Brezin:1979am}
E.~Brezin, C.~Itzykson, J.~Zinn-Justin and J.~B.~Zuber,
\textit{``{Remarks About the Existence of Nonlocal Charges in Two-Dimensional
  Models}''},
\textsf{\doiref{10.1016/0370-2693(79)90263-6}{Phys.~Lett.~B82,~442~(1979)}}.

\bibitem{Wu:1982jt}
Y.-S.~Wu,
\textit{``{Extension of the Hidden Symmetry Algebra in Classical Principal
  Chiral Models}''},
\textsf{\doiref{10.1016/0550-3213(83)90190-6}{Nucl.~Phys.~B211,~160~(1983)}}.

\bibitem{Kazakov:2004qf}
V.~A.~Kazakov, A.~Marshakov, J.~A.~Minahan and K.~Zarembo,
\textit{``{Classical/quantum integrability in AdS/CFT}''},
\textsf{\doiref{10.1088/1126-6708/2004/05/024}{JHEP~0405,~024~(2004)}},
\texttt{\arxivref{hep-th/0402207}{hep-th/0402207}}.

\bibitem{Kazakov:2004nh}
V.~A.~Kazakov and K.~Zarembo,
\textit{``{Classical / quantum integrability in non-compact sector of
  AdS/CFT}''},
\textsf{\doiref{10.1088/1126-6708/2004/10/060}{JHEP~0410,~060~(2004)}},
\texttt{\arxivref{hep-th/0410105}{hep-th/0410105}}.

\bibitem{Beisert:2004ag}
N.~Beisert, V.~A.~Kazakov and K.~Sakai,
\textit{``{Algebraic curve for the SO(6) sector of AdS/CFT}''},
\textsf{\doiref{10.1007/s00220-005-1528-x}{Commun.~Math.~Phys.~263,~611~(2006)}},
\texttt{\arxivref{hep-th/0410253}{hep-th/0410253}}.

\bibitem{SchaferNameki:2004ik}
S.~Schafer-Nameki,
\textit{``{The Algebraic curve of 1-loop planar N=4 SYM}''},
\textsf{\doiref{10.1016/j.nuclphysb.2005.02.034}{Nucl.~Phys.~B714,~3~(2005)}},
\texttt{\arxivref{hep-th/0412254}{hep-th/0412254}}.

\bibitem{Beisert:2005bm}
N.~Beisert, V.~A.~Kazakov, K.~Sakai and K.~Zarembo,
\textit{``{The Algebraic curve of classical superstrings on AdS(5) x S**5}''},
\textsf{\doiref{10.1007/s00220-006-1529-4}{Commun.~Math.~Phys.~263,~659~(2006)}},
\texttt{\arxivref{hep-th/0502226}{hep-th/0502226}}.

\bibitem{Beisert:2012ue}
N.~Beisert and F.~Luecker,
\textit{``{Construction of Lax Connections by Exponentiation}''},
\textsf{\doiref{10.1063/1.4769824}{J.~Math.~Phys.~53,~122304~(2012)}},
\texttt{\arxivref{1207.3325}{arxiv:1207.3325}}.

\bibitem{Dolan:1980kz}
L.~Dolan and A.~Roos,
\textit{``{Nonlocal Currents as Noether Currents}''},
\textsf{\doiref{10.1103/PhysRevD.22.2018}{Phys.~Rev.~D22,~2018~(1980)}}.

\bibitem{Dolan:1981fq}
L.~Dolan,
\textit{``{Kac-Moody Algebra Is Hidden Symmetry of Chiral Models}''},
\textsf{\doiref{10.1103/PhysRevLett.47.1371}{Phys.~Rev.~Lett.~47,~1371~(1981)}}.

\bibitem{Hou:1981hn}
B.-y.~Hou, M.-l.~Ge and Y.-s.~Wu,
\textit{``{Noether Analysis for the Hidden Symmetry Responsible for Infinite
  Set of Nonlocal Currents}''},
\textsf{\doiref{10.1103/PhysRevD.24.2238}{Phys.~Rev.~D24,~2238~(1981)}}.

\bibitem{Schwarz:1995td}
J.~H.~Schwarz,
\textit{``{Classical symmetries of some two-dimensional models}''},
\textsf{\doiref{10.1016/0550-3213(95)00276-X}{Nucl.~Phys.~B447,~137~(1995)}},
\texttt{\arxivref{hep-th/9503078}{hep-th/9503078}}.

\bibitem{Devchand:1981wy}
C.~Devchand and D.~B.~Fairlie,
\textit{``{A Generating Function for Hidden Symmetries of Chiral Models}''},
\textsf{\doiref{10.1016/0550-3213(82)90312-1}{Nucl.~Phys.~B194,~232~(1982)}}.

\bibitem{Luscher:1977rq}
M.~Luscher and K.~Pohlmeyer,
\textit{``{Scattering of Massless Lumps and Nonlocal Charges in the
  Two-Dimensional Classical Nonlinear Sigma Model}''},
\textsf{\doiref{10.1016/0550-3213(78)90049-4}{Nucl.~Phys.~B137,~46~(1978)}}.

\bibitem{Faddeev:1987ph}
L.~D.~Faddeev and L.~A.~Takhtajan,
\textit{``{Hamiltonian Methods in the Theory of Solitons}''},
Springer (1987).

\bibitem{Forger:1991cm}
M.~Forger, J.~Laartz and U.~Schaper,
\textit{``{Current algebra of classical nonlinear sigma models}''},
\textsf{\doiref{10.1007/BF02102634}{Commun.~Math.~Phys.~146,~397~(1992)}},
\texttt{\arxivref{hep-th/9201025}{hep-th/9201025}}.

\bibitem{Dorn:2015bfa}
H.~Dorn,
\textit{``{Wilson loops at strong coupling for curved contours with cusps}''},
\textsf{\doiref{10.1088/1751-8113/49/14/145402}{J.~Phys.~A49,~145402~(2016)}},
\texttt{\arxivref{1509.00222}{arxiv:1509.00222}}.

\bibitem{Polyakov:2000ti}
A.~M.~Polyakov and V.~S.~Rychkov,
\textit{``{Gauge field strings duality and the loop equation}''},
\textsf{\doiref{10.1016/S0550-3213(00)00183-8}{Nucl.~Phys.~B581,~116~(2000)}},
\texttt{\arxivref{hep-th/0002106}{hep-th/0002106}}.

\bibitem{Polyakov:2000jg}
A.~M.~Polyakov and V.~S.~Rychkov,
\textit{``{Loop dynamics and AdS / CFT correspondence}''},
\textsf{\doiref{10.1016/S0550-3213(00)00642-8}{Nucl.~Phys.~B594,~272~(2001)}},
\texttt{\arxivref{hep-th/0005173}{hep-th/0005173}}.

\bibitem{Kruczenski:2002fb}
M.~Kruczenski,
\textit{``{A Note on twist two operators in N=4 SYM and Wilson loops in
  Minkowski signature}''},
\textsf{\doiref{10.1088/1126-6708/2002/12/024}{JHEP~0212,~024~(2002)}},
\texttt{\arxivref{hep-th/0210115}{hep-th/0210115}}.

\bibitem{brakke1992}
K.~A.~Brakke,
\textit{``{The surface evolver}''},
\textsf{Experiment.~Math.~1,~141~(1992)},
\href{http://projecteuclid.org/euclid.em/1048709050}{\texttt{http://projecteuclid.org/euclid.em/1048709050}}.

\bibitem{Kruczenski:2014bla}
M.~Kruczenski,
\textit{``{Wilson loops and minimal area surfaces in hyperbolic space}''},
\textsf{\doiref{10.1007/JHEP11(2014)065}{JHEP~1411,~065~(2014)}},
\texttt{\arxivref{1406.4945}{arxiv:1406.4945}}.

\bibitem{Huang:2016atz}
C.~Huang, Y.~He and M.~Kruczenski,
\textit{``{Minimal area surfaces dual to Wilson loops and the Mathieu
  equation}''},
\textsf{\doiref{10.1007/JHEP08(2016)088}{JHEP~1608,~088~(2016)}},
\texttt{\arxivref{1604.00078}{arxiv:1604.00078}}.

\bibitem{Semenoff:2004qr}
G.~W.~Semenoff and D.~Young,
\textit{``{Wavy Wilson line and AdS / CFT}''},
\textsf{\doiref{10.1142/S0217751X0502077X}{Int.~J.~Mod.~Phys.~A20,~2833~(2005)}},
\texttt{\arxivref{hep-th/0405288}{hep-th/0405288}}.

\bibitem{Gurdogan:2015csr}
{\"O}.~G{\"u}rdo{\u{g}}an and V.~Kazakov,
\textit{``{New Integrable 4D Quantum Field Theories from Strongly Deformed
  Planar $\mathcal N = $ 4 Supersymmetric Yang-Mills Theory}''},
\textsf{\doiref{10.1103/PhysRevLett.117.201602,
  10.1103/PhysRevLett.117.259903}{Phys.~Rev.~Lett.~117,~201602~(2016)}},
\texttt{\arxivref{1512.06704}{arxiv:1512.06704}},
[Addendum: Phys. Rev. Lett. 117, 259903 (2016)].

\bibitem{Alday:2005gi}
L.~F.~Alday, G.~Arutyunov and A.~A.~Tseytlin,
\textit{``{On integrability of classical superstrings in AdS(5) x S**5}''},
\textsf{\doiref{10.1088/1126-6708/2005/07/002}{JHEP~0507,~002~(2005)}},
\texttt{\arxivref{hep-th/0502240}{hep-th/0502240}}.

\bibitem{Zarembo:2010sg}
K.~Zarembo,
\textit{``{Strings on Semisymmetric Superspaces}''},
\textsf{\doiref{10.1007/JHEP05(2010)002}{JHEP~1005,~002~(2010)}},
\texttt{\arxivref{1003.0465}{arxiv:1003.0465}}.

\bibitem{Henneaux:1984mh}
M.~Henneaux and L.~Mezincescu,
\textit{``{A Sigma Model Interpretation of Green-Schwarz Covariant Superstring
  Action}''},
\textsf{\doiref{10.1016/0370-2693(85)90507-6}{Phys.~Lett.~B152,~340~(1985)}}.

\bibitem{Berkovits:1999zq}
N.~Berkovits, M.~Bershadsky, T.~Hauer, S.~Zhukov and B.~Zwiebach,
\textit{``{Superstring theory on AdS(2) x S**2 as a coset supermanifold}''},
\textsf{\doiref{10.1016/S0550-3213(99)00683-5}{Nucl.~Phys.~B567,~61~(2000)}},
\texttt{\arxivref{hep-th/9907200}{hep-th/9907200}}.

\bibitem{Green:1983wt}
M.~B.~Green and J.~H.~Schwarz,
\textit{``{Covariant Description of Superstrings}''},
\textsf{\doiref{10.1016/0370-2693(84)92021-5}{Phys.~Lett.~B136,~367~(1984)}}.

\bibitem{Green:1987sp}
M.~B.~Green, J.~H.~Schwarz and E.~Witten,
\textit{``{Superstring Theory. Vol. 1: Introduction}''},
Cambridge University Press (1987),
Cambridge.

\bibitem{deAzcarraga:1982dhu}
J.~A.~de~Azcarraga and J.~Lukierski,
\textit{``{Supersymmetric Particles with Internal Symmetries and Central
  Charges}''},
\textsf{\doiref{10.1016/0370-2693(82)90417-8}{Phys.~Lett.~B113,~170~(1982)}}.

\bibitem{Siegel:1983hh}
W.~Siegel,
\textit{``{Hidden Local Supersymmetry in the Supersymmetric Particle
  Action}''},
\textsf{\doiref{10.1016/0370-2693(83)90924-3}{Phys.~Lett.~B128,~397~(1983)}}.

\bibitem{McArthur:1999dy}
I.~N.~McArthur,
\textit{``{Kappa symmetry of Green-Schwarz actions in coset superspaces}''},
\textsf{\doiref{10.1016/S0550-3213(99)00800-7}{Nucl.~Phys.~B573,~811~(2000)}},
\texttt{\arxivref{hep-th/9908045}{hep-th/9908045}}.

\bibitem{Roiban:2000yy}
R.~Roiban and W.~Siegel,
\textit{``{Superstrings on AdS(5) x S**5 supertwistor space}''},
\textsf{\doiref{10.1088/1126-6708/2000/11/024}{JHEP~0011,~024~(2000)}},
\texttt{\arxivref{hep-th/0010104}{hep-th/0010104}}.

\bibitem{Hatsuda:2004it}
M.~Hatsuda and K.~Yoshida,
\textit{``{Classical integrability and super Yangian of superstring on AdS(5) x
  S**5}''},
\textsf{\doiref{10.4310/ATMP.2005.v9.n5.a2}{Adv.~Theor.~Math.~Phys.~9,~703~(2005)}},
\texttt{\arxivref{hep-th/0407044}{hep-th/0407044}}.

\bibitem{Hatsuda:2011mt}
M.~Hatsuda and K.~Yoshida,
\textit{``{Super Yangian of superstring on $AdS_5 x S^5$ revisited}''},
\textsf{\doiref{10.4310/ATMP.2011.v15.n5.a6}{Adv.~Theor.~Math.~Phys.~15,~1485~(2011)}},
\texttt{\arxivref{1107.4673}{arxiv:1107.4673}}.

\bibitem{Beisert:2011pn}
N.~Beisert and B.~U.~W.~Schwab,
\textit{``{Bonus Yangian Symmetry for the Planar S-Matrix of N=4 Super
  Yang-Mills}''},
\textsf{\doiref{10.1103/PhysRevLett.106.231602}{Phys.~Rev.~Lett.~106,~231602~(2011)}},
\texttt{\arxivref{1103.0646}{arxiv:1103.0646}}.

\bibitem{Arutyunov:2006yd}
G.~Arutyunov, S.~Frolov and M.~Zamaklar,
\textit{``{The Zamolodchikov-Faddeev algebra for AdS(5) x S**5 superstring}''},
\textsf{\doiref{10.1088/1126-6708/2007/04/002}{JHEP~0704,~002~(2007)}},
\texttt{\arxivref{hep-th/0612229}{hep-th/0612229}}.

\bibitem{Beisert:2007ds}
N.~Beisert,
\textit{``{The S-matrix of AdS / CFT and Yangian symmetry}''},
\textsf{PoS~SOLVAY,~002~(2006)},
\texttt{\arxivref{0704.0400}{arxiv:0704.0400}},
in: Proceedings, Bethe ansatz: 75 years later: Brussels, Belgium, 19-21 October
  2006, p. 002, 2006.

\bibitem{Matsumoto:2007rh}
T.~Matsumoto, S.~Moriyama and A.~Torrielli,
\textit{``{A Secret Symmetry of the AdS/CFT S-matrix}''},
\textsf{\doiref{10.1088/1126-6708/2007/09/099}{JHEP~0709,~099~(2007)}},
\texttt{\arxivref{0708.1285}{arxiv:0708.1285}}.

\bibitem{deLeeuw:2012jf}
M.~de~Leeuw, T.~Matsumoto, S.~Moriyama, V.~Regelskis and A.~Torrielli,
\textit{``{Secret Symmetries in AdS/CFT}''},
\textsf{\doiref{10.1088/0031-8949/86/02/028502}{Phys.~Scripta~02,~028502~(2012)}},
\texttt{\arxivref{1204.2366}{arxiv:1204.2366}}.

\bibitem{Beisert:2014hya}
N.~Beisert and M.~de~Leeuw,
\textit{``{The RTT realization for the deformed $\mathfrak {gl}(2\vert2)$
  Yangian}''},
\textsf{\doiref{10.1088/1751-8113/47/30/305201}{J.~Phys.~A47,~305201~(2014)}},
\texttt{\arxivref{1401.7691}{arxiv:1401.7691}}.

\bibitem{Berkovits:2011kn}
N.~Berkovits and A.~Mikhailov,
\textit{``{Nonlocal Charges for Bonus Yangian Symmetries of
  Super-Yang-Mills}''},
\textsf{\doiref{10.1007/JHEP07(2011)125}{JHEP~1107,~125~(2011)}},
\texttt{\arxivref{1106.2536}{arxiv:1106.2536}}.

\bibitem{Dekel:2016oot}
A.~Dekel,
\textit{``{Dual conformal transformations of smooth holographic Wilson
  loops}''},
\textsf{\doiref{10.1007/JHEP01(2017)085}{JHEP~1701,~085~(2017)}},
\texttt{\arxivref{1610.07179}{arxiv:1610.07179}}.

\bibitem{Berkovits:2008ic}
N.~Berkovits and J.~Maldacena,
\textit{``{Fermionic T-Duality, Dual Superconformal Symmetry, and the
  Amplitude/Wilson Loop Connection}''},
\textsf{\doiref{10.1088/1126-6708/2008/09/062}{JHEP~0809,~062~(2008)}},
\texttt{\arxivref{0807.3196}{arxiv:0807.3196}}.

\bibitem{Beisert:2008iq}
N.~Beisert, R.~Ricci, A.~A.~Tseytlin and M.~Wolf,
\textit{``{Dual Superconformal Symmetry from AdS(5) x S**5 Superstring
  Integrability}''},
\textsf{\doiref{10.1103/PhysRevD.78.126004}{Phys.~Rev.~D78,~126004~(2008)}},
\texttt{\arxivref{0807.3228}{arxiv:0807.3228}}.

\bibitem{Beisert:2009cs}
N.~Beisert,
\textit{``{T-Duality, Dual Conformal Symmetry and Integrability for Strings on
  AdS(5) x S**5}''},
\textsf{\doiref{10.1002/prop.200900060}{Fortsch.~Phys.~57,~329~(2009)}},
\texttt{\arxivref{0903.0609}{arxiv:0903.0609}},
in: \textit{``{Constituents, fundamental forces and symmetries of the universe.
  Proceedings, 4rd EURTN Workshop, Varna, Bulgaria, September 11-17, 2008}''},
329-337p.

\bibitem{Arutyunov:2005nk}
G.~Arutyunov and M.~Zamaklar,
\textit{``{Linking Backlund and monodromy charges for strings on AdS(5) x
  S**5}''},
\textsf{\doiref{10.1088/1126-6708/2005/07/026}{JHEP~0507,~026~(2005)}},
\texttt{\arxivref{hep-th/0504144}{hep-th/0504144}}.

\bibitem{Kruczenski:2003gt}
M.~Kruczenski,
\textit{``{Spin chains and string theory}''},
\textsf{\doiref{10.1103/PhysRevLett.93.161602}{Phys.~Rev.~Lett.~93,~161602~(2004)}},
\texttt{\arxivref{hep-th/0311203}{hep-th/0311203}}.

\bibitem{Delduc:2013qra}
F.~Delduc, M.~Magro and B.~Vicedo,
\textit{``{An integrable deformation of the $\mathrm{AdS}_5 \times
  \mathrm{S}^5$ superstring action}''},
\textsf{\doiref{10.1103/PhysRevLett.112.051601}{Phys.~Rev.~Lett.~112,~051601~(2014)}},
\texttt{\arxivref{1309.5850}{arxiv:1309.5850}}.

\bibitem{Delduc:2016ihq}
F.~Delduc, S.~Lacroix, M.~Magro and B.~Vicedo,
\textit{``{On q-deformed symmetries as Poisson–Lie symmetries and application
  to Yang–Baxter type models}''},
\textsf{\doiref{10.1088/1751-8113/49/41/415402}{J.~Phys.~A49,~415402~(2016)}},
\texttt{\arxivref{1606.01712}{arxiv:1606.01712}}.

\bibitem{Alday:2009yn}
L.~F.~Alday and J.~Maldacena,
\textit{``{Null polygonal Wilson loops and minimal surfaces in Anti-de-Sitter
  space}''},
\textsf{\doiref{10.1088/1126-6708/2009/11/082}{JHEP~0911,~082~(2009)}},
\texttt{\arxivref{0904.0663}{arxiv:0904.0663}}.

\bibitem{Gaiotto:2008cd}
D.~Gaiotto, G.~W.~Moore and A.~Neitzke,
\textit{``{Four-dimensional wall-crossing via three-dimensional field
  theory}''},
\textsf{\doiref{10.1007/s00220-010-1071-2}{Commun.~Math.~Phys.~299,~163~(2010)}},
\texttt{\arxivref{0807.4723}{arxiv:0807.4723}}.

\bibitem{Belitsky:2003sh}
A.~V.~Belitsky, S.~E.~Derkachov, G.~P.~Korchemsky and A.~N.~Manashov,
\textit{``{Superconformal operators in N=4 superYang-Mills theory}''},
\textsf{\doiref{10.1103/PhysRevD.70.045021}{Phys.~Rev.~D70,~045021~(2004)}},
\texttt{\arxivref{hep-th/0311104}{hep-th/0311104}}.

\bibitem{Groeger}
J.~Groeger,
\textit{``{Supersymmetric Wilson Loops in N=4 Super Yang-Mills Theory}''},
Diploma Thesis, Humboldt-Universit{\"a}t zu Berlin (2012).

\end{thebibliography}

\newpage
$\mbox{ }$
\newpage
\phantomsection
\addcontentsline{toc}{chapter}{Acknowledgements}
\section*{Acknowledgements} 

First, I would like to thank my supervisor Jan Plefka for introducing me to this field of theoretical physics, for his guidance of my work in it and for always making himself available to my questions. I am very grateful to Konstantin Zarembo for his guidance in a part of the work on this thesis and for his very valuable advice on numerous questions. I also acknowledge the hospitality of Nordita during my visits there. 

I would also like to thank Gleb Arutyunov and Valentina Forini, who kindly offered to be referees for this thesis. 

Moreover, I thank Thomas Klose and Florian Loebbert for the very interesting collaboration that eventually lead to the description of the master symmetry and Jonas Pollok for the collaboration on the minimal surfaces in superspace. 

Over the course of the last years, I had many interesting discussions with colleagues, from which I have greatly profited. For those that were directly related to my work on this thesis I thank Gleb Arutyunov, Till Bargheer, Dmitri Bykov, Amit Dekel, Harald Dorn, Valentina Forini, Ben Hoare, Vladimir Kazakov, Thomas Klose, Ivan Kostov, Martin Kruczenski, Florian Loebbert, Vladimir Mitev, Didina Serban, Matthias Staudacher, Stijn van Tongeren and Edoardo Vescovi. I especially thank Dennis M{\"u}ller for the many discussions we had on our respective research and related topics. They were invaluable to me and I shall miss them much.

In addition, I thank Dennis M{\"u}ller, Jan Plefka, Stijn van Tongeren and in particular Florian Loebbert for their careful reading of the manuscript and their helpful comments which improved the presentation.

\newpage
\pagestyle{empty}

\section*{Hilfsmittel} 
Diese Arbeit wurde in \LaTeX$\,$ unter Verwendung von MiKT\raisebox{-0.6mm}{E}X 2.9 gesetzt. 
F{\"u}r die Erstellung der Grafiken wurden Mathematica sowie der Open-Source Vektorgrafikedtitor Inkscape verwendet. Zur Unterst{\"u}tzung einiger Rechnungen wurde Mathematica herangezogen. In der Erstellung der Bibliographie wurde BiBTeX verwendet.  

\newpage
\thispagestyle{empty}
\section*{Eigenst\"andigkeitserkl\"arung}
Ich erkl{\"a}re, dass ich die Dissertation selbst{\"a}ndig und nur unter Verwendung der von mir gem{\"a}{\ss} \S 7 Abs. 3 der Promotionsordnung der Mathematisch-Naturwissen\-schaftlichen Fakult{\"a}t, ver{\"o}ffentlicht im Amtlichen Mitteilungsblatt der Humboldt-Universit{\"a}t zu Berlin Nr. 126/2014 am 18.11.2014 angegebenen Hilfsmittel angefertigt habe. \\
\vspace{12mm} \\
\renewcommand{\arraystretch}{1.25}
\begin{tabular}{lp{6em}l}
 \hspace{3cm}   & & \hspace{3cm} \\ \cline{1-1} \cline{3-3} 
 \textit{Ort, Datum}  & & \textit{Hagen M\"unkler}  
\end{tabular}
   
\end{document}